\documentstyle[12pt,epsfig]{article}
\catcode`\@=11
\def\marginnote#1{}
\def\pss#1{\hfill\\[#1\baselineskip]}
\def\vvs#1{\vspace{#1\baselineskip}}
\newcount\hour
\newcount\minute
\newtoks\amorpm
\hour=\time\divide\hour by60
\minute=\time{\multiply\hour by60 \global\advance\minute by-
\hour}

\edef\standardtime{{\ifnum\hour<12 \global\amorpm={am}%
    \else\global\amorpm={pm}\advance\hour by-12 \fi
    \ifnum\hour=0 \hour=12 \fi
    \number\hour:\ifnum\minute<100\fi\number\minute\the\amorpm}}
\edef\militarytime{\number\hour:\ifnum\minute<100\fi\number\minute}
\def\draftlabel#1{{\@bsphack\if@filesw {\let\thepage\relax
  \xdef\@gtempa{\write\@auxout{\string
    \newlabel{#1}{{\@currentlabel}{\thepage}}}}}\@gtempa
    \if@nobreak \ifvmode\nobreak\fi\fi\fi\@esphack}
     \gdef\@eqnlabel{#1}}
\def\@eqnlabel{}
\def\@vacuum{}
\def\draftmarginnote#1{\marginpar{\raggedright\scriptsize\tt#1}}
\def\draft{\oddsidemargin -.5truein
        \def\@oddfoot{\sl preliminary draft \hfil
        \rm\thepage\hfil\sl\today\quad\militarytime}
        \let\@evenfoot\@oddfoot \overfullrule 3pt
        \let\label=\draftlabel
        \let\marginnote=\draftmarginnote
\def\@eqnnum{(\theequation)\rlap{\kern\marginparsep\tt\@eqnlabel}%
\global\let\@eqnlabel\@vacuum}  }
\def\preprint{\twocolumn\sloppy\flushbottom\parindent 1em
        \leftmargini 2em\leftmarginv .5em\leftmarginvi .5em
        \oddsidemargin -.5in    \evensidemargin -.5in
        \columnsep 15mm \footheight 0pt
        \textwidth 250mmin      \topmargin  -.4in
        \headheight 12pt \topskip .4in
        \textheight 175mm
        \footskip 0pt

\def\@oddhead{\thepage\hfil\addtocounter{page}{1}\thepage}
        \let\@evenhead\@oddhead \def\@oddfoot{} \def\@evenfoot{} }
\def\titlepage{\@restonecolfalse\if@twocolumn\@restonecoltrue\onecolumn
     \else \newpage \fi \thispagestyle{empty}\c@page\z@
        \def\thefootnote{\fnsymbol{footnote}} }
\def\endtitlepage{\if@restonecol\twocolumn \else  \fi
        \def\thefootnote{\arabic{footnote}}
        \setcounter{footnote}{0}}  
\catcode`@=12
\relax
\def\be{\begin{equation}}
\def\ee{\end{equation}}

\def \P               {\phi} 
\def\bea{\begin{eqnarray}}
\def\eea{\end{eqnarray}}

\def\simlt{\stackrel{<}{{}_\sim}}
\def\simgt{\stackrel{>}{{}_\sim}}

\def\PLB#1#2#3{{\it Phys.~Lett.} {\bf{B#1}} (19#2) #3}

\def\MPL#1#2#3{{\it Mod.~Phys.~Lett.} {\bf#1} (19#2) #3}

\relax
\def\infb{\mbox{$\hbox{fb}^{-1}$}}
\def\Gcs{\hbox{GeV}}
\def\Z{\mbox{$\hbox{Z}$}}
\def\W{\mbox{$\hbox{W}$}}
\def\A{\mbox{$\hbox{A}$}}
\def\H{\mbox{$\hbox{H}$}}
\def\h{\mbox{$\hbox{h}$}}
\def\e{\mbox{$\hbox{e}$}}

\def\epemto{\mbox{$e^+ e^- \rightarrow$}}
\def\epem{\mbox{$\hbox{e}^+\hbox{e}^-$}}
\def\mpmm{\mbox{$\mu^+\mu^-$}}
\def\tptm{\mbox{$\tau^+\tau^-$}}
\def\lplm{\mbox{$\hbox{l}^+\hbox{l}^-$}}
\def\nnbar{\mbox{$\nu\bar\nu$}}
\def\qqbar{\mbox{$\hbox{q}\bar{\hbox{q}}$}}
\def\bbbar{\mbox{$\hbox{b}\bar{\hbox{b}}$}}

\def\mst11{m_{\;\widetilde{t}_{1}}}

\def\mst22{m_{\;\widetilde{t}_{2}}}
\def\mst12{m_{\;\widetilde{t}_{1,2}}}

\def\msb11{m_{\;\widetilde{b}_{1}}}
\def\msb22{m_{\;\widetilde{b}_{2}}}
\def\msb12{m_{\;\widetilde{b}_{1,2}}}

\def\mtilde2{\widetilde{m}^{2}}

\newcommand{\jnp}{{\it Nucl.~Phys.} {\bf B }}

\newcommand{\jzp}{{\it Z.~Phys.} {\bf C} }
\newcommand{\jpr}{{\it Phys.~Rev.} {\bf D} }

\newcommand{\jprp}{\it Phys.~Rep.} 

\newcommand{\np}[3]{{\it Nucl.~Phys.} {\bf #1} (19#2) #3}
\newcommand{\npb}[3]{{\it Nucl.~Phys.} {\bf B#1} (19#2) #3}
\newcommand{\prl}[3]{{\it Phys.~Rev.~Lett.} {\bf #1} (19#2) #3}
\newcommand{\pl}[3]{{\it Phys.~Lett.} {\bf #1B} (19#2) #3}
\newcommand{\zp}[3]{{\it Z.~Phys.} {\bf C#1} (19#2) #3}
\newcommand{\pr}[3]{{\it Phys.~Rev.} {\bf #1} (19#2) #3}
\newcommand{\prd}[3]{{\it Phys.~Rev.} {\bf D#1} (19#2) #3}
\newcommand{\ptp}[3]{{\it Prog.~Theor.~Phys.} {\bf #1} (19#2) #3}

\newcommand{\prp}[3]{{\it Phys.~Rep.} {\bf #1C} (19#2) #3}
\newcommand{\mpl}[3]{{\it Mod.~Phys.~Lett.} {\bf A#1} (19#2) #3}
\newcommand{\sjp}[3]{{\it Sov.~J.~Part.~Nucl.} {\bf #1} (19#2) #3}
\newcommand{\inc}[3]{{\it Nuovo~Cimento} {\bf #1A} (19#2) #3}
\newcommand{\cpc}[3]{{\it Comput.~Phys.~Commun.} {\bf #1} (19#2) #3}

\newcommand{\lsim}{\raisebox{-0.13cm}{~\shortstack{$<$ \\[-0.07cm] $\sim$}}~}

\newcommand{\beq}{\begin{equation}}
\newcommand{\eeq}{\end{equation}}
\newcommand{\SM}{{\rm SM}}
\newcommand{\MSSM}{{\rm MSSM}}
\newcommand\MSQCD{\overline{{\rm MS}}}
\newcommand{\ba}{\begin{eqnarray}}
\newcommand{\ea}{\end{eqnarray}}
\newcommand{\bano}{\begin{eqnarray*}}
\newcommand{\eano}{\end{eqnarray*}}

\newcommand\tb{\tan\beta}
\newcommand\ctb{\cot\beta}
\newcommand\ra{\rightarrow}
\newcommand\eepm{e^+ e^-}

\def\fmfL(#1,#2,#3)#4{\put(#1,#2){\makebox(0,0)[#3]{#4}}}
\font\eightrm=cmr8

\def\etal{et al., }
\def\LEPI{LEP1}
\def\LEPII{LEP2}
\def\tanb{\mbox{$\tan\beta$}}
\def\inpb{\mbox{$\hbox{pb}^{-1}$}}
\def\infb{\mbox{$\hbox{fb}^{-1}$}}
\def\Gcs{\hbox{GeV}/\mbox{$c^2$}}

\def\Z{\mbox{$\hbox{Z}$}}
\def\W{\mbox{$\hbox{W}$}}
\def\A{\mbox{$\hbox{A}$}}
\def\H{\mbox{$\hbox{H}$}}
\def\Hpm{\mbox{$\hbox{H}^\pm$}}
\def\HpHm{\mbox{$\H^+\H^-$}}
\def\HSM{\mbox{$\hbox{H}_{\hbox{\eightrm SM}}$}}
\def\h{\mbox{$\hbox{h}$}}
\def\e{\mbox{$\hbox{e}$}}

\def\epemto{\mbox{$\hbox{e}^+\hbox{e}^- \to$}}
\def\epem{\mbox{$\hbox{e}^+\hbox{e}^-$}}
\def\mpmm{\mbox{$\mu^+\mu^-$}}
\def\tptm{\mbox{$\tau^+\tau^-$}}
\def\lplm{\mbox{$\ell^+\ell^-$}}
\def\nnbar{\mbox{$\nu\bar\nu$}}
\def\ffbar{\mbox{$\hbox{f}\bar{\hbox{f}}$}}
\def\qqbar{\mbox{$\hbox{q}\bar{\hbox{q}}$}}
\def\bbbar{\mbox{$\hbox{b}\bar{\hbox{b}}$}}
\def\csbar{\mbox{$\hbox{c}\bar{\hbox{s}}$}}
\def\cbars{\mbox{$\bar{\hbox{c}}\hbox{s}$}}
\def\mh{\mbox{$m_{\hbox{\eightrm h}}$}}
\def\mH{\mbox{$m_{\hbox{\eightrm H}}$}}
\def\mHpm{\mbox{$m_{{\hbox{\eightrm H}}^\pm}$}}
\def\mHp{\mbox{$m_{{\hbox{\eightrm H}}^+}$}}
\def\mHm{\mbox{$m_{{\hbox{\eightrm H}}^-}$}}
\def\mA{\mbox{$m_{\hbox{\eightrm A}}$}}
\def\mZ{\mbox{$m_{\hbox{\eightrm Z}}$}}
\def\mW{\mbox{$m_{\hbox{\eightrm W}}$}}

\newcommand{\ptmis}{{ {\rm p} \hspace{-0.53 em} \raisebox{-0.27 ex}{/}_T }}

\begin{document}
\topmargin-0.5cm
\noindent
\vspace*{2cm}
~\pss0
{\large {\bf HIGGS PHYSICS}}\pss3 
{\it Conveners:}\pss1
M.\ Carena and P.M.\ Zerwas\pss2
{\it Working Group:}\pss1
E.~Accomando,
P.~Bagnaia,
A.~Ballestrero,
P.~Bambade,
D.~Bardin,
F.~Berends,
J.~van der Bij, 
T.~Binoth,
G.~Burkart,
F.~de Campos, 
R.~Contri,
G.~Crosetti,
J.~Cuevas Maestro, 
A.~Dabelstein, 
W.~de Boer,
C.~de StJean,
F.~Di Lodovico,
A.~Djouadi, 
V.~Driesen,
M.~Dubinin,
E.~Duchovni,
O.J.P.~Eboli,
R.~Ehret,
U.~Ellwanger,
J.-P.~Ernenwein,
J.-R.~Espinosa,
R.~Faccini, 
M.~Felcini, 
R.~Folman,
H.~Genten,
J.--F.~Grivaz,
E. Gross,
J.~Guy,
H.~Haber,
Cs.~Hajdu, 
S.W.~Ham,
R.~Hempfling,
A.~Hoang,
W.~Hollik,
S.~Hoorelbeke,
K.~Hultqvist, 
P.~Igo--Kemenes,
P.~Janot,
S.~de Jong,
U.~Jost,
J.~Kalinowski,
S.~Katsanevas,
R.~Ker\"anen,
W.~Kilian,
B.R.~Kim,
S.F.~King,
R.~Kleiss,
B.~A.~Kniehl,
M.~Kr\"amer, 
A.~Leike, 
E.~Lund,
V.~Lund,
P.~Lutz,
J.~Marco, 
C.~Mariotti,
J.--P.~Martin, 
C.~Martinez--Rivero,
G.~Mikenberg,
M.R.~Monge,
G.~Montagna, 
O.~Nicrosini,
S.K.~Oh,
P.~Ohmann,
G.~Passarino, 
F.~Piccinini, 
R.~Pittau,
T.~Plehn,
M.~Quiros,
M.~Rausch de Traubenberg,
T.~Riemann,
J.~Rosiek,
V.~Ruhlmann-Kleider,
C.A.~Savoy,
P.~Sherwood, 
S.~Shichanin,
R.~Silvestre, 
A.~Sopczak,
M.~Spira,
J.W.F.~Valle,
D.~Vilanova,
C.E.M.~Wagner,
P.L.~White,
T.~Wlodek,
G.~Wolf,
S.~Yamashita,
and F.~Zwirner. 

\vfill
\newpage

\tableofcontents

\newpage

\section{Synopsis}
1. The understanding of the mechanism responsible for the breakdown of
 the electroweak symmetry is one of $\:$the central $\:$problems 
in $\:$particle
physics. $\:$If $\:$the fundamental $\:$particles  -- leptons, quarks and gauge
bosons -- remain weakly interacting up to high energies, then the sector in
which the electroweak symmetry is broken must contain one or more
fundamental scalar Higgs bosons with masses of the order of the
symmetry breaking scale $v \sim$ 174 GeV. Alternatively, the symmetry
breaking could be generated dynamically by novel strong forces at the
scale $\Lambda \sim $ 1 TeV.  However, no compelling model of this kind
 has yet
been formulated which provides a satisfactory description
of the  fermion sector  and reproduces the high precision electroweak 
measurements.
\vvs1

\noindent 2. The simplest mechanism for the breaking of the electroweak symmetry
is realized in the Standard Model (\SM) \cite{R1}. To accommodate all
observed phenomena, a complex isodoublet scalar field 
is introduced which, through
self-interactions, spontaneously breaks the electroweak symmetry SU(2)$_{\rm L}
\times$U(1)$_{\rm Y}$ down to the electromagnetic U(1)$_{\rm EM}$
symmetry, by acquiring a non--vanishing vacuum
expectation value. After the electroweak symmetry breakdown, the 
interaction of the gauge bosons and fermions
with the isodoublet scalar field generates the masses of these particles
\cite{R2}.  In this process, one
scalar field
remains in the spectrum,
 manifesting itself as the physical  Higgs particle $H$.

The mass of the \SM\ Higgs boson is constrained in two ways. Since the
quartic self-coupling of the Higgs field grows indefinitely with
rising energy, an upper limit on the Higgs mass is obtained by
demanding that the SM particles remain weakly interacting up to a scale
$\Lambda$ \cite{R3}. On the other hand, stringent lower bounds on the
Higgs mass can be derived from the  requirement of stability of the
electroweak vacuum \cite{R3,R6}.  Hence, if the Standard Model
is valid up to scales near the Planck scale, then the  \SM\ Higgs mass is
restricted to the range between $\sim 130$ GeV and $\sim$ 180 GeV, for
a top-quark mass $M_t \sim$ 176 GeV. Moreover, if the Higgs particle
is discovered in the mass range up to the $100$ GeV accessible at
LEP2, this will imply that new physics beyond the Standard Model should exist
at energies below a scale $\Lambda$ of order 10 TeV. [These bounds 
become stronger (weaker) for larger (smaller) values of the top quark mass].

The high precision electroweak data give a slight preference to Higgs
masses of less than 100 GeV, 
despite the fact that the electroweak 
observables depend only 
logarithmically on the Higgs mass through radiative corrections \cite{R7}. 
 They do not, however, exclude values up to
$\sim 700$ GeV at the $2\sigma$ level \cite{R8}, thus sweeping the
entire Higgs mass range of the Standard Model. By searching directly
for the \SM\ Higgs particle, the LEP experiments \cite{R9} have set a
lower bound, $m_H > 65.2$ GeV [95\% CL], on the Higgs mass.

The dominant production mechanism for the \SM\ Higgs boson within the
energy range of LEP2 is the Higgs--strahlung process $\eepm \ra ZH$ in
which the Higgs boson is emitted from a virtual $Z$ boson \cite{R10}.
The cross section monotonically falls from $\sim 1$ pb at $m_H \sim $
65 GeV down to very small values for Higgs masses near the kinematical
threshold. The cross section for the production of Higgs bosons via
$WW$ fusion \cite{R11,R11C} is nearly two orders of magnitude smaller
at LEP2, except at the edge of the phase space for Higgs--strahlung
where both are small.  In the mass range between 60 and 120 GeV, the
dominant decay mode of the \SM\ Higgs particle is $b\bar{b}$
\cite{R11A}. Branching ratios for Higgs decays to $\tau^+ \tau^-,
c\bar{c}$ and  $gg$ final states are suppressed by an order of
magnitude or more.

The experimental search for the \SM\ Higgs boson at LEP2 will be based
primarily on the Higgs--strahlung process. The $Z$ boson can easily be
reconstructed in all charged leptonic and hadronic decay channels
while the Higgs decay mostly leads to $b\bar{b}$ and, less frequently,
to $\tau^+ \tau^-$ final states. Moreover, neutrino decays of the $Z$
boson, augmented by $W$ fusion events, can be exploited in the
experimental analyses.  Higgs events can be searched for with an
average efficiency of about 25\%.  Exploiting micro--vertex detection
for tagging $b$ quarks, the Higgs events can be well discriminated
from the main background process of $ZZ$ production even for a Higgs
mass near the $Z$ mass. When the results of all four LEP experiments
are combined, after accumulating an integrated luminosity $\int{\cal
L} =150$ pb$^{-1}$ per experiment, the \SM\ Higgs boson can be
discovered in the mass range up to $m_{H} \simeq$ 95 GeV at LEP2 for a
total center of mass energy of $\sqrt{s} = 192$ GeV.
\vvs1

\noindent 3. If the Standard Model is embedded in a Grand Unified
Theory (GUT) at high energies, then the natural scale of electroweak
symmetry breaking would be close to the unification scale $M_{\rm GUT}$,
due to the quadratic nature of the radiative corrections to the Higgs
mass.  Supersymmetry \cite{R12} provides a solution to this hierarchy
problem through the cancellation of these quadratic divergences via
the contributions of fermionic and bosonic loops \cite{R13}.
Moreover, the Minimal Supersymmetric extension of the Standard Model
(MSSM) can be derived as an effective theory from supersymmetric Grand
Unified Theories \cite{R13A}, involving not only the strong and
electroweak interactions but gravity as well.  A strong indication for
the realization of this physical picture in nature is the excellent
agreement between the value of the weak mixing angle $\sin^2\theta_W$
predicted by the unification of the gauge couplings, and the measured
value \cite{R13A}-\cite{R13G}.  In particular, if the gauge couplings
are unified in the minimal supersymmetric theory at a scale $M_{\rm
  GUT} = {\cal{O}}(10^{16}$~GeV) and if the mass spectrum of the
supersymmetric particles is of order $m_Z$, then the electroweak
mixing angle is predicted to be $\sin^2{\theta}_W = 0.2336 \pm 0.0017$
in the ${\overline{{\rm MS}}}$ scheme for $\alpha_s = 0.118 \mp
0.006$, to be compared with the experimental result
$\sin^2{\theta}_W^{exp} = 0.2314 \pm 0.0003$.  Threshold effects at
both the low scale of the supersymmetric particle spectrum and at the
high unification scale may drive the prediction for $\sin^2\theta_W$
even closer to its experimental value.

In the past two decades a detailed picture has been developed of the
Minimal Supersymmetric Standard Model. In this extension of the
Standard Model the Higgs sector is built up of two doublets, necessary
to generate masses for up-- and down--type fermions in a
supersymmetric theory, and to render the theory anomaly--free
\cite{R14}.  The Higgs particle spectrum consists of a quintet of
states: two CP--even scalar $(h,H)$, one CP-odd pseudoscalar neutral
$(A)$, and a pair of charged $(H^\pm)$ Higgs bosons \cite{hhg} .

\vfill

\newpage

Since the tree--level quartic Higgs self--couplings in this minimal theory are
determined in terms of  the gauge couplings, the mass of the lightest CP-even
Higgs boson $h$ is constrained very stringently. At tree-level, the
mass $m_h$ has been predicted to be less than the $Z$ mass
\cite{Marc,R15}. Radiative corrections to $m_h^2$ grow as the fourth
power of the top mass and the logarithm of the stop masses.
They shift the upper limit to about $\simlt 150$ GeV
\cite{RC1,ERZ}, depending on the MSSM parameters.

The upper limit on $m_h$ depends on $\tb$, the ratio of the vacuum
expectation values associated with the two neutral scalar Higgs
fields. This parameter can be constrained by additional symmetry
concepts. If the theory is embedded into a grand unified theory, the
$b$ and $\tau$ Yukawa couplings can  be expected to unify at
$M_{\rm GUT}$.  The condition of $b$-$\tau$ Yukawa coupling
unification determines the value of the top-quark Yukawa coupling at
low energies \cite{Inoue}, thus explaining qualitatively the large 
value of the top
quark mass \cite{R13D},\cite{R17}-\cite{R18}.  For the present
experimental range \cite{CDF}, $M_t = 180 \pm 12$ GeV, the condition
of $b$-$\tau$ unification implies either low values of $\tan \beta$, $
1\simlt \tan \beta \simlt 3$, or very large values of $\tan \beta =
{\cal{O}} (m_t/m_b)$ \cite{R17}-\cite{R18}.
In the small $\tan \beta$ regime, the top-quark mass is strongly
attracted to its infrared fixed point \cite{IRold}, implying a strong
correlation between the top-quark mass and $\tan \beta$.  The large
$\tan\beta$ regime is more complex because of possible large radiative
corrections to the $b$ quark mass associated with supersymmetric
particle loops \cite{Hall,wefour}.  For small $\tan \beta$ and $M_t
\lsim 176$ GeV, the upper bound on the mass of the lightest neutral
Higgs particle is reduced to $\sim $ 100 GeV. This mass bound is just
at the edge of the kinematical range accessible at a center of mass
energy of 192 GeV \cite{interim} -- raising 
the prospects of discovering
this Higgs boson at LEP2.

The structure of the Higgs sector in the \MSSM\ at  tree level is
determined by one Higgs mass parameter, which we choose to be $m_A$, and
$\tb$.  The mass of the pseudoscalar Higgs boson $m_A$ may vary
between the present 
experimental lower bound of 45  GeV \cite{R9} and
$\sim$~1~TeV, the heavy neutral scalar mass $m_H$ is in general larger
than $\sim $ 120 GeV, and the mass of the charged Higgs bosons exceeds
$\sim 90$ GeV. Due to the kinematics the primary focus at LEP2 will be
on the light scalar particle $h$ and on the pseudoscalar particle $A$.
In the decoupling limit of large $A$ mass [yielding
large $H, H^\pm$
masses], the Higgs sector becomes \SM\ like and the properties of the
lightest neutral Higgs boson $h$ coincide with the properties of the
Higgs boson $H$ in the Standard Model \cite{R19}.

The processes for producing the Higgs particles $h$ and $A$ at LEP2
are Higgs--strahlung $e^+ e^- \ra Zh$, and associated pair production
$e^+ e^- \ra Ah$ \cite{R19a}. These two processes are complementary.
For small values of $\tb$ the $h$ Higgs boson is produced primarily
through Higgs--strahlung; if kinematically allowed, associated $Ah$
production becomes increasingly important with rising $\tb$. The
typical size of the cross sections is of order 1~pb or slightly below.
The dominant decay modes of the $h,A$ Higgs bosons are decays into $b$
and $\tau$ pairs, if we consider \SM\ particles in the final state
\cite{R11A}. Only near the maximal $h$ mass for a given value of $\tb$
do $c\bar{c}$ and $gg$ decays occur at a level of several percent, in
accordance with the decoupling theorem.  However, there are areas in
the SUSY $[\mu, M_2$] parameter space where Higgs particles can decay
into invisible $\chi_1^0 \chi_1^0$ LSP final states or possibly other
neutralino and chargino final states \cite{R20,shgs3b}. If the LSP
channel is open, the $h$ and $A$ invisible decay branching ratios can
be close to 100\% for small to moderate values of $\tb$. However, the
Higgs boson $h$ can still be found in the Higgs--strahlung process.
The pseudoscalar $A$, produced only in association with $h$, would be
hard to detect in this case since both particles decay into invisible
channels for small $\tb$.

The experimental search for $h$ in the Higgs--strahlung process
follows the lines of the Standard Model, while for associated $Ah$
production $b\bar{b}b\bar{b}$ and $b\bar{b} \tau^+ \tau^-$ final
states can be exploited. Signal events of the $Ah$ type can be
searched for with an efficiency of about 30\%; the background
rejection is somewhat more complicated than for Higgs--strahlung, due
to two unknown particles in the signal final state. For small to
moderate $\tb$, $h$ particles with masses up to $\sim 100$ GeV can be
discovered in the Higgs--strahlung process.  For large $\tb$ the
experimentally accessible limits are typically reduced by about 10
GeV. The pseudoscalar Higgs boson $A$ is accessible for masses up to about
$80$ GeV. [These limits are based on the LEP2 energy of 192 GeV
and an integrated luminosity of $\int{\cal L}=150$ pb$^{-1}$ per
experiment, with all four experiments pooled.]

The supersymmetric theory may be distinguished from the Standard Model
if one of the following conditions occurs: (i) at least two different
Higgs bosons are found; (ii)  precision  measurements of
production cross sections and decay branching ratios of $h$ can
 be performed at a  level of a few per cent; and (iii) genuine
SUSY decay modes are observed. Near the maximum $h$ mass, the
decoupling of the heavy Higgs bosons reduces the \MSSM\ to the SM
Higgs boson except for the SUSY decay modes.
\vvs1

\noindent 4. {\em In summary.} If a neutral scalar Higgs boson is found at LEP2,
new physics beyond the Standard Model should exist at scales 
of order 10 TeV.  In the framework of the Minimal
Supersymmetric extension of the Standard Model, there are good
prospects of discovering the lightest of the neutral scalar Higgs
bosons at LEP2.  Even though this discovery cannot be ensured,
observation or non--observation will have far reaching consequences on
the possible structure of low--energy supersymmetric theories. 
\vvs1

In  section 2 the theoretical analysis and experimental
simulations for the search for the Higgs boson in the Standard Model
are presented. In section 3 the Higgs spectrum and the couplings in
the MSSM as well as the relevant cross sections and branching ratios
are studied.  In addition, the results of the experimental simulations
are thoroughly discussed. Section 4 investigates opportunities of
detecting Higgs particles at LEP2 within non-minimal extensions of the
SM and the MSSM.  In particular, the next--to--minimal extension of
the MSSM with an additional isoscalar Higgs field (NMSSM) is studied.

\section{The Standard--Model Higgs Particle}

\subsection{Mass Bounds}

(i) \underline{Strong interaction limit and vacuum stability.} Within
the Standard Model the value of the Higgs mass $m_H$ cannot be
predicted. The mass $m_H = \sqrt{2 \lambda} v$ is given as a function
of the vacuum expectation value of the Higgs field, $v$ = 174 GeV, and
the quartic coupling $\lambda$ which is a free parameter.  However,
since the quartic coupling grows with rising energy indefinitely, an
upper bound on $m_H$ follows from the requirement that the theory be
valid up to the scale $M_{Planck}$ or up to a given cut-off scale
$\Lambda$ below $M_{Planck}$ \cite{R3}. The scale $\Lambda$ could be
identified with the scale at which a Landau pole develops. However, 
in the following the upper
bound on $m_H$ shall be defined by the requirement $\lambda(\Lambda)/
4 \pi \leq 1$ so that $\Lambda$ characterizes the energy where the
system becomes strongly interacting. [This scale is very close to the scale
associated with the 
Landau pole  in practice.]  The upper bound on $m_H$ depends mildly on
the top-quark mass through the impact of the top-quark Yukawa coupling
on the running of the quartic coupling $\lambda$,
\be 
\frac{d \lambda}{d t} = \frac{6}{16 \pi^2} \left( \lambda^2 
                        + \lambda h_t^2 - h_t^4 \right) 
                        +   {\rm elw. \; corrections}
\label{eq:betalambda} 
\ee
with $t= \ln (Q^2/\Lambda^2)$.  The first two terms inside the
parentheses are crucial in driving the quartic coupling to its
perturbative limit.  On the other hand, the requirement of vacuum
stability in the SM imposes a lower bound on the Higgs boson mass,
which depends crucially on the top-quark mass as well as on the
cut-off $\Lambda$ \cite{R3,R6}.  Again, the dependence of this lower
bound on $M_t$ is due to the effect of the top-quark Yukawa coupling
on the quartic coupling of the Higgs potential [third term inside the
parentheses of eq.(\ref{eq:betalambda})], which drives $\lambda$ to
negative values at large scales, thus destabilizing the standard
electroweak vacuum.

\begin{figure}[htb]
\vspace{-1cm}
\centerline{
\epsfig{figure=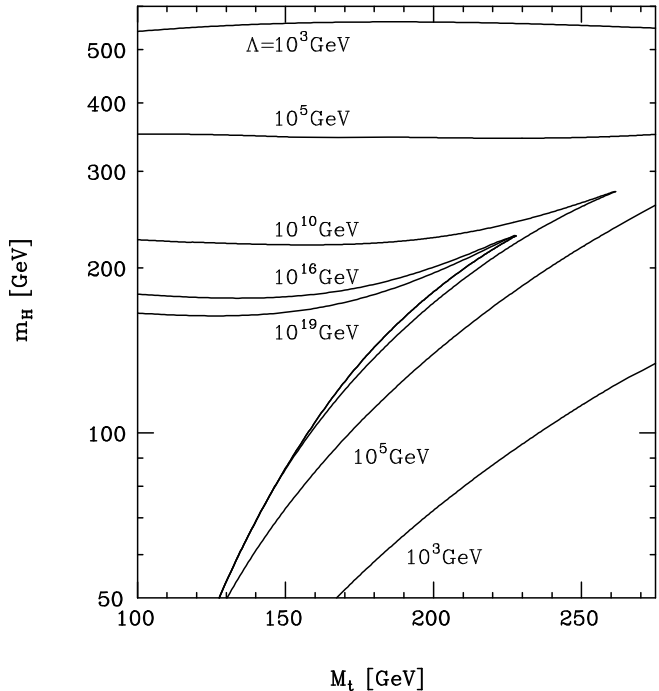,height=15cm,angle=90}}
\vspace{-1.5cm}
\caption{\it Strong interaction and stability bounds on the SM Higgs 
  boson mass. $\Lambda$ denotes the energy scale where the particles
  become strongly interacting.}
\label{mariano}
\end{figure}

Fig.\ref{mariano} shows the perturbativity and stability bounds on the
Higgs boson mass of the SM for different values of the cut-off
$\Lambda$ at which new physics is expected.  From the point of view of
LEP physics, the upper bound on the SM Higgs boson mass does not pose
any relevant restriction. The lower bound on $m_H$, instead, needs to
be carefully considered.  To define the conditions for vacuum
stability in the SM and to derive the lower bounds on $m_H$ as a
function of $M_t$, it is necessary to study the Higgs potential for
large values of the Higgs field $\phi$ and to determine under which
conditions it develops an additional minimum deeper than the
electroweak minimum.
The renormalization group improved effective potential of the SM is
given by
\be
V_{eff.} = V_0 + V_1 \simeq -m^2(t) \P^2(t) 
           + \frac{\lambda(\phi)}{2} \P^4(t)
\label{eq:pot}
\ee
where $V_0$ and $V_1$ are the tree--level potential and the one--loop
correction, respectively.  A rigorous analysis of the structure of the
potential has been done in Ref.\cite{R6}. Quite generally it follows
that the stability bound on $m_H$ is defined, for a given value of
$M_t$, as the lower value of $m_H$ for which $\lambda(\phi) \geq$ 0
for any value of $\phi$ below the scale $\Lambda$ at which new physics
beyond the SM should appear. From eq.(\ref{eq:betalambda}) it is clear
that the stability condition of the effective potential demands new
physics at lower scales for larger values of $M_t$ and smaller values
of $m_H$.

From Fig.\ref{mariano} it follows that for $M_t$ = 175 GeV and $m_H
< 100$~GeV [i.e. in the LEP2 regime] new physics should appear
below the scale $\Lambda \sim$ a few to 100~TeV. The dependence on
the top-quark mass however is noticeable. A lower value, $M_t \simeq$
160 GeV, would relax the previous requirement to $\Lambda \sim 10^3 $
TeV, while a heavier value $ M_t \simeq$ 190 GeV would demand new
physics at an energy scale as low as 2~TeV.
 
The previous bounds on the scale at which new physics should appear
can be relaxed if the possibility of a metastable vacuum is taken into
account \cite{meta}. In fact, if the effective potential of the SM has
a non-standard stable minimum deeper than the standard minimum, the
decay of the electroweak minimum by thermal fluctuations or quantum
tunnelling to the stable minimum must be suppressed.  In this case,
the lower bounds on $m_H$ follow from requiring that no transition at
any finite temperature occurs, so that all space remains in the
metastable electroweak vacuum. In practice, if the metastability
arguments are taken into account, the lower bounds on $m_H$ become
gradually weaker. They seem to disappear if the cut-off of the theory
is at the TeV scale; however, the calculations are technically not
reliable in this energy regime. Moreover, the metastability bounds
depend on several cosmological assumptions which may be relaxed in
several ways.

\begin{figure}[htb]
\centerline{
\epsfig{figure=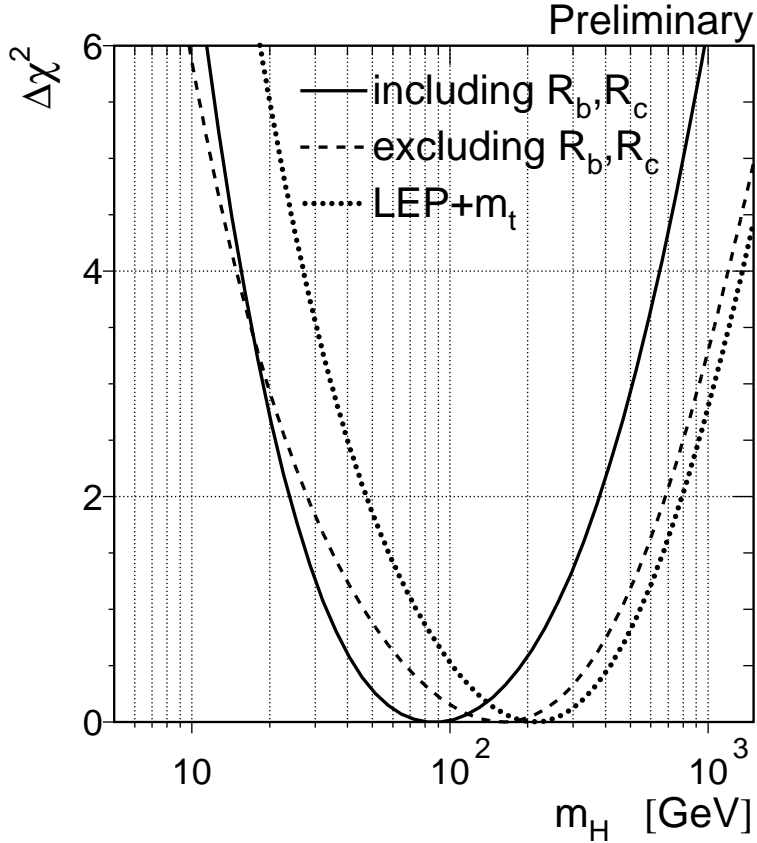,height=12cm,angle=0}}
\vspace{-1.cm}
\caption{\it $\Delta\chi^{2}=\chi^2-\chi^2_{min}$ vs $m_H$ curves.
  Continuous line: based on all LEP, SLD, $p\bar{p}$ and $\nu$N data;
  dashed line: as before, but excluding the LEP+SLD measurements of
  $R_b$ and $R_c$; dotted line: LEP data including measurements of
  $R_b$ and $R_c$. In all cases, the direct measurement of $M_t$ at
  the TEVATRON is included.  }
\label{mhfit}
\end{figure}

\noindent 
(ii) \underline{Estimate of the Higgs mass from electroweak data.}
Indirect evidence for a light Higgs boson comes from the
high--precision measurements at LEP \cite{R8} and elsewhere. Indeed,
the fact that the SM is renormalizable only after including the top
and Higgs particles in the loop corrections shows that the electroweak
observables should be sensitive to these particle masses.  Although
the sensitivity to the Higgs mass is only logarithmic, while the
sensitivity to the top-quark mass is quadratic, the increasing
precision of present experiments makes it possible to derive $\chi^2$
curves as a function of $m_H$.  Several groups \cite{R8} have
performed an analysis of $m_H$ by means of a global fit to the electroweak
data, including low and high energy data. In the light of the recent
direct determination of $M_t$, the results favor a light Higgs boson.
With all LEP, SLD, $p\bar{p}$ and $\nu$N data included, a central
value for $m_H$ around 80 GeV and $M_t~\sim$~170~GeV is obtained \cite
{R8}.  However, the recently reported LEP values of $R_b \equiv
\Gamma_{Z \rightarrow b \bar{b}}/ \Gamma_{Z \rightarrow hadrons} $ and
$R_c \equiv \Gamma_{Z \rightarrow c \bar{c}}/ \Gamma_{Z \rightarrow
  hadrons} $ which are more than 2 standard deviations away from the
SM predictions, and the left-right asymmetries of SLD which still lead
to a 2$\sigma $ discrepancy in $\sin^2 \theta_W$ compared with LEP
analyses, have drastic effects on the SM fits.  Fig.\ref{mhfit} shows
$\Delta \chi^2 = \chi^2 - \chi^2_{min}$ as a function of $m_H$; the
curve is rather flat at the minimum due to the mild logarithmic
dependence of the observables on $m_H$.  It should be noticed in this
context that the bounds on $m_H$ become very weak if $R_b$, $R_c$
and/or the left-right asymmetries are excluded from the data.


\vfill

\newpage
                       
{\it In summary.} It is clear that the indirect bounds on $m_H$ cannot
assure the existence of a light Higgs boson at the reach of LEP2.
However, the fact that the best fit to the present high-precision data
tends to prefer a light SM Higgs boson, indicates that this particle
may be found either at LEP or LHC.  On the other hand, the stability
bounds imply that if the Higgs boson is light, new physics beyond the
Standard Model should appear at relatively low energies in the TeV
regime.

\subsection{Production and Decay Processes}

The main mechanism for the production of Higgs particles in $e^+e^-$
collisions at LEP2 energies is the radiation off the virtual $Z$-boson
line \cite{R10},
\begin{equation}\label{smh:hst}
  \mbox{Higgs-strahlung :} \quad e^{+}e^{-}\to ZH\hspace*{4mm}
\end{equation}
The fusion process~\cite{R11,R11C,R11B} in which the Higgs bosons are
formed in $WW$ collisions, the virtual $W$'s radiated off the
electrons and positrons,
\begin{equation}\label{smh:wwf}
  \mbox{WW fusion\hspace*{8mm} :}\quad e^{+}e^{-}\to \bar{\nu}_{e}\nu_{e}H
\end{equation}
has a considerably smaller cross section at LEP energies. It is 
suppressed by an additional power of the electroweak coupling with
respect to the Higgs-strahlung process, becoming competitive only at
the edge of phase space in (\ref{smh:hst}), where the $Z$ boson turns
virtual. In this corner, however, both cross sections are small and
the experimentally accessible mass parameter space will be extended
only slightly by the fusion channel.
\begin{figure}[hbt]
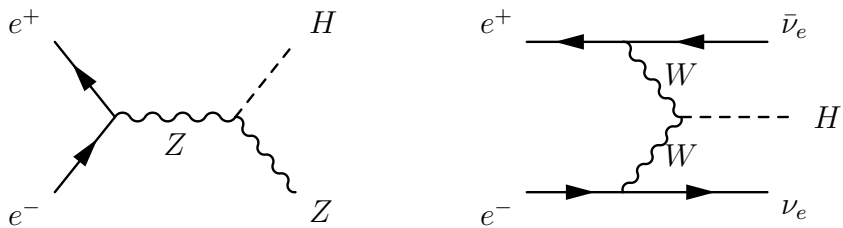

\begin{center}
\begin{picture}(40,20)
  \put(0,0){\epsfig{file=smh-graphs.1}}
  \input{smh-graphs.t1}
\end{picture}
\hspace{2cm}
\begin{picture}(40,20)
  \put(0,0){\epsfig{file=smh-graphs.2}}
  \input{smh-graphs.t2}
\end{picture}
\end{center}
\caption{\it Higgs-strahlung and $WW$ fusion of the SM Higgs boson.}
\end{figure}

\subsubsection{Higgs-strahlung} 
The \underline{cross section} for the Higgs-strahlung process can be
written in the following compact form:
\begin{equation}\label{smh:equ3}
  \sigma(e^{+}e^{-}\to ZH) =
  \frac{G_{F}^{2}m_{Z}^{4}}{96 \pi s}\,\left( v_{e}^{2}+a_{e}^{2}\right)
  \,\lambda^{\frac{1}{2}}\,
  \frac{\lambda + 12m_{Z}^{2}/s}{(1-m_{Z}^{2}/s)^{2}}
\end{equation}
where $\sqrt{s}$ denotes the center-of-mass energy, and $a_e=-1$,
$v_e=-1+4\,s_w^{2}$ are the $Z$ charges of the electron; $\lambda =
(1-m_{H}^{2}/s-m_{Z}^{2}/s)^{2}-4m_{H}^{2} m_{Z}^{2}/s^{2}$ is the
usual two-particle phase space function.  The radiative corrections to
the cross section are well under control.  The genuine electroweak
corrections \cite{smh5} are small at the LEP energy, less than 1.5\%
(for a recent review see Ref.\cite{smh6}). By contrast, photon
radiation \cite{smh7} affects the cross section in a significant
way. The bulk of the corrections, real and virtual contributions due
to photons and $e^+e^-$ pairs, can be accounted for by convoluting the
Born cross section in eq.(\ref{smh:equ3}) with the radiator function $G(x)$,
\begin{equation}
  \langle\sigma\rangle
  =\int_{x_{H}}^{1} dx\,G (x)\,\sigma (xs)
\end{equation}
with $x_{H}=m_{H}^{2}/s$.  The radiator function is known to order
$\alpha^2$, including the exponentiation of the infrared sensitive
part, 
\begin{equation}
  G(x) =\beta\,(1-x)^{\beta-1}\,\delta_{V+S}\,+\,\delta_{H}(x)
\end{equation}
where $\delta_{V+S}$ and $\delta_{H}$ are polynomials in $\log s/m_{e}^{2}$
and $\beta=\frac{2\alpha}{\pi}\,[\log s/m_{e}^{2} -1]$. $\delta_{V+S}$
accounts for virtual and soft photon effects, $\delta_{H}$ for hard
photon radiation.  The $\delta$'s are given in Ref.\cite{smh7}.

\begin{figure}[p]
\begin{center}
  \epsfig{file=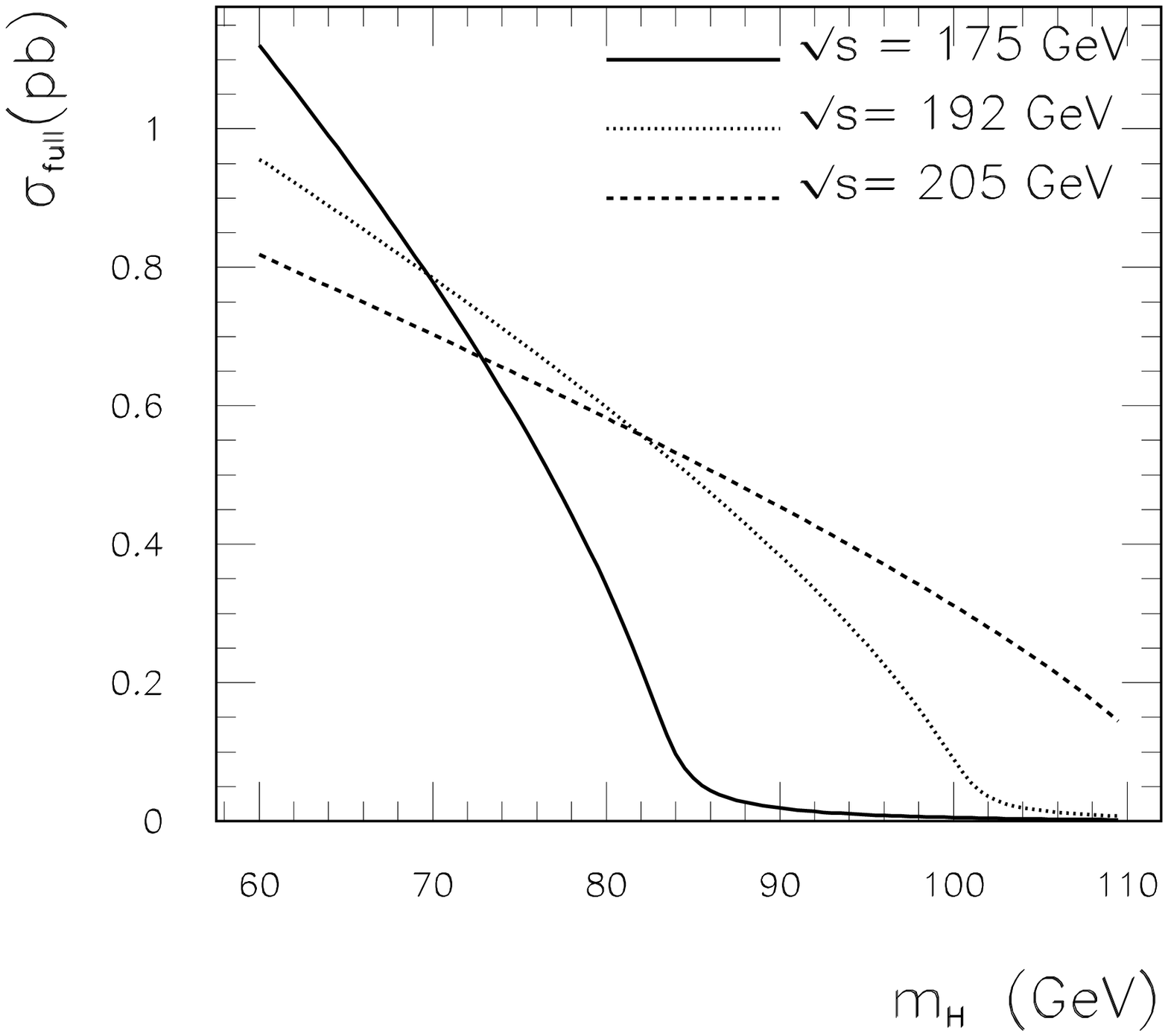,
          bbllx=0, bblly=0, bburx=575, bbury=455,
          height=10cm, clip=}
\end{center}
\vspace*{-8mm}
\caption{\it The cross section for Higgs-strahlung as a function of
  the Higgs mass for three repre\-sentative energy values [QED and
  electroweak radiative corrections included].}
\label{smh:xsec-hst}
\vspace*{8mm}
\begin{center}
  \epsfig{file=smh-sigma192.eps}
\end{center}
\vspace*{-5mm}
\caption{\it Higgs-strahlung (dashed) and $WW$ fusion
  (long-dashed) processes for Higgs production in the cross-over
  region [without radiative corrections].  The solid line shows the
  total cross section for both processes including the (dotted line)
  interference term.}
\label{smh:xsec-192}
\end{figure}

The cross-section for Higgs-strahlung is shown in
Fig.\ref{smh:xsec-hst} for the three representative energy values
$\sqrt{s}=175$, $192$ and $205$ GeV as a function of the
Higgs-mass \cite{smh8}.  The curves include all genuine electroweak
and QED corrections introduced above. The $Z$ boson in the final state
is allowed to be off-shell, so that the tails of the curves extend beyond
the on-shell limit $m_H = \sqrt{s} - m_Z$. [The Higgs boson is so
narrow, $\Gamma_{H}<3$ MeV for $m_{H}<100$ GeV, that the particle need
not be taken off-shell.] From a value of order $0.3$ to $1$ pb at
$m_H\sim\sqrt{s}-110$ GeV, the cross section falls
steadily, reaching a level of less than 0.05 pb at the mass 
$m_H\sim\sqrt{s}-90$~GeV.

\begin{figure}
\begin{center}
\begin{picture}(170,75)
\put(0,0){\epsfig{file=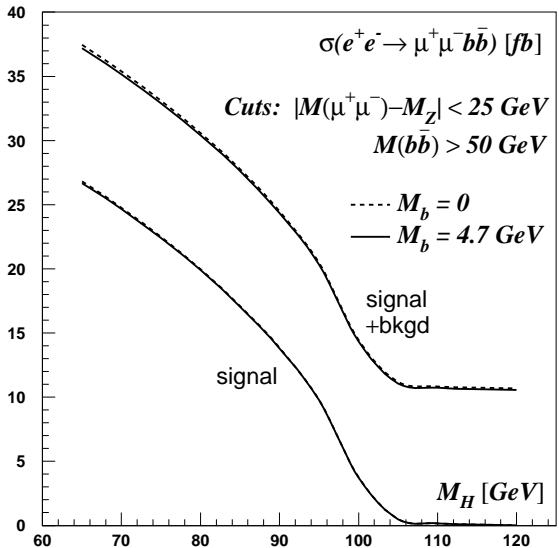,%
        bbllx=90pt, bblly=260pt, bburx=490pt, bbury=610pt,%
        height=7cm, clip=}}
\put(85,0){\parbox[b]{85mm}{%
\caption{\it Comparison of the Higgs signal with the background in the
  $\mu^+\mu^-b\bar b$ final state for zero and non-zero quark mass.}
\label{smh:fig-shi}}}
\end{picture}
\end{center}
\end{figure}

Since the Higgs particle decays predominantly to $b\bar{b}$ and
$\tau^{+}\tau^{-}$ pairs, the observed final state consists of four
fermions. Among the possible final states, the channel
$\mu^{+}\mu^{-}b\bar{b}$, the $\mu$ pair being generated by the $Z$
decay, has a particularly simple structure. Background events of this
type are generated by double vector-boson production
$e^{+}e^{-}\to\,Z^{*}Z^{*},\ Z^{*}\gamma^{*}$ and $\gamma^{*}
\gamma^{*}$ with the virtual $Z^{*},\ \gamma^{*}$ decaying to
$\mu^{+}\mu^{-}$ and $b\bar{b}$; $Z$ final states generate by far the
dominant contribution.  Since these processes are suppressed by one
and two additional powers of the electroweak coupling compared with
the signal [except for $m_H\sim m_Z$], the background can be
controlled fairly easily up to the kinematical limit of the Higgs
signal. This is demonstrated in
Tables~\ref{smh:tab-ee-mumubb_a}/\ref{smh:tab-ee-mumubb_b} and
Fig.~\ref{smh:fig-shi} where signal and background cross sections for the
process $e^{+}e^{-}\to\,\mu^{+}\mu^{-}b\bar{b}$ are compared for three
Higgs masses at $\sqrt{s}=192$ GeV. The invariant $\mu^{+}\mu^{-}$
mass is restricted to $m_{Z}\pm 25$ GeV and the invariant $b\bar{b}$
mass is cut at $m(b\bar{b})> 50$ GeV.  The following conclusions can
be drawn from the tables and the figure: (i) The signal-to-background
ratio decreases steadily with rising Higgs mass from a value of about
three near $m_{H}=65$ GeV; (ii) The initial state QED radiative
corrections are large, varying between 10 and 20\%; (iii) The cross
sections are lowered by taking non-zero $b$ quark masses into account,
but only marginally at a level of less than 1\%. Since massless
fermions are coupled to spin-vectors in $Z^{*}$ decays but to
spin-scalars in Higgs decays, signal and background amplitudes do not
interfere as long as $b$ quark masses are neglected.
\vfill\newpage

\begin{table}[hbt]
\caption{\it The process $e^+e^-\to \mu^+\mu^-b\protect\bar b$ at
  $\protect\sqrt{s} = 192$ GeV. No initial state radiation is
  included.  The cross sections are given in fb.}
\label{smh:tab-ee-mumubb_a}
 {\footnotesize
\begin{center}
\begin{tabular}{|c|c|c|c|c|}
\hline 
\rule[-2mm]{0mm}{6mm}
$m_H$ [GeV] & 65 & 90 & 115 & $\infty$ \\
\hline
\rule{0mm}{4.5mm}
CompHEP${}_0$     & 37.264(58) & 24.395(46) & 10.696(13) & 10.634(13) \\
CompHEP${}_{4.7}$ & 37.147(58) & 24.279(46) & 10.580(13) & 10.518(13) \\
EXCALIBUR         & ---        & ---        & ---        & 10.6398(15) \\
FERMISV           & ---        & ---        & ---        &  9.49(23) \\
GENTLE${}_0$      & 37.3975(37)& 24.4727(25)& 10.7022(11)& 10.6401(11) \\
HIGGSPV           & 37.393(27) & 24.490(21) & 10.694(16) & 10.65(05) \\
HZHA/PYTHIA       & 36.79(13)  & 23.53(13)  & 10.28(13)  & 10.22(13) \\
WPHACT${}_{4.7}$  & ---        & ---        & ---  & 10.5243$\pm$0.24E-02 \\
WPHACT${}_0$      & 37.39896$\pm$0.64E-02 & 24.47269$\pm$0.40E-02
 & 10.70272$\pm$0.24E-02 & 10.64070$\pm$0.24E-02 \\
WTO               & 37.40994$\pm$0.32E-02 & 24.47653$\pm$0.42E-02
 & 10.70360$\pm$0.21E-02 & 10.64157$\pm$0.21E-02 \\[1mm]
\hline
\end{tabular}
\end{center}
\caption{\it The process $e^+e^-\to \mu^+\mu^-b\protect\bar b$ at
  $\protect\sqrt{s} = 192$ GeV. Initial state radiation included,
  cross sections in fb.}
\label{smh:tab-ee-mumubb_b}
\begin{center}
\begin{tabular}{|c|c|c|c|c|}
\hline
\rule[-2mm]{0mm}{6mm}
$m_H$ [GeV] & 65 & 90 & 115 & $\infty$ \\
\hline
\rule{0mm}{4.5mm}
EXCALIBUR         & ---        & ---         & ---         & 8.4306(29) \\
FERMISV           & ---        & ---         & ---         & 7.90(27) \\
GENTLE${}_0$      & 33.7575(34)& 19.4717(19) & 8.47729(85) & 8.43290(84) \\
HIGGSPV           & 33.759(12) & 19.480(09)  & 8.483(05)   & 8.44(05) \\
HZHA/PYTHIA       & 33.48(11)  & 18.91(11)   & 8.31(11)    & 8.27(11) \\
WPHACT            & 33.75217$\pm$0.16E-01 & 19.46923$\pm$0.91E-02 
 & 8.47665$\pm$0.57E-02 & 8.43236$\pm$0.57E-02 \\
WTO               & 33.77741$\pm$0.10E-01 & 19.48562$\pm$0.83E-02
 & 8.48511$\pm$0.78E-02 & 8.44090$\pm$0.78E-02 \\[1mm]
\hline
\end{tabular}
\end{center}
}
\end{table}

The \underline{angular distribution} of the $Z/H$ bosons in the 
Higgs-strahlung process is sensitive to the spin-parity quantum 
numbers $J^P=0^{+}$ of the Higgs particle. At high energies the 
$Z$ boson is produced in a
state of longitudinal polarization according to the equivalence
theorem so that the angular distribution approaches asymptotically the
$\sin^{2}\theta$ law, where $\theta$ is the polar
angle between the $Z/H$ flight direction and the $e^+e^-$ beam axis. At
non-asymptotic energies the distribution is shoaled~\cite{smh9},
\begin{equation}
  \frac{d \sigma}{d \cos\theta}\sim \lambda \, 
  \sin^{2}\theta\,+\,8\,m_{Z}^{2}/s
\end{equation}
becoming independent of $\theta$ at the threshold.  Were a
pseudoscalar particle produced in association with the $Z$, the
angular distribution would be given by $\sim\,(1+\cos^{2}\theta)$, 
independent of the energy; the $Z$ polarization would be transverse 
in this case. Thus, the angular distribution is sensitive to the 
assignment of spin-parity quantum numbers to the Higgs particle. The
coefficients of the $\sin^{2}\theta$ term and the constant term are
independent and could be modified separately by additional effective
$ZZH$, $\gamma ZH$ couplings or $eeZH$ contact terms induced by
interactions outside the Standard Model~\cite{smh10}.

\subsubsection{The $WW$ Fusion Process} 
The final state in which the Higgs
particle is produced in association with neutrinos
\begin{equation}
  e^{+}e^{-}\to H + \nu\bar{\nu}
\end{equation}
is built up by two different mechanisms, Higgs-strahlung with $Z$
decays to the three types of neutrinos and $WW$
fusion~\cite{R11,R11C,R11B,smh12,boos}.  For $\nu_{e}\bar{\nu}_{e}$ final states the
two amplitudes interfere. At $e^+e^-$ energies above the $HZ$
threshold for on-shell $Z$, Higgs-strahlung is by far the dominant
process, while below the $HZ$ threshold the fusion process becomes
dominant.  Correspondingly, the interference term is most important
near the threshold where the cross-over between the two mechanisms
occurs. The cross section for Higgs-strahlung above the $HZ$ threshold
is of order $g_{W}^{4}$ while below the threshold it is
suppressed by the additional electroweak vertex as well as by the
off-shell $Z$ propagator. The fusion cross-section is of order
$g_{W}^{6}$ and therefore small at LEP energies where no $\log
s/m_{H}^{2}$ enhancement factors are effective.%
\footnote{The cross-section for $ZZ$ fusion is reduced by another
  order of magnitude since the leptonic NC couplings are considerably
  smaller than the CC couplings.}
The cross section for $WW $ fusion can be expressed in a compact
form~\cite{smh12}:
\begin{equation}
  \sigma(e^{+}e^{-}\to \nu_{e}\bar{\nu}_{e}H) 
  = \frac{G_{F}^{3}m_{W}^{4}}{4\sqrt{2}\pi^{3}}\,\int_{x_{H}}^{1}\,dx\,
    \int_{x}^{1}\,\frac{dy\,F(x,y)}{[1+(y-x)/x_{W}]^{2}}
\end{equation}
\begin{equation}
  F(x,y)
  = \left[ \frac{2x}{y^{3}}-\frac{1+3x}{y^{2}}+\frac{2+x}{y}-1\right]\,
    \left[\vphantom{\frac{2x}{y^{3}}}\frac{z}{1+z}-\log(1+z)\right]\,
    +\,\frac{x}{y^{3}}\frac{z^{2}(1-y)}{1+z}\nonumber
\end{equation}
with $x_{H}=m_{H}^{2}/s,\ x_{W}=m_{W}^{2}/s $ and
$z=y(x-x_{H})/(xx_{W})$.  The more involved analytic form of the
interference term between fusion and Higgs-strahlung~\cite{R11B} is
given in the Appendix~\ref{app:HSWW}.

The size of the various contributions to the cross section for the
final state $e^{+}e^{-}\to H+\mbox{neutrinos}$ is shown in
Fig.\ref{smh:xsec-192} at $\sqrt{s}=192$ GeV. The Higgs-strahlung
includes all three neutrinos in the final state. The nominal threshold
value of the Higgs mass for on-shell $Z$ production in Higgs-strahlung
is $m_{H}=101$~GeV at $\sqrt{s}=192$~GeV. A few GeV above this mass
value the fusion mechanism becomes dominant while the Higgs-strahlung
becomes rapidly more important for smaller Higgs masses. In the
cross-over range, the cross-sections for fusion, Higgs-strahlung and
the interference term are of the same size. With a cross section of
the order of 0.01 pb only a few events can be generated in the
cross-over region for the integrated luminosity at LEP.
Dedicated
efforts are therefore needed to explore this domain experimentally and 
to extract the signal from the event sample 
$e^{+}e^{-}\to b\bar{b} + \mbox{neutrinos}$, which includes several 
background channels.
Nevertheless, $WW$ fusion can extend the Higgs mass range that can be
explored at LEP2 by a few (perhaps very valuable) GeV.

\subsubsection{Higgs Decays} 
\begin{figure}[hbt]
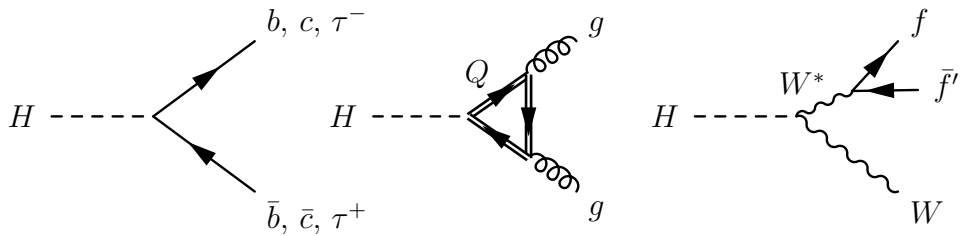

\begin{center}
\begin{picture}(30,20)
  \put(0,0){\epsfig{file=smh-graphs.3}}
  \input{smh-graphs.t3}
\end{picture}
\hspace{1cm}
\begin{picture}(30,20)
  \put(0,0){\epsfig{file=smh-graphs.4}}
  \input{smh-graphs.t4}
\end{picture}
\hspace{1cm}
\begin{picture}(30,20)
  \put(0,0){\epsfig{file=smh-graphs.5}}
  \input{smh-graphs.t5}
\end{picture}
\end{center}
\caption{\it The main decay modes of Higgs particles in the LEP2 mass
  range.}
\label{smh:decays}
\end{figure}

The Higgs decay width is predicted in the Standard
Model to be very narrow, being less than 3 MeV for $m_{H}$ less than 100
GeV.  The width of the particle can therefore not be resolved
experimentally. The main decay modes (Fig.\ref{smh:decays}), relevant
in the LEP2 Higgs mass range, are in the following channels~\cite{R11A,smh8}:
\begin{equation}
  \begin{array}{lll}
  \mbox{quark decays}&:& H\ \to \ b\bar{b} \mbox{ and }\ c\bar{c}\\
  \mbox{lepton decay}&:& H\ \to \ \tau^{+}\tau^{-}\\
  \mbox{gluon decay}&:& H\ \to \ gg\\ 
  \mbox{$W$ boson decay\hskip-.8em}&:&H\ \to \ WW^*
  \end{array}
\end{equation}
The $b\bar b$ decays are by far the leading decay mode, followed by
$\tau$, charm, and gluon decays at a level of less than 10\%. Only at
the upper end of the mass range do decays of the Higgs particle to $W$
pairs start playing an increasingly important role.

The theoretical analysis of the Higgs decay branching ratios is not
only important for the prediction of signatures to define the
experimental search techniques.  In addition, once the Higgs boson is
discovered, the measurement of the branching ratios will be necessary
to determine its couplings to other particles.  This will allow
us to explore the physical nature of the Higgs particle and to encircle the
Higgs mechanism as the mechanism for generating the
masses of the fundamental particles.  In fact, 
the strength of the Yukawa coupling of the Higgs boson to
fermions,
$g_{ffH}=[\sqrt{2}\,G_{F}]^{1/2}m_{f}$, and the couplings to the
electroweak $V=W,Z$ gauge bosons,
$g_{VVH}=2\,[\sqrt{2}\,G_{F}]^{1/2}m_{V}^{2}$, both grow with the
masses of the particles. While the latter can be measured through the
production of Higgs particles in the Higgs-strahlung and $WW$ fusion
processes, fermionic couplings can be measured at LEP only through
decay branching ratios.

\noindent 
\underline{Higgs decay to fermions.}
The partial width of the Higgs decay to 
$\tau^{+}\tau^{-}$ pairs is given by~\cite{smh14}
\begin{equation}
  \Gamma (H\to \tau^{+}\tau^{-})=\frac{G_{F}m_{\tau}^{2}}
  {4\sqrt{2}\pi}m_{H}
\end{equation}
For the decay into $b\bar{b}$ and $c\bar{c}$ quark pairs, 
QCD radiative corrections \cite{smh15} must be included which are known 
up to order $\alpha_s^{2}$ [in the $\delta'_t$ term up to order $\alpha_s^3$],
\begin{equation}\label{partial-widths}
  \Gamma (H\to q\bar{q})
  = \frac{3G_F}{4\sqrt2\pi}\,m_q^2(m_H)\,m_H
  \left[1 + 5.67\left(\frac{\alpha_s}{\pi}\right)
  + (35.94 - 1.36N_F+\delta_t+\delta'_t)\left(\frac{\alpha_s}{\pi}\right)^2 
\right]
\end{equation}    
$\delta_t$ accounts for the top-quark triangle coupled to the
$q\bar q$ final state in second order by 2-gluon $s$-channel exchange 
\cite{smh15a}, $
  \delta_t = 1.57 - {\textstyle \frac23}\log(m_H^2/M_t^2) 
  + {\textstyle \frac19}\log^2(m_q^2(m_H)/m_H^2) $, 
while $\delta'_t$ accounts for Higgs decays to two gluons with one 
gluon split into a $q\bar{q}$ pair~\cite{R11A}, discussed in detail 
below. 
The strong coupling $\alpha_s$ is to be evaluated at the scale $m_{H}$, 
and $N_{F}=5$ is the number of active flavors [all quantities defined 
in the $\MSQCD$ scheme]. The bulk of the QCD corrections 
can be absorbed into the running quark masses 
evaluated at the scale $m_{H}$, 
\begin{equation}
  m_{q}(m_{H}) 
  = m_{q}(M_{q})\,\left[ \frac{\alpha_s (m_{H})}
       {\alpha_s (M_{q})}\right]^{\frac{12}{33-2N_{F}}}
    \frac{1+c_{1}\,[\alpha_s(m_{H})/\pi ]\,
          +\,c_{2}[\alpha_s(m_{H})/\pi ]^{2}}
         {1+c_{1}\,[\alpha_s(M_{q})/\pi ]\,
          +\,c_{2}[\alpha_s(M_{q})/\pi ]^{2}}
\end{equation}
In the case of bottom (charm) quarks, the coefficients $c_{1}$ and
$c_{2}$ are 1.17 (1.01) and 1.50 (1.39), respectively.  Since the
relation between the pole mass $M_{c}$ of the charm quark and the
$\MSQCD$ mass $m_{c}\,(M_{c})$ evaluated at the pole mass is badly
convergent, the running quark masses $m_{q}(M_{q})$ lend themselves as
the basic mass parameters in practice.  They have been extracted
directly from QCD sum rules evaluated in a consistent ${\cal
O}(\alpha_s )$ expansion \cite{smh16}. Typical values of the running
$b$, $c$ masses at the scale $\mu = 100$ GeV, which is of the order of
the Higgs mass, are displayed in Table~\ref{smh:masses}.  The
evolution has been performed for the QCD coupling $\alpha_s
(m_{Z})=0.118\pm0.006$. The large uncertainty in the running charm
mass is a consequence of the small scale at which the evolution starts
and where the errors of the QCD coupling are very large.  In any case 
the value of the $c$ mass, relevant for the prediction of the $c$
branching ratio of the Higgs particle, is reduced to about 600 MeV.
\begin{table}[htbp]
\caption{\it The running $b,c$ quark masses in the $\MSQCD$
  scheme at the scale $\mu = 100$ GeV. The initial values
  $m_{Q}(M_{Q})$ of the evolution are extracted from QCD sum rules;
  the pole masses $M_{Q}^{pt2}$ are defined by the ${\cal
    O}(\alpha_{s})$ relation with the running masses
  $m_{Q}(M_{Q}^{pt2})= M_{Q}^{pt2}/[1+4\alpha_{s}/3\pi]$.}
\label{smh:masses}

{\footnotesize \begin{center}
\begin{tabular}{|c|c|cc|c|} \hline
\rule[-2mm]{0mm}{6mm}
&$ \alpha_{s}(m_{Z}) $
& $ m_{Q}(M_{Q})$\ & \ $M_{Q}\,=\,M_{Q}^{pt2} $
& \ $m_{Q}\,(\mu\,=\ 100$ GeV)  \\
\hline
\rule{0mm}{4.5mm}
$b$ & $0.112$ & $(4.26 \pm 0.02)$ GeV & $(4.62 \pm 0.02)$ GeV 
& $(3.04 \pm 0.02)$ GeV \\
    & $0.118$ & $(4.23 \pm 0.02)$ GeV & $(4.62 \pm 0.02)$ GeV
& $(2.92 \pm 0.02)$ GeV \\
    & $0.124$ & $(4.19 \pm 0.02)$ GeV & $(4.62 \pm 0.02)$ GeV
& $(2.80 \pm 0.02)$ GeV \\[1mm]
\hline
\rule{0mm}{4.5mm}
$c$ & $0.112$ & $(1.25 \pm 0.03)$ GeV & $(1.42 \pm 0.03)$ GeV
& $(0.69 \pm 0.02)$ GeV \\
    & $0.118$ & $(1.23 \pm 0.03)$ GeV & $(1.42 \pm 0.03)$ GeV
& $(0.62 \pm 0.02)$ GeV \\
    & $0.124$ & $(1.19 \pm 0.03)$ GeV & $(1.42 \pm 0.03)$ GeV
& $(0.53 \pm 0.02)$ GeV \\[1mm]
\hline
\end{tabular}
\end{center}}
\vspace*{-5mm}
\end{table}

An additional mechanism for $b$, $c$ quark decays of the Higgs particle 
\cite{R11A} is provided by the gluon decay mechanism where virtual 
gluons split into $b\bar{b},\ c\bar{c}$ pairs, $H\ \to\ gg^{*}\ 
\to\ gb\bar{b},\
gc\bar{c}$. These contributions add to the QCD corrected partial widths
(\ref{partial-widths}) in which the $b$, $c$ quarks are coupled to the
Higgs boson directly. As 
long as quark masses are neglected in the final states, the two
amplitudes do not interfere. In this approximation, the contributions
of the splitting channels are obtained by taking the differences of
the widths $H\ \to\ gg(g), q\bar{q}g$ between $N_{F}=5$ and 4 for $b$,
and $N_{F}=4$ and 3 for $c$ final states, given below in
eq.(\ref{smh:gluonic-width}). The $b/\bar{b}$ and the $c/\bar{c}$
quarks are in general emitted into two different parts of the phase
space for the two mechanisms; for the direct process the flight
directions tend to be opposite, while by contrast for gluon splitting 
they are parallel.

The electroweak radiative corrections to fermionic Higgs decays
are well under control \cite{smh17,smh6}. If the Born formulae 
are parametrized in terms of the Fermi coupling $G_{F}$, the 
corrections are free of large logarithms associated with light 
fermion loops. For $b$, $c$, $\tau$ decays the electroweak corrections 
are of the order of one percent.

\noindent 
\underline{Higgs decays to gluons and light quarks.}
In the Standard Model, gluonic Higgs decays $H\to gg$ are primarily
mediated by top-quark loops \cite{smh18}.  Since in the LEP2 range
Higgs masses are much below the top threshold, the gluonic width can
be cast into the approximate form~\cite{smh19}
\begin{equation}
\label{smh:gluonic-width}
  \Gamma (H\ \to\ gg(g), q\bar{q}g) 
  = \frac{G_{F}\,\alpha_s^{2}(m_{H})}{36\sqrt{2}\,\pi^{3}}m_{H}^{3}
    \,\left[\,1+\left(\frac{95}{4}-\frac{7}{6}N_{F}\right)
    \frac{\alpha_s(m_{H})}{\pi}\right]
\end{equation}
The QCD corrections, which include the splitting of virtual gluons
into $gg$ and $q\bar{q}$ final states, are very important; they nearly
double the partial width.

It is physically meaningful to separate the gluon and light-quark
decays of the Higgs boson \cite{R11A} from the $b$, $c$ decays which
add to the $b$, $c$ decays through direct coupling to the Higgs boson.
In this case, the partial width $\Gamma (H\to \mbox{gluons} +
\mbox{light quarks})$ is obtained from (\ref{smh:gluonic-width}) by
choosing $N_{F}=3$ for the light $u$, $d$, $s$ quarks and by
evaluating the running QCD coupling at $m_{H}$ for three flavors only
[corresponding to $\Lambda_{\MSQCD}^{(3)} = 378^{+105}_{-92}$ MeV for
$\alpha_s^{(5)}(m_{Z})=0.118\pm0.006$].

\noindent 
\underline{Higgs decay to virtual W bosons.}  The
channel $H\to WW^{*}\to 4\ \mbox{fermions}$ becomes relevant for Higgs
masses $m_{H}\ >\ m_{W}$ when one of the W bosons can be produced
on-shell. The partial width for this final state is given by
\begin{equation}
  \Gamma (H\ \to\ WW^{*}) 
  = \frac{3\,G_{F}^{2}\,m_{W}^{4}} {16\,\pi^{3}}m_{H}\,R(x)
\end{equation}
\begin{displaymath}
  R(x)= \frac{3(1-8x+20x^2)}{(4x-1)^{1/2}}
  {\rm arccos}\left(\frac{3x-1}{2x^{3/2}}\right)
  - \frac{1-x}{2x}(2-13x+47x^2) - \frac32(1-6x+4x^2)\log x
\end{displaymath}
with $x=m_{W}^{2}/m_{H}^{2}$. Due to the larger $Z$ mass and the
reduced NC couplings compared with $W$ mass and the CC couplings,
respectively, decays to $ZZ^{*}$ final states are suppressed by one
order of magnitude.

\begin{figure}[tb]
\begin{center}
\begin{picture}(170,75)
\put(0,75){\epsfig{file=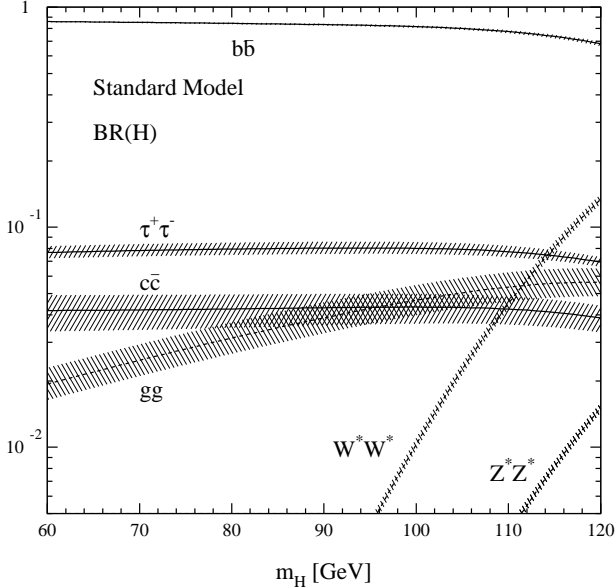,%
        bbllx=45pt, bblly=200pt, bburx=545pt, bbury=695pt,%
        height=8cm, angle=-90, clip=}}
\put(90,5){\parbox[b]{80mm}{%
\caption{\it Branching ratios for the Higgs decays in the Standard
  Model.  The bands include the uncertainties due to the errors in the
  quark masses and the QCD coupling.}
\label{smh:BR}}}
\end{picture}
\end{center}
\end{figure}

\noindent
\underline{Summary of the branching ratios.}
The numerical results for the branching ratios are displayed in
Fig.\ref{smh:BR}, taking into account all QCD and electroweak
corrections available so far. Separately shown are the branching
ratios for $\tau$'s, $c$, $b$ quarks, gluons plus light quarks, and
electroweak gauge bosons. The analyses have been performed for the
following set of parameters: $\alpha_s^{(5)}(m_{Z})=0.118\pm 0.006$,
$t$ pole mass $M_{t}=176\pm 11\ {\rm GeV}$, and the $\MSQCD$ masses
$m_b(M_b)$ and $m_c(M_c)$ as listed in
Table~\ref{smh:masses}.  The dominant error in the predictions is due
to the uncertainty in $\alpha_s$ which migrates to the running quark
masses at the high energy scales.

Despite the uncertainties, the hierarchy of the Higgs decay modes is
clearly preserved. The $\tau^+\tau^-$ branching ratio is more than an
order of magnitude smaller than the $b\bar b$ branching ratio,
following from the ratio of the two masses squared and the color
factor. Since the charm quark mass is small at the scale of the Higgs
mass, the ratio of $BR_{c}$ to $BR_{b}$ is reduced to about $0.04$,
i.e.\ more than would have been expected na\"{\i}vely.

Thus, the measurements of the production cross sections and of the
decay branching ratios enable us to explore experimentally the
physical nature of the Higgs boson and the origin of mass through the
Higgs mechanism.

\vfill\newpage

\subsection{The Experimental Search for the SM Higgs Particle}
\label{sec:exp}

Selection algorithms were developed by the four LEP experiments \cite{PJ} 
towards the Higgs production {\it via} the Higgs-strahlung process,
for the following event topologies:
\begin{itemize}
\item[{\it (i)}]   the four-jet channel, $(\Z \to \qqbar)$
$(\HSM \to \bbbar)$;
\item[{\it (ii)}]  the missing energy channel, $(\Z \to \nnbar)$
$(\HSM \to \bbbar)$;
\item[{\it (iii)}] the leptonic channel, $(\Z \to \epem, \mpmm)$
$(\HSM \to$ anything$)$;
\item[{\it (iv)}]  the \tptm\qqbar\ channel,
$(\Z \to \tptm)$ $(\HSM \to$ hadrons$)$ and vice-versa;
\end{itemize}
altogether amounting to more than 90\% of the possible final states
in the \LEPII\ mass range.

All important background processes were included in the simulations.
Whenever possible, the corresponding cross-sections were computed
and events were generated using PYTHIA 5.7~\cite{pythia}. 
The \Z\nnbar\ process being not simulated in PYTHIA, the corresponding results 
were derived from a Monte Carlo generator based on Ref.\cite{mele}. The most 
relevant cross-sections are indicated in Table~\ref{table:cross} for
the three different center-of-mass energies at which the studies were
carried out.
Events from the Higgs-strahlung process were generated using either
PYTHIA (DELPHI, L3, OPAL), the HZGEN generator~\cite{hzgen} (DELPHI,
for the $\H\Z\ \to\ \bbbar\nu_e\bar\nu_e$ final state) or the
HZHA generator~\cite{hzha} (ALEPH, for all signal
final states), and the signal cross-section and Higgs boson decay
branching ratios were determined from Ref.\cite{smh8}, or directly from
the HZHA program in the case of ALEPH.

\begin{table}[htbp]
\caption{\it
  The cross-sections for the most relevant background processes, in
  pb.  Whenever a Z is indicated, the cross-section also includes the
  $\gamma^\ast$ contribution. The $\gamma\gamma \to \ffbar$
  cross-section is given for a fermion pair mass in excess of
  30~\Gcs.}
\label{table:cross}
\begin{center}
\begin{tabular}{|l||c|c|c|} \hline\hline
          & 175 GeV & 192 GeV & 205 GeV  \\ \hline
\epemto\ \ffbar    &  173.4  &  135.5  & 116.5    \\
\epemto\ \W\W      &  14.63  &   17.74 &  18.07   \\
\epemto\ \Z\Z      &  0.45   &    1.20 &   1.43   \\
\epemto\ \Z\epem   &  2.75   &    2.93 &   3.05   \\
\epemto\ \W e$\nu$ &  0.68   &    0.90 &   1.10   \\
\epemto\ \Z\nnbar  &  0.011  &   0.015 &   0.020  \\
$\gamma\gamma \to \ffbar$
          & 22.3    &   24.9  &  26.3    \\
\hline\hline
\end{tabular}
\end{center}
\end{table}

The selection efficiencies and  the background rejection capabilities
were evaluated after a simulation of each of the four LEP detectors.
Fully simulated events were produced by DELPHI \cite{delphione,delphinote}, 
L3 \cite{l3note} and OPAL \cite{opalnote} for all the background 
processes and for the signal at several Higgs boson masses,
including all the detector upgrades foreseen for the \LEPII\ running. A fast,
but reasonably detailed simulation was used in ALEPH \cite{moriond} instead, 
with the current detector design (in particular, the gain expected from
the installation of a new vertex detector was conservatively ignored),
but it was checked in the four-jet topology and in the missing energy
channel, at $\sqrt{s} = 175$~GeV and with $\mH = 70$~GeV, that this
fast simulation faithfully reproduces the predictions of the full
simulation chain both for the background rejection and for the signal
selection, up to an accuracy at the percent level.
\vvs1

\noindent {\bf a) Search in the Four-jet Topology}
\pss{0.5}
\begin{picture}(158,70)
\put(0,65){\parbox[t]{105mm}{
The four-jet topology arises when the Z decays into a pair of quarks,
in 70\% of the cases, and the Higgs boson decays into hadrons, in more
than 90\% of the cases. This topology represents therefore by far the most
abundant final state (occurring in $\sim 65\%$ of the cases) produced by the
Higgs-strahlung process. However, the search in this channel is affected
by a large background consisting of multijet events from \epemto\ \qqbar,
\W\W\ and \Z\Z\ production. For instance, at $\sqrt{s}=192$~GeV, and
for an integrated luminosity of 150~\inpb, approximately 1500~\qqbar, 
1000~\W\W\ and 80~\Z\Z\ events have at least four jets with all jet-jet 
invariant masses in excess of 10~\Gcs, while only 40 \HSM\Z\ events are 
expected if $\mH = 90$~\Gcs.}}
\put(110,6){\epsfxsize60mm\epsfbox{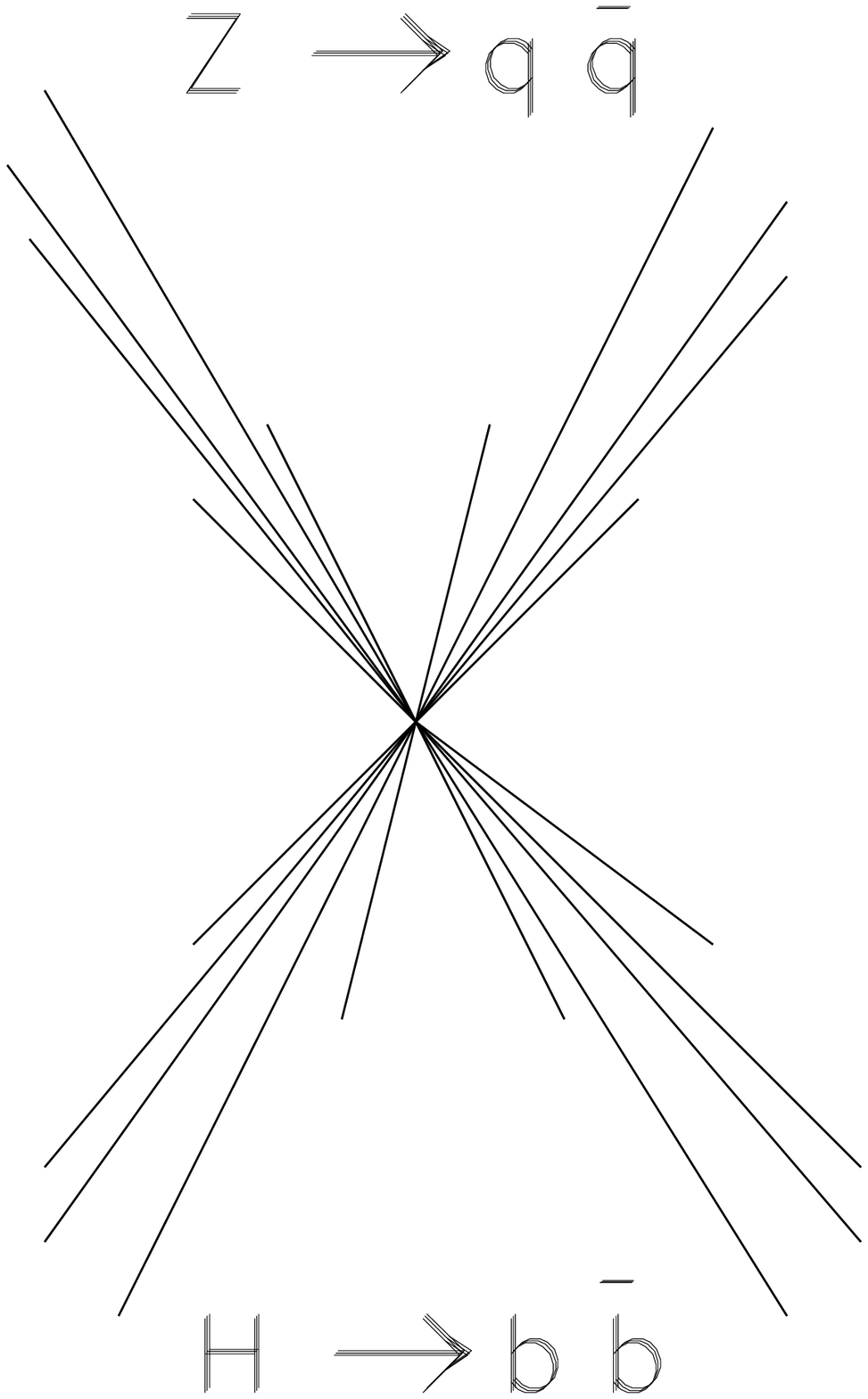}}
\end{picture}

The selection procedures developed by the four collaborations to improve the
signal-to-noise ratio are very close to each other. After a preselection
aimed at selecting four-jets events, either from global events properties
or directly from a jet algorithm such as the DURHAM or JADE algorithms,
the four-jet energies and momenta are subjected to a kinematical fit with the
four constraints resulting from  the energy-momentum conservation, in order
to improve  the Higgs boson mass resolution beyond the detector resolution.
Events consistent with the \epemto\ \W\W\ hypothesis, {\it i.e.}
events in which two pairs of jets have an invariant mass close to
\mW,  are rejected. Only events in which the mass of one pair of jets  is
consistent with \mZ\ are kept, and they are fitted again with 
the Z mass constraint in addition. This last step improves again the 
Higgs boson mass resolution, which is found to be  between 2.5 and 
3.5~\Gcs\ by the four LEP experiments.

\begin{figure}[htbp]
\begin{picture}(120,77)
\put(66,35){(a)}
\put(93,35){(b)}
\put(0,-5){\epsfxsize83mm\epsfbox{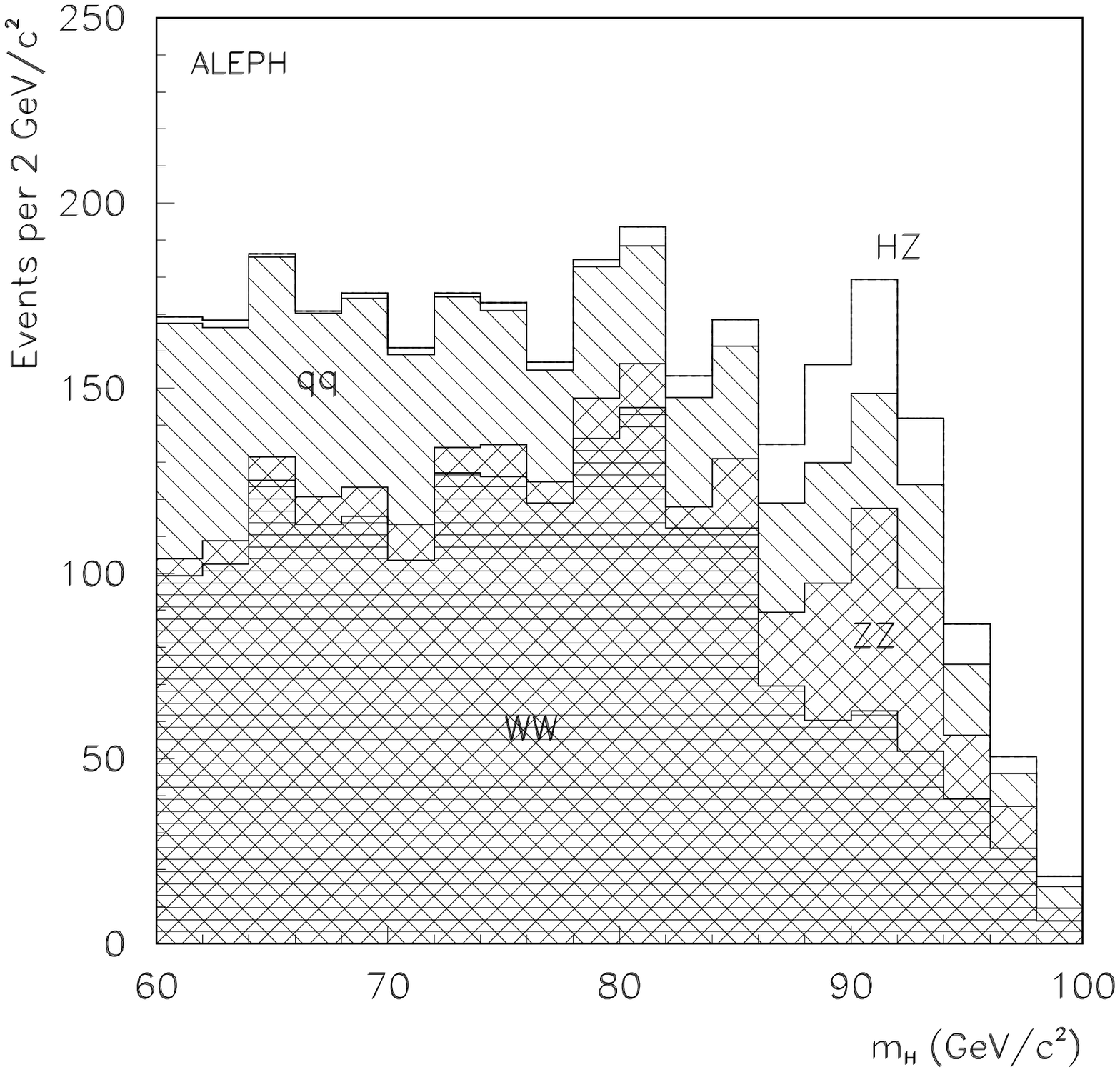}}
\put(80,-5){\epsfxsize83mm\epsfbox{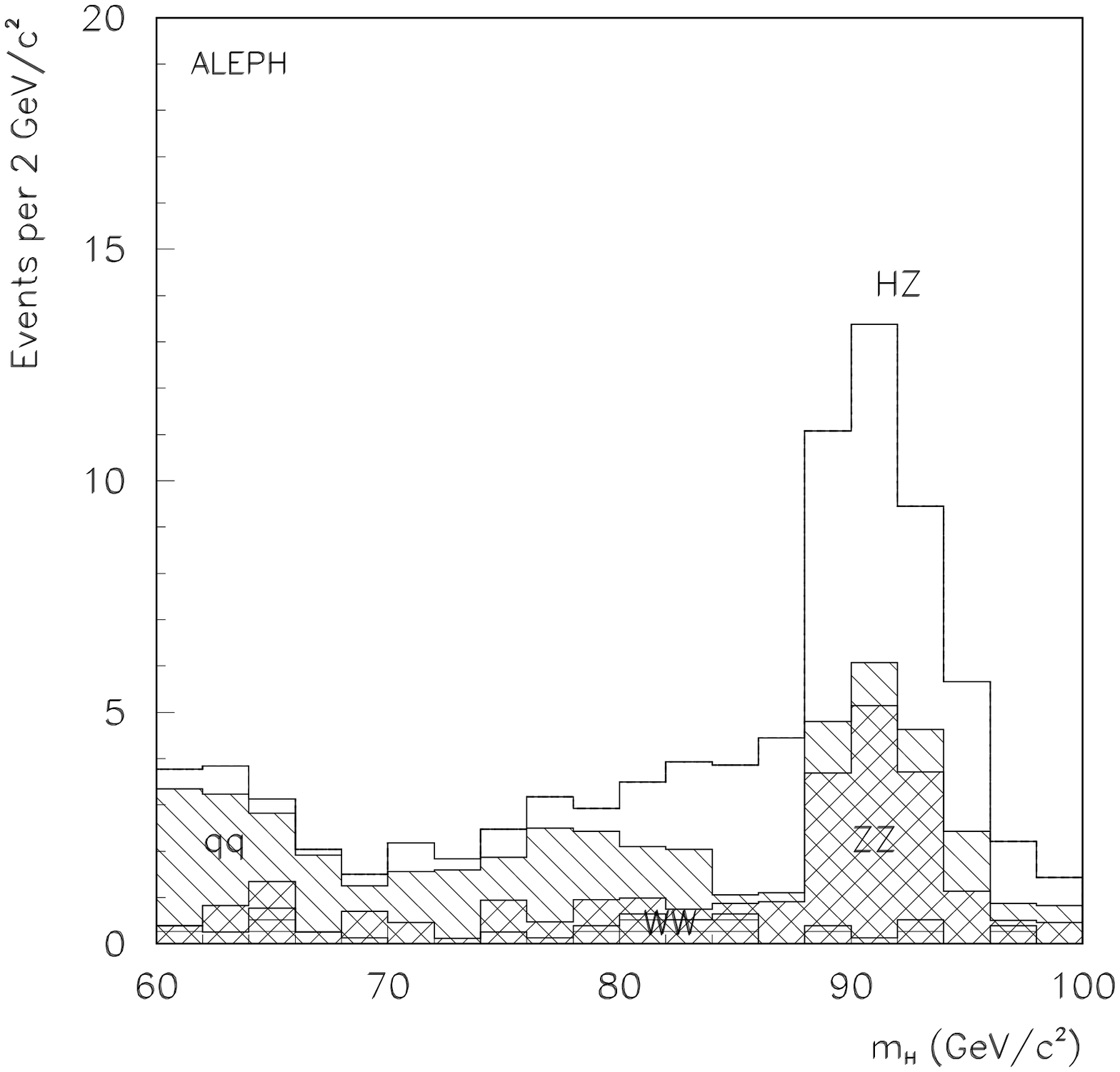}}
\end{picture}
\caption{\it Distribution of the fitted Higgs boson mass as
  obtained from the ALEPH simulation, in the four-jet topology before
  (a) and after (b) a b-tagging requirement is applied, at 192~GeV,
  with 500~\inpb\ and for $\mH = 90$~\Gcs.}
\label{fig:hqq}
\end{figure}

However, these requirements do not suffice to reduce the background
contamination to an adequate level. This is illustrated in Fig.\ref{fig:hqq}a
where the distribution of the fitted Higgs boson mass ({\it i.e.} the mass
of the pair of jets recoiling against the pair consistent with a Z) is shown,
for the signal ($\mH = 90$~\Gcs) and for the backgrounds, at $\sqrt{s} =
192$~GeV and for a luminosity of 500~\inpb\, as obtained from the ALEPH
simulation at this level of the analysis.

\begin{figure}[htbp]
\begin{picture}(120,77)
\put(66,35){(a)}
\put(93,35){(b)}
\put(0,-5){\epsfxsize83mm\epsfbox{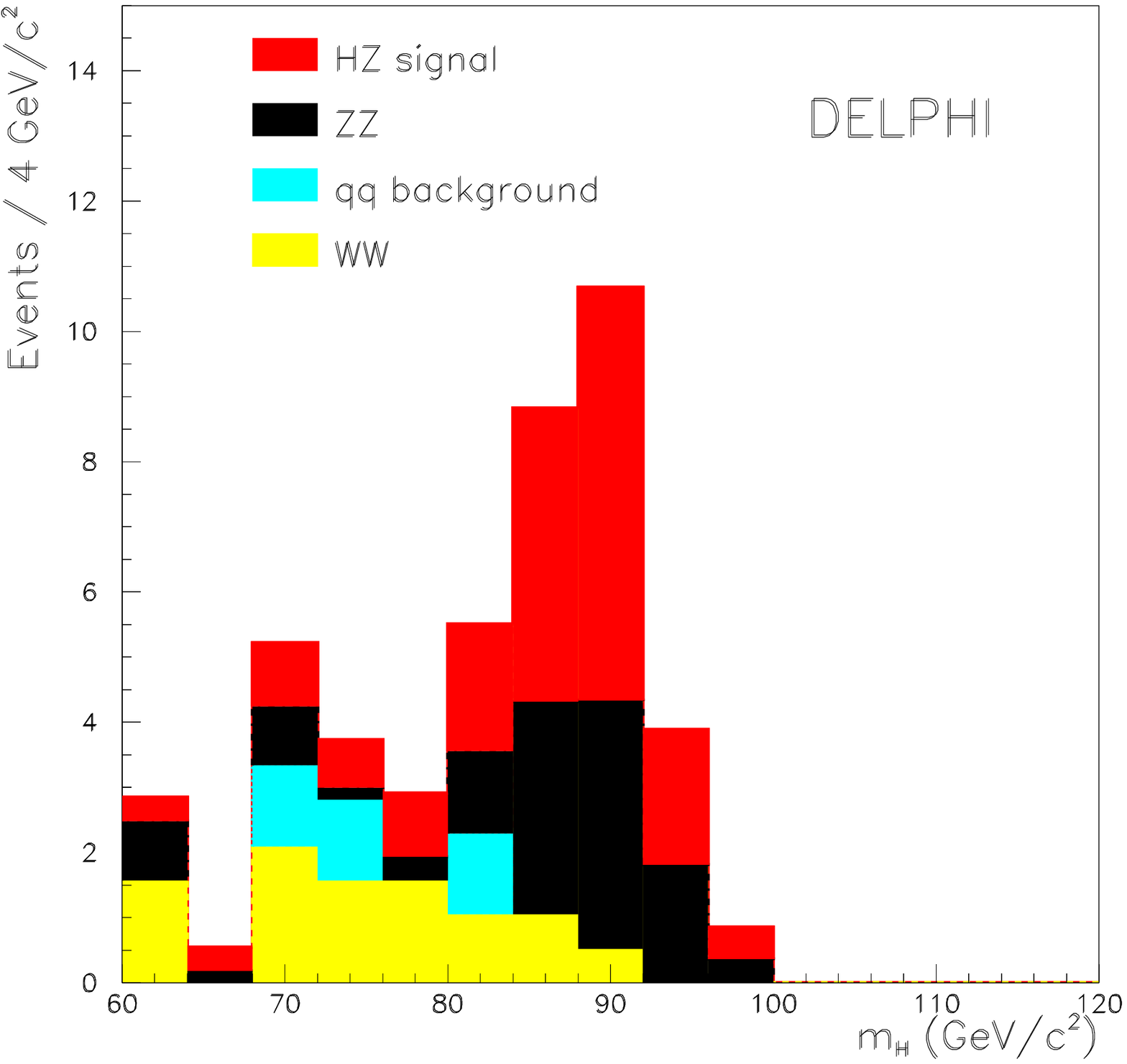}}
\put(80,-5){\epsfxsize83mm\epsfbox{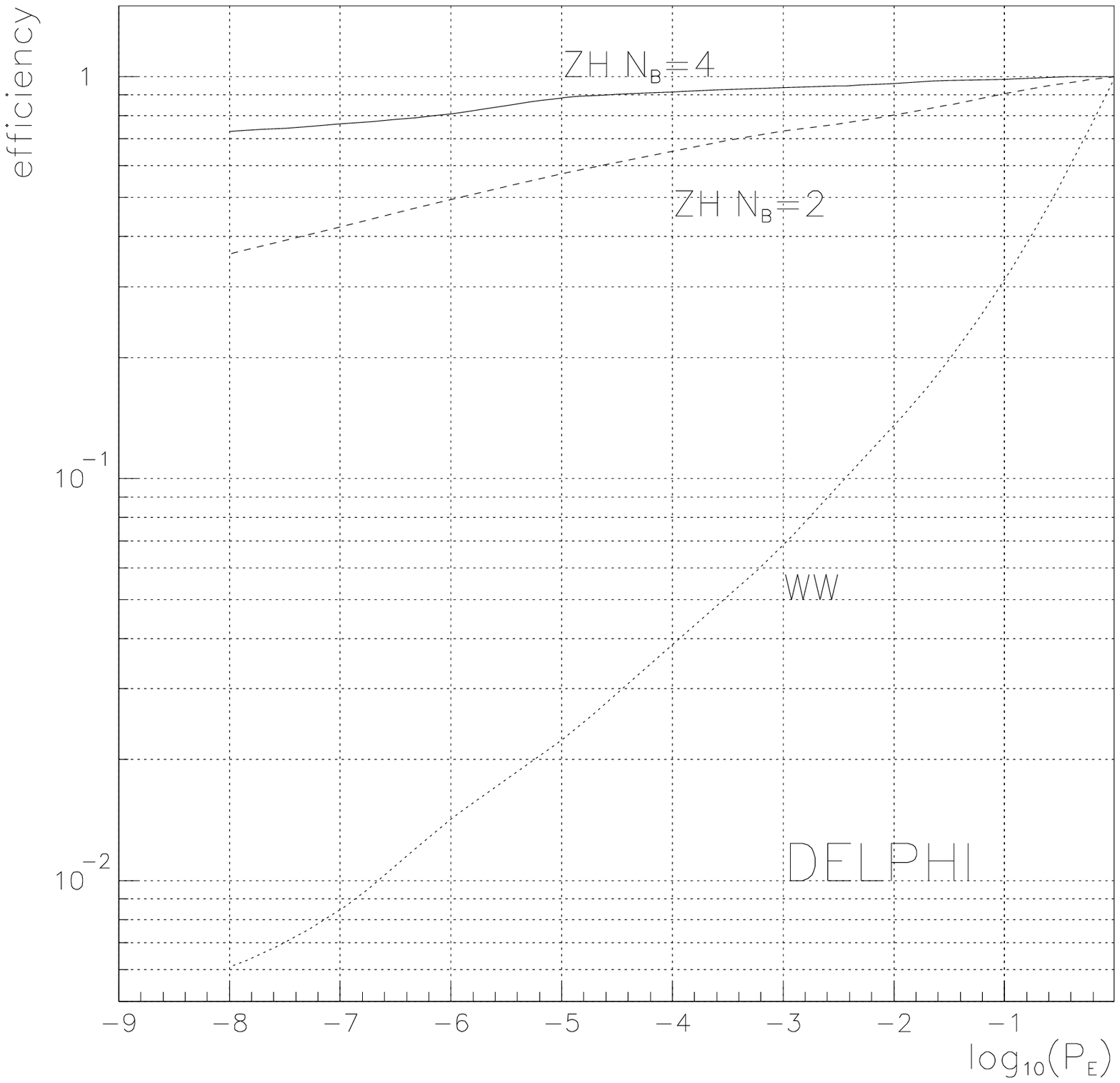}}
\end{picture}
\caption{\it 
  (a) Distribution of the fitted Higgs boson mass as obtained from the
  DELPHI experiment, after a b-tagging requirement is applied, at
  192~GeV, with 300~\inpb\ and for $\mH = 90$~\Gcs, and (b) Evolution
  of the b-tagging efficiency as a function of the cut on $\cal P$
  when applied to four jet events, with four, two or zero b-jets.}
\label{fig:delhqq}
\end{figure}

\begin{figure}[htbp]
\begin{picture}(120,77)
\put(66,50){(a)}
\put(103,50){(b)}
\put(0,-10){\epsfxsize83mm\epsfbox{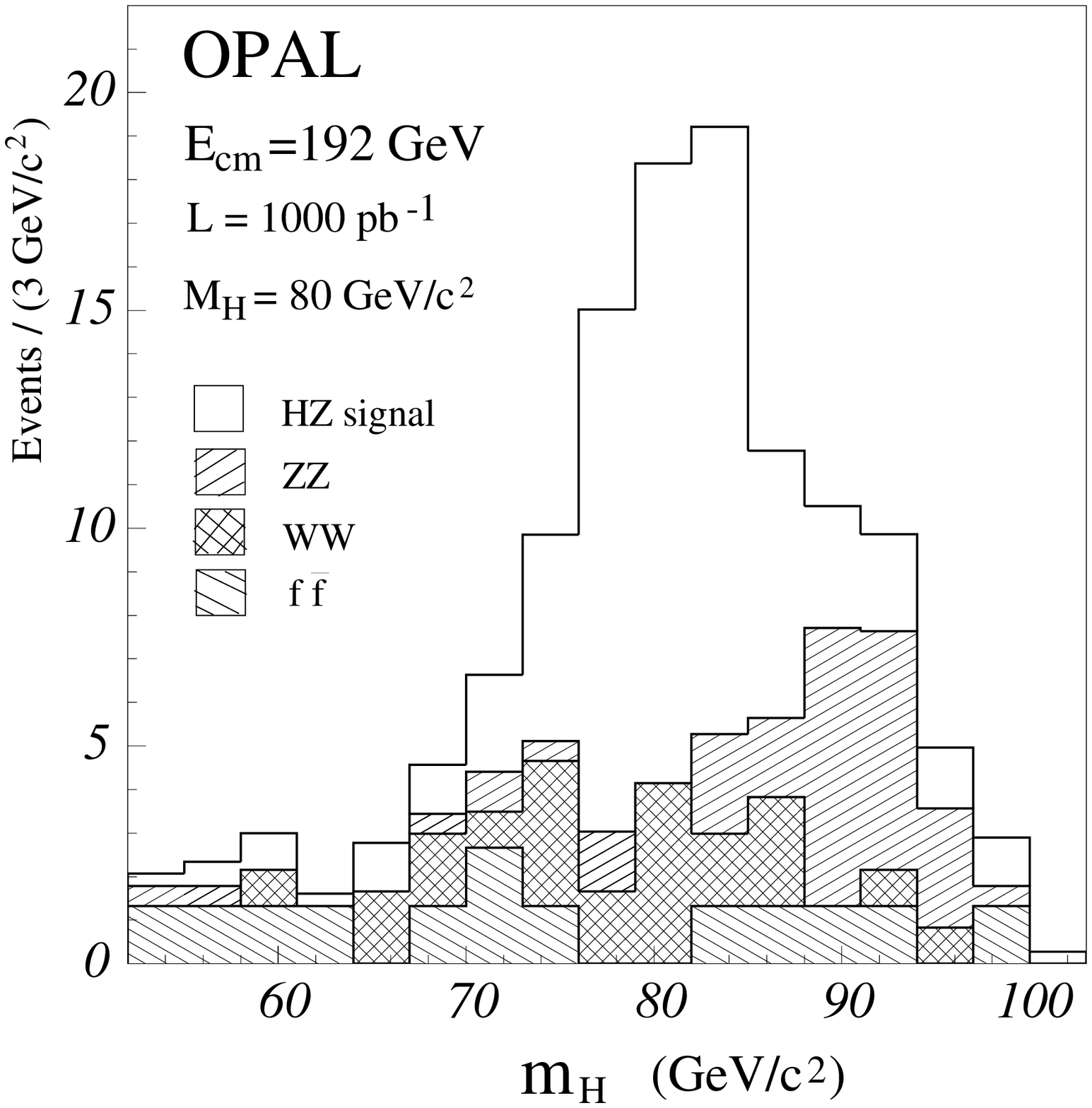}}
\put(80,0){\epsfxsize93mm\epsfbox{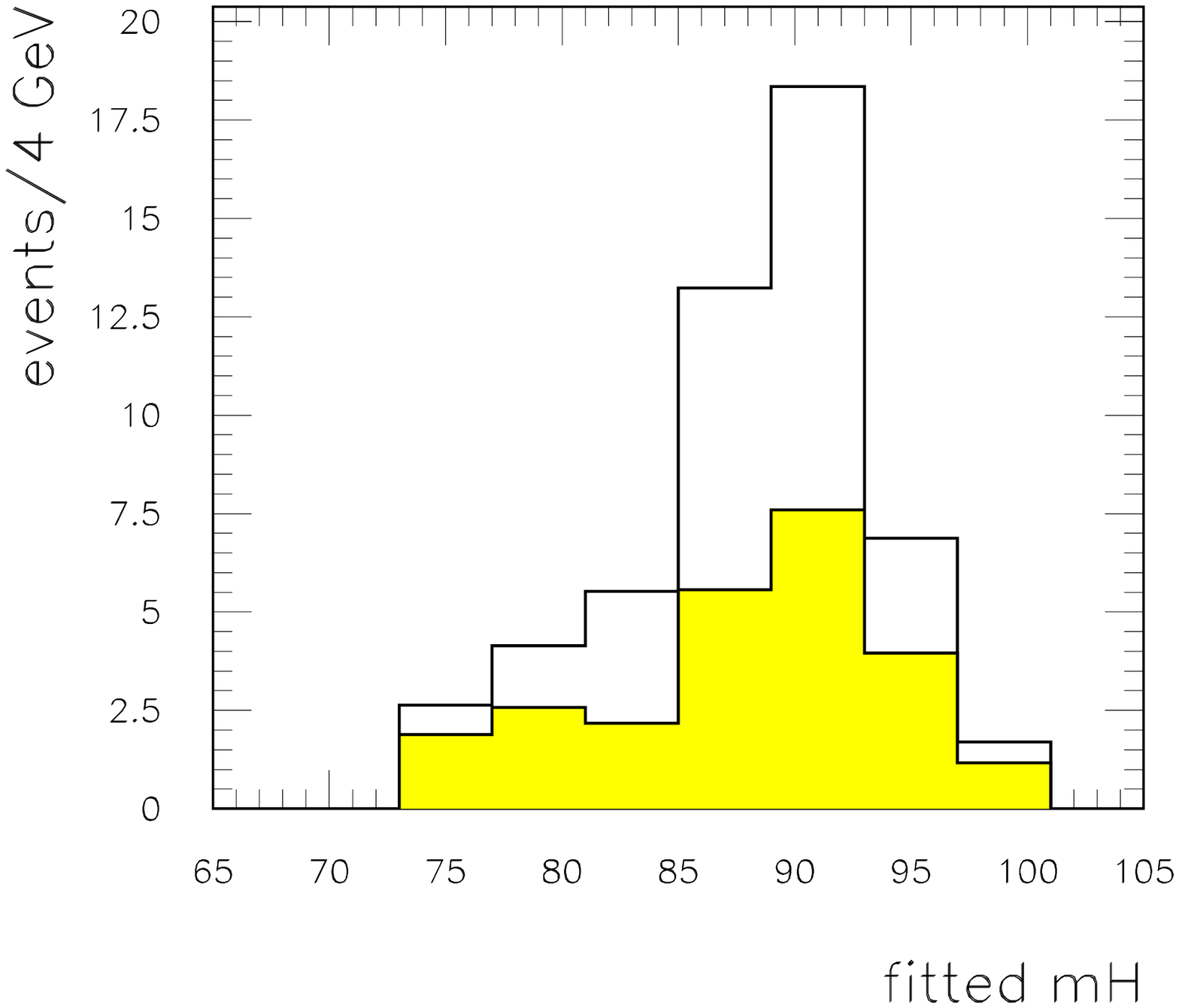}}
\end{picture}
\caption{\it 
  Distribution of the fitted Higgs boson mass as obtained from the
  OPAL experiment, in the four-jet channel (a) and in the missing
  energy channel (b), at a centre-of mass energy of 192~GeV,
  normalized to a luminosity of 1000~\inpb\ and for $\mH = 80$ and
  90~\Gcs, respectively. The signal (in white) is shown on top of the
  background (shaded histogram).}
\label{fig:ophnn}
\end{figure}

The high branching of the Higgs boson into \bbbar\ must then be taken
advantage of to further reduce the background, by requiring that the
jets associated to the Higgs boson be identified as b-jets. This is
done by means of a microvertex detector, either by counting
the charged particle tracks with large impact parameters, or by evaluating
the probability $\cal P$ that these tracks come from the main interaction
point~\cite{brown}, or by directly reconstructing secondary decay
vertices~\cite{mattison}. Shown in Fig.\ref{fig:hqq}b is the resulting
Higgs boson mass distribution after such a b-tagging requirement is applied.
The same distribution as seen by DELPHI is shown in Fig.\ref{fig:delhqq}, 
together with the efficiency of the DELPHI lifetime b-tagging requirement
applied to four-jet events, in which four b-jets, two b-jets or 
no b-jets are present, as a function of the logarithm of the probability
$\cal P$. The OPAL result in this topology is shown in Fig.\ref{fig:ophnn}a.
Due to the recent vertex detector installation, the L3 b-tagging 
algorithm is not yet fully optimized and its performance is thus expected to 
improve in the future.

\vskip -.5cm

\vskip -.5cm
\begin{table}[htbp]
\setlength{\tabcolsep}{1.1pc}
\caption{\it 
  Accepted cross-sections (in fb) for the signal and the backgrounds,
  as expected by ALEPH, DELPHI, L3, OPAL, for $\mH = 90~\Gcs$ at 192
  GeV, in the four-jet topology.}
\label{tab:hqq}
\begin{center}
\begin{tabular}{rrrrr}
\hline
\hline
 Experiment & ALEPH  & DELPHI  &   L3   &   OPAL   \\
\hline
 Signal     &   58   &    43   &  43    &   46     \\
 Background &   33   &    33   &  47    &   26     \\
\hline
\hline
\end{tabular}
\end{center}
\end{table}

\vskip -.0cm
The numbers of background and signal events expected to be selected by
ALEPH, DELPHI, L3, and OPAL in a window of $\pm 2 \sigma$ around the
reconstructed Higgs boson mass are shown in Table~\ref{tab:hqq} for a Higgs
boson mass of 90~\Gcs\ and at a center-of mass energy of 192~GeV. 
\vvs1

\noindent {\bf b) Search in the Missing Energy Channel}\pss{0.5}
\begin{picture}(158,60)
\put(56,55){\parbox[t]{114mm}{
The topology of interest here, arising in 18\% of the cases,
is an acoplanar pair of b-quark jets with mass \mH, accompanied by large
missing energy and large missing mass, close to the Z mass. The
background, with the exception of the $\Z\Z\to\bbbar\nnbar$
or the $\Z\nnbar$ with $\Z\to\bbbar$ processes, either has no missing energy
($\epemto\ \qqbar$ with no initial state radiation, $\W\W, \Z\Z \to$
four-jets), or no missing mass and isolated particles
($\epemto\ \qqbar(\gamma)$, $\W\W \to \ell\nu + $ two jets, $\Z\epem$),
or no missing transverse momentum and small acoplanarity angle
($\epemto\ \qqbar(\gamma\gamma)$, $\gamma\gamma \to \qqbar$), or light
quark jets ($\epemto\ (\e)\nu W$, $\W\W\to\tau\nu + $ two jets,
$\Z\Z\to\qqbar\nnbar$).}}
\put(0,5){\epsfxsize50mm\epsfbox{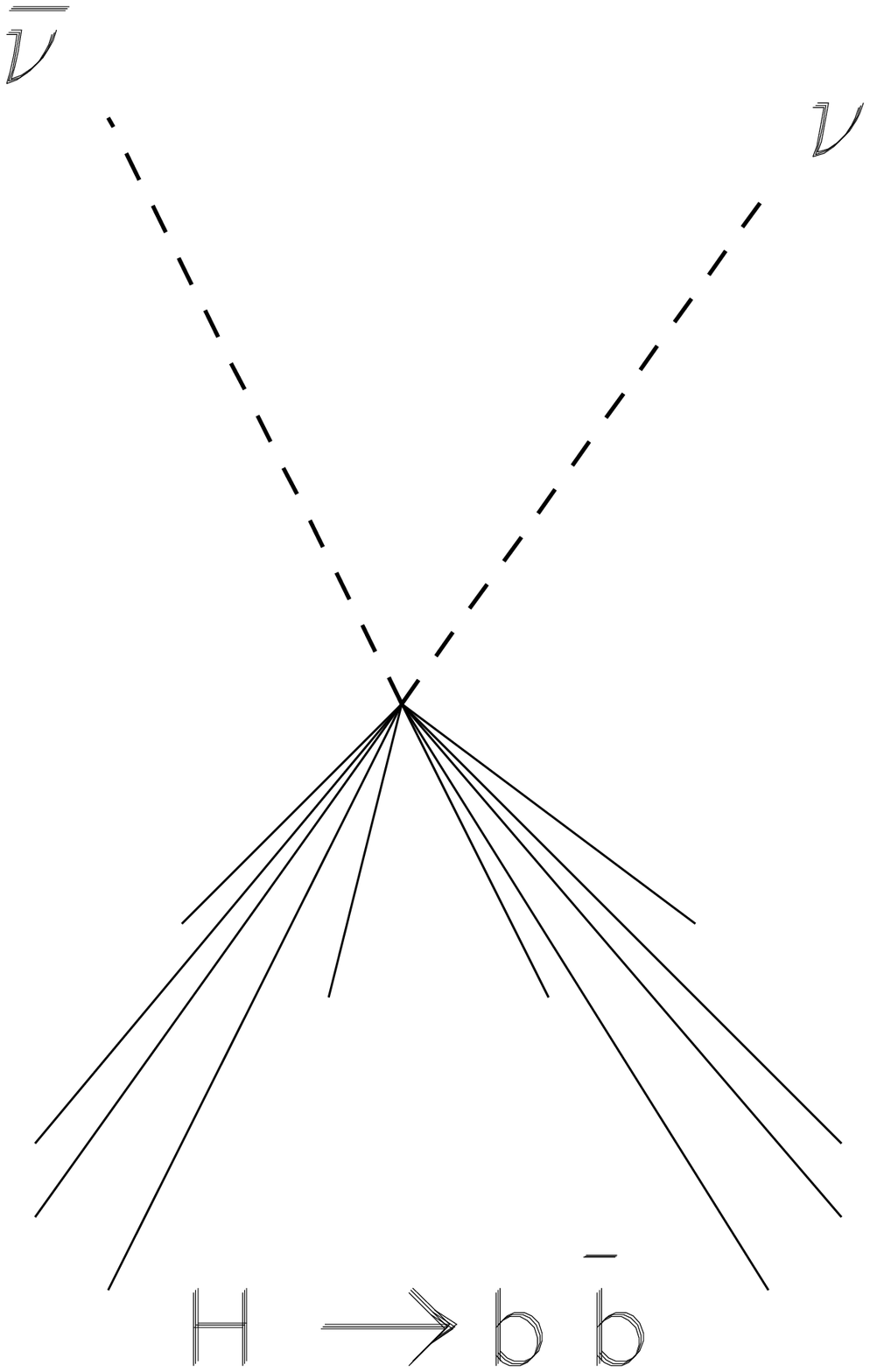}}
\end{picture}

The four collaborations developed a selection procedure with a sequence
of criteria, based on these differences between signal and background,
including a b-tagging requirement. The mass of the Higgs boson can be
either rescaled or fitted by constraining the missing mass to equal
the Z mass, allowing mass resolutions from 3.5 to 5~\Gcs\ to be achieved.
The mass distribution obtained by OPAL in this channel,
for a Higgs boson mass of 90~\Gcs\ and at a center-of mass energy of 192~GeV,
is shown in Fig.\ref{fig:ophnn}b.

This selection procedure was supplemented in DELPHI by an alternative
multi-variate probabilistic method, confirming (or slightly
improving) the first analysis results. The contribution
of the $t$-channel \W\W\ fusion to the \H\nnbar\ final topology
was also estimated by DELPHI with the recently released HZGEN
event generator which includes both the Higgs-strahlung and the
\W\W\ fusion diagrams together with their interference. As can be
naively expected, the relative gain is only sizeable above the \H\Z\
kinematical threshold, and amounts to 28\% for a 100~\Gcs\ Higgs boson at
192~GeV, corresponding to 0.25 additional events expected for an
integrated luminosity of 300~\inpb.

\begin{table}[htbp]
\setlength{\tabcolsep}{1.1pc}
\caption{\it
  Accepted cross-sections (in fb) for the signal and the backgrounds,
  as expected by ALEPH, DELPHI, L3, OPAL, for $\mH = 90~\Gcs$ at 192
  GeV, in the missing energy channel.}
\label{tab:hnn}
\begin{center}
\begin{tabular}{rrrrr}
\hline
\hline
 Experiment & ALEPH  & DELPHI  &   L3   &   OPAL   \\
\hline
 Signal     &   24   &    24   &  9     &   25     \\
 Background &   13   &    17   &  11    &   20     \\
\hline
\hline
\end{tabular}
\end{center}
\end{table}

The numbers of background and signal events expected to be selected by
ALEPH, DELPHI, L3, and OPAL in a window of $\pm 2 \sigma$ around the
reconstructed Higgs boson mass are shown in Table~\ref{tab:hnn} for a Higgs
boson mass of 90~\Gcs\ and at a center-of mass energy of 192~GeV.
\vvs1

\noindent {\bf c) Search in the Leptonic Channel}\pss{0.5}
\begin{picture}(158,65)
\put(0,60){\parbox[t]{105mm}{
Although occurring in only 6.7\% of the cases, this topology can be
selected in a simple way by requiring the presence of a high mass
pair of energetic, isolated, and thus well identifiable leptons (e or
$\mu$) in association with a high multiplicity hadronic system.
The process \epemto\ \Z\Z\ where one of the Z bosons decays into
a lepton pair and the other into \qqbar\ and, to a much lesser extent,
the \epemto\ \Z\epem\ process, constitute the only irreducible background
sources. A mild b-tagging requirement can also be applied, especially
when $\mH\sim\mZ$, to improve the signal-to-noise ratio. Selection
efficiencies varying from 50 to 80\% were achieved by the four LEP
experiments.
}}
\put(110,6){\epsfxsize60mm\epsfbox{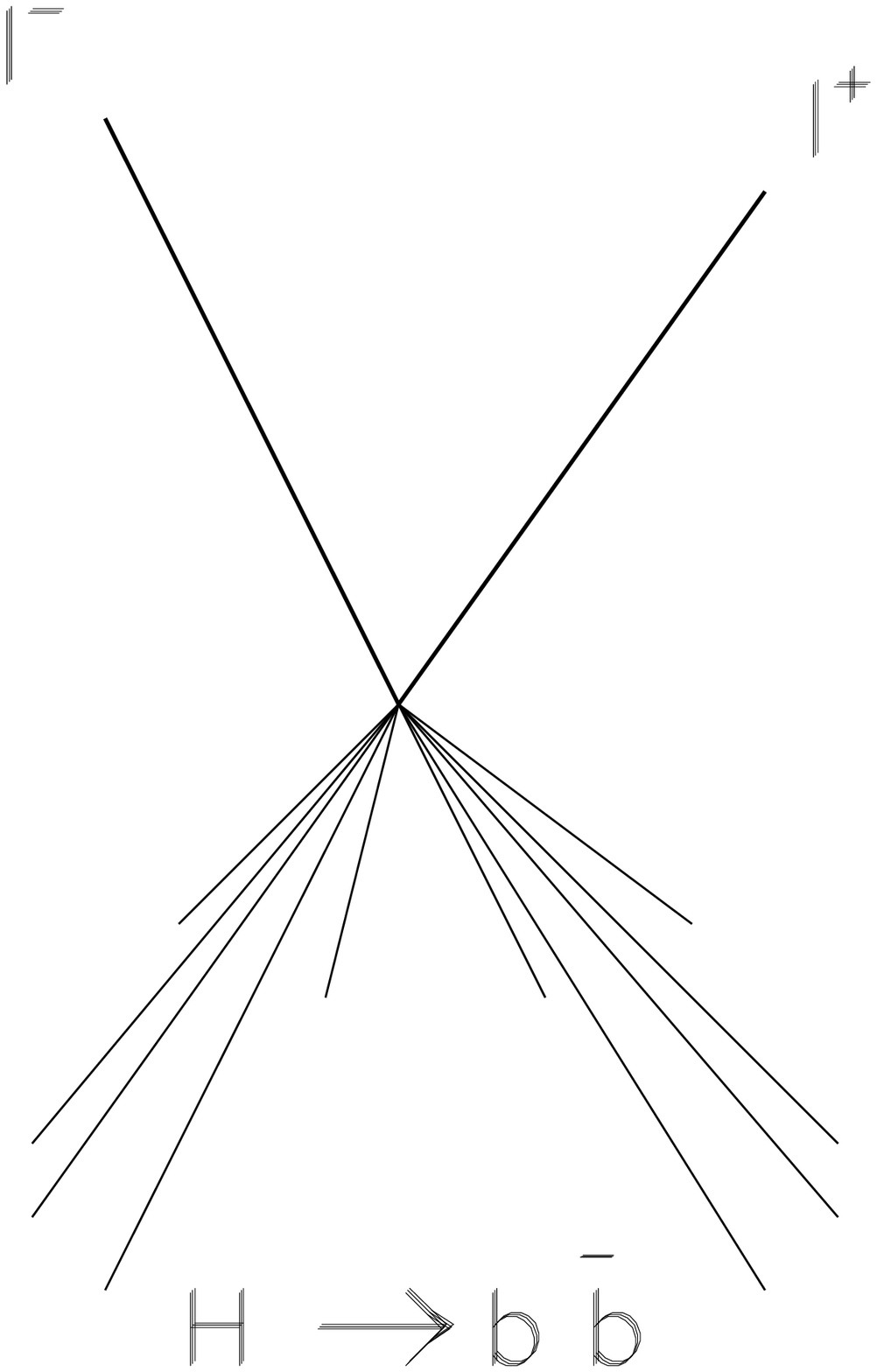}}
\end{picture}

\begin{figure}[htbp]
\begin{picture}(120,120)
\put(20,-5){\epsfxsize120mm\epsfbox{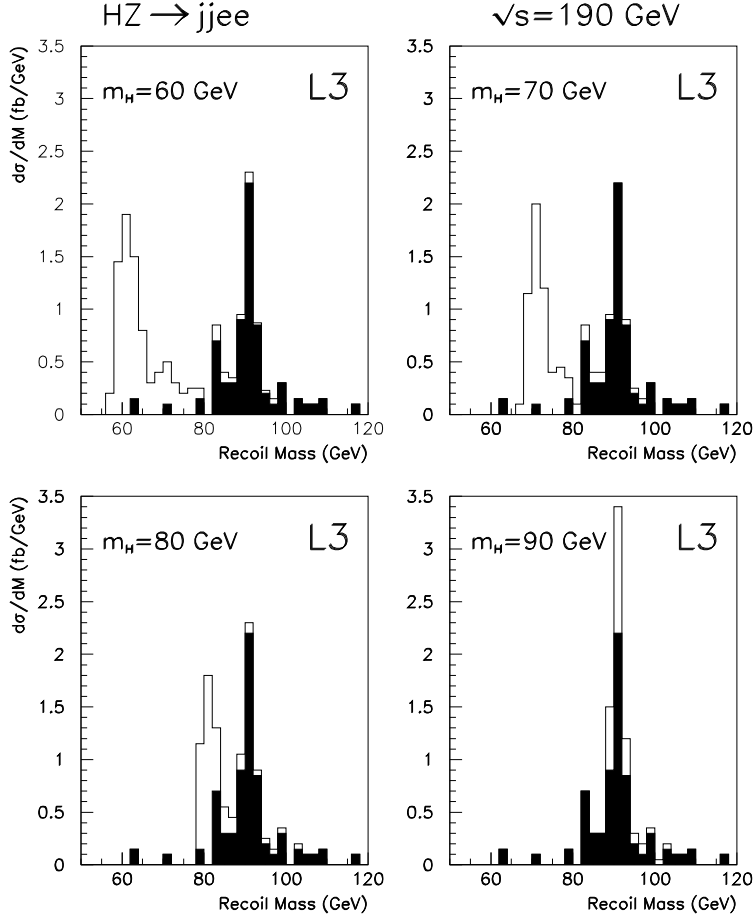}}
\end{picture}
\caption{\it 
  Distribution of mass recoiling to the lepton pair as obtained from
  the L3 experiment, in the \H\epem\ channel, at a center-of mass
  energy of 192~GeV, normalized to a luminosity of 1000~\inpb\ and for
  $\mH = 60, 70, 80, 90$~\Gcs. The signal (in white) is shown on top
  of the \Z\Z\ background (in black).}
\label{fig:l3hll}
\end{figure}

In addition to these high efficiencies, the mass of the Higgs boson
can be determined with a very good resolution (typically better than 2~\Gcs)
either as the mass recoiling to the lepton pair with the mass of the 
pair constrained to the Z mass, or with  a full fitting procedure using 
the energies and the directions of the
leptons and of the Higgs decay products, the energy-momentum conservation and
the Z mass constraint. As shown in Fig.\ref{fig:l3hll} from L3, this
drastically reduces the \Z\Z\ background contamination, except if 
$\mH\sim\mZ$ when the two mass peaks merge together.

The numbers of background and signal events expected to be selected by
ALEPH, DELPHI, L3, and OPAL in a window of $\pm 2 \sigma$ around the
reconstructed Higgs boson mass are shown in Table~\ref{tab:hll} for a Higgs
boson mass of 90~\Gcs\ and at a centre-of mass energy of 192~GeV.

\vvs1

\noindent {\bf d) Search in the \tptm\qqbar\ Channel}\pss{0.5}
\begin{picture}(158,60)
\put(56,55){\parbox[t]{114mm}{
At present, only ALEPH \cite{moriond} and DELPHI \cite{lutz} have 
investigated this topology, occurring in 9\% of the cases when
$(\Z \to \tptm)$ $(\HSM \to$ hadrons$)$ (3\%) or when
$(\HSM \to \tptm)$ $(\Z \to$ hadrons$)$ (6\%). It is characterized by
two energetic, isolated {\it taus}, defined as 1- or 3-prong slim jets,
with masses compatible with $m_\tau$, not identified as an electron or a
muon pair, and associated to a high multiplicity hadronic system.
After a selection of this topology either by successive topological cuts
(ALEPH) or by a single multi-dimensional cut (DELPHI), a fit
to the four-body final state hypothesis with the energy-momentum
conservation constraint is performed  to reject most of the backgrounds.
}}
\put(0,5){\epsfxsize50mm\epsfbox{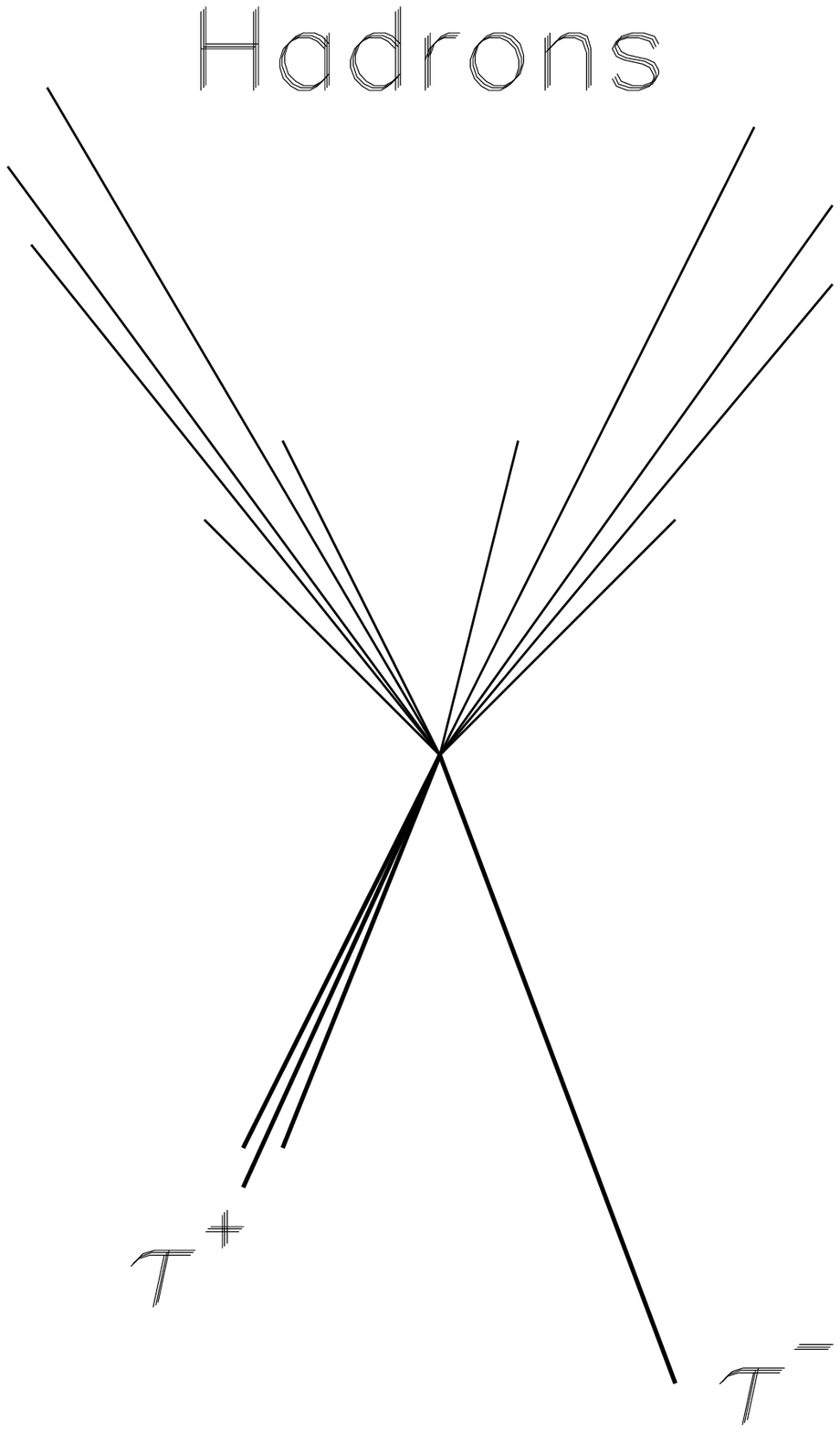}}
\end{picture}

\begin{table}[htbp]
\setlength{\tabcolsep}{1.1pc}
\caption{\it 
  Accepted cross-sections (in fb) for the signal and the backgrounds,
  as expected by ALEPH, DELPHI, L3 and OPAL, for $\mH = 90~\Gcs$ at 192 GeV,
  in the leptonic channel. }
\label{tab:hll}
\begin{center}
\begin{tabular}{rrrrr}
\hline
\hline
 Experiment & ALEPH  & DELPHI  &   L3   &   OPAL   \\
\hline
 Signal     &   12   &   11   &  7     &   6.5     \\
 Background &   12   &   24   &  10    &   9.4      \\
\hline
\hline
\end{tabular}
\end{center}
\end{table}

The typical efficiency for such an analysis is 20 to 30\%, corresponding
to 6 to 8 signal events expected for a 90~\Gcs\ Higgs boson with 1~\infb\
at 192 GeV, and the \tptm\ and the hadronic mass
resolutions amount to approximately 3~\Gcs. These resolutions can be further
improved by fitting the final state to the \H\Z\ hypothesis, with \mH\ free
and \mZ\ constrained. As in the leptonic channel, the only really irreducible
background source is the process $\epemto\ \Z\Z$ when one of the Zs decays into a
$\tau$ pair and the other hadronically. The existence of the Higgs boson would
then be observed as an accumulation around (\mH, \mZ) in the folded
two-dimensional distribution of these masses. A signal-to-noise ratio between
1 and 2 can be achieved when $\mH \sim \mZ$.
It could be further improved by a factor of two with a b-tagging
requirement, at the expense of a drastic efficiency loss, since two
thirds of these events (when $\H\to\tptm$) do not contain b-quarks.
\vvs1

\noindent {\bf Summary: Numbers of Events Expected}\pss{0.5}
Tables~\ref{tab:175}, \ref{tab:192} and \ref{tab:205} summarize the 
results of the standard model Higgs boson search, with 
the total numbers of signal and background events expected by
each experiment given for several Higgs boson masses, at $\sqrt{s} =
175$, 192 and 205~GeV, respectively. The uncertainties are due to
the limited Monte Carlo statistics. No systematic uncertainties (due for
instance to the simulation of the b-tagging efficiency) are included.

\begin{table}[htbp]
\setlength{\tabcolsep}{0.82pc}
\caption{\it 
  Accepted cross-sections (in fb) expected for the signal and the
  background, for various Higgs boson masses, at a center-of-mass
  energy of 175~GeV.}
\label{tab:175}
\begin{center}
\begin{tabular}{lrrrrrr} \hline\hline
\mH~(\Gcs)  &  $60$  & $65$ & $70$ & $75$ & $80$ & $85$  \\
 \hline
ALEPH & & & & & & \\
Signal   & $275\pm5$ & $234\pm4$ & $168\pm4$ & $115\pm3$ & $61\pm2$   &
 $7\pm1$               \\
Background & $51\pm7$ & $45\pm7$ & $38\pm6$ & $31\pm6$ & $24\pm5$  &
   $24\pm5$             \\       \hline

DELPHI & & & & & & \\
Signal &  $210\pm13$  & $180\pm11$ & $147\pm9$ &
 $109\pm7$ & $64\pm4$ & $ 7\pm1$
                \\
Background &$25\pm4$  & $25\pm4$ & $28\pm5$ & $28\pm5$
& $28\pm5$ & $ 15\pm3$
                \\   \hline
L3 & & & & & & \\
Signal &$167\pm10$  &$ 142\pm9$  & $119\pm7$ & $ 88\pm5$ & $49\pm3$
& $ 7\pm3$
                \\
Background &$79\pm11$  &$ 83\pm10$  & $87\pm9$ & $65\pm7$ & $44\pm5$
& $44\pm5$
                \\   \hline
OPAL & & & & & & \\
Signal & $188\pm9$  &$160\pm8$  & $128\pm6$ & $98\pm7$ &
$56\pm4$ & $6\pm1$
                \\
Background & $27\pm5$  &$27\pm5$  & $26\pm4$ & $27\pm5$ &
$17\pm4$ & $ 7\pm3 $
                \\   \hline \hline
ALL & & & & & & \\
Signal & $ 840\pm20$ & $ 715\pm17$ & $ 561\pm13$ & $ 410\pm12$ &
$ 229\pm 6$ & $ 27\pm3$
   \\
Background & $ 182\pm14$ & $ 180\pm13$ & $ 179\pm13$ & $ 151\pm11$
 &$112\pm9 $  & $ 89\pm8$
   \\ \hline\hline
\end{tabular}
\end{center}
\end{table}

\begin{table}[htbp]
\setlength{\tabcolsep}{0.82pc}
\caption{\it 
  Accepted cross-sections (in fb) expected for the signal and the
  background, for various Higgs boson masses, at a center-of-mass
  energy of 192~GeV.}
\label{tab:192}
\begin{center}
\begin{tabular}{llrrrrr} \hline \hline
& \mH~(\Gcs) & $80$  & $85$ & $90$ & $95$ & $100$   \\
 \hline
ALEPH &
Signal
 & $125\pm3$ & $115\pm3$ & $103\pm3$ & $64\pm2$ & $13\pm2$
                \\
& Background & $33\pm6$ & $48\pm7$ & $63\pm8$ & $57\pm7$ & $51\pm7$
                \\       \hline

DELPHI &
Signal &  $99\pm4$  & $108\pm5$ & $85\pm4$ &
 $60\pm4$ & $13\pm2$
                \\
& Background &$42\pm5$  & $68\pm5$ & $79\pm5$ & $50\pm3$ & $25\pm2$
                \\   \hline
L3 &
Signal &$93\pm5$  &$ 77\pm4$  & $64\pm3$ & $ 45\pm3$ & $9\pm1$
                \\
& Background &$66\pm6$  &$ 67\pm5$  & $68\pm5$ & $ 44\pm5$ & $19\pm3$
                \\   \hline
OPAL &
Signal   &$98\pm4$  & $ 81\pm3$ & $72\pm3$ & $ 40\pm2$ & $ 13\pm2$
                \\
& Background   &$28\pm4$  & $ 37\pm4$ & $46\pm5$ & $36\pm5$ & $ 26\pm5 $
                \\   \hline \hline
ALL &
Signal &$414\pm8 $ & $ 381\pm8$ & $323\pm6 $ & $ 209\pm6$ & $ 47\pm4$
   \\
& Background & $169\pm10 $ & $ 220\pm1$ & $255\pm12 $ & $ 187\pm11$ &
$ 121\pm9$
   \\ \hline\hline
\end{tabular}
\end{center}
\end{table}

\begin{table}[htbp]
\setlength{\tabcolsep}{0.82pc}
\caption{\it 
  Accepted cross-sections (in fb) expected for the signal and the
  background, for various Higgs boson masses, at a center-of-mass
  energy of 205~GeV.}
\label{tab:205}
\begin{center}
\begin{tabular}{lrrrrrr} \hline\hline
\mH~(\Gcs)  & $80$  & $90$ & $100$ & $105$ & $110$ & $115$  \\
\hline
ALEPH & & & & & & \\
Signal  & $118\pm3$ & $90\pm3$ & $63\pm2$ & $46\pm2$ & $32\pm3$
   & $5\pm1$
                \\
Background & $48\pm7$ & $82\pm9$ & $28\pm5$ & $24\pm5$ & $20\pm5$  &
 $20\pm5$               \\       \hline

DELPHI & & & & & & \\
Signal &  $78\pm4$  & $84\pm4$ & $66\pm4$ &
 $48\pm3$ & $23\pm2$ & $ 3\pm1$
                \\
Background &$56\pm6$  & $66\pm6$ & $52\pm6$ & $26\pm4$
& $13\pm3$ & $ 8\pm2$
                \\   \hline
L3 & & & & & & \\
Signal &$88\pm6$  &$68\pm4$  & $51\pm3$ & $ 38\pm3$ & $ 22\pm3$
& $ 4\pm1$
                \\
Background &$70\pm6$  &$94\pm7$  & $59\pm6$ & $ 30\pm6$ & $ 20\pm6$
& $ 20\pm6$
                \\   \hline
OPAL & & & & & & \\
Signal & $55\pm2$ &$48\pm2$  & $39\pm2$ & $26\pm2$ &
$13\pm1$ & $4\pm0.3$
                \\
Background & $25\pm4$ &$45\pm4$  & $26\pm4$ & $20\pm4$ &
$15\pm4$ & $ 15\pm4 $
                \\   \hline \hline
 ALL & & & & & & \\
Signal & $ 339\pm8$ & $287\pm7 $ & $220\pm6 $ & $ 158\pm5$ &
$89\pm4 $  & $ 16\pm1$
   \\
Background & $ 198\pm12$ & $ 288\pm13$ & $166\pm11 $ & $ 101\pm10$ &
$ 68\pm9$  & $ 62\pm9$
   \\ \hline\hline
\end{tabular}
\end{center}
\end{table}

\subsection{Discovery and Exclusion Limits}
\label{sec:lim}

Based on the simulations described in Section~\ref{sec:exp}, it is
possible to derive the exclusion and discovery limits of the standard
model Higgs boson as a function of the luminosity for the three
center-of-mass energies specified earlier. The contours are defined
at $5\sigma$ for the discovery in the case of the existence
of the Higgs boson and at 95\%~C.L. for the exclusion limits in the
case of negative searches, with the specifications described in
Appendix~\ref{sec:app}.

\begin{figure}[htbp]
\begin{picture}(160,195)
\put(20,130){\epsfxsize130mm\epsfbox{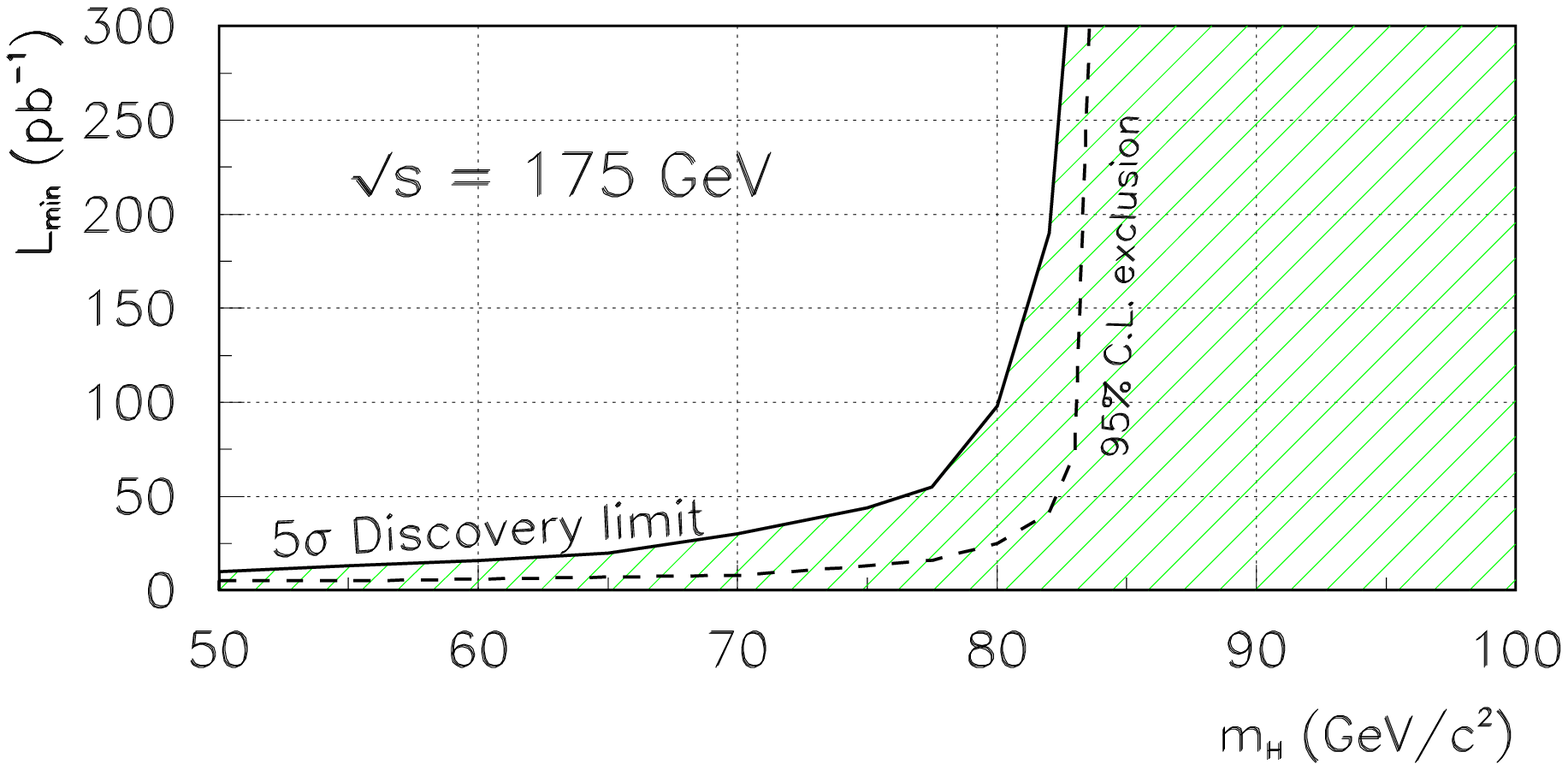}}
\put(20,65){\epsfxsize130mm\epsfbox{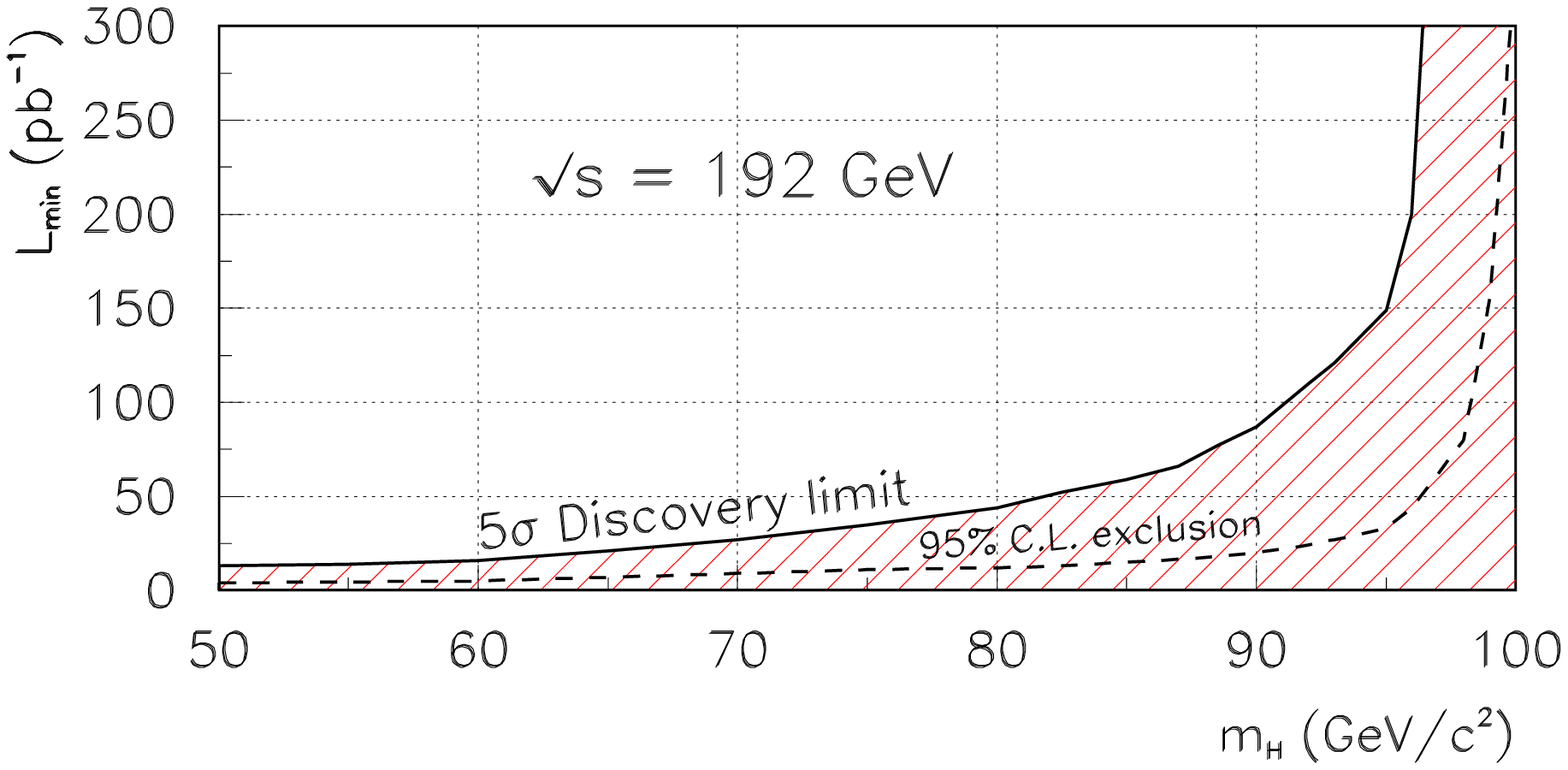}}
\put(20,0){\epsfxsize130mm\epsfbox{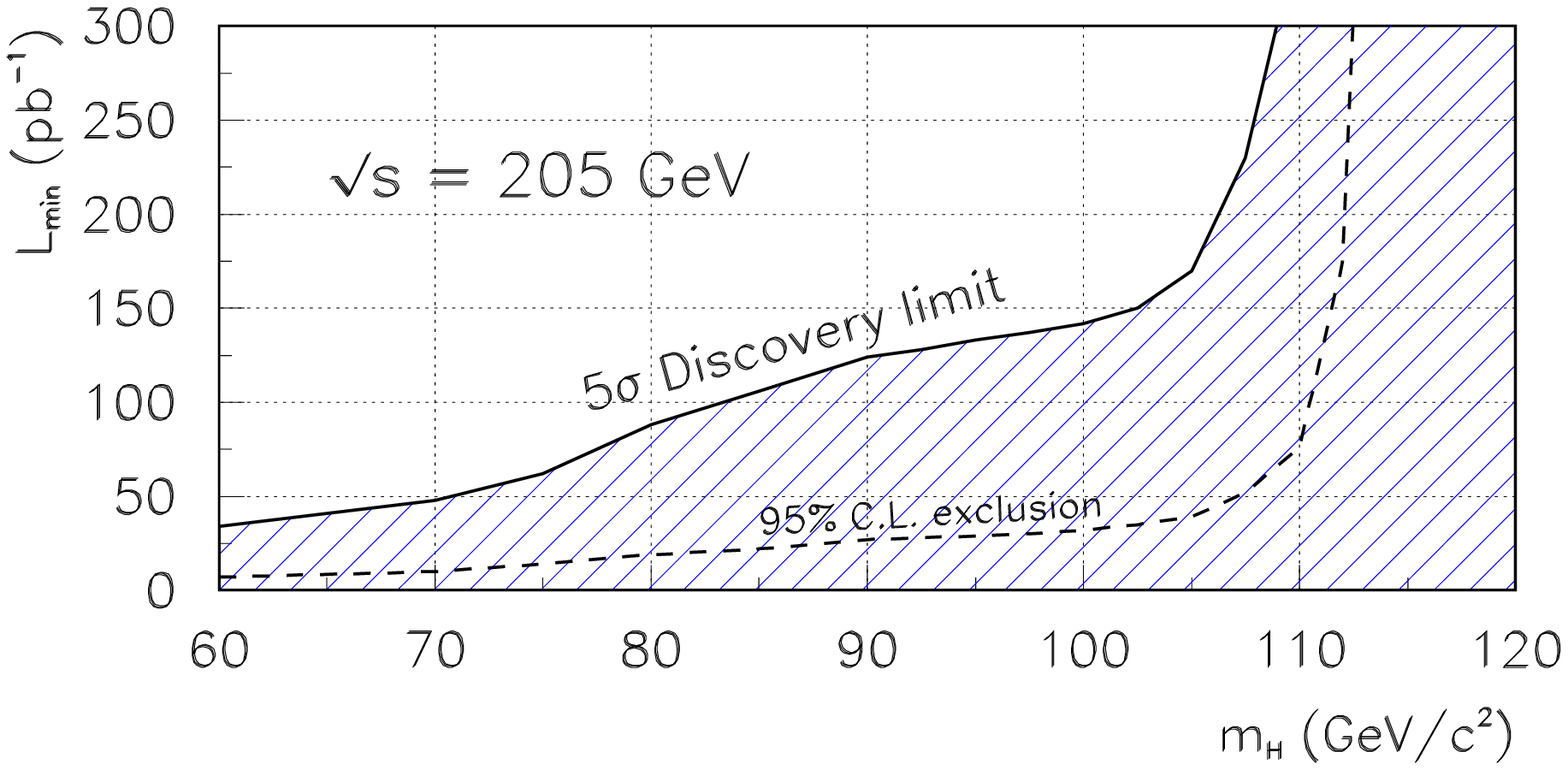}}
\end{picture}
\caption{\it 
  Minimum luminosity needed per experiment, in \inpb, for a combined
  $5\sigma$ discovery (full line) or a 95\% C.L.  exclusion (dashed
  line) as a function of the Higgs boson mass, at the three
  center-of-mass energies.}
\label{fig:lumin}
\end{figure}

In Table~\ref{tab:lumin}, the minimum integrated luminosities needed
to exclude or discover a given Higgs boson mass at the center-of-mass energies
$\sqrt{s}=175$, 192 and 205 GeV are given for the combination of all channels
for each of the four experiments separately, as well as for the combination of
all channels for the four LEP experiments together. The results of the
combination of the four experiments are graphically shown in
Fig.\ref{fig:lumin}, and summarized in Table~\ref{tab:SM}.

\begin{table}[htbp]
\setlength{\tabcolsep}{0.8pc}
\caption{\it
  Minimum luminosity needed, in \inpb, by ALEPH, DELPHI, L3, OPAL, and
  for a simple combination of the four experiments, at the three
  center-of-mass energies and for various Higgs boson masses.  The
  first number holds for the 95\% C.L. exclusion, the second one for
  the $5\sigma$ discovery.}
\label{tab:lumin}
\begin{center}
$\sqrt{s}=175$ GeV

\begin{tabular}{rrrrrr}
\hline\hline
 Experiment &$\mH = 60$&    65    &    70    &    75    &    80    \\
\hline
 ALEPH      &  12:34   &  18:49   &  25:76   &  36:126  &  80:316  \\
 DELPHI     &  16:48   &  18:51   &  31:87   &  40:140  &  78:335  \\
 L3         &  29:127  &  39:180  &  56:244  &  75:334  & 152:727  \\
 OPAL       &  17:56   &  20:75   &  34:96   &  44:161  &  74:294  \\
\hline
 All        &   6:15   &   6:19   &   8:28   &  10:41   &  21:90   \\
\hline\hline
\\
\end{tabular}

$\sqrt{s}=192$ GeV

\begin{tabular}{rrrrr}
\hline\hline
 Experiment & \mH = 80 &    85    &    90    &    95    \\
\hline
 ALEPH      &  33:117  &  42:166  &  59:238  & 103: 510  \\
 DELPHI     &  50:195  &  50:231  &  80:388  & 118: 529  \\
 L3         &  64:306  &  90:426  & 118:596  & 172: 832  \\
 OPAL       &  43:157  &  60:251  &  85:360  & 182: 825 \\
\hline
 All        &  12:44   &  15:60   &  20:87   &  33:149  \\
\hline\hline
\\
\end{tabular}

$\sqrt{s}=205$ GeV

\begin{tabular}{rrrrrr}
\hline\hline
 Experiment & \mH = 80 &    90    &   100    &   105    &   110   \\
\hline
 ALEPH      &  41:157  &  80:369  &  76:327  & 119: 504 & 186: 870 \\
 DELPHI     &  75:356  &  78:372  &  97:462  & 114: 507 & 283:1296 \\
 L3         &  74:342  & 142:704  & 162:817  & 164: 897 & 409:2103 \\
 OPAL       &  87:372  & 149:735  & 151:719  & 267:1284 & 680:3500 \\
\hline
 All        &  16:66   &  25:119  &  30:125  &  38:158  &  72:339 \\
\hline\hline
\end{tabular}

\end{center}
\end{table}

\begin{table}[htb]
\setlength{\tabcolsep}{0.85pc}
\caption{\it
  Maximal Higgs boson masses that can be excluded or discovered with a
  given integrated luminosity $L_{min}$ per experiment at the three
  representative energy values of 175, 192 and 205~GeV, when the four
  LEP experiments are combined.}
\label{tab:SM}
\begin{center}
\begin{tabular}{|c|cc|cc|}
\hline\hline
\rule[0mm]{0mm}{3ex}
  &   Exclusion: &  &  Discovery: &   \\
\rule[0mm]{0mm}{3ex}
$\sqrt{s}$ (GeV) & \mH~(\Gcs) & $L_{min}$ (\inpb) & \mH~(\Gcs) & $L_{min}$ (\inpb) \\
\hline 
\rule[0mm]{0mm}{3ex} 175 & 83 & 75  & 82 & 150 \\
\rule[0mm]{0mm}{3ex} 192 & 98 &  150  & 95 & 150 \\
\rule[0mm]{0mm}{3ex} 205 & 112 & 200  & 108 & 300  \\
\hline\hline
\end{tabular}
\end{center}
\end{table}

Combining the four LEP experiments, the required minimal integrated
luminosity per experiment
to discover or exclude a certain Higgs boson mass at a given
center-of-mass
energy is reduced to approximately a fourth of the average
minimal integrated luminosity of each individual experiment.
This implies that the maximal value of the Higgs boson mass
will be reached at a given energy for luminosities
which can be naturally  expected at \LEPII.
The following conclusions can be drawn from detailed analyses of
the figures and tables.

\begin{itemize}

\item[{\it (i)}] At a center-of-mass energy of 175 GeV, the maximum integrated luminosity
needed is of the order of 150~\inpb\ and this allows  the discovery of a
Higgs boson with  a maximum mass of about 82~\Gcs.
Indeed, combining the four experiments it follows  that raising the
luminosity leads only to a marginal increase of the exclusion and discovery
limits, which are very close to each other.

\item[{\it (ii)}] At 192 GeV it is again sufficient to have an integrated luminosity of
about 150~\inpb, in this case  to discover a Higgs boson with mass up to 95~\Gcs.
Increasing the center-of-mass energy from 175 to 192 GeV leads to
a significant extension in the discovery range of the Higgs boson mass.
It is of great interest to observe that at $\sqrt{s} = 192$~GeV
a 95~\Gcs\ Higgs boson mass can be excluded at the 95\% confidence level with
an integrated luminosity as low as 33~\inpb\ while with
150~\inpb\  a Higgs boson mass close to 100~\Gcs\ can be excluded.

\item[{\it (iii)}] This development continues up to 205 GeV, where a luminosity
as low as 70~\inpb\ is sufficient to exclude Higgs boson masses up to about 110~\Gcs,
and  a 5$\sigma$ discovery of a Higgs boson with a mass of order 105~\Gcs\
requires an  integrated luminosity of $\sim 160$~\inpb.
More luminosity is needed in this case, since the cross section of the
 irreducible ZZ background increases. With an integrated luminosity of
$\sim 300$~\inpb\ a Higgs boson mass close to 110 GeV can be discovered.

\end{itemize}

If each experiment is considered separately,
the 5$\sigma$ discovery limit
for an integrated luminosity of 500~\inpb\ is, on average, approximately
given by $\mH =$ 82 (95) (103)~\Gcs\ for $\sqrt{s}$= 175 (192) (205) GeV.
 Similar results may be obtained by combining the four experiments  for an
 integrated luminosity per experiment of about 150~\inpb.
For the combined exclusion limits, the maximum
value of \mH\ at $\sqrt{s}$= 175, 192, 205~GeV is reached for a luminosity
 per experiment of about 75, 150, 200~\inpb. A further increase
in luminosity is not very useful in case of negative searches. Clearly, energy rather than
luminosity is the crucial parameter to improve the range of masses which can be reached
at \LEPII.
\vvs1

\subsection{The LHC Connection}

It has been shown in  section \ref{sec:lim} that LEP2 can cover
the SM Higgs mass range up to $82 \rm{\;GeV}$ at a total energy of
$\sqrt{s} = 175 \rm{\;GeV}$ while the Higgs mass discovery limit
increases to
$\sim 95 \rm{\;GeV}$ for a total energy of $192 \rm{\;GeV}$. Since this mass
range contains the lower limit at which the SM Higgs particle
can be searched for at the LHC, the upper limit of the LEP2 energy is
quite crucial for the overlap in the discovery regions
of the two accelerators.

Low--mass Higgs particles are produced at the LHC predominantly in
gluon--gluon collisions~\cite{glashow,s1a} or in Higgs--strahlung processes 
\cite{richter,s2},
\begin{eqnarray*}
pp & \rightarrow & H \rightarrow \gamma\gamma \\
pp & \rightarrow & WH,ZH,t\bar{t}H \rightarrow \ell+\gamma\gamma
\quad\mbox{and}\quad\ell + b\bar{b}
\end{eqnarray*}
with the Higgs boson emitted from a virtual $W$ boson or from a top
quark. In the gluon--fusion process the Higgs particle is searched for
as a resonance in the $\gamma\gamma$ decay channel which comes with a
branching ratio of order $10^{-3}$. Even though large samples of Higgs
particles can be generated in this mass range, the
signal--to--background ratio is only a few percent and the rejection
of jet background events which are eight orders of magnitude more
frequent, is a very difficult experimental task.  Excellent energy
resolution and particle identification is needed \cite{atlas} to
tackle this problem. It has been shown in detailed experimental
simulations that the significance $S/\sqrt{B}$ of the Higgs signal is
expected to rise in this channel from a value $\sim 2.5$ at $m_{H} =
80 \rm{\;GeV}$ to a value $\sim 4.5$ at $m_{H} = 100 \rm{\;GeV}$ for
$\int{\cal L} = 3\times 10^4$ ${\rm pb}^{-1}$ if ATLAS and CMS
analyses are combined.

In the Higgs--strahlung process, the events can be tagged by leptonic
decays of the $W/Z$ bosons or the $t$ quark to trigger the experiment and
to reduce the jet background. In these subsamples the Higgs boson can
be searched for in the $b\bar{b}$ decay mode with a branching ratio
close to unity. This method is based on $b$ tagging by micro--vertex
detection which is anticipated to be an excellent tool of the LHC
detectors.  After suitable cuts in the transverse momenta of the
isolated lepton and the $b$ jets, a peak is looked for in the
invariant $M(b\bar{b})$ mass. The experimental significance
$S/\sqrt{B}$ of this method is biggest for small Higgs masses.
For $\int{\cal L}= 3\times 10^4$ ${\rm pb}^{-1}$ and ATLAS/CMS combined,
experimental simulations of the $[W]_l b\bar{b}$ sample suggest that
$S/\sqrt{B}$ falls from $\sim 8$ at $m_{H} = 80 \rm{\;GeV}$ down to
$\sim 6$ at $m_{H} = 100 \rm{\;GeV}$.  It is not yet clear how the search
can be extended to higher luminosities where the layers in the 
micro--vertex detectors closest to the beams may not survive, thus
reducing significantly the $b$--tagging performance of the experiments.

Combining the prospective signals from the $\gamma\gamma$ and the
$[W]_l b\bar{b}$ analyses, an overall significance of 7 to 8 may be
reached for Higgs masses below $100 \rm{\;GeV}$, based on a low
integrated luminosity of $\int\!{\cal L}=3\!\times\!10^4
\rm{\;pb}^{-1}$ within three years.  Raising the integrated
luminosity 
to $\int\!{\cal L} = 10^5
\rm{\;pb}^{-1}$ increases the discovery significance to almost $9$ for
$80<m_H<100$ GeV~\cite{gian}.
\vvs2

\vfill

\newpage

\section{The Higgs Particles in the Minimal Supersymmetric Standard Model}

The Minimal Supersymmetric Standard Model leads to clear and distinct
experimental signatures in the Higgs sector. Two Higgs doublets, $H_1$
and $H_2$, must be present, in order to give masses to the up and down
quarks and leptons, and to cancel the gauge anomalies induced through
the  Higgs superpartners. In the supersymmetric limit, the Higgs
potential is fully determined as a function of the gauge couplings and
the supersymmetric mass parameter $\mu$. The breakdown of
supersymmetry is associated with the introduction of soft
supersymmetry breaking parameters, which are essential to yield a
proper electroweak symmetry breaking. In the broken phase, the ratio
of the Higgs vacuum expectation values, $\tan\beta = v_2/v_1$, appears
as a new parameter, which can be related to the other parameters of
the theory by minimizing the Higgs potential.

The physical Higgs spectrum of the MSSM
contains two CP-even and one CP-odd neutral Higgs bosons, $h/H$ and $A$, 
respectively, and a charged Higgs boson pair $H^\pm$ \cite{hhg}. 
The  tree--level  Higgs spectrum is   determined by 
 the weak gauge boson masses, 
the CP--odd Higgs mass, $m_A$, and $\tan\beta$.
It is only through radiative corrections
that the other parameters of the model affect  the Higgs mass
spectrum.  
The dominant radiative corrections  to the Higgs masses 
grow as the fourth power of the top-quark mass and they are logarithmically 
dependent on the  sparticle spectrum. 
The mass of the heavy Higgs doublet is controlled by the CP-odd
Higgs mass  and, for large values of $m_A$, the effective low
energy theory contains only one Higgs doublet, which couples to fermions
in the standard way.  In a first approximation,
the Higgs masses may be calculated by assuming that
all sparticles acquire masses of order of the characteristic supersymmetry
breaking scale $M_S$ which, based on naturalness arguments, should
be below a few TeV. The low--energy effective theory below $M_S$ is
a general two--Higgs doublet model, with couplings which can be calculated
as a function of the other parameters of the theory. Under these conditions,
 a general upper bound on the lightest CP-even Higgs boson mass is derived for
values of the CP-odd Higgs mass of order $M_S$. 
For smaller values of
$m_A$, a more stringent upper bound is obtained. 
In the following, we shall discuss in detail the different methods
to compute the Higgs spectrum in the MSSM and the bounds which can be
derived in each case.
\vvs1

\subsection{Higgs Mass Spectrum and Couplings}

\subsubsection{Tree--level Mass Bounds} 

The masses of the
Higgs bosons   at tree level are determined as a function of $m_A$,
$\tan\beta$ and the gauge boson masses as follows,
\begin{equation}
m_{h,H}^2 = \frac{1}{2} \left[ m_A^2 + m_Z^2 \mp
\sqrt{ \left(m_A^2 + m_Z^2 \right)^2 - 4 m_Z^2 m_A^2 
\cos^2 2 \beta} \right] ,
\end{equation}
\begin{equation}
m_{H^{\pm}}^2 = m_A^2 + m_W^2.
\end{equation}
The 
mass of the lightest MSSM neutral Higgs particle is bounded 
to be smaller than the $Z$ mass \cite{Marc,R15},
\begin{equation}
m_h^{\rm tree} \leq
m_Z |\cos 2 \beta|,
\end{equation} 
 and it approaches  this upper bound 
for large values of $m_A$. 
The bound is modified by  radiative corrections, which raise
the upper limit on the lightest CP-even Higgs mass to values close
to 150 GeV.

\subsubsection{Radiative Corrections to the Higgs Masses}
\label{sec:massesMSSM}

The one- and partial two-loop radiative corrections to the Higgs mass
spectrum in the MSSM have been calculated.
Computations implying a variety of  different approximations, 
which may be distinguished according to their level of refinement, exist. 
In general, the radiative corrections to the Higgs masses 
are  large and positive, being dominated by the contributions of the
third-generation  quark superfields. Since the upper bound on $m_h$
determines the limit for the  detectability of the Higgs boson at LEP2, it is
interesting to discuss the different methods in some detail. 
\vvs1

\noindent {\bf a) Diagrammatic Approach.}
Order by order, a  precise method of computation
 of the radiative corrections 
to the Higgs masses is  the full diagrammatic approach. At
the one-loop level such calculations have been pursued by several 
authors \cite{Marc,RC1,fulld}. Complete expressions, including all
supersymmetric particle contributions
are available \cite{FDCFOR}.
 The resulting Higgs masses are defined as the location
of the pole in the Higgs propagator. 
In order to obtain a more  accurate estimate of the Higgs spectrum in
the diagrammatic approach,  the two-loop effects must be evaluated.
A first step in this direction was 
performed in Ref.\cite{RC2}
for the case of large values of the CP-odd Higgs boson mass, large
 $\tan\beta$, and degenerate squark masses. It was shown
that these corrections may be quite significant, of order 10--15 GeV, 
underlining the need for a careful treatment of the two-loop
effects on the Higgs mass spectrum. 
\vvs1

\noindent {\bf b) Effective Potential Methods.}
The leading corrections to 
the Higgs mass spectrum in the MSSM can be computed in  a very simple way 
by means of effective potential methods \cite{ERZ,BERZ}. 
If all the contributions from the
MSSM particles  are included, the results
within this scheme differ
from those of the full  diagrammatic approach in that the
Higgs masses are evaluated at zero momentum.  
In order to simplify the calculations, it is possible to consider
only the contributions of the third-generation quark superfields,
neglecting all weak gauge coupling effects  in the one-loop
expressions \cite{ERZ}.
This treatment of the effective potential has the
virtue of displaying, in a compact way, the full dependence of the
one-loop radiative corrections on the stop/sbottom
masses and mixing angles. 
For a given squark spectrum,  the numerical 
results obtained in this case differ by only  a few GeV from the results 
obtained within the full one--loop diagrammatic approach.
This reflects the smallness of  the one-loop contributions
from  superfields other than  top and bottom. Moreover, it shows that
the one-loop vacuum polarization effects relating the  Higgs pole masses to the
running masses calculated through the  effective potential approach  
 are in general  small.
The effective potential computation
 can be improved by including 
the dependence of the stop and sbottom spectrum  on the
weak gauge couplings \cite{B}.
In the limit $m_{{\tilde t}_1}$,  $m_{{\tilde t}_2}$, $m_A \gg m_Z$, 
where $m_{\tilde{t}_{1,2}}$ are the two stop mass eigenvalues, the 
expression of the lightest Higgs mass takes a  simple form, 
\be
 \label{hlmass}
m_h^2= m_Z^2\cos^2 2\beta+(\Delta m_{h}^2)_{\rm 1LL}+
(\Delta m_{h}^2)_{\rm mix}
\ee
where
\begin{equation}
(\Delta m_h^2)_{\rm 1LL}= \frac{3 m_t^4}{ 4 \pi^2 v^2}
 \ln\left(\frac{ m_{{\tilde t}_1} m_{{\tilde t}_2}}
{m_t^2}\right)\left[1+{\cal O}\left(
{m_W^2\over m_t^2}\right)\right]
\label{mhlonell}
\end{equation}
and
\be
(\Delta m_h^2)_{\rm mix}= {3 m_t^4 {\tilde A}_t^2 \over 8 \pi^2 v^2}
\left[2h(m^2_{{\tilde t}_1},m^2_{{\tilde t}_2})
+{\tilde A}_t^2 f(m^2_{{\tilde t}_1},m^2_{{\tilde t}_2})\right]
\left[1+{\cal O}\left({m_W^2\over m_t^2}\right)\right]
\label{mhlmix}
\ee
In eq.(\ref{mhlmix}), $\tilde{A}_t =  A_t-\mu\cot\beta$ and 
the functions $h$ and $f$ are given by
\be
\label{hfeq}
h(a,b) = {1\over a-b}\ln\left({a\over b}\right) \;\;\;\;\; \mbox{and} 
\;\;\;\;\; 
f(a,b) \equiv {1\over (a-b)^2}\left[2-{a+b\over a-b}\ln\left({a\over b}
\right)\right] 
\ee
The above expression is particularly interesting since it provides
the upper bound on $m_h$ for a given stop spectrum.
Including two--loop effects
remains, however, a necessary further step to obtain a correct quantitative
estimate of the Higgs mass. 
\vvs1

\noindent {\bf c) Renormalization Group Improvement 
of the Radiatively Corrected Higgs Sector.}
The most important two--loop effects may be included 
by performing a renormalization group improvement of the 
effective potential, while
taking into account, in a proper way, the effect of the decoupling of
the heavy third--generation squarks. This program can be easily carried out
in the case of a large CP-odd Higgs boson mass and 
degenerate squarks \cite{lambdar}.  Since only one
Higgs doublet survives at low energies, 
the lightest CP--even Higgs mass may be calculated through the renormalization 
group evolution of the  effective quartic coupling, 
assuming that the heavy sparticles  decouple
at a common scale $M_{S}$.  The one-loop renormalization
group evolution of the quartic couplings 
 includes two-loop effects through the resummation of the one-loop result. The 
general result is, however, scale dependent
but this  dependence is reduced by taking into account
the two-loop renormalization group improvement of the
one--loop effective potential \cite{RC2s,RC2p}. The
vacuum expectation value of the Higgs field and the renormalized 
Higgs mass scale
(approximately) with the appropriate one--loop anomalous dimension factors
within this approximation. The scale dependence of the Higgs mass is
cancelled by adding the one-loop vacuum polarization
effects, necessary to define the Higgs pole mass.
For the case of small stop  mixing and large
values of $\tan\beta$, the Higgs spectrum evaluated through this
method agrees with the  diagrammatic
computation  at the two--loop level \cite{RC2,RC2p}.
\vvs1

\begin{figure}[htp]
\vspace{-1cm}
\centerline{
\epsfig{figure=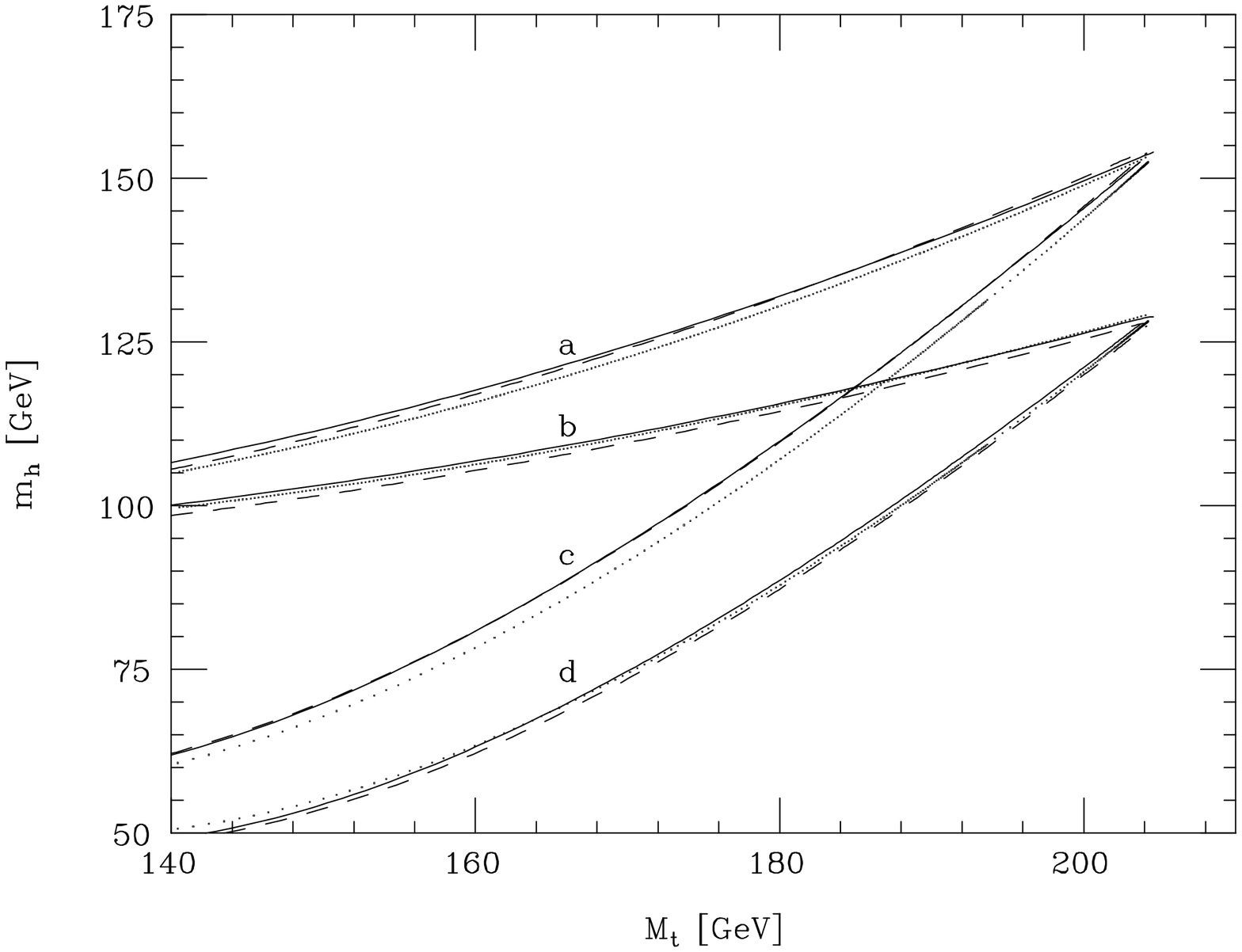,height=13cm,angle=90} }
\vspace{-1.3cm}
\caption[0]{\it 
  The lightest Higgs mass as a function of the physical top-quark
  mass, for $M_S$ = 1 TeV, evaluated in the limit of large $m_A$, as
  obtained from the two--loop RG improved effective potential (solid
  lines), the one--loop improved RG evolution (dashed lines) and the
  analytical approximation, eq.(\ref{mhsm}) (dotted-lines). The four
  sets of lines correspond to a) $\tan \beta$ = 15 with maximal squark
  mixing, b) $\tan \beta$ = 15 with zero-squark mixing, c) the minimal
  value of $\tan \beta$ allowed by perturbativity constraints for the
  given value of $M_t$ (IR fixed point) for maximal mixing and d)
  $\tan \beta$ the same as in c) for zero mixing.}
\label{fig:mhmt}
\vvs2
\centerline{
\epsfig{figure=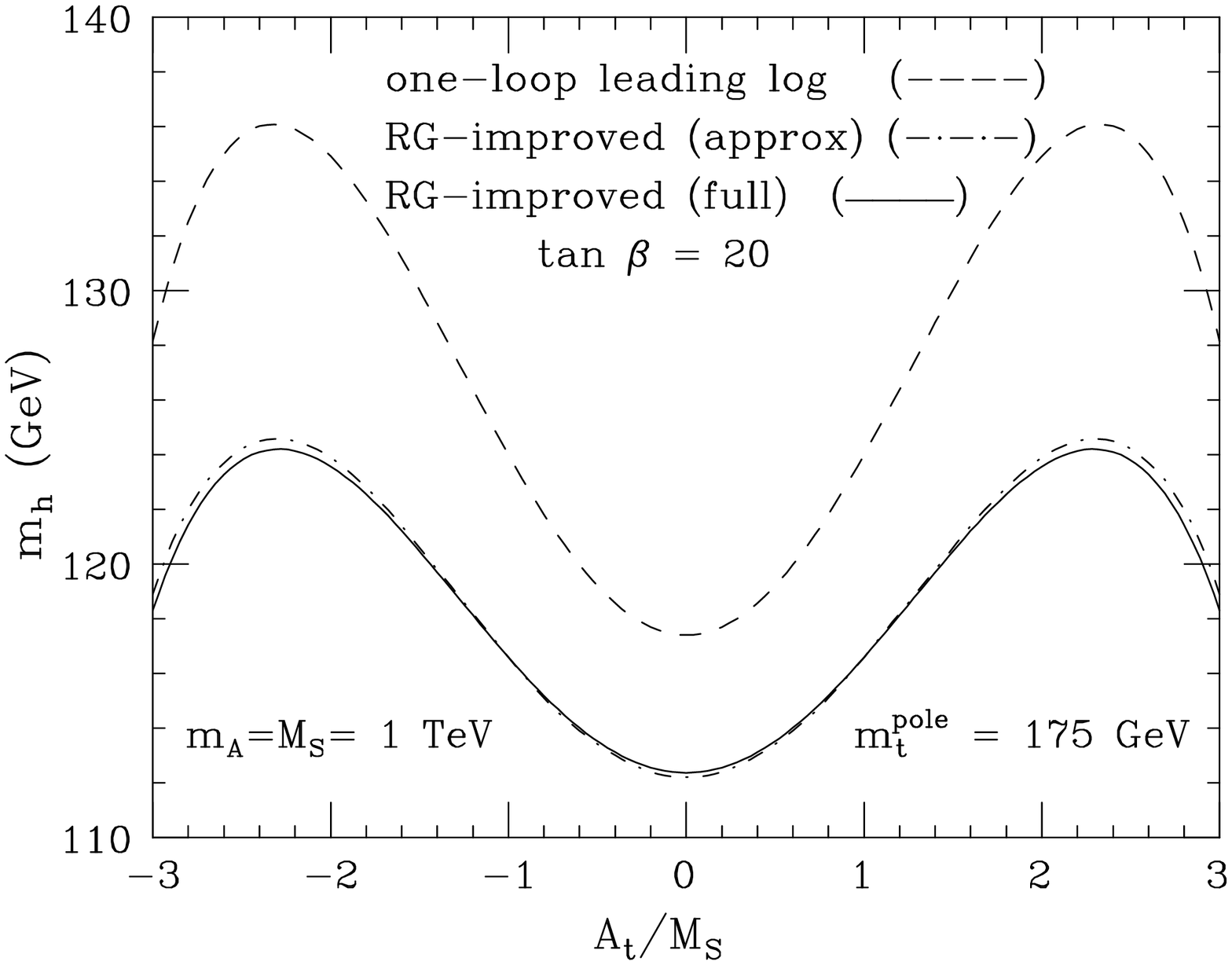,height=8.0cm,angle=90}
\hfill
\epsfig{figure=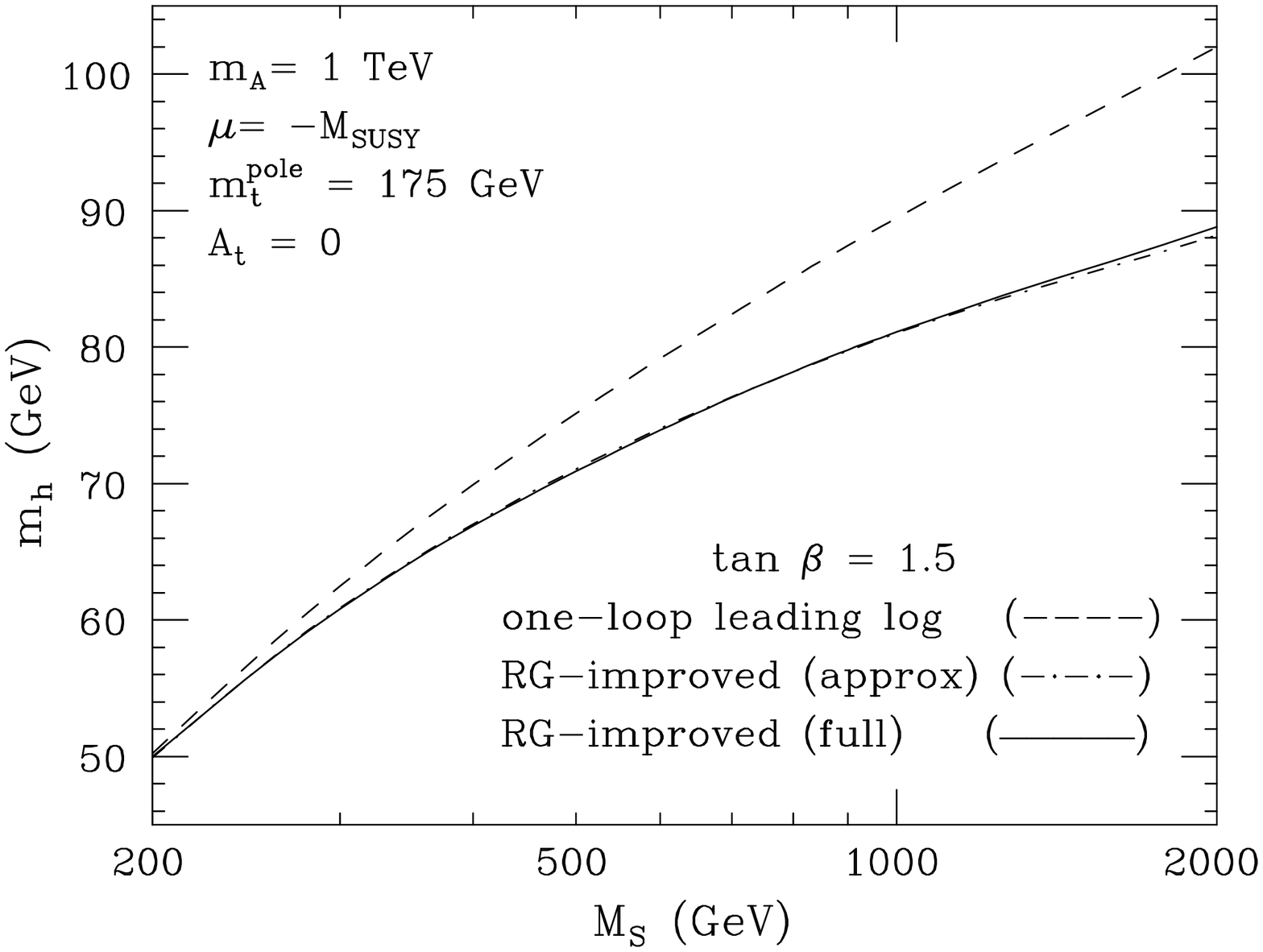,height=8.0cm,angle=90}
}
\caption{\it
  The radiatively corrected light CP-even Higgs mass is plotted as a
  function of the MSSM parameters.  The one-loop leading-log
  computation is compared with the RG-improved result which was
  obtained by numerical analysis and by using the simple analytic
  result given in eq.(\protect\ref{hlmassimproved}).}
\label{fig2}
\end{figure}

\noindent \underline {Analytical Expression for the Lightest CP-even Higgs Mass.}
The two-loop RG improvement of the one--loop
effective potential
includes two--loop effects in two different ways: through the 
resummation of one--loop effects and through
genuine two--loop  effects. Numerically, the latter are 
small compared to the resummation effects \cite{EQ1}.
Once an appropriate  scale  of order of the
top-quark mass is adopted, the results of the one--loop RG
  improvement of the tree--level effective potential including the proper
threshold effects of squark decoupling,
 are in excellent agreement with the pole Higgs masses 
computed by the two-loop RG improvement of the one-loop effective 
potential \cite{RC2p,RC4}.  This holds,  
for large values of the CP-odd Higgs mass, for any value of $\tan\beta$ 
and the squark mixing angles. 
Based on  this result, an analytical approximation
may be obtained \cite{RC4} 
which reproduces the dominant two-loop
results \cite{RC2p} within an error of less than 2 GeV. Fig.\ref{fig:mhmt}
 shows the agreement of 
the one--loop and
two--loop results for  $m_h$ evaluated at 
the appropriate scale $M_t$, and the accuracy 
of the analytical approximation.
In the $\overline{MS}$ scheme, the pole top-quark mass 
$M_t$ must be related to the on-shell running mass 
$m_t \equiv m_t(M_t)$ 
by taking into account the corresponding one-loop QCD correction factor
\begin{equation}
m_t=\frac{M_t}{1+\frac{4}{3\pi}\alpha_s(M_t)}
\label{mtMt}
\end{equation}
 Top Yukawa effects have been neglected in eq.(\ref{mtMt}), 
since they are essentially   cancelled
by the two--loop QCD effects. Observe that eq.(\ref{mtMt}) gives the
correct relation between the running and the pole top-quark masses only
if the leading-log contributions to the running mass, associated with
the decoupling of the heavy sparticles, are properly taken
into account and the sparticles are sufficiently heavy so that the
finite corrections become small \cite{Donini}. 
The analytical approximation to the one-loop
renormalization-group improved result, including two-loop
leading-log effects, is given by \cite{RC4} 
\begin{eqnarray}
\label{upperb}
m_h^2& = & m_Z^2\cos^2 2\beta\left( 1-\frac{3}{8\pi^2}\frac{m_t^2}
{v^2}\ t\right) \nonumber \\
\label{mhsm}
& & + \frac{3}{4\pi^2}\frac{m_t^4}{v^2}\left[ \frac{1}{2}\tilde{X}_t + t
+\frac{1}{16\pi^2}\left(\frac{3}{2}\frac{m_t^2}{v^2}-32\pi\alpha_s
\right)\left(\tilde{X}_t t+t^2\right) \right]
\end{eqnarray}
where the angle $\beta$ is defined  at the scale $m_A = M_{S}$ and 
$ t=\log (M_{S}^2/M_t^2) $. $\tilde{A}_t $ is defined above and 
\begin{equation}
\label{stopmix}
\tilde{X}_{t}  = \frac{2 \tilde{A}_t^2}{M_{S}^2}
\left(1 - \frac{\tilde{A}_t^2}{12 M_{S}^2} \right)
\end{equation}
Furthermore, 
$\alpha_s(M_t)=\alpha_s(m_Z)/ \left(1+\frac{b_3}{4\pi}\alpha_s(m_Z)
\log(M_t^2/m_Z^2) \right)$
with $b_3 = 11 - 2 N_F/3$  being the one-loop QCD beta function and
$N_F$  the number of quark flavours [$N_F= 5$ at scales below $M_t$].   
The supersymmetric scale $M_S$ is defined as
$M_S = \sqrt{(m_{\tilde{t}_1}^2 + m_{\tilde{t}_2}^2)/2}$.  
For simplicity, all supersymmetric particle masses
are assumed to be of order $M_{S}$.
Notice that eq.(\ref{mhsm}) includes
the leading $D$-term correction ${\cal O}(m_Z^2 m_t^2)$ \cite{B}. 

A similar analytical result to eq.(\ref{mhsm}) 
has been obtained in Ref.\cite{HHH}.  In this approximation
the two-loop leading-logarithmic 
contributions to $m_{h}^2$ are incorporated by replacing
$m_t$ in eq.(\ref{hlmass}) 
by the running top quark mass evaluated at appropriately chosen
scales.  For $m_{\tilde t_1}\approx m_{\tilde t_2}\equiv
M_S$ the  result is:
\begin{equation} \label{hlmassimproved}
m_{h}^2= m_Z^2\cos^2 2\beta+(\Delta m_{h}^2)_{\rm
1LL}(m_t(\mu_t))+(\Delta m_{h}^2)_{\rm mix}(m_t(M_S))
\end{equation}
where $\mu_t\equiv\sqrt{M_t M_S}$;
the running top-quark mass is given by
\begin{equation}
m_t(\mu) = m_t \left[
1-\left({\alpha_s\over\pi}-{3\alpha_t\over 16\pi}\right)\,
\ln\left({\mu^2\over M_t^2}\right)\right]
\label{mtrun}
\end{equation}
with $\alpha_t = h_t^2/ 4\pi$.
All couplings on the right hand side of eq.(\ref{mtrun})
are evaluated at $M_t$.  The requirement that the two stop mass
eigenstates be close to each other allows an expansion of
the functions $h$ and $f$, eq.(\ref{hfeq}),   
in powers of $m_t \tilde{A}_t/M_S^2$. The resulting expression 
for the Higgs mass is
equivalent to the one obtained by performing
an expansion of the effective potential
in powers of the Higgs field~$\phi$. Keeping only operators up to order 
four in the effective potential,
eq.(\ref{hlmassimproved})  reproduces
the expression of eq.(\ref{upperb}). This comparison
holds up to small differences associated with the 
treatment of the effects due to 
the weak gauge couplings in the one-loop effective potential,
and with the inclusion of the top Yukawa effects in the 
relation between the pole and running top-quark mass \cite{RC4,HHH}.
A more detailed treatment of the dependence of the Higgs mass on the
weak gauge couplings may be also found in Ref.\cite{CQW}.

Fig.\ref{fig2} shows the comparison between the
results of the analytical approximation, 
eq.(\ref{mtrun}), and
the one-loop RG improvement to the full one--loop leading--log
diagrammatic calculation.  In general, the
prescription given in eq.(\ref{mtrun}) reproduces the full
one-loop RG-improved Higgs masses to within 2 GeV for top-squark
masses of 2~TeV or below.  The dashed line in the figure shows the
result that would be obtained by 
ignoring the RG-improvement; it reflects  the relevance of the
two-loop effects in the evaluation of the Higgs mass.
\vvs1

\noindent\underline {The Case $m_A \simlt M_S$.}
A similar RG improvement of the effective potential
method to the one already discussed
can be applied to calculate all the masses and couplings 
in the more general case  of a light CP-odd 
Higgs boson $m_A\simlt M_{S}$. As above,  the finite
one-loop threshold corrections to the quartic couplings
at the scale $M_S$ at which the heavy 
squarks decouple \cite{HH} are also taken into account. 
The effective theory below the scale $M_{S}$ \cite{RC3,HH} is a 
two-Higgs 
doublet model where the tree--level quartic couplings can be written in terms
of dimensionless parameters $\lambda_i$, $i=1,\dots ,\ 7$, whose tree--level
values are functions of the gauge couplings. The  one-loop
threshold corrections $\Delta\lambda_i$, $i=1,\dots ,\ 7$, are expressed as
functions of the supersymmetric Higgs mass $\mu$ and the soft supersymmetry
breaking parameters $A_t$, $A_b$ and $M_S$~\cite{HH}. 
An analytical  approximation, which reproduces the previous one--loop 
RG improved results for all values of $\tan \beta$ and $m_A$, 
can also be derived. For example, generalizations of Eq. (\ref{hlmass})
can be found in Refs.\cite{RC4,HHH}.
The CP-even light and heavy Higgs masses 
and the charged Higgs mass are given as functions of
$\tan \beta$, $M_{S}$, $A_t$, $A_b$, $\mu$, the CP-odd Higgs mass $m_A$
and the physical top-quark mass $M_t$ related to the on-shell running
mass $m_t$ through eq.(\ref{mtMt}).  The analytical expressions for the
masses and mixing angle of the Higgs sector as a function of the parameters
$\lambda_i$ are 
presented in Appendix \ref{app:H1}.
These expressions are the analogue of eq.(\ref{mhsm}) for the case
in which two-Higgs doublets survive at low energies. 
Effects of the bottom Yukawa coupling, which may become large for values  
$\tan\beta \simeq m_t/m_b$  ($m_b$ being the running
bottom mass at the scale $M_t$), are also included. 
A subroutine implementing this method is available \cite{subh}.
\vvs1
\begin{table}[htbp]
\caption{\it MSSM Higgs couplings relevant at LEP2.}
\label{table:couplings}
\begin{center}
\begin{tabular}{||c|c||}\hline
Vertex &  Coupling \\ \hline 
  &   \\
$\{h,H\}W_{\mu}W_{\nu}$ &${\displaystyle i2(G_F \sqrt{2})^{1/2}m_W^2
g_{\mu\nu} \{\sin(\beta-\alpha),
\cos(\alpha-\beta)\} }$\\\  & \\
$\{h,H\}Z_{\mu}Z_{\nu}$ & ${\displaystyle i2(G_F \sqrt{2})^{1/2}\frac{m_W^2}
{\cos^2\theta_W}g_{\mu\nu} \{\sin(\beta-\alpha),\cos(\alpha-\beta)\} }$\\\
 &  \\
$\{h,H,A\}u\overline{u}$ & ${\displaystyle -i
\frac{(\sqrt{2} G_F)^{1/2} m_u}{\sin \beta}
\{\cos\alpha,\sin\alpha,-i\gamma_5\cos\beta\}  }$\\\
 & \\
$\{h,H,A\}d\overline{d}$ & ${\displaystyle -i
\frac{(\sqrt{2} G_F)^{1/2} m_d}{\cos\beta}
\{-\sin\alpha,\cos\alpha,-i\gamma_5\sin\beta\}  }$\\\
& \\
$\{h,H\} A Z_{\mu}$ & ${\displaystyle - (\sqrt{2} G_F)^{1/2} m_Z (p + k)_{\mu}
\{\cos(\beta - \alpha),-\sin(\beta-\alpha)\} }$\\  &  \\
\hline
\end{tabular}
\end{center}
\end{table}

\begin{figure}[p]
\begin{center}
\epsfig{file=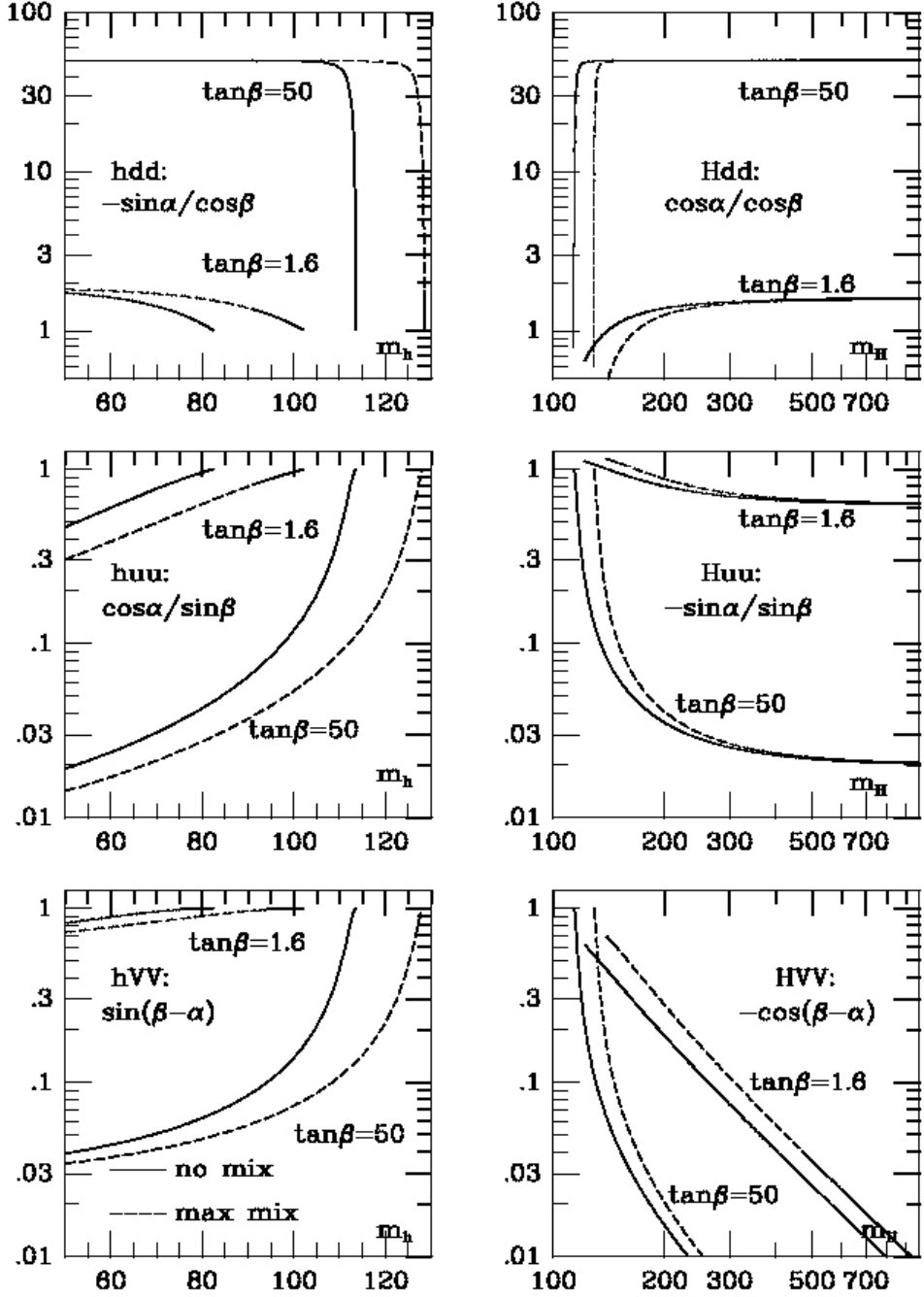, 
          bbllx=55, bblly=150, bburx=505, bbury=790,
          height=19.5cm, clip=}
\end{center}
\caption[0]{\it 
  MSSM Higgs couplings normalized to the SM couplings $g^{SM}_{Hff} =
  [ \sqrt{2} G_F]^{1/2} m_f$ and $g^{SM}_{HVV} = 2 [\sqrt{2}
  G_F]^{1/2} m_V^2$.}
\label{shgs:fig-couplings}
\end{figure}

\noindent \underline{Couplings}.
Notice that the radiatively corrected quartic couplings
$\lambda_i$, $i=1,\dots, 7$, and hence  the corresponding
 value of the Higgs mixing
angle $\alpha$ as given in Appendix \ref{app:H1},
 permit us to   evaluate  all radiatively corrected Higgs couplings. For
instance, the   Yukawa and gauge Higgs couplings  relevant for LEP2
energies  are listed in  Table \ref{table:couplings} 
[$p_{\mu}$ ($k_{\mu}$) is the incoming (outgoing) CP-odd 
(CP-even) Higgs momentum].  
The size of the couplings of the
two scalar Higgs bosons to fermions and a gauge boson
are shown in Fig.\ref{shgs:fig-couplings} \cite{DKZ}.
For fermions the charged Higgs particles couple to
mixtures of scalar and pseudoscalar currents, with components
proportional to $m_u\ctb$ and $m_d\tb$ for the two $\pm$
chiralities.  The couplings to left(right)-handed ingoing $u$ quarks
are given by $g_{H^+\bar du_{L(R)} } = [ \sqrt{2} G_F]^{1/2} 
m_u\ctb$ $(m_d\tb)$.  For large $\tb$ the down--type mass defines
the size of the coupling; for small to moderate $\tb$ it is 
defined by  the up--type
mass.
Furthermore, the trilinear Higgs
couplings can be explicitly written as functions of 
$\lambda_i$, $\alpha$ and $\beta$~\cite{HH,HHN}. 

\noindent {\bf d) Renormalization Group Improvement of
the Effective Potential: 
General Third--Generation Squark Mass Parameters.}
The above one--loop RG improved treatment of the effective
potential relies on the definition of an effective supersymmetric threshold 
scale,  $M_{S}^2 = (m^2_{\tilde{t}_1} + m^2_{\tilde{t}_2})/2$.
Its validity is therefore  restricted to the case
of small differences between the squark mass eigenvalues.
Quantitatively, the method is valid if
$(m^2_{\tilde{t}_1} - m^2_{\tilde{t}_2})/
 (m^2_{\tilde{t}_1} + m^2_{\tilde{t}_2}) \leq 0.5$. Furthermore,
 all the RG Higgs analyses performed in the literature, besides 
Ref.\cite{CQW}, rely on the expansion of the effective potential up to 
operators of dimension four. However, to safely neglect higher dimensional
operators, the conditions $2|M_t A_t| \leq  M_S^2$
 and $2|M_t \mu| \leq M_S^2$  
must be fulfilled.

\vfill

\newpage

The case of large splitting in the stop sector is particularly interesting 
in the light of recent measurements of $R_b \equiv
\Gamma (Z \rightarrow b \bar{b})/ \Gamma (Z \rightarrow {\rm hadrons})$,
whose discrepancy of more than 3 standard deviations with the SM prediction 
can be ameliorated in the presence of a light higgsino together with a
light right--handed stop (see the discussion in 
the chapter on New Particles) \cite{RbmA}-\cite{johnE}.
 The left--handed stop must instead remain reasonably
 heavy to avoid undesirable contributions to the $W$ mass and the $Z$ 
leptonic width. 
It is hence important  to generalize  the results previously  obtained
by using the
renormalization group improved 
one-loop effective potential, to the case
of general values of the left-- and  right--handed squark masses 
 and mixing parameters, $m_Q$, $m_U$, $m_D$, $A_t$ and $A_b$, respectively.
In this case the contribution of higher
dimensional operators to the effective potential must be properly
taken into account; hence, the naive treatment in terms of quartic 
couplings is  no longer appropriate.

In Ref.\cite{CQW},  a method has been developed 
for the neutral Higgs sector of the theory, in  which 
 each stop and sbottom particle is decoupled at its corresponding mass scale. 
Threshold effects, associated with the
decoupling of the heavy sparticles, 
are frozen at the decoupling scales; they 
evolve, in the squared mass matrix, with the anomalous dimensions of the
Higgs fields. The threshold effects achieve a complete
matching of the effective potential for scales above and below the 
decoupling scales, and include all higher order (non--renormalizable) terms 
arising from the whole MSSM effective potential. The dominant leading--log
contributions in the expressions of the  
renormalized Higgs quartic couplings  involve
the scale dependent contributions to the effective potential and are treated
in the same way as in the RG improved approach described above.
The way to proceed in evaluating  the CP-even Higgs mass values and
mixing angle $\alpha$ is explained in detail in Ref.\cite{CQW}.
A subroutine implementing the method is  available \cite{subhpole}.
This approach  makes contact with the computation of the Higgs masses 
by means of the effective potential performed in Ref.\cite{ERZ}.
Moreover, it
reproduces the results of Ref.\cite{RC4} for
 small mass splitting of the squark masses. This comparison holds up
to a tiny difference coming from the inclusion 
of the small dependence of the one-loop radiative corrections 
on the weak couplings and the vacuum polarization effects. 
Indeed, in Ref.\cite{CQW} the definition of pole Higgs masses
is introduced by including 
the most relevant vacuum polarization effects.
The gaugino corrections, which  are relatively small, 
have been also included by incorporating (only) the one-loop 
leading logarithmic contributions.


\subsubsection{Results}

\begin{figure}[htp]
\vspace{-1cm}
\centerline{
\epsfig{figure=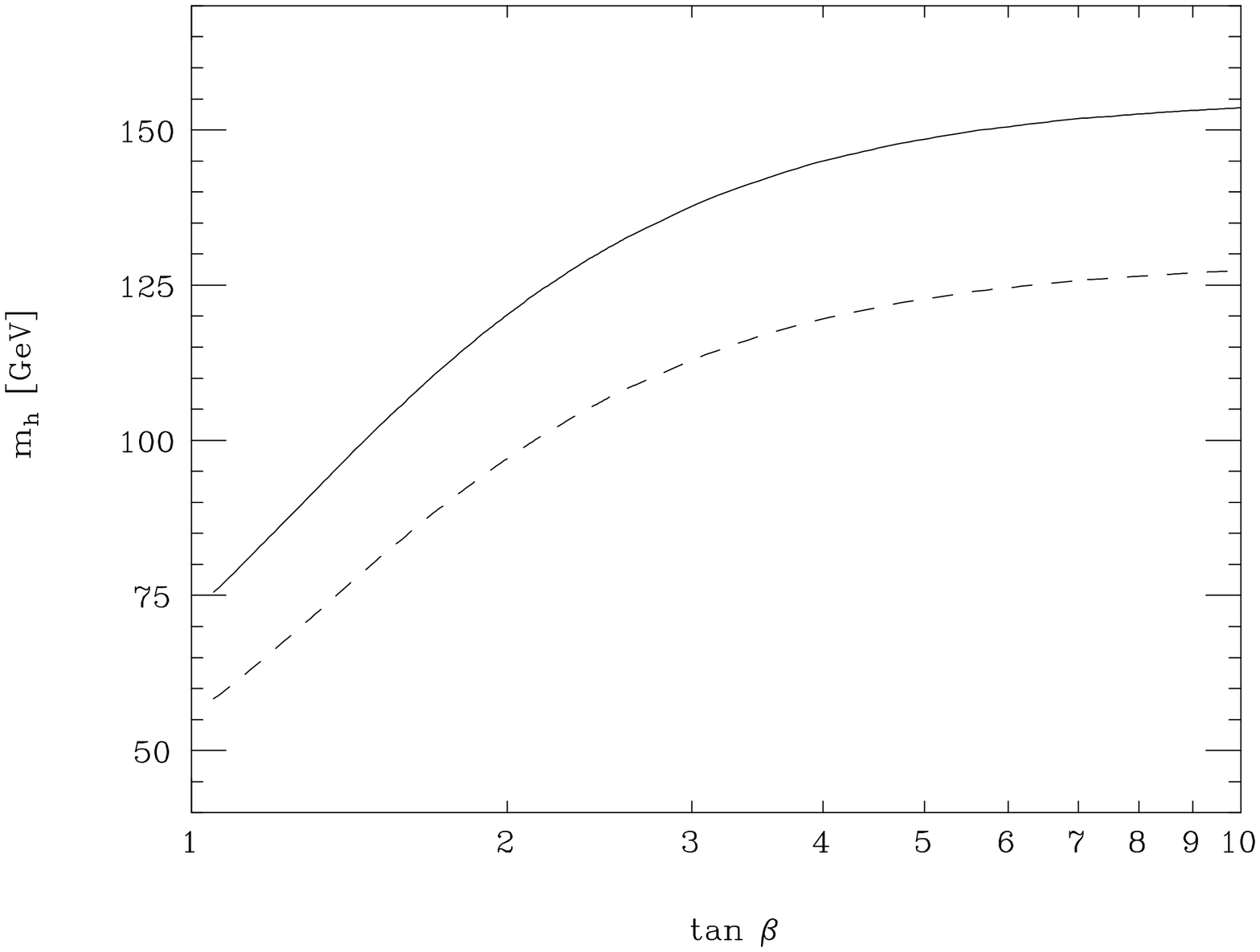,height=13cm,angle=90} }
\vspace{-1.3cm}
\caption[0]{\it
  Upper limit on the mass of the lightest neutral Higgs boson mass
  $m_h$ as a function of $\tan\beta$ for zero mixing (dashed line) and
  for the maximal impact of mixing in the stop sector (solid line);
  $M_S= $ 1 TeV.}
\label{fig:mhtb}
\vvs2
\centerline{
\epsfig{figure=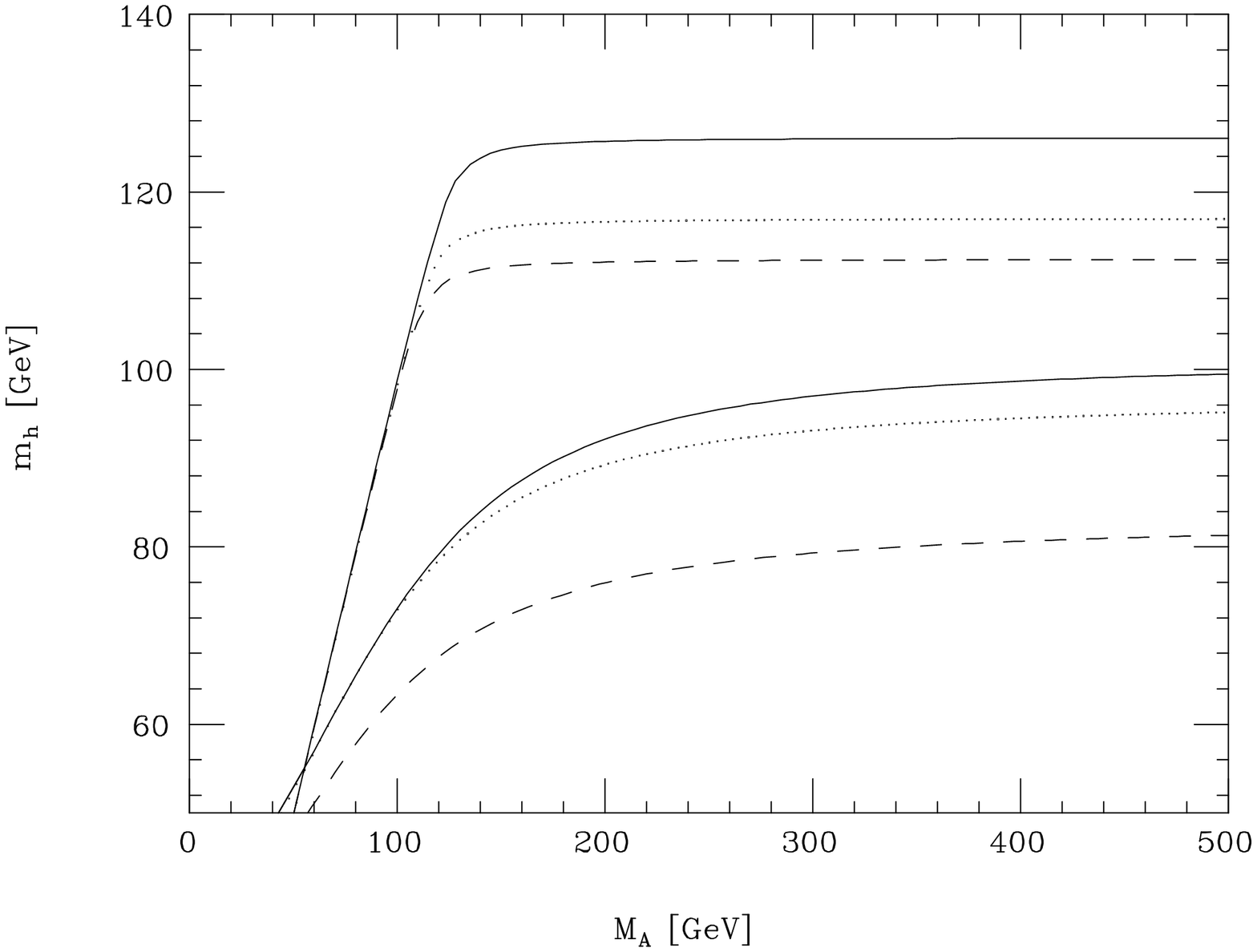,height=13cm,angle=90} }
\vspace{-1.5cm}
\caption[0]{\it
  Lightest neutral Higgs boson $h$ in the MSSM as a function of $m_A$
  for zero mixing (dashed line), for intermediate mixing (dotted line)
  and for the maximal impact of mixing in the stop sector (solid
  line); for two values of $\tan \beta$ = 1.6 (lower set), 15 (upper
  set): $M_S= $ 1 TeV and $M_t$ = 175 GeV.}
\label{fig:amutalk}
\end{figure}

\begin{figure}[htp]
\vspace{-1.3cm}
\centerline{
\epsfig{figure=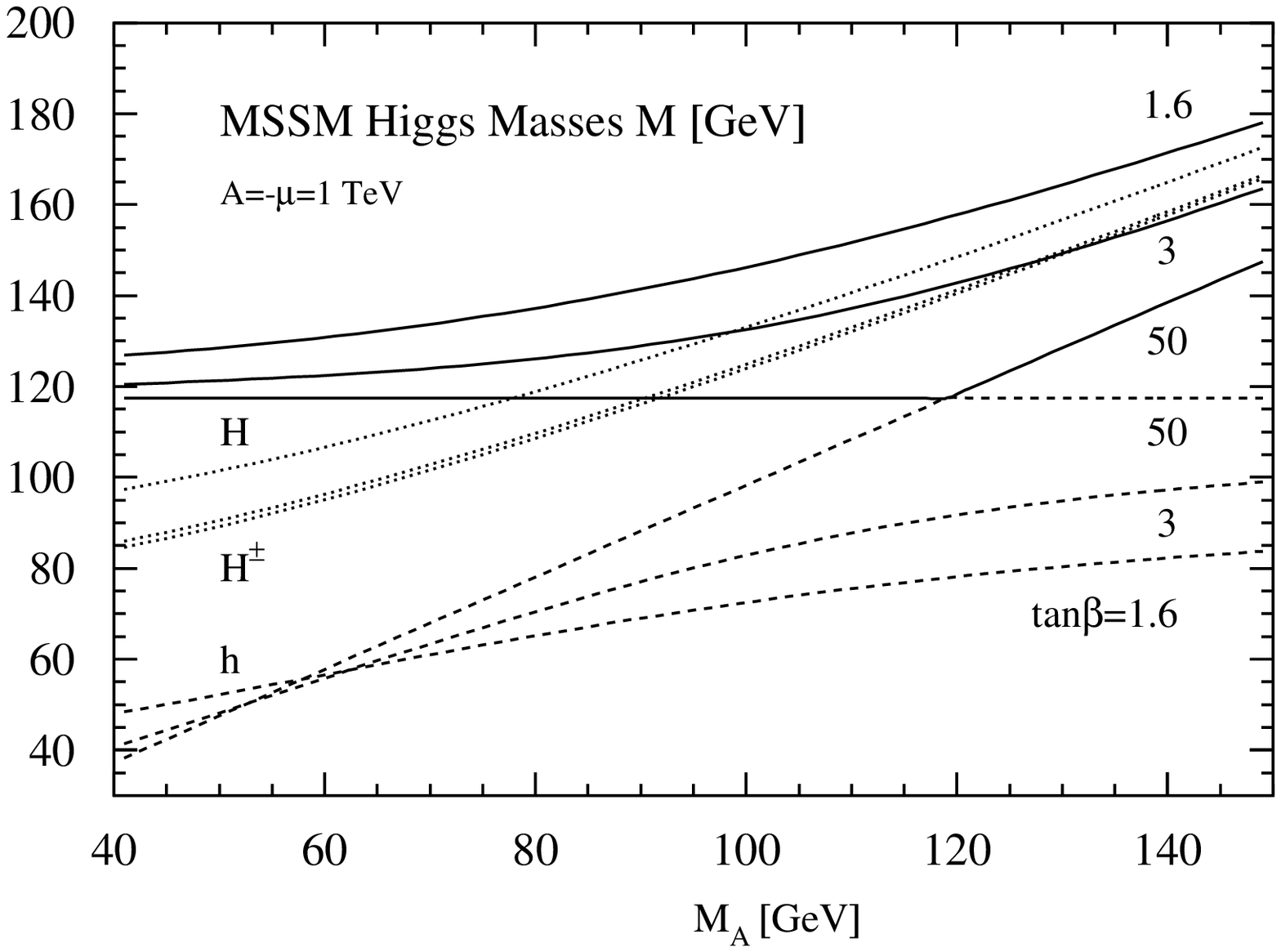,height=11cm,angle=0} }
\vspace{-1.cm}
\caption[0]{\it
  Lightest CP--even Higgs boson mass (dashed line), heaviest CP--even
  Higgs mass (solid) and charged Higgs mass (dotted line) in the MSSM
  as a function of $m_A$ for $A=-\mu = M_S = $ 1 TeV, $M_t$ = 175 GeV
  and different values of $\tan \beta$}
\label{fig:masses2}
\end{figure}

The lightest CP--even Higgs mass  is a 
monotonically increasing function of $m_A$, which in the low $\tan \beta$
regime converges to its maximal value for $m_A \simgt$ 300 GeV.   
In Fig.\ref{fig:mhtb} the upper
limits on the lightest CP-even
Higgs mass $m_h$ [realized in the large $m_A$ limit] 
are shown as a function of
$\tan\beta$. Since the radiative corrections to the
Higgs mass  depend on the fourth power of the top mass,
 the maximal top-quark
mass compatible with perturbation theory up to the
GUT scale has been adopted for each value of $\tan \beta$. 
Apart from the natural choice of the mixing mass--parameters 
and the scale $M_S$,
this result is the most general upper limit on
$m_h$ for a given value of $\tan \beta$ in the MSSM.
The variation of  the upper bound on $m_h$ as a function of $M_t$ is 
shown by the solid line (a) of Fig.\ref{fig:mhmt}.
In  Fig.\ref{fig:amutalk} the mass $m_h$ is plotted for different
values of the mixing parameters $A =A_t$ and $\mu$.  In fact, 
$A \simeq |\mu| \ll M_S$ yields the case of
negligible  squark mixing, while $A = \sqrt{6} M_S$,  $|\mu| \ll M_S$
characterizes the case of large mixing [i.e.\  the impact of stop
 mixing in the radiative corrections is maximized]; $A = -\mu = M_S$ 
 yields moderate mixing for large $\tan \beta$ while the mixing effect 
is close to maximal for low $\tan \beta$. 
 In  Fig.\ref{fig:masses2} we show the masses of the two CP--even
Higgs bosons and of the charged Higgs boson as a function of $m_A$
for the case $A=- \mu = M_S$ = 1 TeV, $M_t= $ 175 GeV and different
values of $\tan \beta$. The peculiar behavior of $m_h$ and $m_H$
for large $\tan \beta$ will be explained in the following.

In general, 
for very {\it large values of $\tan\beta$}
and values of $\mu$, $A_t$ and $A_b$ of order or
smaller than $M_S$, the mixing
in the Higgs sector is negligible and the CP-even Higgs mass
eigenstates are approximately given by $H_1$ and $H_2$. As a result,
the properties of h and H mainly depend on the value of $m_A$.
For $m_A^2 >   2 v^2 \lambda_2 \equiv m_Z^2 + {\rm rad. corr.}$, one 
approaches the decoupling limit and 
 the relations
$\sin\alpha  \simeq  - \cos\beta$ and
$\cos\alpha  \simeq  \sin\beta$ hold.
Hence, the  CP-even Higgs mass eigenstates are given by
$ h   \simeq  \sin\beta H_2 + \cos\beta H_1 \simeq H_2$ and
$H   \simeq  \sin\beta H_1 - \cos\beta H_2 \simeq H_1$.
In this case the lightest 
CP-even Higgs couples
to up (down) fermions as
\begin{equation}
\label{eq:uuh}
h u \bar{u} \rightarrow h_u \sin\beta \;\;\;\; {\rm and } \;\;\;\;
 h d \bar{d} \rightarrow h_d \cos\beta 
\end{equation}  
 where  $h_u \sin\beta$  
$(h_d \cos\beta)$ is  the SM coupling $h_u^{SM}$ ($h_d^{SM}$) [Observe that
$h_f^{SM} = g_{Hff}^{SM} \sqrt{2}$, with $f=u,d$].
The heaviest
 CP-even Higgs boson, instead,  couples in highly non-standard way
to fermions,
\begin{equation}
\label{eq:Huu}
H u \bar{u} \rightarrow h_u \cos\beta = h^{SM}_u \cot \beta 
 \;\;\;\; {\rm and } \;\;\;\; H d \bar{d} \rightarrow
 h_d \sin\beta = h^{SM}_d \tan \beta 
\end{equation}
so that the coupling to up (down) quarks is highly suppressed 
(enhanced) with respect to the coupling in the Standard Model.     
For $m_A^2 <  2 v^2 \lambda_2$  instead,
$
\sin\alpha  \simeq - \sin\beta$ and
$\cos\alpha  \simeq  \cos\beta$.
Hence, the  CP-even Higgs mass eigenstates are given by
$h   \simeq  \cos\beta H_2 + \sin\beta H_1 \simeq H_1$ and 
$H   \simeq  -\sin\beta H_2 + \cos\beta H_1 \simeq - H_2$.
In this case the situation is interchanged; $h$ has the  non-standard
 type of couplings to fermions, eq.(\ref{eq:Huu}),
 and $H$ has the SM couplings,
eq.(\ref{eq:uuh}).

The values of the CP-even Higgs masses depend on  the size of the
$H_2$ or $H_1$ component. 
When the Higgs is predominantly $H_2$, 
its mass is given by eq.(\ref{upperb}) for $|\cos 2 \beta| = 1$, 
neglecting the small bottom--quark Yukawa effects. 
When the Higgs is predominantly $H_1$, instead, its mass is given by
$m_A$. Hence, the mass of the lightest Higgs boson is  given by
$m_h^2 \simeq m_A^2$ (and non-standard couplings to fermions) if
$m_A^2 \leq 2 v^2 \lambda_2 $, and it is given  by eq.(\ref{upperb})  
for larger $m_A$, for which the couplings to fermions are SM-like. 
The complementary situation occurs for $H$ and this can clearly be observed
in Fig.\ref{fig:masses2}.

\begin{figure}[htp]
\vspace{-1cm}
\centerline{
\epsfig{figure=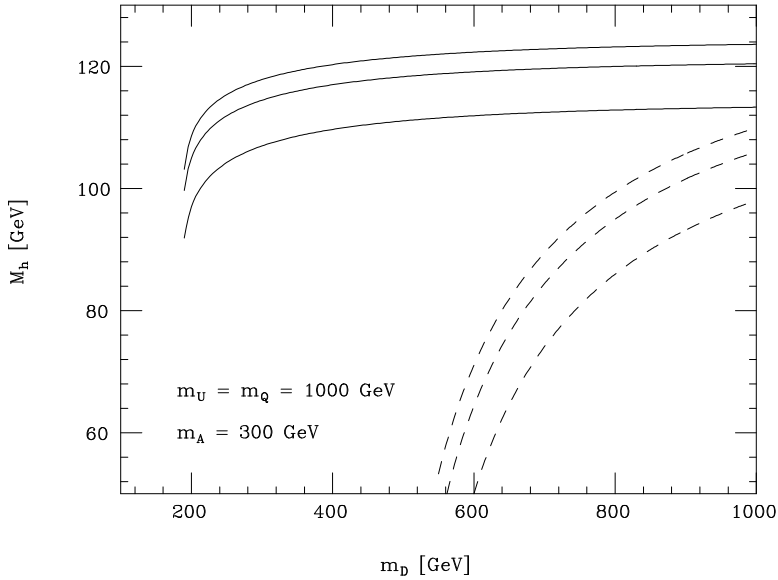,height=13cm,angle=90} }
\vspace{-1.5cm}
\caption[0]{\it 
  Plot of the pole Higgs mass $M_h$ as a function of $m_D$, for $M_t$
  = 175 GeV, $ \tan \beta $ = 60, $A_b = 0$, $m_U = m_Q $ = 1 TeV,
  $A_t = $0. 1.5, 2.4 TeV (from bottom to top) and $\mu = $ 1 TeV
  (solid curves), $\mu = $ 2 TeV (dashed curves).}
\label{figef9}
\end{figure}

The effects of the bottom quark are
only relevant in the limit of large $\mu$ parameters. For 
values of $\mu$ larger than $M_{S}$ relevant  corrections,
which are  dependent on the bottom mass, enter the 
Higgs mass formulae. This can
be easily understood
in the case $m_Q = m_U = m_D = M_S$,
by studying the dependence of $\lambda_2$ 
on the supersymmetric Yukawa coupling $h_b$
[see appendix \ref{app:H1}]. For values of $h_b$ of
order of $h_t$, or equivalently for $\tan\beta \simeq m_t/m_b$, 
$\lambda_2$ depends significantly on the fourth power
of the $\mu$ parameter. These radiative corrections are negative,
lowering the mass of the CP-even Higgs associated
with the $H_2$ doublet. Fig.\ref{figef9} shows the case of large $m_A$.

For large values of $\mu$ and small values of
$m_A$, the charged Higgs mass also receives  large negative 
radiative corrections, which grow as the fourth power of the
$\mu$ parameter. Hence, large negative corrections to the
charged Higgs mass may be obtained. Such large values of
$\mu$, however, may be in conflict with the stability of the
ordinary vacuum state.

\subsubsection{%
Additional Constraints: $b$-$\tau$ Unification and Infrared
Fixed Point Structure}

The MSSM can be derived as an effective theory in the framework
of supersymmetric grand unified theories. In addition to the
unification of gauge couplings,
the unification of the $b$ and $\tau$ Yukawa couplings,  $h_b(M_{GUT}) 
= h_{\tau}(M_{GUT})$,
 appears naturally in most grand unified scenarios.
 Given this additional constraint,
the experimental values of the
$b$ and $\tau$ masses at low energies determine the value of $M_t$ as a
function of $\tan \beta$ \cite{R13D,R17,Yukawa}.
In fact, for the present experimental range of the top-quark mass
$M_t = 180 \pm 12 $ GeV~\cite{CDF}, 
the condition of $b$-$\tau$ unification implies either low values of
$\tan \beta$, $ 1\leq \tan \beta \leq 3$, or very large values 
of $\tan \beta = {\cal{O}} (m_t/m_b) \simeq 50$ \cite{R13D,R17}-\cite{R18}.
To accommodate $b$-$\tau$ unification,
large values of the top Yukawa coupling
are necessary in order to compensate for the effects of the  
renormalization by strong interactions in the running of the 
bottom Yukawa coupling. Large 
values of $h_t^2(M_{GUT})/4 \pi \simeq$ 0.1--1 
ensure the attraction towards the infrared (IR) 
fixed point solution for the top
quark mass \cite{IRold}. 
The strength of the strong gauge coupling as well as 
the experimentally allowed range of the bottom mass play a decisive role
in this behavior \cite{Yukawa}-\cite{R18}. 
In the low $\tan \beta$ case,
for the presently allowed values of the electroweak parameters and of the
 bottom mass and for values of $\alpha_s(m_Z) \simgt 0.115$,  
$b$-$\tau$ unification implies that the top-quark mass
must be within ten percent of its infrared fixed point 
values.
A mild 
relaxation of exact unification [0.85-0.9 $ \leq 
h_b/h_{\tau}|_{M_{GUT}} \leq$ 1.15]
still preserves this feature, especially for 
values of $M_b \leq 4.95$ GeV. In the large $\tan \beta$ region,
$h_b$ is ${\cal{O}}(h_t)$ and
 the infrared fixed point attraction, 
within the context of b-$\tau$ Yukawa coupling unification,  is much weaker.

The top-quark mass is also predicted to be close to
its infrared-fixed point value
in string scenarios, 
in which the top-quark Yukawa
coupling is determined by minimizing the effective potential with respect to
moduli fields~\cite{fabio}.
Quite generally, the fixed point solution, $h_t=h_t^{IR}$,
is obtained for large values of the top Yukawa 
coupling  at high energy scales, which however remain in the
 perturbative regime.
Within the framework of grand unification, one obtains 
$(h_t^{IR})^2/ 4 \pi \simeq
(8/9) \alpha_s(m_Z)$ for $M_{GUT} \simeq 10^{16}$ GeV, 
and the running top-quark mass tends to its
infrared fixed point value $m_t^{IR} = h_t^{IR} v \sin \beta$.
 Hence, relating the running top-quark mass $m_t$
 with the pole
top-quark mass $M_t$ by taking into account
the appropriate QCD corrections  we arrive in the low 
$\tan \beta$ regime at \cite{CW1}, 
\bea 
M_t^{IR} &\simeq & \sin \beta \; 
 [1+ 2\left(\alpha_s(m_Z)-0.12 \right)] 
\left[1+\frac{4 \alpha_s(m_Z)}{3 \pi} +                                         {\cal{O}}(\alpha_s^2) \right] \times  
196~{\rm GeV}
\label{eq:IR}
\eea
The strong $M_t$--$\tan \beta$ correlation associates
with each value of $M_t$ at the infrared fixed point
 the lowest value of $\tan\beta$ 
consistent with the validity of perturbation theory up to scales
of order $M_{GUT}$. 
If  the physical top-quark mass  is  in
the range 160--190 GeV, the values of $\tan\beta$ are restricted
to the interval
between 1 and 3. This is in agreement with the results from
b-$\tau$ Yukawa unification. 

The infrared fixed point solution can also be analysed 
in the large $\tan \beta$ case, where the effects
 of the bottom Yukawa coupling need to be taken into account
in the RG evolution as well. For instance, if the
values of the supersymmetric Yukawa couplings of the bottom and top quarks
are very close to each other, $m_t(M_t) \simeq m_b(M_t) \tan \beta$,
the  infrared fixed point prediction for the top-quark mass
is reduced by a factor $\sqrt{6/7}$
with respect to  eq.(\ref{eq:IR}) \cite{COPW,fabio1}. Still,   
the values of $M_t$ predicted in this regime are
about  190 GeV. 
\vvs1

After the above  general discussions we shall describe
their consequences for the Higgs sector:\\[2mm]
(i) The infrared fixed point structure in the low $\tan\beta$
region  have far-reaching consequences
for the lightest CP-even  Higgs mass  in the MSSM \cite{Diaz}-\cite{COPW}. 
Indeed, for $\tan\beta$ larger than one, the lowest tree--level Higgs mass
is obtained  at the lowest value of $\tan\beta$. Hence, in any theory
consistent with perturbative unification, the fixed point solution
is associated with the lowest value of the tree--level mass consistent
with the theory. Even after including radiative corrections,
the upper bound on the Higgs mass is considerably reduced at the
fixed point solution: for a top mass of 175 GeV, the upper limit
of the Higgs mass
is less than 100 GeV, while for $M_t = 160$ GeV, it is even
less than 80 GeV (see Fig.\ref{fig:mhmt}). 
Hence, 
if the infrared  fixed point solution for the top-quark with 
$M_t \simlt 175$ GeV  
is realized in nature, the lightest CP-even Higgs mass must be accessible  
at LEP2 for $\sqrt{s} =$ 192 GeV \cite{interim,RC4}. 
Fig.\ref{fig:mhmt} shows also that for $M_t $= 175 GeV
the upper bound on the lightest Higgs mass in the case of $b$-$\tau$
Yukawa coupling unification 
is nearly 25 GeV smaller than the
unrestricted MSSM limit.

\begin{figure}[htp]
\vspace{-1cm}
\centerline{
\epsfig{figure=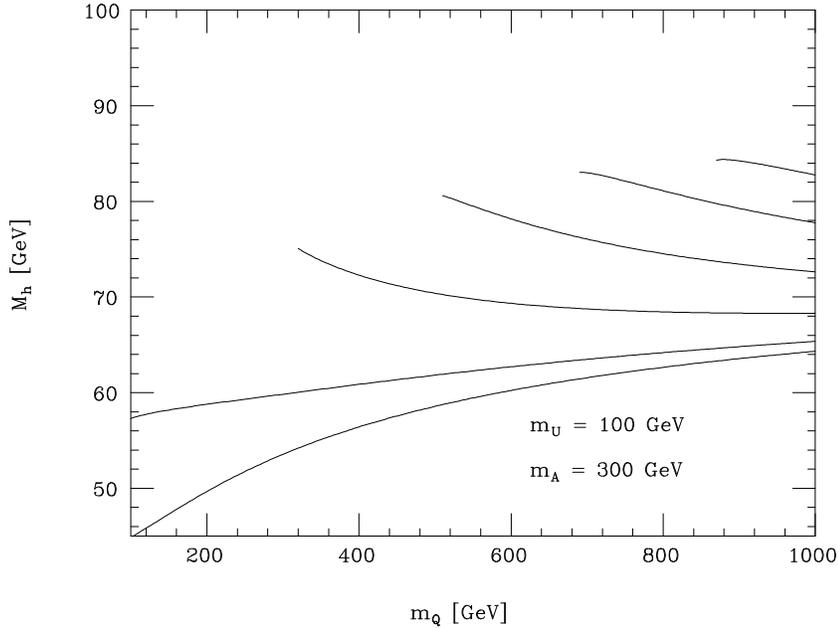,height=14cm,angle=90} }
\vspace{-1.5cm}
\caption[0]{\it 
  Plot of the pole Higgs mass $M_h$ as a function of $m_Q$, for $M_t$
  = 175 GeV, $ \tan \beta $ = 1.6, $\mu = A_b = 0$, $m_U = m_D $ = 1
  TeV and $m_A = 300~GeV$. The lines denote different values of the
  $A_t$ parameter. Starting from below at $m_Q$ = 1 TeV, $A_t = $ 0,
  0.2, 0.4, 0.6, 0.8, 1 TeV, respectively.}
\label{figef7}
\end{figure}

The present  data indicate that the
value of $R_b = \Gamma_b/\Gamma_h$ is more than $3\sigma$
above the SM prediction for this quantity. Large positive
radiative corrections to $R_b$ are always associated with large values
of the Yukawa couplings; they are therefore 
maximized at the infrared fixed 
point solution \cite{Gordy,CW1}.
 Moreover, presicion measurements also provide information
about the structure of the soft supersymmetry breaking terms: Low values
of the right--handed \mbox{SUSY} breaking stop mass $m_U$ and of the 
\mbox{SUSY} mass parameter $\mu$ are preferred, while
the left--handed stop mass parameter $m_Q$ must be
larger than $m_U$. For a fixed large value of $m_Q$, the upper bound on the
Higgs mass is significantly lower in the case $m_U \ll m_Q$ than in
the case $m_U \simeq m_Q$. 
Fig\ref{figef7}  shows the Higgs mass as a function of $m_Q$ for 
$m_U = 100$ GeV, $m_A = 300$ GeV and $\tan\beta$ 
consistent with the fixed point solution for $M_t = 175$ GeV, 
for different values of the mixing mass parameter
$A_t$.  Even for the largest value of $A_t$ physically acceptable
[i.e.\ $m_{\tilde{t}}$ above the experimental
lower bound], the Higgs mass remains below 85 GeV. Hence, for the values of the
supersymmetry breaking mass terms preferred by the precision electroweak
data,  associated with
a light right-handed stop, lower values of $m_h$ than na\"{\i}vely 
expected are
obtained.

 Furthermore, the most general upper bounds on $m_h$ at the infrared fixed
point are valid for  very large values of the
mixing parameters in the squark sector [$A_t$, $A_b$ and $\mu$] 
which in general are hard to
realize.
Requiring radiative breaking of the electroweak
symmetry [yet no colour breaking],
and imposing the boundary conditions from
experimental SUSY mass limits, the range of upper values of $m_h$  is
reduced further \cite{CW1}.
 In the general framework of
 supergravity models, various analyses have been performed in the literature
 to study the spectrum of a constrained 
 MSSM at different levels of refinement~\cite{CMSSM}.


(ii) The condition of
$b$-$\tau$ Yukawa coupling unification is also consistent with
the values of the top-quark mass measured at the Tevatron 
for very large values of $\tan\beta \simeq m_t/m_b$.
There are, however, large uncertainties in this sector associated with 
one--loop supersymmetric corrections to the bottom mass. 
These radiative corrections are strongly
 dependent on the structure of the supersymmetric spectrum and induce 
 strong 
 variations in the predictions for the top-quark mass and $\tan \beta$,
 once the unification of the $b$ and $\tau$ Yukawa couplings is implemented.
Nevertheless, the large $\tan \beta$ regime 
with unification of the $b$-$\tau$ Yukawa couplings,
although more model dependent,
provides an interesting framework for Higgs particle searches at LEP2.

Large positive radiative corrections to $R_b$ can also be obtained for large
values of $\tan\beta$, since the supersymmetric
bottom--quark Yukawa coupling is enhanced in this regime. 
Indeed, the value of $R_b$ can be significantly increased
 if the CP-odd Higgs mass is below
70 GeV \cite{RbmA}-\cite{Joan}. 
This is a result of the large positive one-loop  corrections 
associated with the neutral CP-odd Higgs scalar sector of the 
theory. 
 Low values of the
CP-odd Higgs mass, $m_A \simeq m_Z$, 
 imply that both the lightest
CP-even and the CP-odd Higgs masses would be at the reach of LEP2.
The charged Higgs mass is approximately determined through the 
CP-odd Higgs mass value, $m_{H^{\pm}}^2 \simeq m_A^2 + m_W^2$, and 
hence, strong constraints on $m_A$ are obtained from the
charged Higgs contributions to ${\rm BR}(b \rightarrow s \gamma)$.
Even conservatively taking into account the QCD uncertainties
associated with the branching ratio 
${\rm BR}(b \rightarrow s \gamma)$ [i.e.\ assuming e.g. 40$\%$ QCD 
uncertainties], 
the $b \rightarrow s \gamma$ decay rate becomes larger than the
presently allowed experimental values \cite{bsgexp} for  $m_{H^{\pm}}
\simlt 130$ GeV,    unless the
supersymmetric particle contributions suppress the 
charged Higgs enhancement of the decay rate.
The most important supersymmetric
contributions to this rare bottom decay
come from the chargino-stop one-loop diagram \cite{bsga}. 
The chargino contribution to the $b \rightarrow s \gamma$
decay amplitude depends on the soft supersymmetry breaking
mass parameter $A_t$ and on the supersymmetric mass parameter
$\mu$. For very large values of $\tan\beta$, it is
given by 
\be
A_{\tilde{\chi}^+} \simeq \frac{m_t^2}{m_{\tilde{t}}^2}
\frac{A_t \mu}{m_{\tilde{t}}^2} \tan\beta \; G\left(
\frac{m_{\tilde{t}}^2}{\mu^2}\right)
\ee
where $G(x)$ is a function with values of order unity
when the characteristic stop mass $m_{\tilde{t}}$
is of order $\mu$, and it grows as
$\mu$ decreases. For positive (negative)
values of $A_t \times \mu$ the chargino contributions are of
the same (opposite) sign as the charged--Higgs contributions. Hence,
to partially cancel the light charged--Higgs contributions and render
the  $b \rightarrow s \gamma$
decay rate acceptable, negative values for $A_t \times \mu$ are required.
This requirement
has direct implications on the corrections to the bottom mass 
mentioned above
 and  puts strong constraints on models with unification of the
Yukawa couplings \cite{wefour,we5}.

\subsubsection{MSSM Parameters}

In the experimental simulations, we have chosen as
 the two basic parameters of the Higgs sector 
 the mass $m_A$ of the pseudoscalar Higgs boson within
the limits  40 GeV $\leq m_A \leq$ 400 GeV,
and the  angle $\beta$ within the bounds $
1 \leq \tan\beta
\leq m_t(M_t)/m_b(M_t) \simeq 60 $.
 The upper limit on $m_A$ is introduced
merely for convenience,
since the variation of $m_h$ with $m_A$ becomes negligible for values of
$m_A \geq $ 200--250 GeV.  The upper value of $\tan \beta$ is chosen
such that the bottom Yukawa coupling 
remains in the perturbative regime for scales
below the grand unification scale. 
A given value of $\tan\beta$ implies an upper limit on the
top mass for which the theory can be extended perturbatively up to
the GUT scale.
For $\tan \beta$ = 1 this upper limit is already close to 150 GeV so that
lower values of $\tan \beta$ would be inconsistent with values of the top
quark mass in the experimental range.
In the examples we shall discuss, we have chosen:
\vspace*{-.5\baselineskip}
\begin{itemize}
\item[(i)]   Top mass, $M_t = 175 \pm 25$ GeV;
\item[(ii)]  SUSY scale, $M_S = 10^3$ GeV;
\item[(iii)] SUSY Higgs mass parameter $\mu$ and
             soft SUSY breaking parameter $ A_t = A_b = A$:\\
             $A = 0$  \mbox{and} $|\mu| \ll M_S\;\;$ 
             \hspace*{8.25mm} [no mixing];\\
             $A = \sqrt{6} M_S$ and $|\mu| \ll M_S\;\;$ [maximal mixing];\\
             $A = M_S = -\mu\;\;$ \hspace*{19.75mm} [``typical'' mixing].
\end{itemize}
\newpage
~\\
We have taken $M_S$ of order 1 TeV to include the effects of possibly
large radiative corrections to the lightest CP-even Higgs mass. In the
same way the choice of the soft SUSY breaking parameter $A$ and of the
SUSY mass parameter $\mu$ is motivated.  The central top mass value is
close to the central value measured at the Tevatron \cite{CDF}. The
upper and lower bounds are extreme, roughly corresponding to the $\pm
2 \sigma$ limits of the CDF measurement.  Although the central value
for $M_t$ extracted from the LEP precision measurements in the MSSM
for large masses of the SUSY particles would be somewhat lower than
the central value observed in the Tevatron events, the lower values
are still consistent at the $2 \sigma$ level.  
\vvs1

\subsection{Production and Decay Modes of MSSM Higgs Particles}

\subsubsection{Higgs Production}

The main production mechanisms of the \underline {neutral Higgs
bosons} $h$ and $A$ in the \MSSM\ at LEP2 energies are 
through the following processes~\cite{shgs1}:
\begin{equation}\label{eq_prodmech}
\begin{array}{ll}
\mbox{Higgs--strahlung:}& e^{+}e^{-}\to Z\,h\\
\mbox{Associated pair production:}&e^{+}e^{-}\to A\,h
\end{array}
\end{equation}
The fusion processes, similar to the Standard Model, play only a
marginal role at the kinematical limit of the Higgs-strahlung process
for the production of the CP-even Higgs boson $h$. The CP-odd Higgs
boson $A$ cannot be produced in Higgs-strahlung and in fusion
processes to leading order.

The production of the heavy CP-even Higgs particle $H$ is very
difficult at LEP energies. In the tiny corner of parameter space, for
moderate to large $\tb$, where associated $A H$ production would be
allowed kinematically, the production cross section
is suppressed by the small coefficient $\sin^{2}(\beta
-\alpha )$, due to the $ZAH$ coupling discussed earlier, and the 
threshold P-wave factor. For $\tb = 3\,(50)$,
$m_{A}=60\;{\rm GeV}$, $m_{H}=123\,(117)\; {\rm GeV}$, it is
$4\,(0.001)\;{\rm fb}$.

The cross sections (\ref{eq_prodmech}) may be expressed in terms of
the cross section $\sigma_\SM$ for Higgs-strahlung in the Standard
Model in the following way \cite{shgs1,shgs3b}:
\begin{eqnarray}
\sigma (e^{+}e^{-}\to Z\,h)&=&\sin^{2}(\beta -\alpha)\,
 \sigma_\SM\\
\sigma (e^{+}e^{-}\to A\,h)&=&\cos^{2}(\beta -\alpha)\,{\bar \lambda}
\,\sigma_\SM
\end{eqnarray}
The factor ${\bar \lambda }=\lambda^{3/2}_{Ah}/ \{\lambda^{1/2}_{Zh}
\,[ 12m_{Z}^{2}/s\,+\, \lambda_{Zh} ]\}$ accounts for the correct
suppression of the P-wave cross section near the thresholds.
[$\lambda_{ij}=(1-(m_i+m_j)^2/s)(1-(m_i-m_j)^2/s)$ 
is the usual momentum factor of the two
particle phase space.]  The cross section for $WW$ fusion of $h$ is
reduced by the same factor $\sin^2(\beta -\alpha)$ as is the cross section for
Higgs--strahlung.

\begin{figure}[p]
\begin{center}
  \epsfig{file=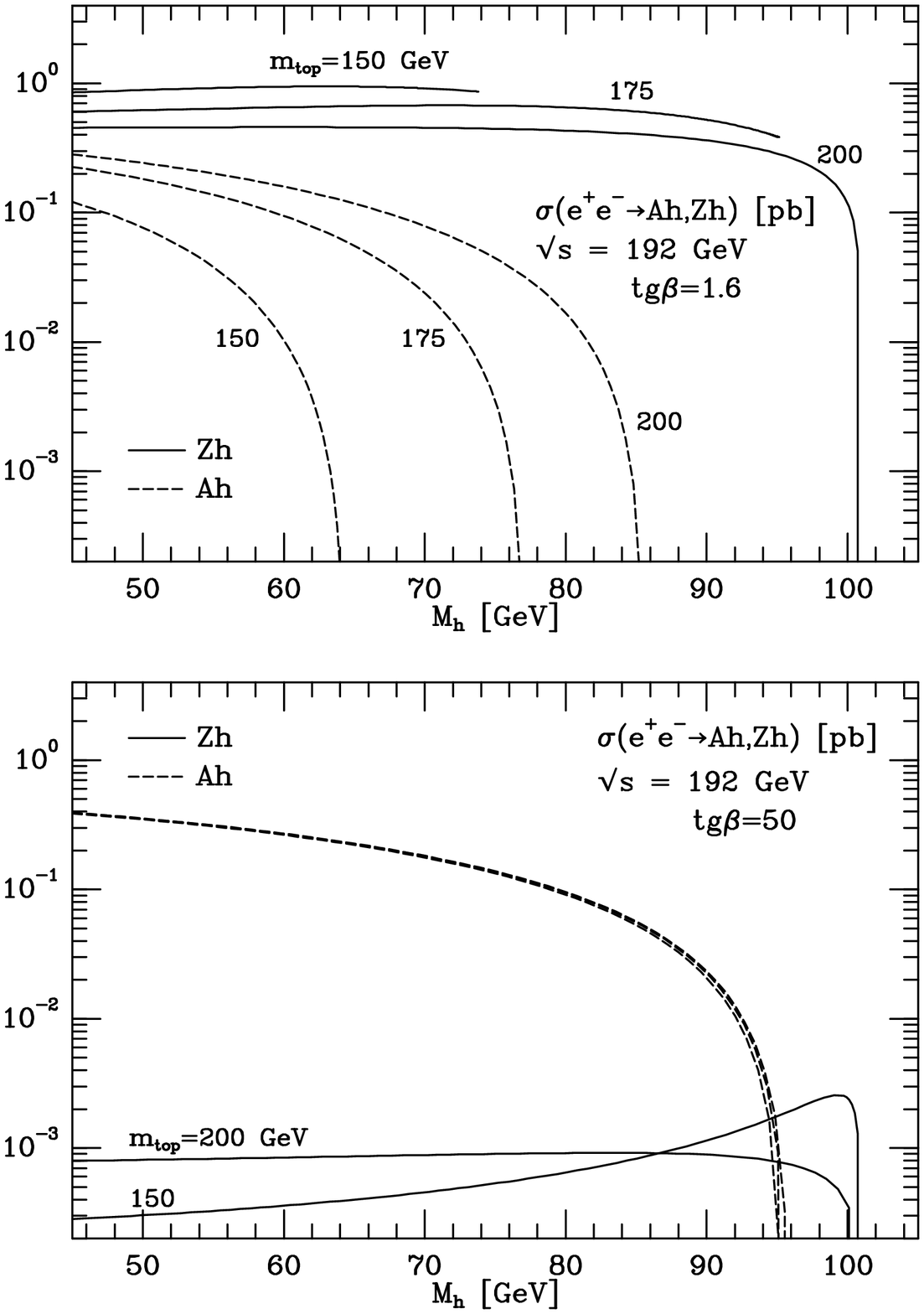,
          bbllx=45, bblly=130, bburx=505, bbury=780,
          height=19cm,clip=}
\end{center}
\caption{\it 
  The cross sections for Higgs-strahlung $Zh$ and associated pair
  production $Ah$ in the MSSM for two values of \/ $\tb=1.6$ and $50$
  and the top mass $M_t=175\pm 25$ GeV.}
\label{shgs:Zh-Ah}
\end{figure}

The cross sections for Higgs--strahlung $Zh$ and associated pair
production $Ah$ are complementary to each other, coming either with
the coefficient $\sin^{2}(\beta -\alpha)$ or $\cos^{2}(\beta
-\alpha)$. The cross sections are shown for two representative values
of $\tb=1.6$ and 50 in Fig.\ref{shgs:Zh-Ah}. The top-quark mass is
varied, as usual, between $175\pm25$ GeV.  Since the upper limit on
$m_h$ depends strongly on $M_t$ for small values of $\tb$ where the
tree-level mass is small, the endpoints of the curves are shifted
upwards significantly with increasing top mass.  For large values of
$\tb$, on the other hand, the dependence of the upper bound of $m_h$
on the radiative corrections is weaker for rising top mass due to the
large value of $m_h$ at the tree level.  The supersymmetric coefficient
$\cos^{2}(\beta -\alpha)$ is  nearly independent of the top
mass and it is very close to unity, so that the spread between the
curves is negligible; the coefficient $\sin^{2}(\beta -\alpha)$ is
correspondingly small.  In this large $\tb$ case, the curves terminate
at the kinematical limit before $m_h^{\rm max}$ can be reached, in
contrast to the small $\tb$ case.  For large $\tb$ the non-zero widths
of the particles are taken into account.

For small $\tb$, Higgs--strahlung $Zh$ provides the largest
production cross section while the cross section for $Ah$ associated
pair production is much smaller. With $\sigma_{Zh}\sim {\cal O}
(0.5\;{\rm pb})$ at $\sqrt{s}=192$ GeV, this mechanism gives rise to a
large sample of Higgs particles. For large $\tb$, associated
$Ah$ production is the dominant mechanism with rates similar to the
previous case. 

\begin{figure}
\begin{center}
  \epsfig{file=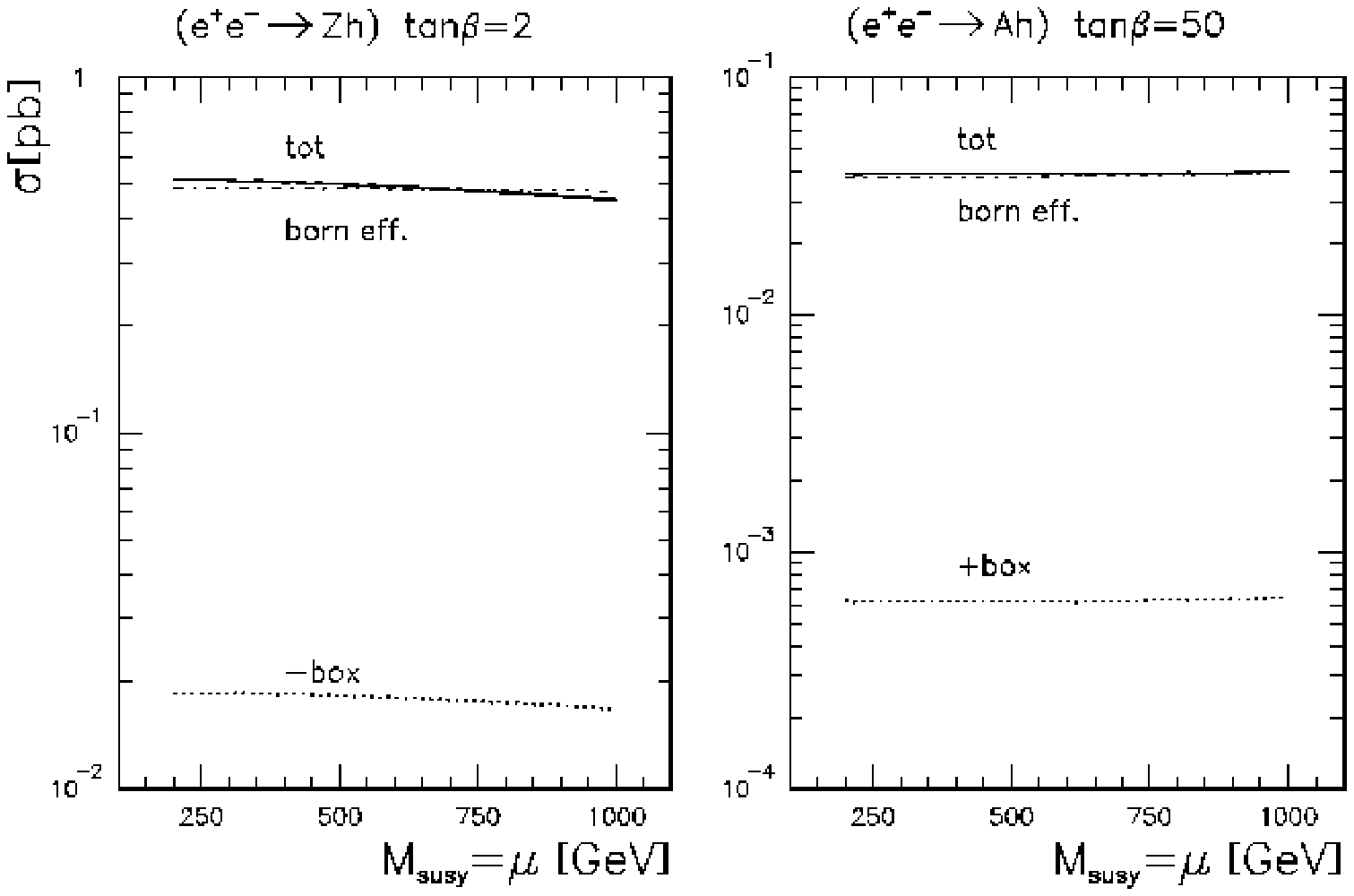,
                    bbllx=40, bblly=260, bburx=525, bbury=555,
                    height=8cm, clip=}
\end{center}
\caption[0]
{\it{ 
  The box contributions to the cross sections for Higgs-strahlung $Zh$
  and associated pair production $Ah$ for $\sqrt{s} = 192 $ GeV;
 $m_A=90$ GeV and a slepton mass  $m_{\tilde\ell}=100$ GeV.}
\label{shgs:Hollik}}
\end{figure}

The predictions for the cross sections $e^+e^-\to Zh$ and $Ah$
presented above have been based on the improved effective potential
approximation which takes into account heavy (s)quark effects on Higgs
masses, mixings and couplings.  It turns out {\em a posteriori\/} that
this scheme is quite accurate.  Indeed, the box contributions to the
cross sections are fairly small~\cite{shgs5}.  This is demonstrated in
Fig.\ref{shgs:Hollik} where the box contributions are compared with
the Born term, defined for the effective value $\tan 2\alpha = -
(m_Z^2+m_A^2)\tb/(m_Z^2+m_A^2\tan^2\beta - m_h^2/\cos^2\beta)$.
The leading part of the box contributions is generated by the
two-Higgs doublet diagrams while the contributions of the genuine SUSY
particles are very small.

The angular distributions are of the standard form~\cite{smh9} for
Higgs--strahlung and spin--zero pair production,
\begin{eqnarray}
\frac{d \sigma}{d \cos \theta}&\sim&\left\{
\begin{array}{lll}
\lambda\,\sin^{2}\theta\,+\,8m_{Z}^{2}/s &\mbox{for}& e^{+}e^{-}\to Zh\\
\sin^{2}\theta&\mbox{for}& e^{+}e^{-}\to Ah
\end{array}
\right.
\end{eqnarray}

Since the main decay mode of scalar and pseudoscalar Higgs particles
are $b{\bar b}$ decays in the \MSSM, it is interesting to study the
4--fermion process $e^{+}e^{-}\to b{\bar b}b{\bar b}$ in greater
detail.  The final state includes the signal $Zh \to (b{\bar
b})_Z(b{\bar b})_h$ in the Higgs--strahlung process, and the signal
$Ah \to (b{\bar b})_A(b{\bar b})_h$ for associated pair production.
The main component of the background is $e^{+}e^{-}\to Z^*Z^*$
production followed by $Z^*\to b{\bar b}$ decays. These cross sections
have been evaluated for a cut on the invariant $b{\bar b}$ mass of
$m(b{\bar b})>20$ GeV. The results are shown for a variety of
combinations ($m_{A}$, $\tb$) in Table~\ref{shgs:ee-bbbb} for
$\sqrt{s}=192$ GeV.\\
\begin{table}[hbt]
\caption{\it 
  The process $e^+e^-\to b{\bar b} b {\bar b} $ at
  $\protect\sqrt{s}=192$ GeV. Cross sections in fb.}
\label{shgs:ee-bbbb}
{\footnotesize
\begin{center}
\begin{tabular}{|c|c|c|c|c|c|c|}
\hline
\rule[0mm]{0mm}{3ex} 
&($m_A$ [GeV], $\tb$) & (75,30) & (400,30) & (75,1.75) & (400,1.75) 
& $\infty$\\
\hline
\rule[0mm]{0mm}{3ex} 
no  &EXCALIBUR  & --- & --- & --- & --- & 25.933(10)   \\
\rule[0mm]{0mm}{3ex} 
ISR &HZHA/PYTHIA & 135.17(61) & 23.286(58) & 163.36(75) & 74.04(31) &
22.816(50) \\
\hline
\rule[0mm]{0mm}{3ex} 
with &EXCALIBUR  & --- & --- & --- & --- & 23.045(23)   \\
\rule[0mm]{0mm}{3ex} 
ISR  &HZHA/PYTHIA & 118.60(58) & 18.761(87) & 151.75(75) & 57.74(28) &
18.384(80) \\
\hline
\end{tabular}
\end{center}
}
\end{table}
\begin{figure}[hbt]
\begin{center}
  \epsfig{file=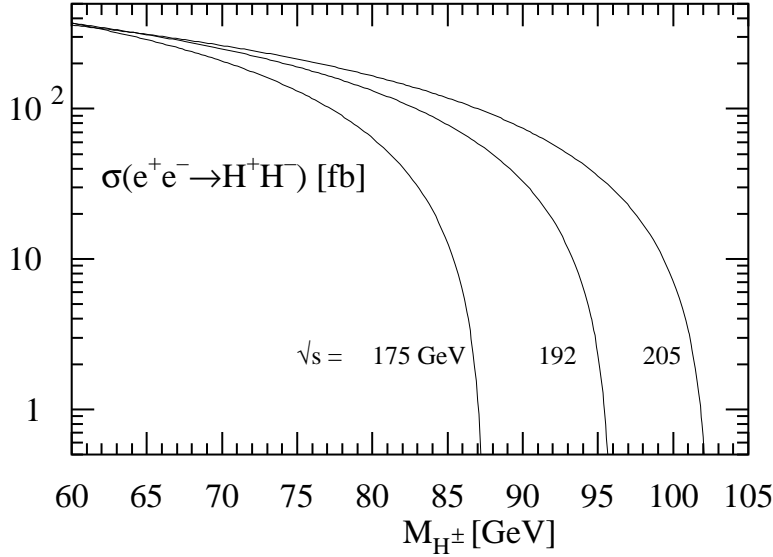,
          bbllx=60, bblly=60, bburx=350, bbury=260}
\end{center}
\caption{\it The cross section for charged Higgs boson production.}
\label{shgs:cxnpm}
\end{figure}

The cross section for the production of \underline{charged Higgs bosons}
\begin{equation}
e^{+}e^{-}\to H^{+}H^{-}
\end{equation}
is built up by $s$-channel $\gamma$ and $Z$
exchanges~\cite{shgs3b,shgs6}.  It depends only on the charged Higgs
mass and no extra parameter,
\begin{equation}
\sigma (e^{+}e^{-}\to H^{+}H^{-}) =
\frac{2 G_{F}^{2}m_{W}^{4}s_W^4 }{3\pi s }
\left[ 1\,+\,\frac{2{\hat v}_{e}{\hat v}_{H}}{1-m_{Z}^{2}/s}\,
+\,\frac{({\hat a}_{e}^{2}+{\hat v}_{e}^{2})\,{\hat v}_{H}^{2}}
{(1-m_{Z}^{2}/s)^{2}}\right] \beta^{3}_H
\end{equation}
where the rescaled $Z$ charges are defined by $\hat{a}_{e}=-1/4c_Ws_W$
and $\hat{v}_{e} =(-1+4s_W^2)/4c_Ws_W$ and
$\hat{v}_{H}=(-1+2s_{W}^{2})/2c_{W}s_{W}$ 
[note that $s^2_W = \sin^2\theta_W$]; $\beta_H =
(1-4m_{H^{\pm}}^{2}/s)^{1/2}$ is the velocity of the Higgs
particles. The cross section is shown in Fig.\ref{shgs:cxnpm} as a
function of the charged Higgs mass for the three representative LEP2
energy values $\sqrt{s} = 175$, $192$ and $205$ GeV.  Within the MSSM
the present lower limit of the charged Higgs boson mass is about $85$
GeV so that only a small window is left for LEP2. Even
though the cross section is not particularly small for
$m_{H^{\pm}}\sim 80$ to 90 GeV, the signal is very hard to extract
from the overwhelming background of $WW$ pair production in this mass
range.  The analysis of cascade decays $H^\pm\to W^*h,
W^*A$~\cite{DKZ} can ameliorate the prospects of detecting the charged
Higgs boson in this mass range for small $\tan\beta$.

\subsubsection{Decay Modes of the \MSSM\ Higgs Particles}
\underline {Decays to \SM\ particles.}
For $\tan\beta > 1$ the lightest CP-even neutral 
Higgs boson $h$ decays almost
exclusively to fermion pairs if the mass $m_{h}$ is less than 100 GeV.
Near the upper limit of $m_{h}$ for a given $\tb$, i.e. in the
decoupling region, the decay pattern becomes \SM--like. Fermion pairs
are also the dominant decay mode of the pseudoscalar Higgs boson $A$.
The partial decay widths of all the neutral Higgs bosons $\Phi$ into
fermions are given by
\begin{equation}
\Gamma (\Phi \to f{\bar f}) = N_{c}\,\frac{G_{F}m_{f}^{2}}{4\sqrt{2}
\pi}\,g_{\Phi ff}^{2}m_{\Phi}\left[1+
\frac{17}{3}\frac{\alpha_s}{\pi}\right] 
\end{equation}
in the limit $m_{\Phi}^2\gg m_{f}^{2}$.  The couplings $g_{\Phi ff}$
have been defined in Table~\ref{table:couplings}. The small additional ${\cal
O}(\alpha_s^2)$ contributions have been summarized in
Ref.\cite{R11A}.  As anticipated from chirality arguments, the
widths, including the QCD radiative corrections, do not depend on the
parity of the state apart from the overall coupling $g_{\Phi ff}$ in
the limit of large Higgs masses. For quark decays, $m_{f}$ has to be
chosen as the running quark mass evaluated at the scale $m_{\Phi}$.
The electroweak corrections are incorporated at a sufficient level of
accuracy by adopting the effective potential approximation for the
couplings~\cite{dab}.

The partial width for charged Higgs decays to quark pairs is
obtained from
\begin{equation}
\Gamma (H^{\pm}\to U{\bar D}) = \frac{3G_{F}m_{H^{\pm}}}
{4\sqrt2\pi}\,|V_{UD}|^{2}\,\left[\cot^2\beta\, m_U^2
+\tan^2\beta\,m_D^2 \right]
\left[1 + \frac{17}{3}\frac{\alpha_s}{\pi}\right]
\end{equation}
This formula is valid if either the first or the second term is
dominant.  The up and down quark masses $m_{U,D}$ are defined again at
the mass scale of the charged Higgs boson.

Since the $b$ quark couplings to the Higgs bosons are in general
strongly enhanced and the $t$ quark couplings suppressed in the \MSSM\
[cf.\ Fig.\ref{shgs:fig-couplings}], $b$ loops may contribute
significantly to the $gg$ coupling so that the approximation
$m_{Q}^{2}\gg m_{\Phi}^{2}$ cannot be applied any more in general.
Nevertheless, it turns out {\it a posteriori\/} that this remains 
 an excellent approximation for the QCD corrections. The CP-even and
CP-odd Higgs decays to gluons and light quarks \cite{R11A} are given
by the expressions
\begin{equation}
\Gamma (\Phi \to gg(g),q{\bar q}g) =
\frac{G_{F}\alpha_s^2 m_{\Phi}^{3}(9/4)}
{16\sqrt{2}\pi^3}\Big|\sum_{t,b}A_{Q}^{\Phi}\Big|^{2}
\,\left\{1+\left[ \frac{95}{4}\left(\frac{97}{4}\right)-
\frac{7}{6}N_{F}\right]\frac{\alpha_s}{\pi}\right\}
\end{equation}
where the parentheses refer to the pseudoscalar particle. The form
factors are defined by
\begin{equation}
A_{Q}^{h,A} =
g_{Q}^{h,A}\times\, \tau\,\left[1+(1-\tau)\,
f(\tau)\right] \quad\mbox{and}\quad A_{Q}^{A}=g_{Q}^{A}\tau f(\tau )
\end{equation}
with $f(\tau )=\arcsin^{2}(1/\sqrt{\tau})$ for $\tau \ge 1$ and
$-\frac{1}{4}\,[\log (1+\sqrt{1-\tau})/(1-\sqrt{1-\tau}) -i\pi]^{2}$
for $\tau < 1$. The parameter $\tau =4m_{Q}^{2}/ m_{\Phi}^{2}$ is
defined by the pole mass of the heavy loop quark $Q$.  In the same way
as $\alpha_s (m_{\Phi})$, the coefficient of the QCD corrections must
be evaluated for $N_{F}=3$ if gluons and only light quarks are
considered in the final state [cf.\ the \SM\ section for details].

At the edge of the mass range accessible at LEP2, the CP-even Higgs
boson $h$ can decay into virtual gauge boson pairs
$W^{*}W/Z^{*}Z$. The widths are the same as in the Standard Model, yet
suppressed by the \MSSM\ coefficient $\sin^{2}(\beta -\alpha )$.

\noindent
(ii) \underline{ Cascade decays.} 
A variety of cascade decays could in principle play a role in some
ranges of the \MSSM\ parameter space accessible at LEP2, if
sufficiently large samples of heavy Higgs bosons were generated.
However, 
for the typical set of  parameters discussed in this report,
these decay modes are not very important in general and details may be
traced back from \cite{DKZ,shgs8}. The only exception are
the cascade decays of the charged Higgs bosons \cite{DKZ} for small
to moderate $\tb$,
\begin{eqnarray}
 \Gamma(H^+ \to hW^{+*} \to hf\bar f')
 &=& \frac{9G_F^2m_W^4}{8\pi^3}\cos^2(\beta-\alpha)\,
        m_{H^\pm}G_{hW}\\
 \Gamma(H^+\to AW^{+*}\to Af\bar f')
 &=& \frac{9G_F^2m_W^4}{8\pi^3}m_{H^\pm}G_{AW}
\end{eqnarray}
The coefficients $G$ depend on the mass ratios of the particles involved,
\begin{eqnarray}
G_{i j} &=& \frac{1}{4} \left\{ 2 (-1+\kappa_j -
\kappa_i) \sqrt{ \lambda_{ij} } \ \left[ \frac{\pi}{2}+ \arctan
\left( \frac{ \kappa_j (1- \kappa_j +\kappa_i) -\lambda_{ij}} {(1-
\kappa_i) \sqrt{\lambda_{ij}}} \right) \right] \right. \nonumber \\
&& \qquad \left. +\,(\lambda_{ij} - 2\kappa_i) \log(\kappa_i) + \frac{1}{3}
(1-\kappa_i) \left[ 5(1+\kappa_i) -4 \kappa_j -\frac{2}{\kappa_j}
\lambda_{ij} \right] \right\} 
\end{eqnarray}
with $\kappa_i = m_i^2 / m_{H^\pm}^2$ and $\lambda_{ij} = -1 +2
\kappa_i + 2 \kappa_j - (\kappa_i - \kappa_j)^2$. These decay modes
are important for sufficiently light $h/A$ Higgs bosons. If they are
allowed, in particular $H^{\pm} \to AW^*$, they reduce
the $\tau \nu_\tau$ branching ratio considerably, and they overrule
$cs$ decays as the second most important decay channel of the charged
Higgs bosons.

\begin{figure}[p]
\begin{center}
\epsfig{file=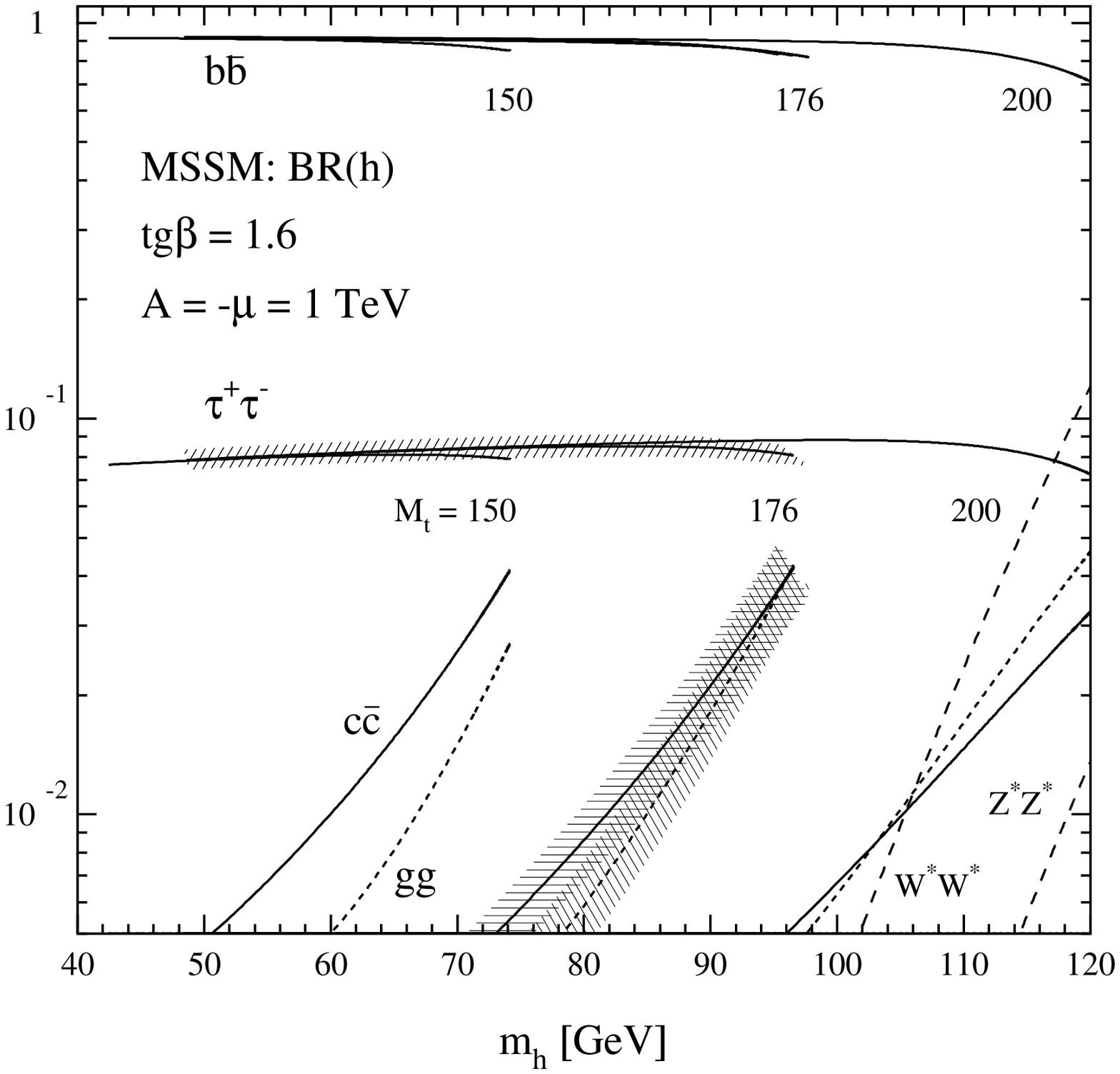,%
        bbllx=45pt, bblly=185pt, bburx=545pt, bbury=695pt,%
        height=8.4cm, angle=-90, clip=}
\epsfig{file=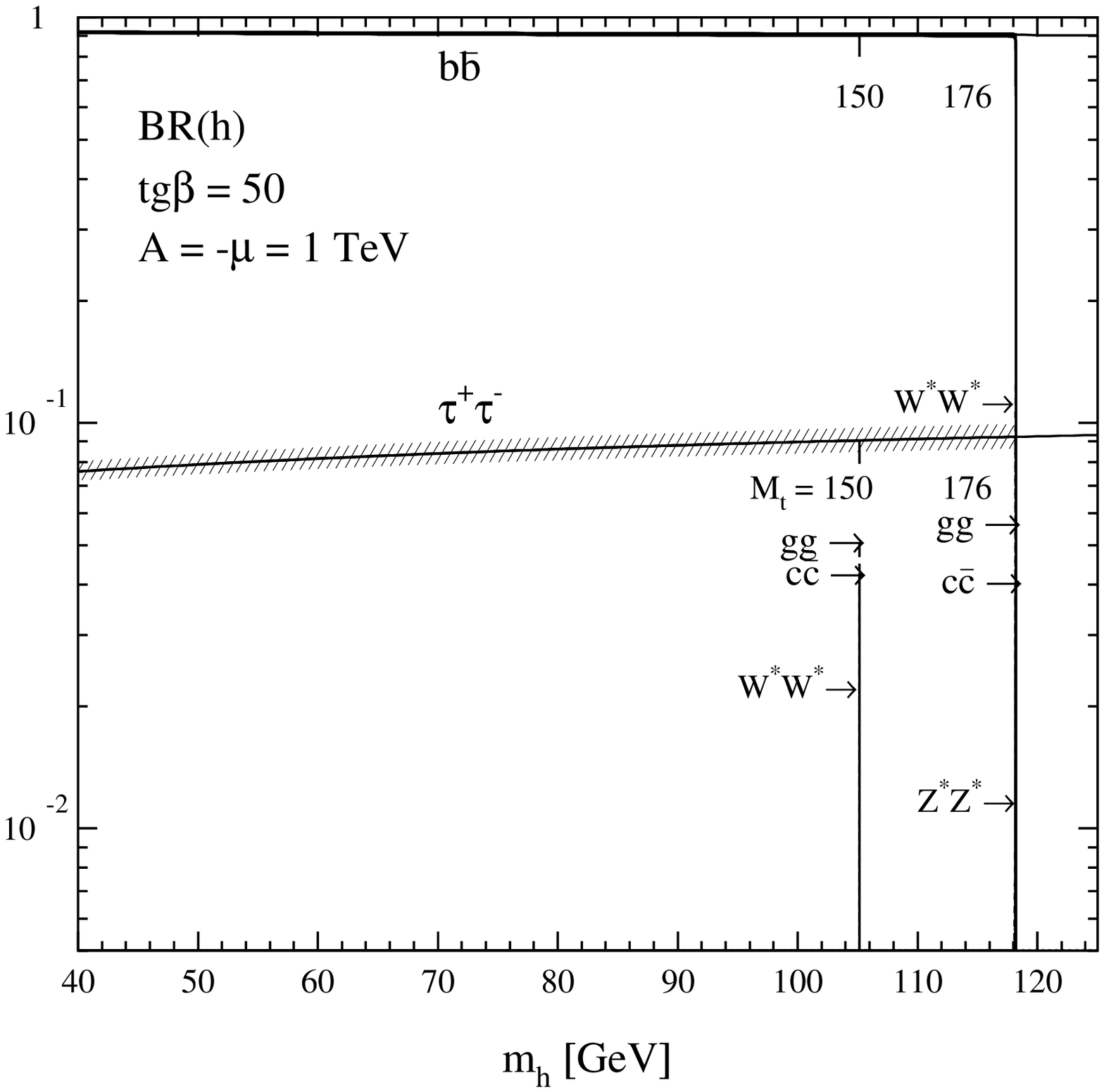,%
        bbllx=45pt, bblly=185pt, bburx=545pt, bbury=695pt,%
        height=8.4cm, angle=-90, clip=}
\end{center}
\begin{center}
\epsfig{file=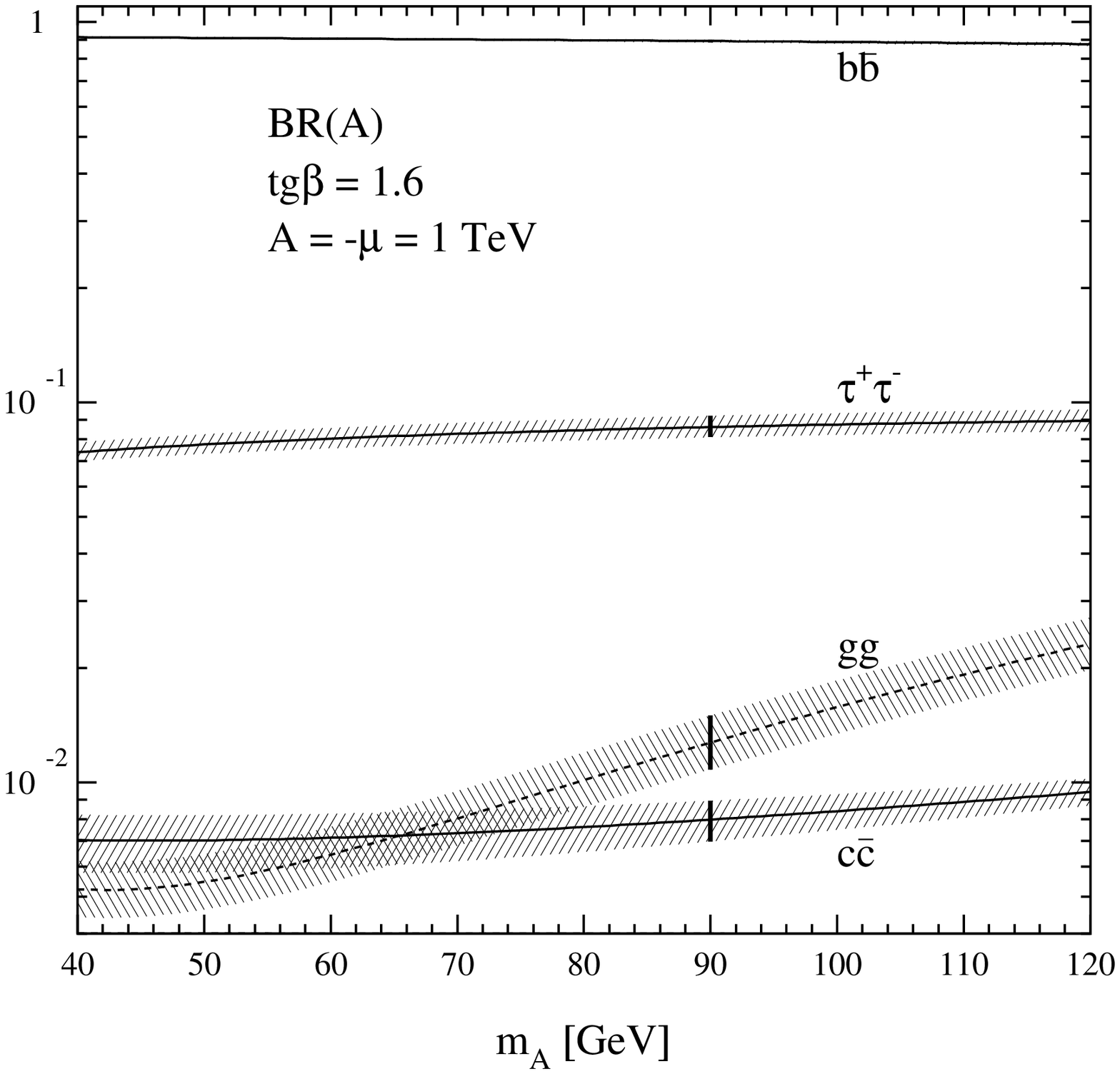,%
        bbllx=45pt, bblly=185pt, bburx=545pt, bbury=695pt,%
        height=8.4cm, angle=-90, clip=}
\epsfig{file=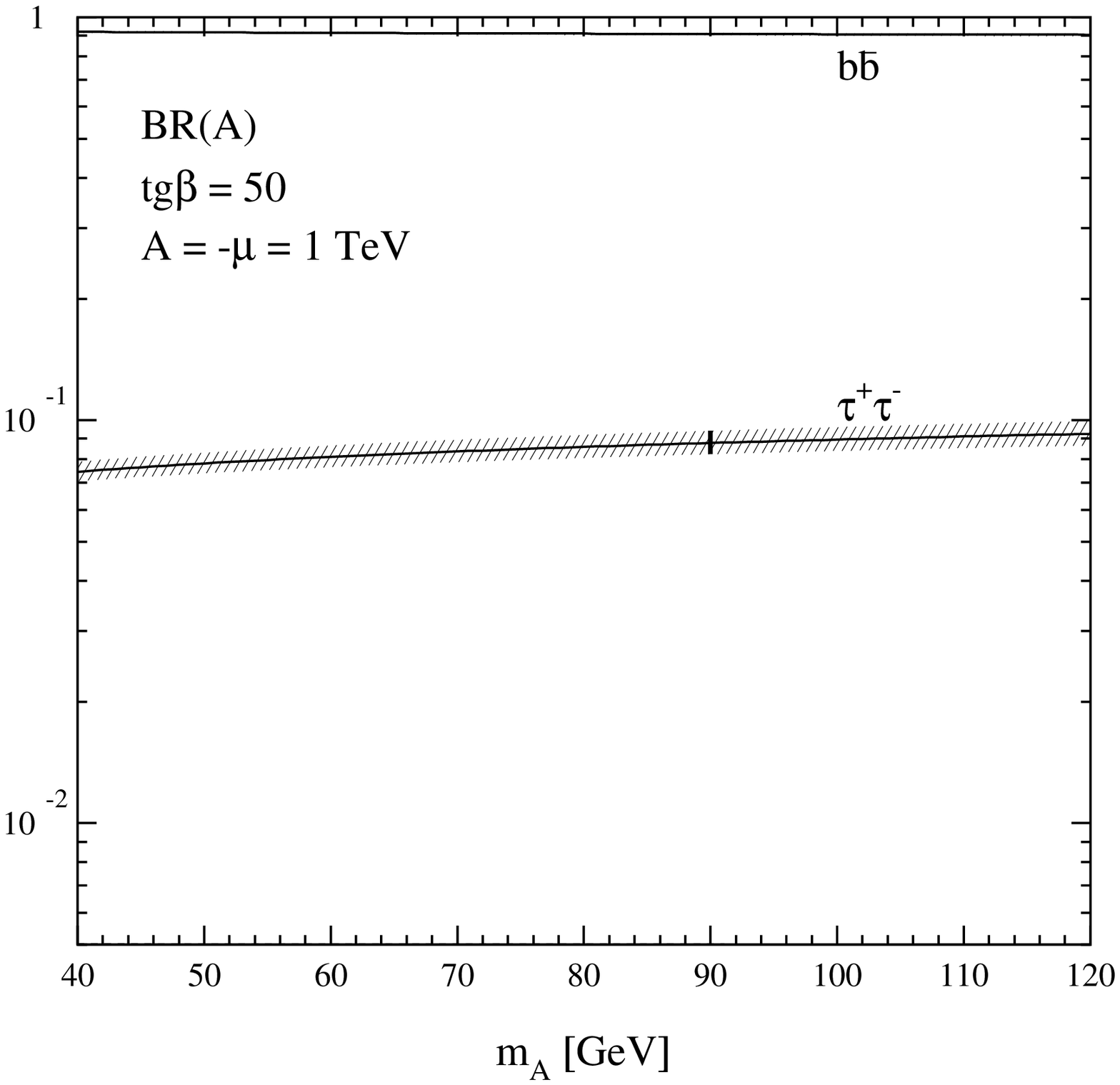,%
        bbllx=45pt, bblly=185pt, bburx=545pt, bbury=695pt,%
        height=8.4cm, angle=-90, clip=}
\end{center}
\caption{\it Decay branching ratios of the MSSM Higgs bosons
  $h,A,H^\pm$ into SM particles and cascade decays. The bands characterize
the uncertainties in the predictions, except those due to the top mass,
which are indicated by bars.}
\label{shgs:br}
\end{figure}

\noindent
\underline {Summary of the branching ratios.}
The branching ratios for the standard quark/lepton/gauge boson and the 
cascade decay modes discussed above, are shown for 
``typical mixing'' and two
representative values $\tb = 1.6$ and $50$ in Fig.\ref{shgs:br}. Unless 
otherwise specified the
top mass in Fig.\ref{shgs:br} has been fixed to  $M_t = 176$ GeV.  
Increasing the top mass
shifts the upper end of the $b$ and $\tau$ curves upwards while the
$cc$ and $gg$ curves are transferred nearly parallel, a consequence of the 
larger Higgs mass values.  The effect of varying $\alpha_s = 0.118 \pm
0.006$ is indicated by the hatched bands. The curves labeled $bb$ and
$cc$ correspond to all mechanisms generating inclusive $b, c$ quarks in the
final states, while the curve labeled $gg$ includes gluons and light
quarks. At $\tb = 1.6$, interesting cascade decays are predicted
for moderately small charged Higgs masses, $H^\pm \to AW^*$
and $hW^*$; they affect the experimental search
techniques also in the $\tau \nu_\tau$ channel by reducing this
important decay branching ratio. For large $\tb$, $b
\overline{b}$ and $\tau^+ \tau^-$ decays are overwhelming except in
the decoupling regime near the upper limit of the $h$ mass.

\begin{figure}[p]
\begin{center}
\epsfig{file=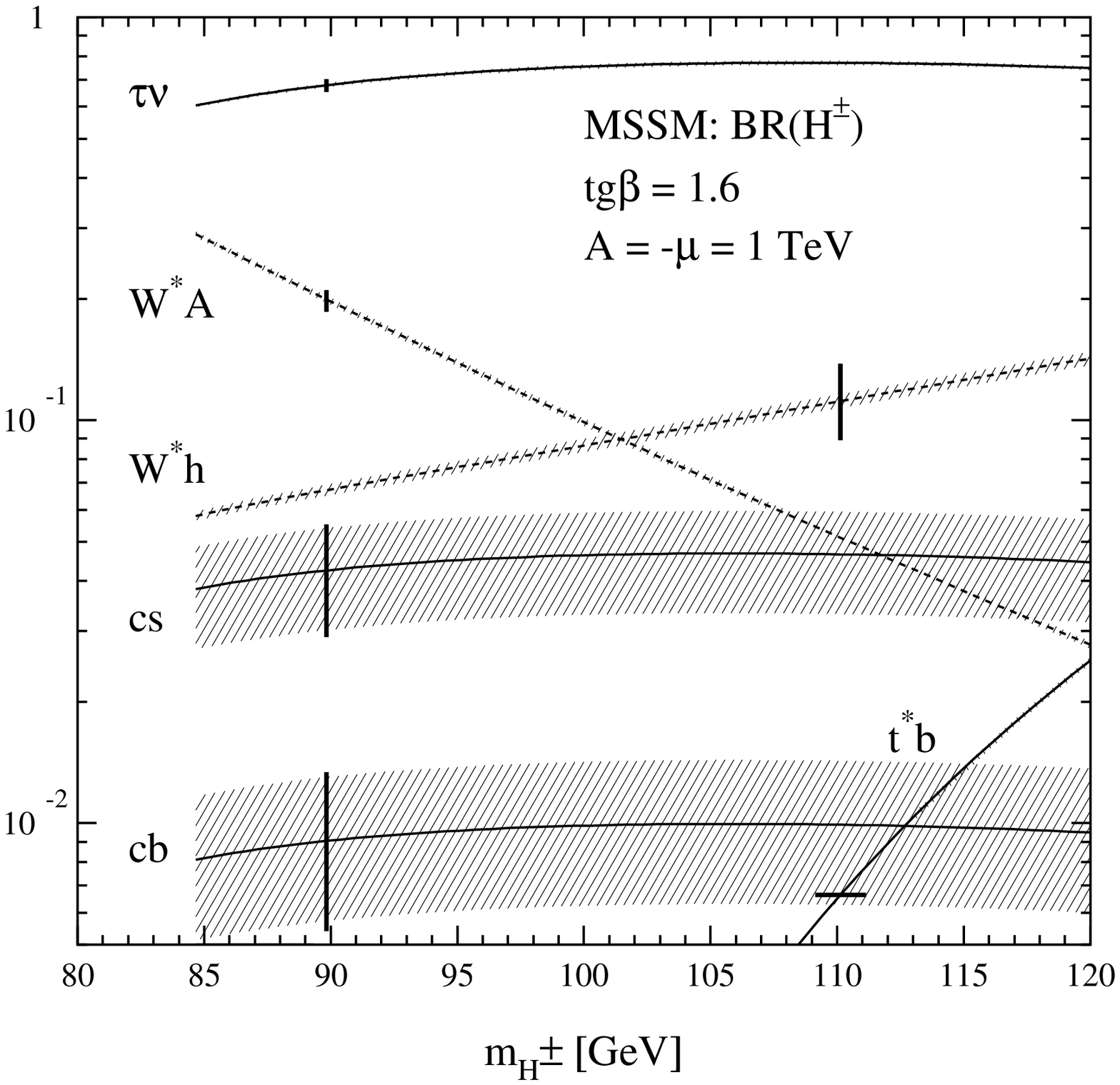,%
        bbllx=45pt, bblly=185pt, bburx=545pt, bbury=695pt,%
        height=8.4cm, angle=-90, clip=}
\epsfig{file=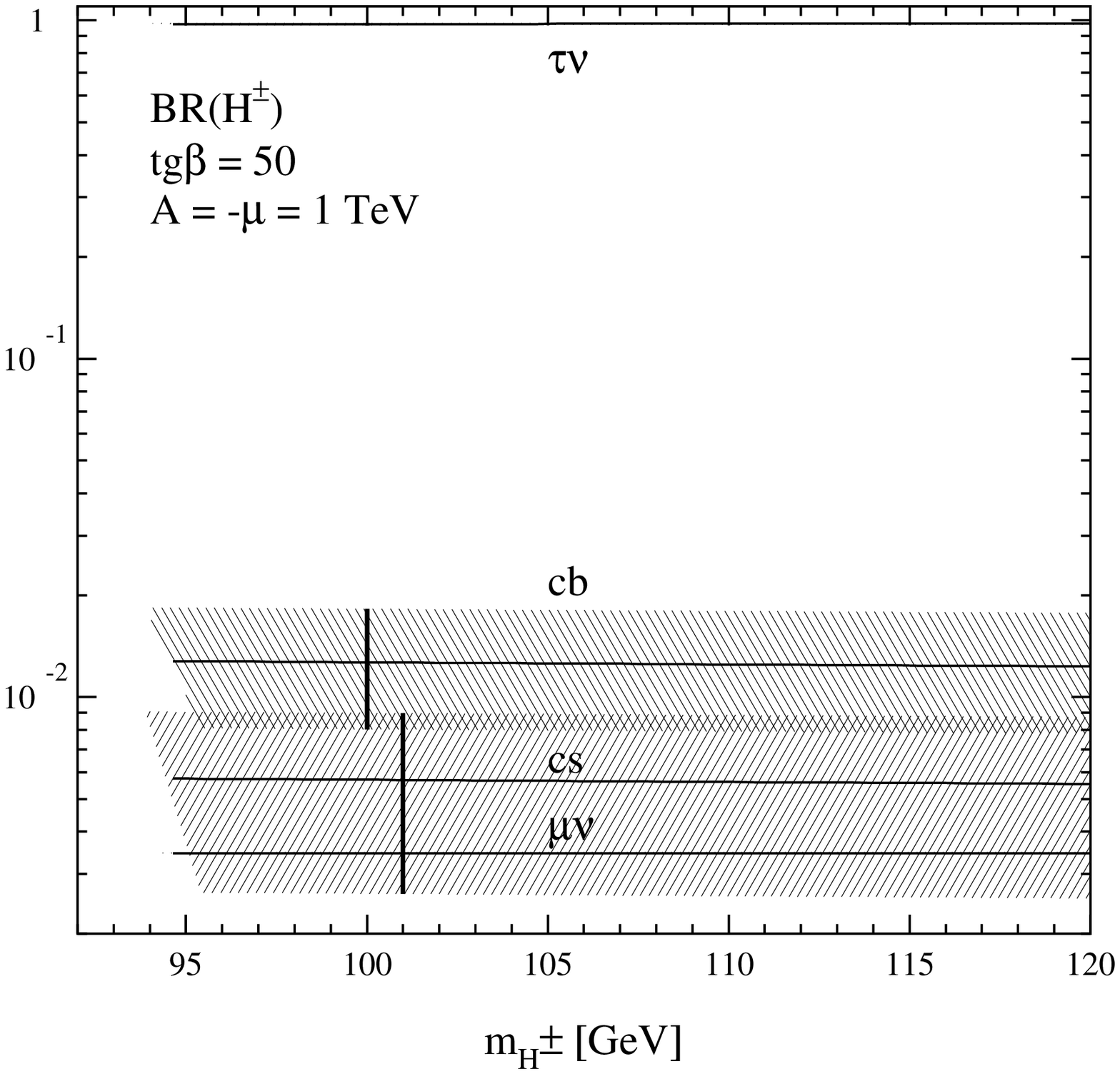,%
        bbllx=45pt, bblly=185pt, bburx=545pt, bbury=695pt,%
        height=8.4cm, angle=-90, clip=}
\end{center}
\vspace*{-7mm}
\begin{center} Figure~\ref{shgs:br}: {\it (cont'd).}
\end{center}
\begin{center}
\epsfig{file=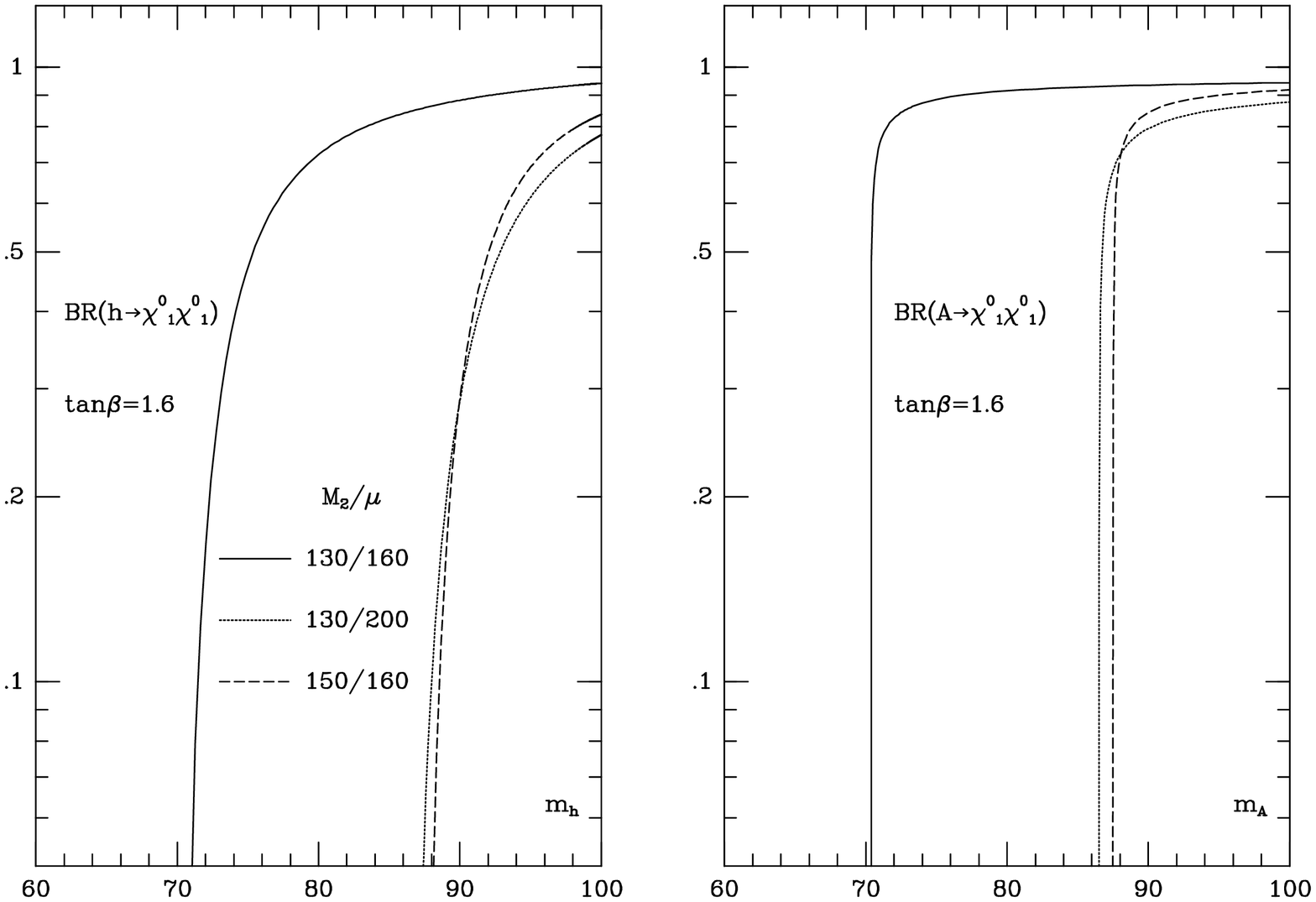,%
        bbllx=80pt, bblly=100pt, bburx=525pt, bbury=745pt,%
        height=14cm, angle=-90, clip=}
\end{center}
\vspace*{-2mm}
\caption{\it Branching ratios for $h,A$ decays into pairs of the
  lightest neutralino for a set of $SU_2$ gaugino and higgsino mass
  parameters not excluded by LEP1/1.5.  If $\chi_1^0$ is the LSP and
  if $R$ parity is unbroken, these decays lead to invisible final
  states.}
\label{shgs:br-typical}
\end{figure}

Predictions for decays of the heavy CP--even Higgs boson $H$ are
discussed in great detail in Ref.\cite{R11A}.

\noindent
\underline {Neutralino decays.} 
For $h$, $A$ Higgs masses up to $\sim 120$ GeV, there are still
windows open for decays into pairs of light
neutralinos~\cite{shgs3b,shgs9}. These windows have been left by LEP1
and they cannot be closed by LEP1.5 either. The decay channels of
interest are
\begin{equation}
  h,\,A \to \chi_1^0\chi_1^0
\end{equation}
for small to moderate $\tan\beta$.

Masses and couplings of the states $\chi_i$ depend on $\tan\beta$, the
$SU_2$ gaugino mass parameter $M_2$, and the Higgs mass parameter
$\mu$. [We assume the relation
$M_1={\textstyle\frac53}\tan^2\theta_WM_2 \simeq
{\textstyle\frac12}M_2$.] For large $M_2$ and small (positive) $\mu$
values, the lightest neutralino $\chi_1^0$ is predominantly built up
by the higgsino component while for large values of $\mu$ the light
$\chi_1^0$ state is predominantly gaugino-like.  The couplings of
$\chi_1^0$ to $h$ and $A$ are given in terms of the neutralino mixing
matrix $Z$ by
\begin{eqnarray*}
  \kappa_h &=& \left(Z_{12} - \tan\theta_W Z_{11}\right)
  \left(\sin\alpha\,Z_{13} + \cos\alpha\,Z_{14}\right)\\
  \kappa_A &=& \left(Z_{12} - \tan\theta_W Z_{11}\right)
  \left(-\sin\beta\,Z_{13} + \cos\beta\,Z_{14}\right)
\end{eqnarray*}
[see e.g.~\cite{shgs10}].
Since $Z_{11}/Z_{12}$ correspond to the gaugino components of
$\chi_1^0$ while $Z_{13}/Z_{14}$ correspond to the higgsino
components, the couplings $h\chi_1^0\chi_1^0$ and $A\chi_1^0\chi_1^0$
can only be non-negligible if the state $\chi_1^0$ incorporates both
components at a significant level.  Moderate values of $M_2$, $\mu$
are therefore the favorable domain for LSP decays.  The widths for the
$h$, $A$ decays into $\chi_1^0\chi_1^0$ pairs can be written as

\vfill

\newpage

\begin{eqnarray}
  \Gamma(h\to\chi_1^0\chi_1^0)
  &=& \frac{G_Fm_W^2}{2\sqrt2\pi}\kappa_h^2m_h\beta_\chi^3\\
  \Gamma(A\to\chi_1^0\chi_1^0)
  &=& \frac{G_Fm_W^2}{2\sqrt2\pi}\kappa_A^2m_A\beta_\chi
\end{eqnarray}

These decays of the Higgs particles are invisible if $\chi_1^0$ is
stable. The Higgs particle $h$ can still be observed in the
Higgs-strahlung process through the recoiling $Z$ at small to moderate
$\tb$ where this production channel is dominant.  However, the
pseudoscalar Higgs boson $A$ cannot be detected if, produced in
associated $Ah$ production, both particles decay into invisible
$\chi_1^0\chi_1^0$ channels for small to moderate values of $\tb$.

Since for large $\tb$ the $b\bar b$ decays of the $h$ and $A$ Higgs
bosons are overwhelming, $\tb$ needs to be small to moderate for
$\chi_1^0$ decays to be relevant.  Typical examples of large branching
ratios for $h, A$ Higgs decays to LSP pairs are shown in
Fig.\ref{shgs:br-typical} for a set of $SU_2$ gaugino and higgsino
mass parameters $M_2$ and $\mu$. The LSP masses can be read off the
threshold values.  The branching ratios are large whenever the LSP
decay channels are open for $\mu > 0$.  For $\mu < 0$ the LSP decays
play a less prominent role; only in a small window close to $\mu\sim -M_2/2$ 
are the couplings large enough to allow for invisible $h$ and $A$
decays~\cite{shgs9}.

\newpage

\subsection{The Experimental Search for the Neutral Higgs Bosons}
\label{sec:mssm}

\subsubsection{Searches in the Higgs-strahlung Process}

For the $\epemto\ \h\Z$ process, as well as for $\epemto\ \H\Z$ when
kinematically allowed, all the analyses \cite{PJ} developed for the standard
model Higgs boson and  presented in Section~\ref{sec:exp}
can be used with no modifications and with a similar efficiency,
provided that the Higgs boson decays into supersymmetric particles, such
as charginos and neutralinos, are not open. As soon as the decay
into a pair of LSPs (Lightest Supersymmetric Particle) $\h\to\chi\chi$ is
allowed, it may even become dominant therefore rendering the existing analyses
ineffective.

Two new selection algorithms were developed by ALEPH to take care
of this particular situation where the Higgs boson would decay invisibly,
for the following events topologies:
\begin{itemize}
\item[{\it (i)}]   the acoplanar lepton pair topology, $(\Z \to \epem, \mpmm)$
$(\h \to \chi\chi)$;
\item[{\it (ii)}]  the acoplanar jet topology, $(\Z \to \qqbar)$
$(\h \to \chi\chi)$.
\end{itemize}
Only minor modifications to the selection procedure would be needed to extend
the validity of these analyses to ``almost invisible'' Higgs boson decays, such
as $\h \to \chi^\prime \chi$ or $\chi^+\chi^-$ when the mass difference between
the LSP and the next-to-LSP is small.
\vvs1

\noindent {\bf a) Search in the Acoplanar Lepton Pair Topology}\pss{0.5}
\label{sec:acopll}
\begin{picture}(158,63)
\put(0,58){\parbox[t]{105mm}{
The acoplanar lepton pair topology arises when the Z decays into a pair of leptons
and the Higgs boson \h\ decays invisibly into a pair of neutralinos. Events can be
selected by requiring a high mass \epem\ or \mpmm\ pair, compatible with 
the \Z\ hypothesis and with large missing energy and missing mass. Events
from $\epemto\ \lplm(\gamma)$, $\Z\epem$ or $\gamma\gamma \to \lplm$
are characterized by a large missing momentum along the beam direction
and a small acoplanarity angle, and can therefore easily be rejected. The only
irreducible background sources are $\epemto\ \W\W \to \e\nu\e\nu, \mu\nu\mu\nu$,
$\epemto \Z\Z \to \epem\nnbar, \mpmm\nnbar$, and to a lesser extent
$\epemto\ \Z\nnbar$.
}}
\put(110,1){\epsfxsize60mm\epsfbox{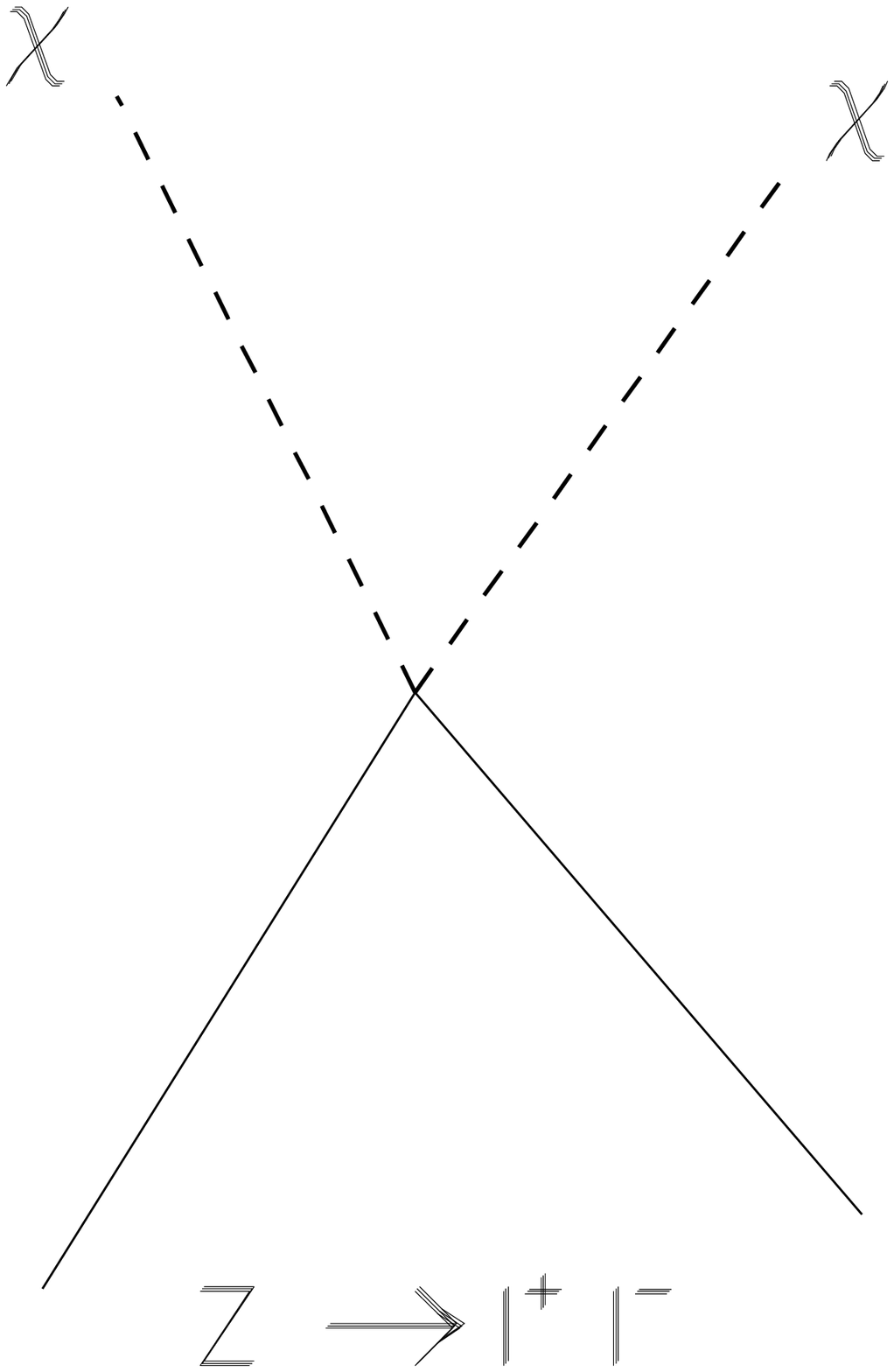}}
\end{picture}

\begin{figure}[htbp]
\begin{picture}(120,75)
\put(0,-5){\epsfxsize83mm\epsfbox{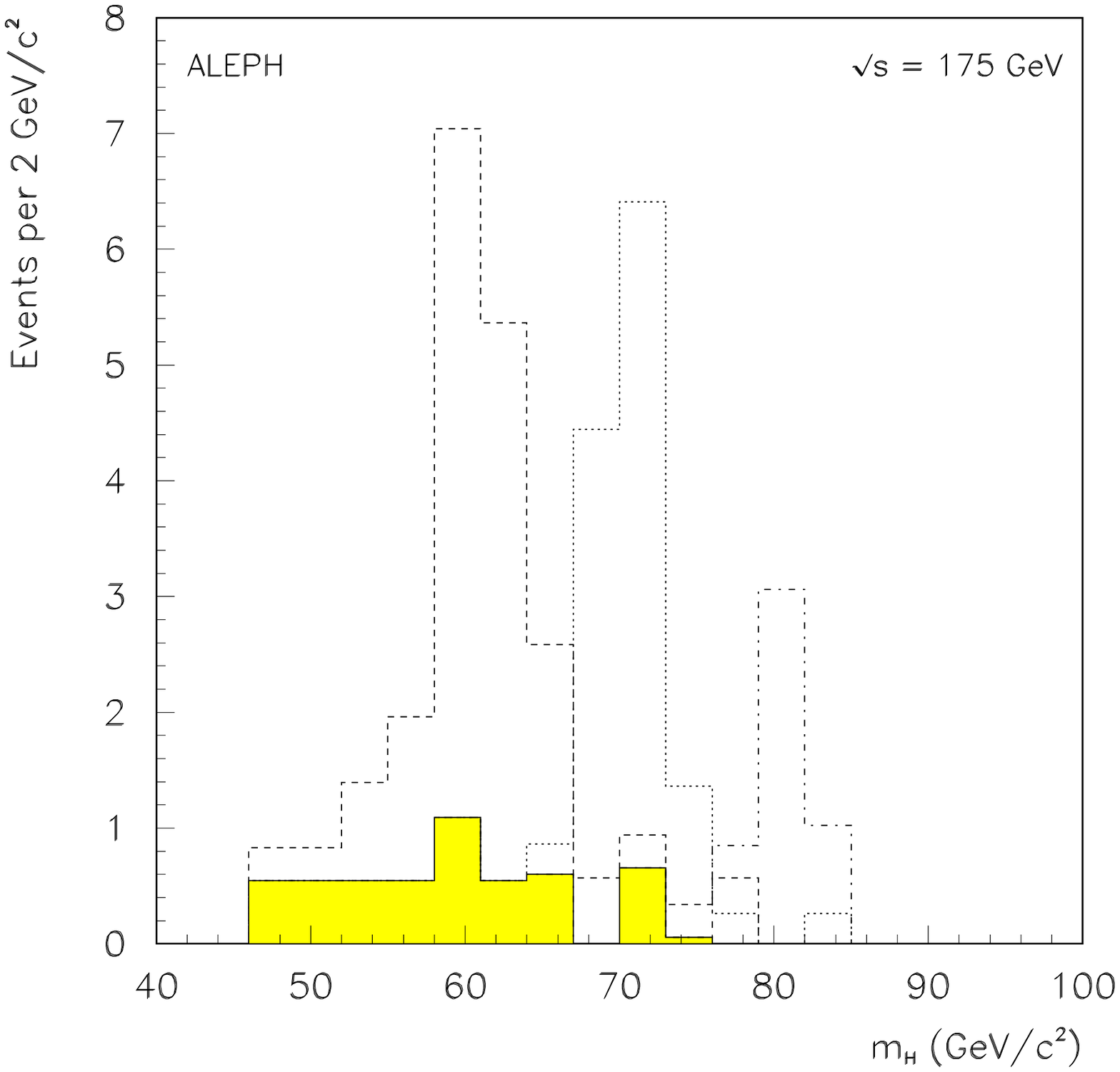}}
\put(80,-5){\epsfxsize83mm\epsfbox{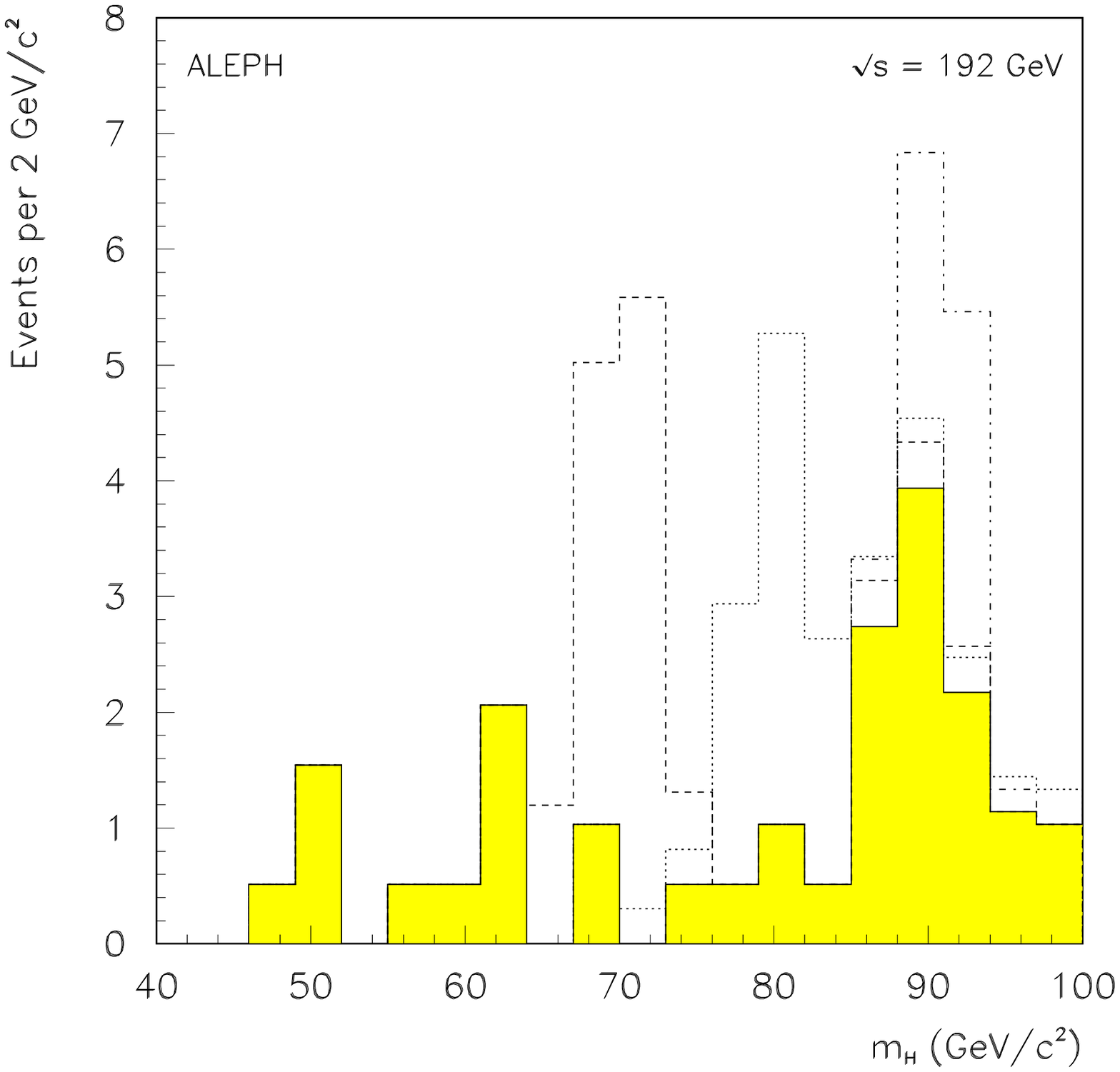}}
\end{picture}
\caption{\it 
  Distribution of the missing mass recoiling against the \epem\ or
  \mpmm\ pair, in the acoplanar lepton pair topology, as obtained from
  the ALEPH simulation at 175~GeV (left) and 192~GeV (right), with an
  integrated luminosity of 500~\inpb.  The signal (in white) is shown
  on top of the background (shaded histogram), with Higgs boson masses
  of 60 (dashed), 70 (dotted) and 80 (dash-dotted)~\Gcs\ at 175~GeV,
  and 70 (dashed), 80 (dotted) and 90 (dash-dotted)~\Gcs\ at 192~GeV.}
\label{fig:acopll}
\end{figure}

An efficiency of 45 to 50\% was achieved independently of \mh. The lepton momenta
were subsequently fitted to the Z mass hypothesis, and the missing
mass calculated from the energy-momentum conservation as recoiling against
the lepton pair, with a typical resolution of 2~\Gcs. Shown in Fig.\ref{fig:acopll}
are the mass distributions obtained with 500~\inpb\ at 175 and 192~GeV, for several
Higgs boson masses. At 192~GeV, and for $\mh = 90~\Gcs$, the numbers of signal and
background events expected in a window of $\pm 2 \sigma$ around the reconstructed
Higgs boson mass are 6.2 and 6.1 respectively, assuming a 100\% branching fraction
into invisible final states.
\vvs1

\noindent {\bf b) Search in the Acoplanar Jet Topology}\pss{0.5}
\label{sec:acopqq}
\begin{picture}(158,58)
\put(56,53){\parbox[t]{114mm}{
In order to add to the numbers of signal events expected from the acoplanar
lepton pair topology, the hadronic decays of the \Z\ were also investigated.
A very similar selection procedure as in Section~\ref{sec:acopll}a) 
was developed,
in which the two hadronic jets played the role of the leptons. A similar
selection efficiency was achieved, but with a much higher background 
from $\epemto\ \W\W$ and $(\e)\nu\W$ in particular, due 
to the much worse jet-jet
invariant mass resolution. A b-tagging requirement may or may not be applied
to reduce the background (at the expense of a 80\% loss of efficiency), with
almost no consequences on the minimum luminosity needed for the discovery or the
exclusion.
}}
\put(0,4){\epsfxsize50mm\epsfbox{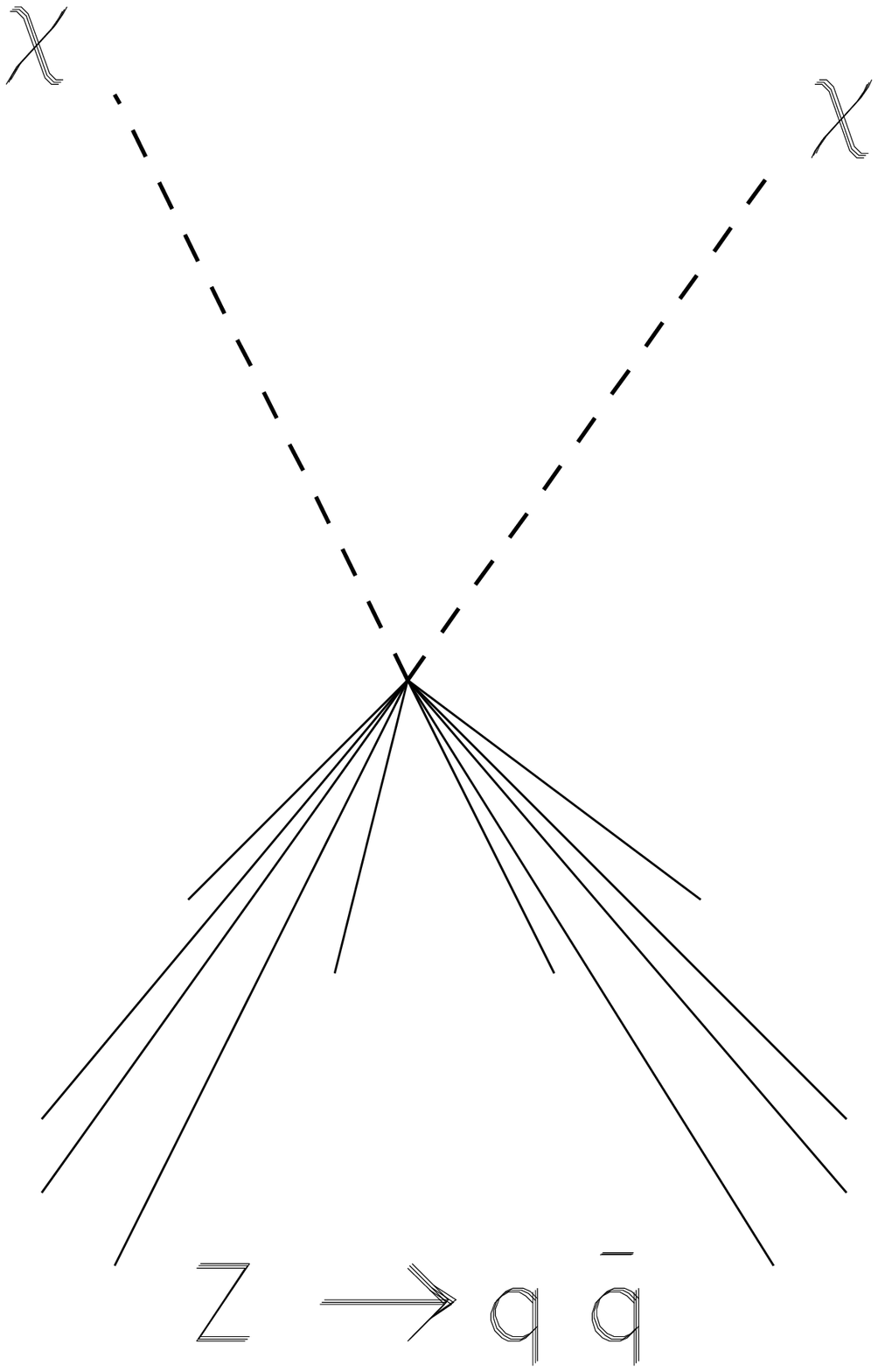}}
\end{picture}

\begin{figure}[htbp]
\begin{picture}(120,77)
\put(0,-5){\epsfxsize83mm\epsfbox{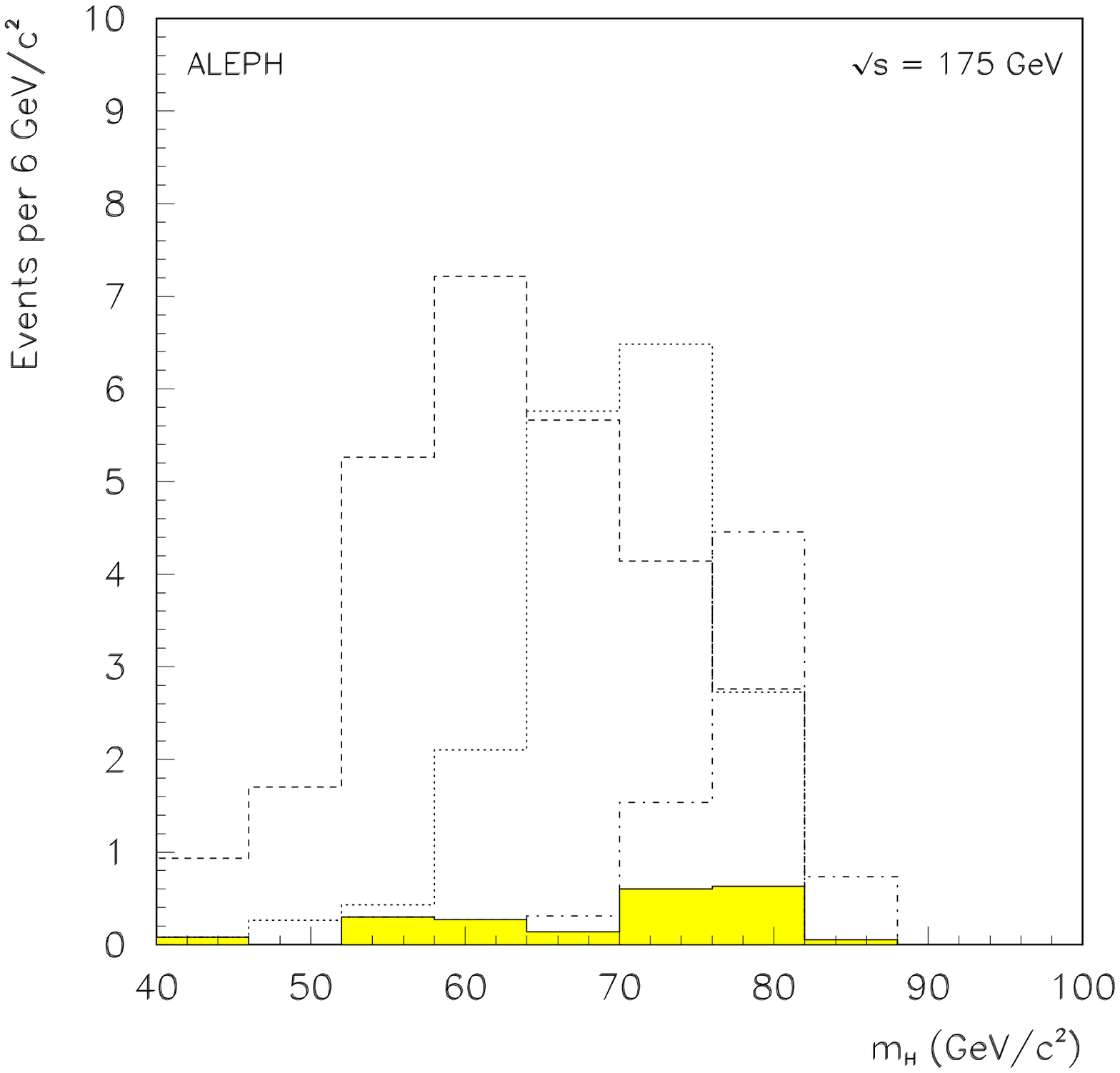}}
\put(80,-5){\epsfxsize83mm\epsfbox{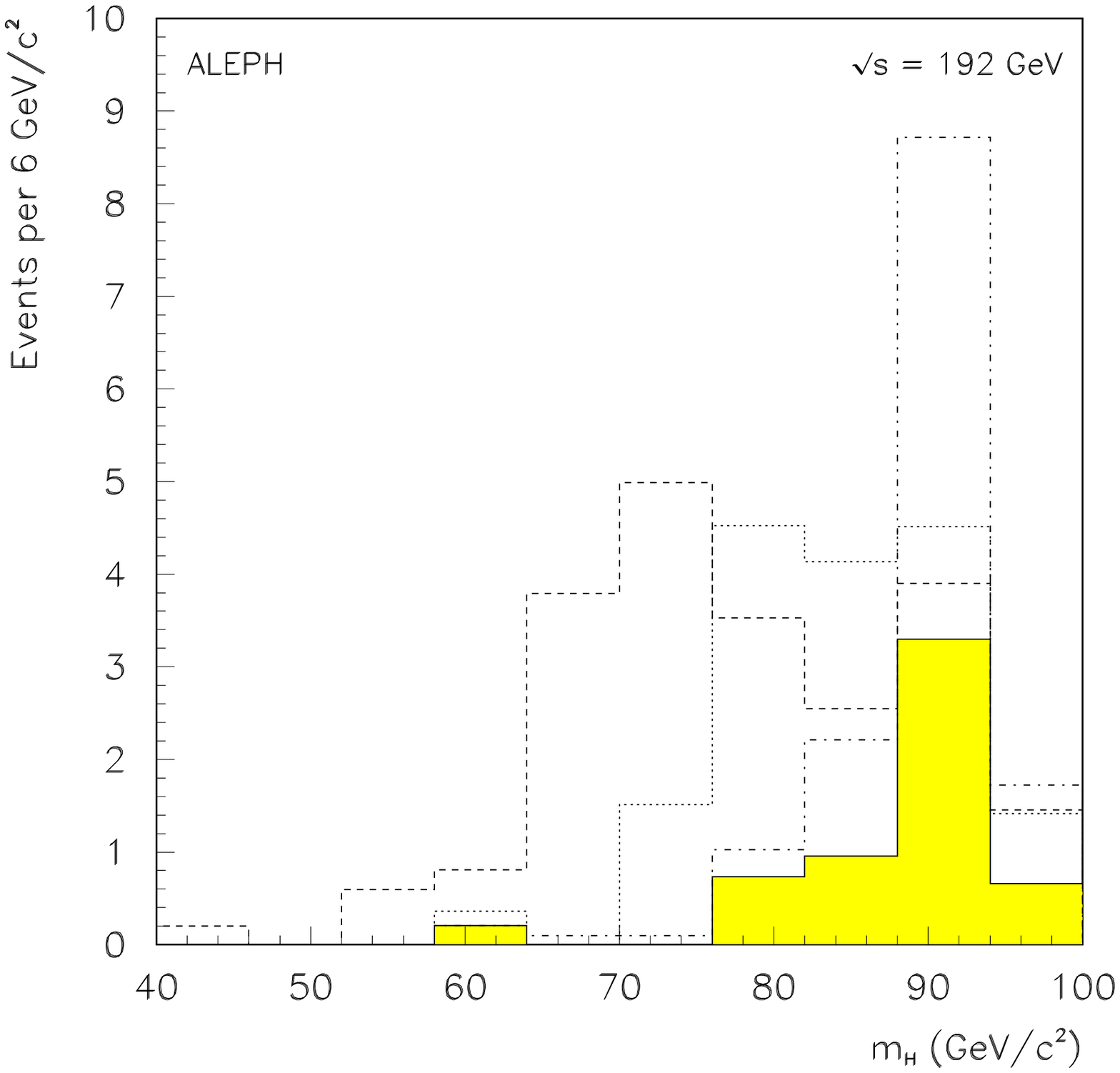}}
\end{picture}
\caption{\it 
  Same as in Fig.\ref{fig:acopll}, for the acoplanar jet topology,
  after a tight b-tagging (optional) requirement is applied.}
\label{fig:acopqq}
\end{figure}

Shown in Fig.\ref{fig:acopqq} are the mass distributions obtained
in the same configurations as in Fig.\ref{fig:acopll}, when a tight b-tagging
criterion is applied.  At 192~GeV, and for $\mh = 90~\Gcs$, the numbers of signal and
background events expected  are 7.7 and 4.9 respectively, assuming a 100\%
branching fraction into invisible final states.

\subsubsection{Search
 in Associated Pair-production \epemto\ \A\h}

The \epemto\ \h\A\ associated production leads to two main final states,
\bbbar\bbbar\ in 83\% of the cases and \tptm\qqbar\ in 16\% of the cases,
if supersymmetric decays are absent.
\vvs1

\noindent {\bf a) The \bbbar\bbbar\ Topology}\pss{0.5}
This four-jet final state is similar to the four-jet topology arising from
the Higgs-strahlung process, and a similar selection procedure can therefore
be applied. Here, the \Z\ mass constraint cannot be used, and the requirement
of incompatibility with a \W\W\ final state must be removed to retain a sizeable
efficiency for the case $\mh=\mA\sim 80~\Gcs$. However, since the b-quark content
is much higher in this four-b-jet topology than in the Higgs-strahlung process, a
much tighter b-tagging criterion can be applied. In terms of background
rejection, this may even over-compensate the removal of the two previous
requirements while keeping a high efficiency, varying between 10 and 35\%.

\begin{figure}[htbp]
\begin{picture}(120,75)
\put(66,35){(a)}
\put(102,35){(b)}
\put(0,-5){\epsfxsize77mm\epsfbox{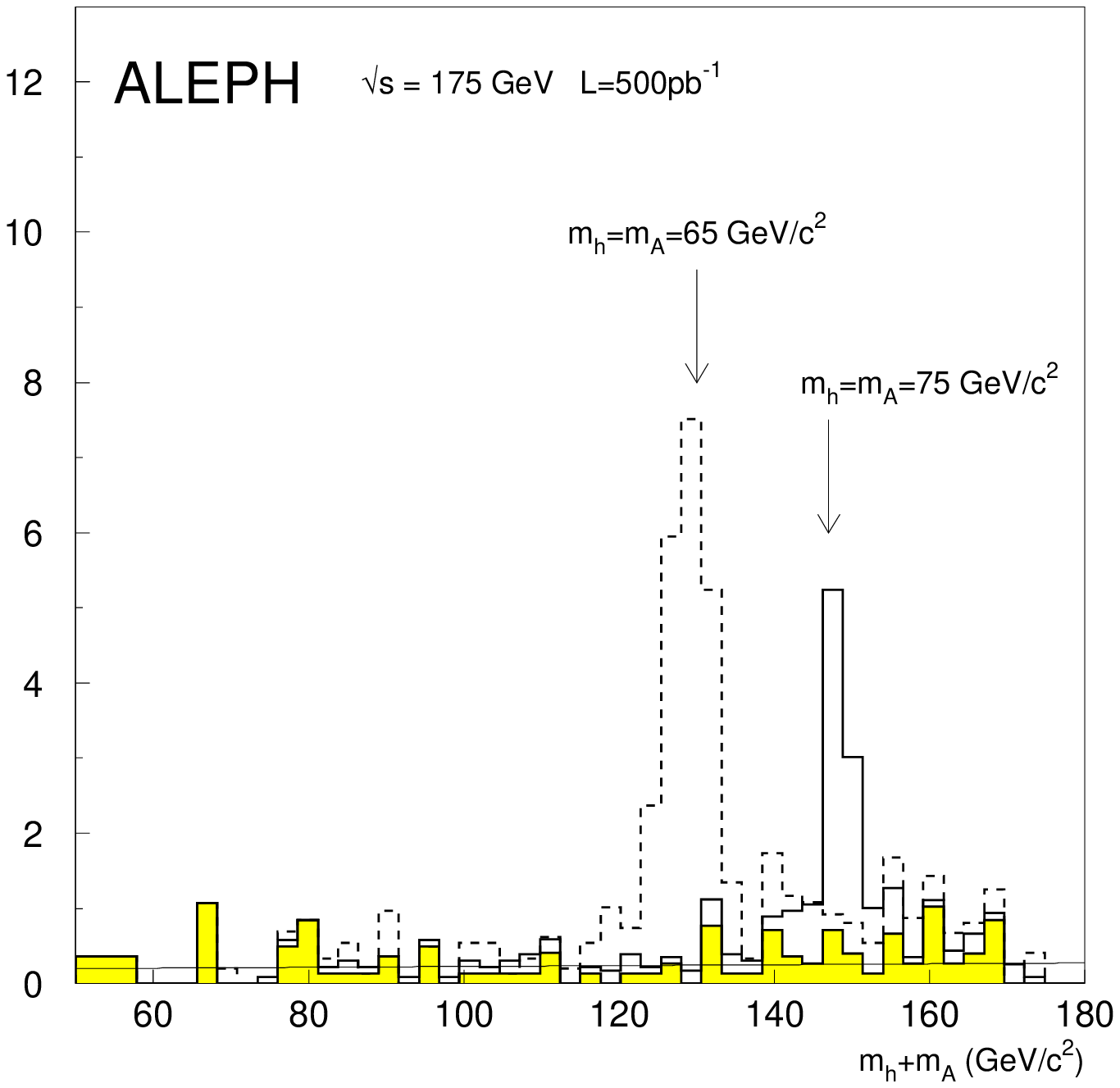}}
\put(83,-18){\epsfxsize85mm\epsfbox{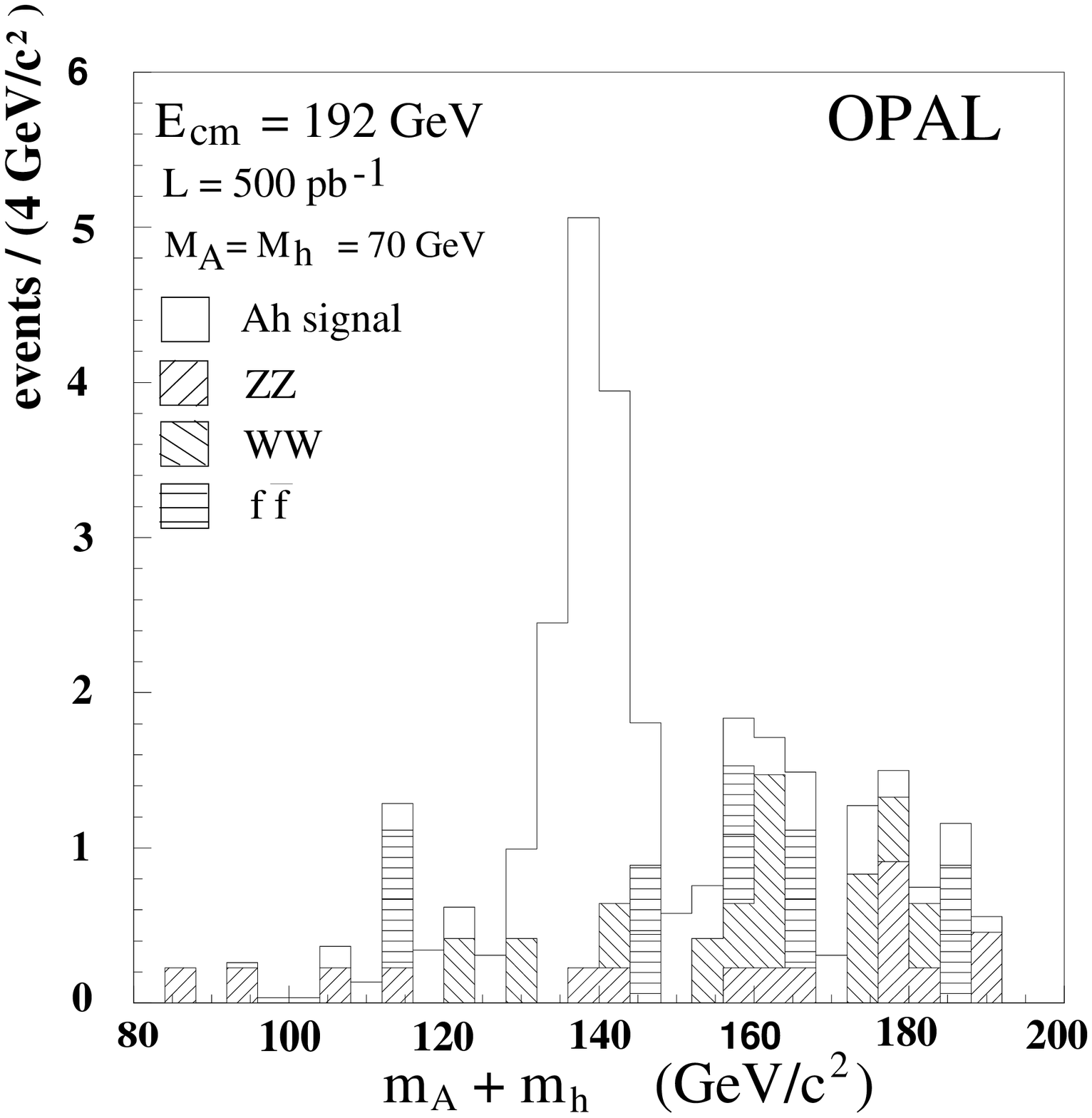}}
\end{picture}
\caption{\it 
  Mass distributions obtained in the $\h\A\ \to \bbbar\bbbar$
  topology.  (a) $\mh+\mA$ from ALEPH (175~GeV, 500~\inpb); and (b)
  $\mh+\mA$ from OPAL (192~GeV, 500~\inpb).}
\label{fig:ha4b}
\end{figure}

\begin{figure}[htbp]
\begin{picture}(120,85)
\put(66,35){(a)}
\put(102,35){(b)}
\put(3,2){\epsfxsize83mm\epsfbox{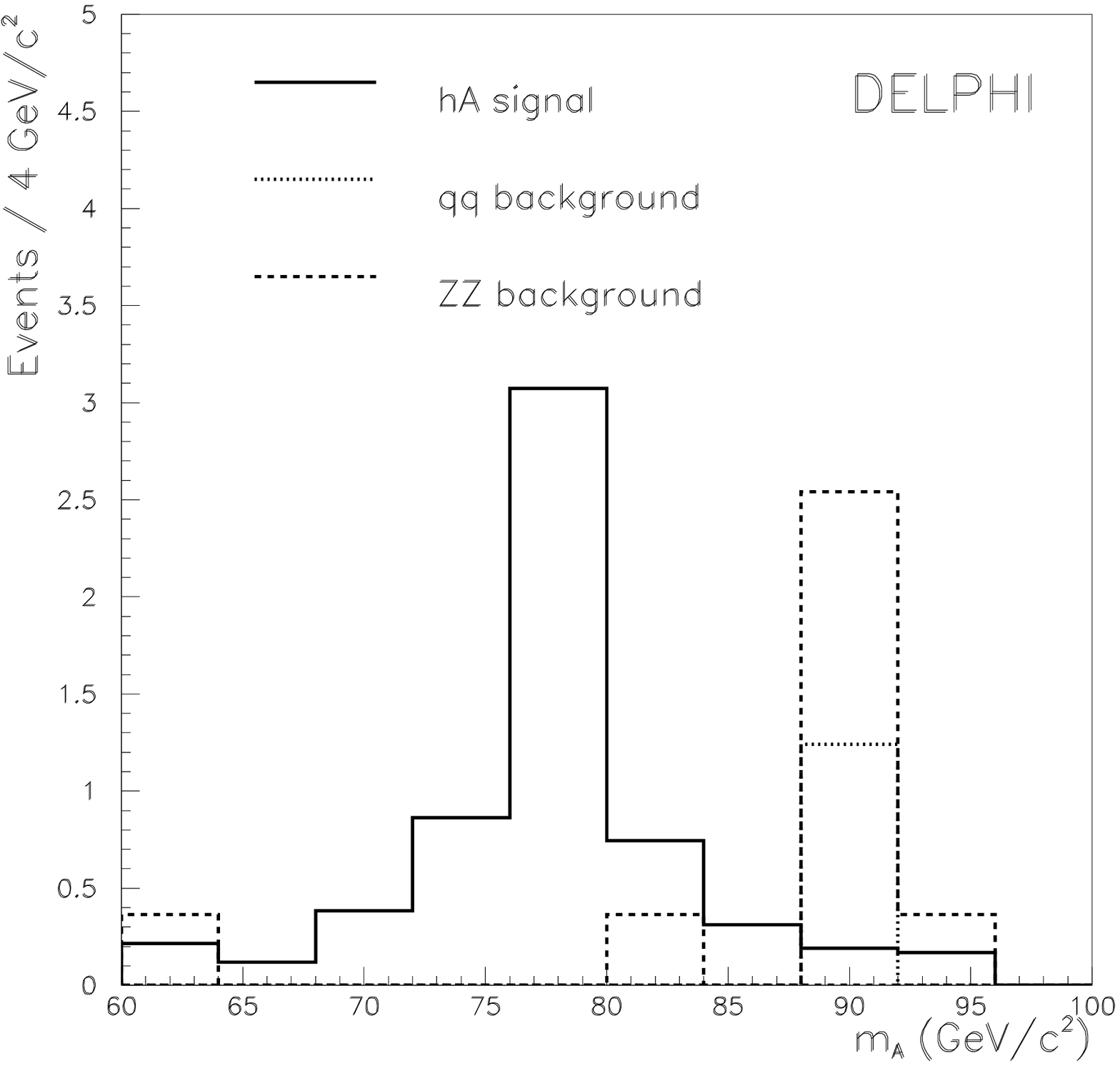}}
\put(94,-5){\epsfxsize77mm\epsfbox{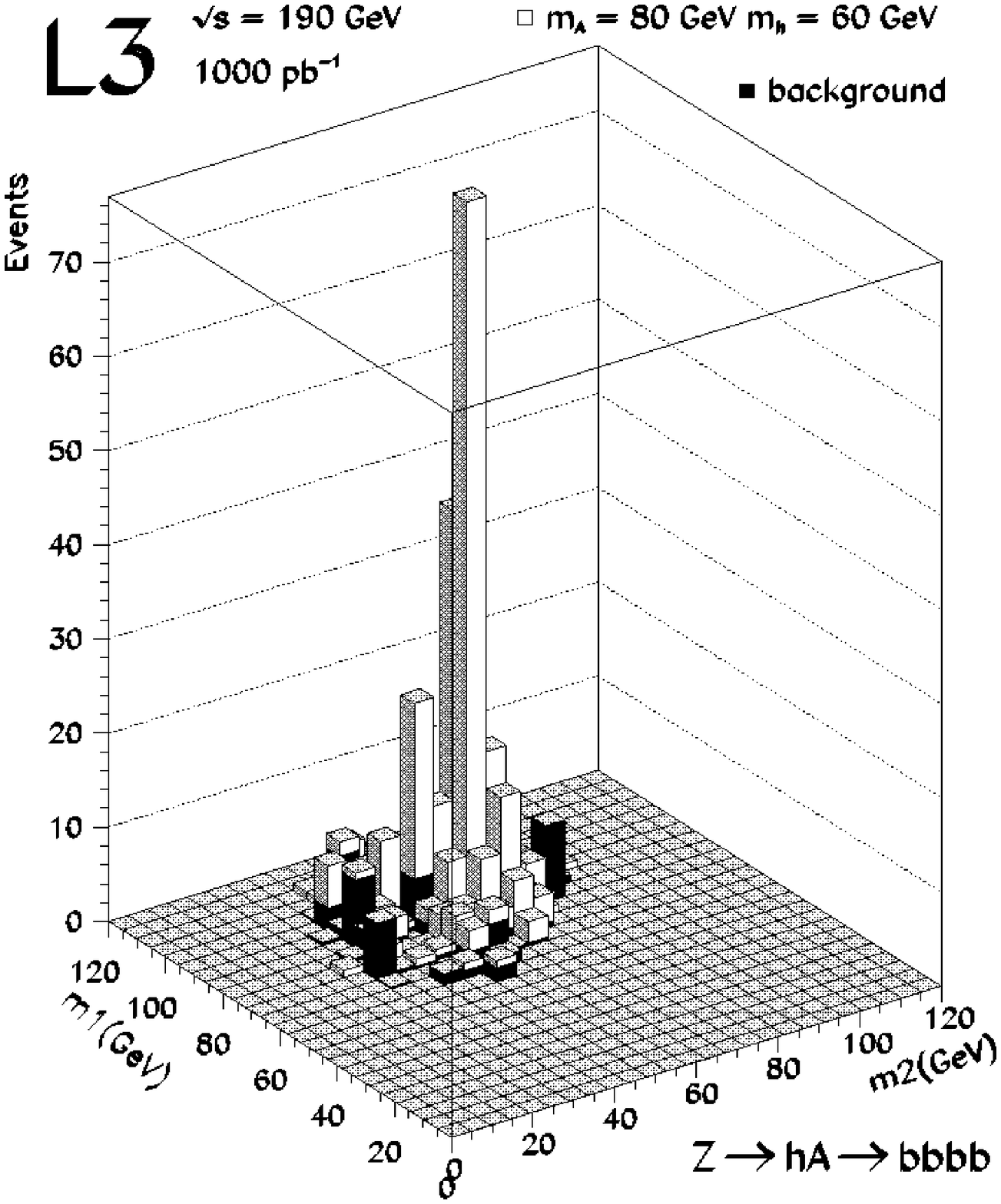}}
\end{picture}
\caption{\it 
  Mass distribution obtained in the $\h\A\ \to \bbbar\bbbar$ topology,
  (a) $\mh=\mA$ from DELPHI (192~GeV, 300~\inpb); and (b) \mh\ {\it
    vs} \mA\ from L3, assuming a signal-cross-section of 0.5~pb
  (190~GeV, 1~\infb).}
\label{fig:hal3}
\end{figure}

The four-jet energies and directions can then be fitted to satisfy the
total energy-momentum conservation constraint in order to achieve a
good mass resolution. Shown in Fig.\ref{fig:ha4b}a 
are the distributions of the sum of the fitted \mh\ and \mA\ 
values as obtained in the ALEPH detector with an integrated luminosity of 
500~\inpb\ at 175~GeV, for $\tanb=10$ and $\mA = 65$ and 75~\Gcs. 
The same distributions as seen by OPAL are shown in Fig.\ref{fig:ha4b}b
with 500~\inpb\ taken at 192 GeV, for $\mh = \mA = 70~\Gcs$.
Since for large \tanb\ values the \h\ and \A\ masses are expected to be close 
to each other, the mass resolution
is expected to improve by imposing this mass equality in the fit procedure.
This was done by DELPHI for $\mh=\mA=79~\Gcs$ at 192~GeV, and the result is
shown in Fig.\ref{fig:hal3}a for an integrated luminosity of 300~\inpb.
Finally, the distribution of \mh\ {\it vs} \mA\ that would be obtained in L3
in the mass configuration (60~\Gcs, 80~\Gcs) if the signal cross-section
amounted to 0.5~pb is shown in Fig.\ref{fig:hal3}b at 190~GeV, for an
integrated luminosity of 1~\infb.
\vvs1

\noindent {\bf b) The \tptm\bbbar\ Topology}\pss{0.5}
For this topology, the same analysis as for the Higgs-strahlung process was used
by ALEPH and DELPHI (with the exception that the very last fit intended to improve
the \tptm\ and hadronic mass resolution with the \mZ\ constraint does not apply).
The background level is already very low, except when \mh\ and \mA\ are close
to \mZ\ in which case the \Z\Z\ background can be reduced by tagging b-quarks.
In this configuration, however, the signal is expected to have a very low
cross-section except at the highest possible center-of-mass energy,
$\sqrt{s} = 205$~GeV. Altogether, when added to the preceding one, this analysis
increases the selection efficiency of the \h\A\ channel by about 20\%.
\vvs1

\noindent {\bf c) The Case \h\ or $\A\to\chi\chi$}\pss{0.5}
If either \h\ or \A\ decays predominantly into $\chi\chi$, the relevant topology
becomes that of an acoplanar jet pair, as already described in
Section~\ref{sec:acopqq}b). However, the pair of jets is actually a pair of
b-quarks in that case, therefore improving the selection efficiency
of a b-tagging criterion with respect to the $\epemto\ \h\Z$
configuration.

In the unfortunate situation where both \h\ and \A\ predominantly decay
into a pair of LSPs, the resulting final state becomes totally invisible
and cannot be found at \LEPII. However, in that case, there is a fair
chance to discover the lightest supersymmetric particle {\it via}
a direct neutralino search.

\subsection{Discovery and Exclusion Limits}

Using the definitions of Appendix~\ref{sec:app}, a minimum signal cross-section
was inferred for the $\epemto\ \h\A$ process from the expected number of
background events, both for the discovery and the exclusion. Since the
background mass distribution is mostly uniform over the (\mh,\mA) plane,
it turns out that this minimum signal cross-section does not depend on
\mh\ and \mA. For instance, at 192 GeV and with an integrated luminosity of
150~\inpb, a cross-section in excess of 65 (30)~fb can be discovered (excluded)
in the $\epemto\ \h\A$ channel
when supersymmetric decays are closed.

\begin{figure}[p]
\centerline{\epsfig{file=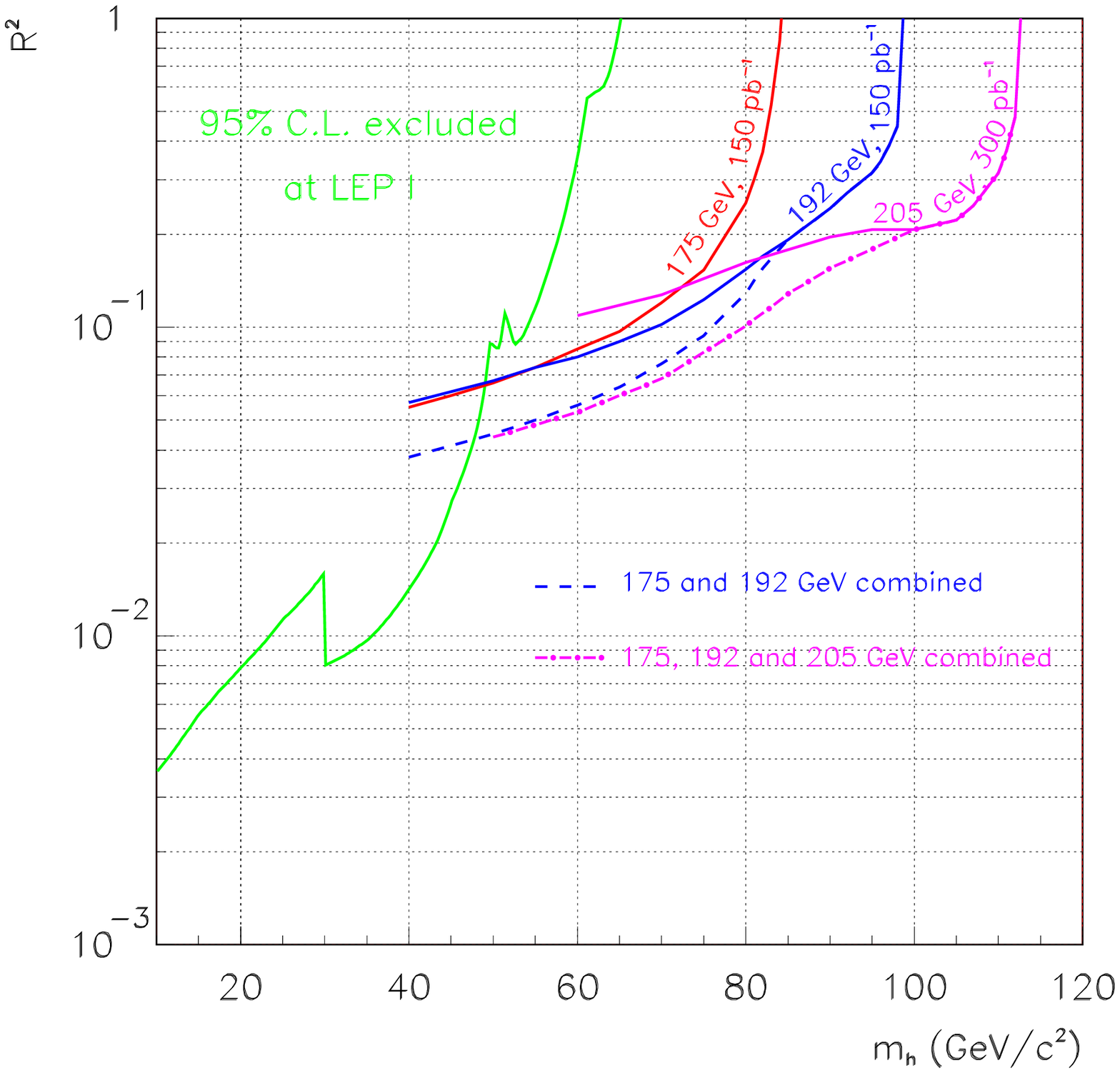,%
        height=9cm, clip=}}
\caption{\it 
  95\% C.L. upper limits on $R^2$, as a function of \mh\, for a
  visible Higgs boson, for the three center-of-mass energies: $R^2 =
  \sin^2(\beta-\alpha) \times BR(\h \to $ any visible final state) for
  the \LEPI\ part of the curve and $R^2 =\sin^2(\beta-\alpha) \times
  BR(\h \to \bbbar)$ for the \LEPII\ part of the curve.}
\label{fig:sbavis}
\vvs{0.9}
\centerline{\epsfig{file=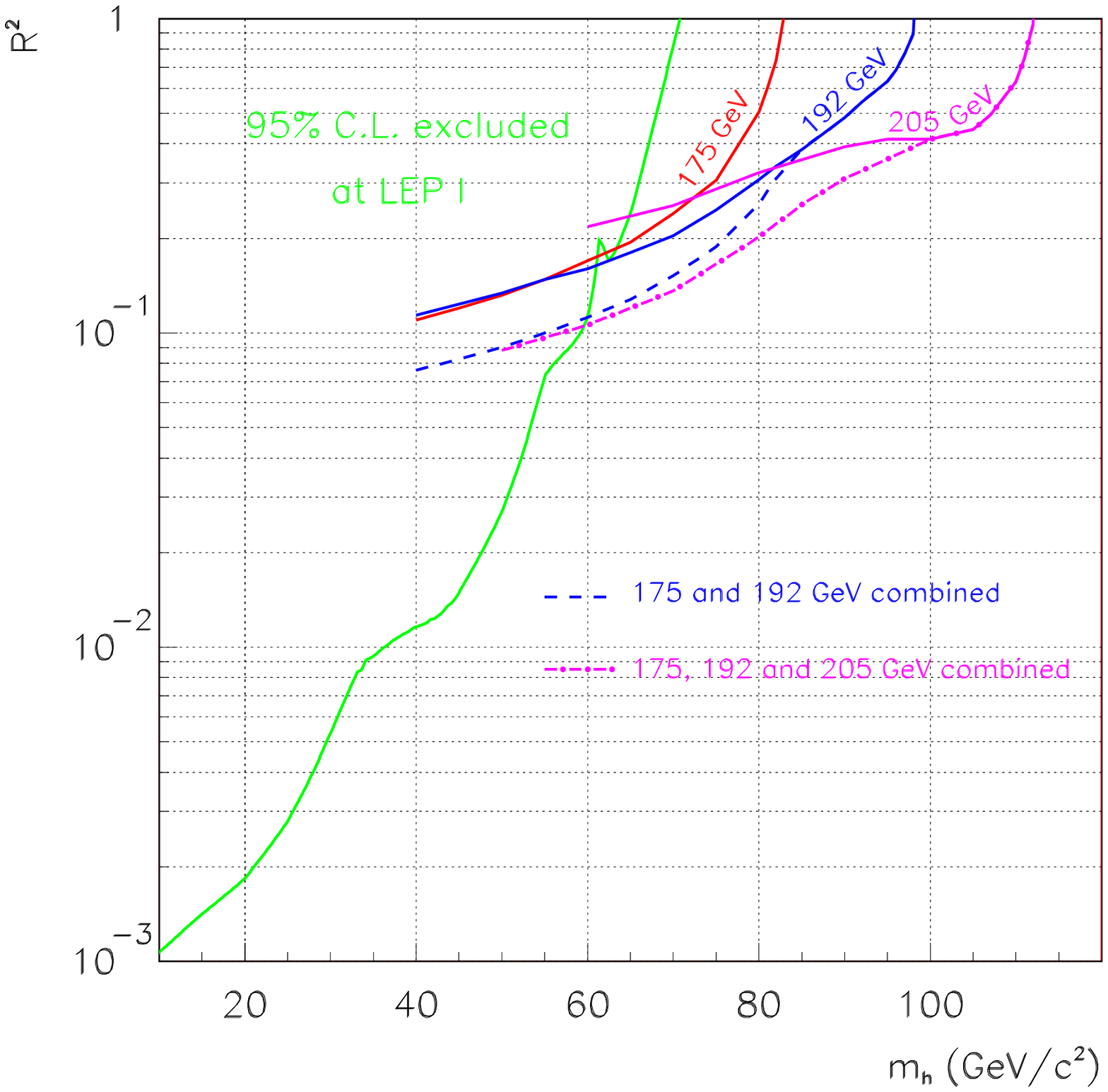,%
        height=9cm, clip=}}
\caption{\it 
  95\% C.L. upper limits on $R^2$, as a function of \mh\, for an
  invisible Higgs boson, for the three center-of-mass energies: $R^2
  =\sin^2(\beta-\alpha) \times BR(\h \to$ any invisible final state).}
\label{fig:sbainv}
\end{figure}

For the Higgs-strahlung process, the total cross-section is reduced
with respect to the standard model expectation by a factor denoted
$\sin^2(\beta-\alpha)$ in the MSSM. The number of events expected is also directly
affected by the branching  ratio of the h decay into \bbbar. For each \mh, a
minimum value for $R^2 \equiv \sin^2(\beta-\alpha) \times BR(\h\to\bbbar)$
was inferred for the discovery and the exclusion.
The result, which is model-independent, is shown in Fig.\ref{fig:sbavis} for the
three center-of-mass energies 175, 192 and 205~GeV, with integrated luminosities
of 150, 150 and 300~\inpb, respectively \cite{PJ}. 
The interpretation of the negative
searches at \LEPI\ (from Ref.\cite{haber}) is also shown in this plot.

Since this limit becomes irrelevant when supersymmetric decays are dominant, it
is supplemented by the 95\% C.L. upper limit on $R^2 \equiv \sin^2(\beta-\alpha)
\times BR(\h\to$ invisible final states), as shown in
Fig.\ref{fig:sbainv}.\footnote{Since this analysis was done in ALEPH only,
it was assumed that the other experiments would contribute in the same proportions
as for the visible decays to perform the combination.}
At \LEPII, this limit is worse by about a factor of two than in the case 
where \h\ predominantly decays into \bbbar, while it was better at \LEPI.
In order to make easier the use of these curves to test other models, such 
as non-minimal supersymmetric extensions of the standard model, 
the exclusion and
discovery limits on $R^2  = \sin^2(\beta-\alpha) \times BR(\h\to\bbbar)$ 
are presented
in Table~\ref{tab:sba}, for the three center-of-mass energies, and when 
combining
the results at 175 and 192 GeV, on the one hand, and with the 205 GeV
results in addition, on the other.

\begin{table}[htbp]
\setlength{\tabcolsep}{0.7pc}
\caption{\it
  Minimum value for $R^2 = \sin^2(\beta-\alpha) \times
  BR(\h\to\bbbar)$ at the three center-of-mass energies, with
  integrated luminosities of 150, 150 and 300~\inpb, respectively, and
  for various Higgs boson masses.  Also shown is the combination of
  the 175 GeV and 192 GeV results, with an integrated luminosity of
  150~\inpb\ taken at each energy, and the overall combination of the
  175, 192 and 205 GeV results, assuming 300~\inpb\ at 205 GeV. The
  combination of several center-of-mass energies have not been used in
  Fig.\ref{fig:neuf} and~\ref{fig:dix}.}
\label{tab:sba}
\begin{center}
$\sqrt{s}=175$ GeV

\vskip .25cm
\begin{tabular}{lrrrrrrrrr}
\hline\hline
\mH (\Gcs) & 40    &  50   &  60   &  70   &  75   &   80  &  82   &  83   & 84   \\
\hline
Exclusion  & 0.055 & 0.066 & 0.085 & 0.120 & 0.153 & 0.252 & 0.369 & 0.525 & 0.842 \\
Discovery  & 0.136 & 0.164 & 0.207 & 0.303 & 0.381 & 0.620 & 0.963 &  ---  &  ---  \\
\hline\hline
\\
\end{tabular}

\vskip .25cm
$\sqrt{s}=192$ GeV

\vskip .25cm
\begin{tabular}{lrrrrrrrrr}
\hline\hline
\mH (\Gcs) &  40   &  50   &  60   &  70   &  80   &  90   &  95   &  97   &  98   \\
\hline
Exclusion  & 0.057 & 0.067 & 0.080 & 0.103 & 0.154 & 0.242 & 0.316 & 0.387 & 0.447 \\
Discovery  & 0.142 & 0.163 & 0.197 & 0.249 & 0.385 & 0.584 & 0.781 &  ---  &  ---  \\
\hline\hline
\\
\end{tabular}

\vskip .25cm
Combination of 175 and 192 GeV

\vskip .25cm
\begin{tabular}{lrrrrrrrrr}
\hline\hline
\mH (\Gcs) &  40   &  50   &  60   &  70   &  80   &  90   &  95   &  97   &  98   \\
\hline
Exclusion  & 0.038 & 0.045 & 0.056 & 0.076 & 0.129 & 0.242 & 0.316 & 0.387 & 0.447 \\
Discovery  & 0.095 & 0.113 & 0.139 & 0.190 & 0.317 & 0.584 & 0.781 &  ---  &  ---  \\
\hline\hline
\\
\end{tabular}

\vskip .25cm
$\sqrt{s}=205$ GeV

\vskip .25cm
\begin{tabular}{lrrrrrrrrr}
\hline\hline
\mH (\Gcs) & 60    &  70   &  80   &  90   &  100  &  105  & 110   & 111   & 112   \\
\hline
Exclusion  & 0.109 & 0.127 & 0.162 & 0.196 & 0.207 & 0.222 & 0.315 & 0.375 & 0.481 \\
Discovery  & 0.270 & 0.315 & 0.404 & 0.490 & 0.510 & 0.539 & 0.785 &  ---  &  ---  \\
\hline\hline
\\
\end{tabular}

\vskip .25cm
Combination of 175, 192 and 205 GeV

\vskip .25cm
\begin{tabular}{lrrrrrrrrr}
\hline\hline
\mH (\Gcs) &  50   &  60   &  70   &  80   &  90   & 100   & 105   & 110   & 112   \\
\hline
Exclusion  & 0.044 & 0.053 & 0.068 & 0.101 & 0.155 & 0.207 & 0.222 & 0.315 & 0.481 \\
Discovery  & 0.111 & 0.132 & 0.168 & 0.252 & 0.370 & 0.510 & 0.539 & 0.785 &  ---  \\
\hline\hline
\\
\end{tabular}

\end{center}
\end{table}

To interpret these results in the MSSM framework, both the $\epemto\ \h\A$ cross-section
and the $\sin^2(\beta-\alpha)$ value were computed and compared
to the above minimum values in a systematic scan of the
(\mA, \tanb)  plane, for $M_{t} = 175 \pm 25~\Gcs$ and for the
three different stop mixing configurations, {\it (i)} No mixing:
$A_t = 0$ and $|\mu| \ll M_S$;
{\it (ii)} Typical mixing: $ A_t = M_S$ and $\mu = -M_S$
 (these values of $A_t$ and $\mu$ give a moderate impact of the stop mixing
for  large \tanb\ but a mixing effect close to maximal
 if \tanb\ is small);
and {\it (iii)} Maximal mixing: $A_t = \sqrt{6} M_S$ and
$|\mu| \ll M_S$, with $M_S =1$ TeV.

The results of the above  analysis are summarized in a series of figures
(Figs.\ref{fig:neuf} and \ref{fig:dix}), which
 display the areas in the $(\mA,\tanb)$
and $(\mh,\tanb)$ planes that can
be covered for a given energy of \LEPII, $\sqrt{s} = 175, 192$ and
205 GeV at the integrated luminosities 150, 150 and 300~\inpb,
respectively.
The  top-quark mass  and
the stop mixing parameters are varied as specified above. The figures are
obtained by combining the four LEP experiments.
At $\sqrt{s}$= 175 GeV, an increase in luminosity
beyond 150~\inpb\  does not improve the
potential of the machine for the  discovery of the light Higgs boson in any
 relevant way. The analogous conclusion was reached in the standard model case.
At $\sqrt{s}$= 192 GeV, still 150~\inpb\ of integrated luminosity per
experiment are sufficient to make proper use of the discovery potential of the
machine, and a larger luminosity of $\simeq 300~\inpb$
gives only a slight improvement in the upper bound on the Higgs boson mass
which can be discovered or excluded. Although
for any center-of-mass energy the variation from
150--200~\inpb\ to 300--400~\inpb\  of integrated luminosity yields a gain of at most
2 to 3~\Gcs\ in the maximal Higgs boson mass that can be reached, at 
$\sqrt{s}$ = 205 GeV the results are presented for 300~\inpb\ since for 
this energy value the increase in luminosity translates into a quite impressive  
coverage of  \tanb.

\begin{figure}[p]
\begin{picture}(140,190)
\put(25,90){\epsfxsize100mm\epsfbox{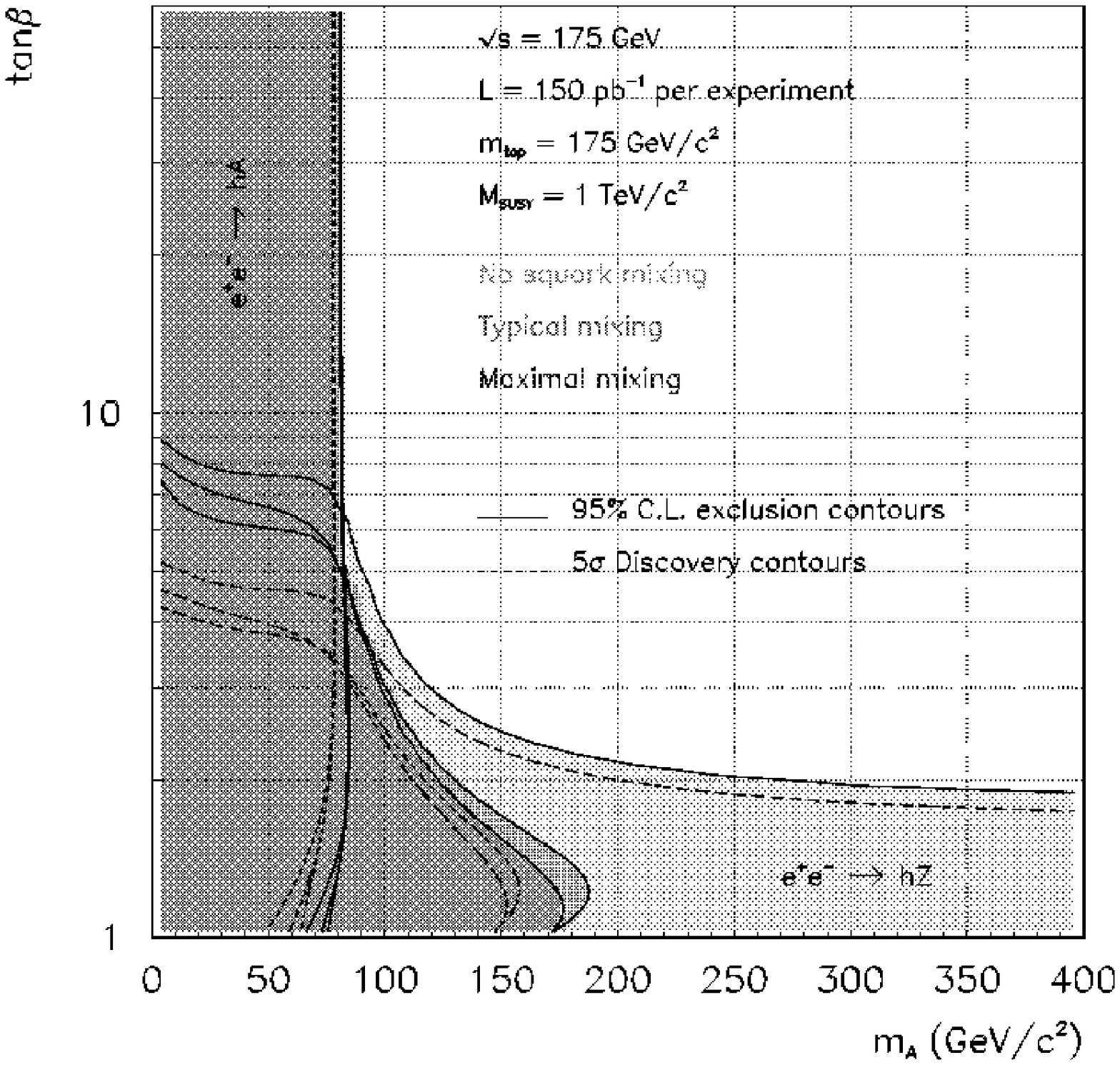}}
\put(15,140){(a)}
\put(25,-5){\epsfxsize100mm\epsfbox{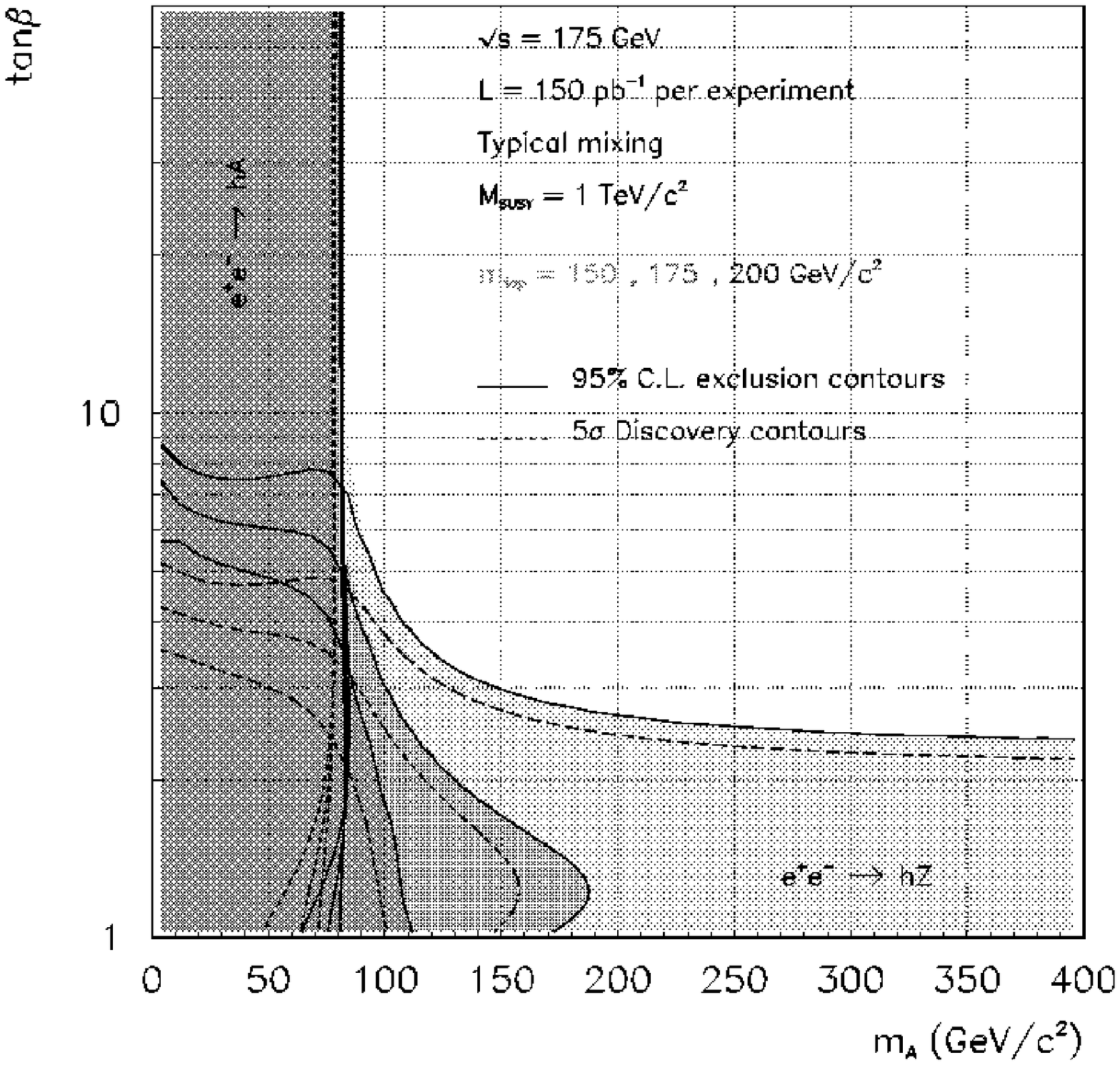}}
\put(15,45){(b)}
\end{picture}
\caption{\it 
  Exclusion and discovery limits in the $[\mA, \tanb]$ plane for each
  of the center-of-mass energies, varying the values of the stop
  mixing parameters as specified in the text (a, c and e) and varying
  the values of $M_t$= 150, 175, 200~\Gcs\ for $A=-\mu = M_S= 1$ TeV
  (b, d and f).}
\label{fig:neuf}
\end{figure}

\begin{figure}[p]
\begin{picture}(140,190)
\put(25,90){\epsfxsize100mm\epsfbox{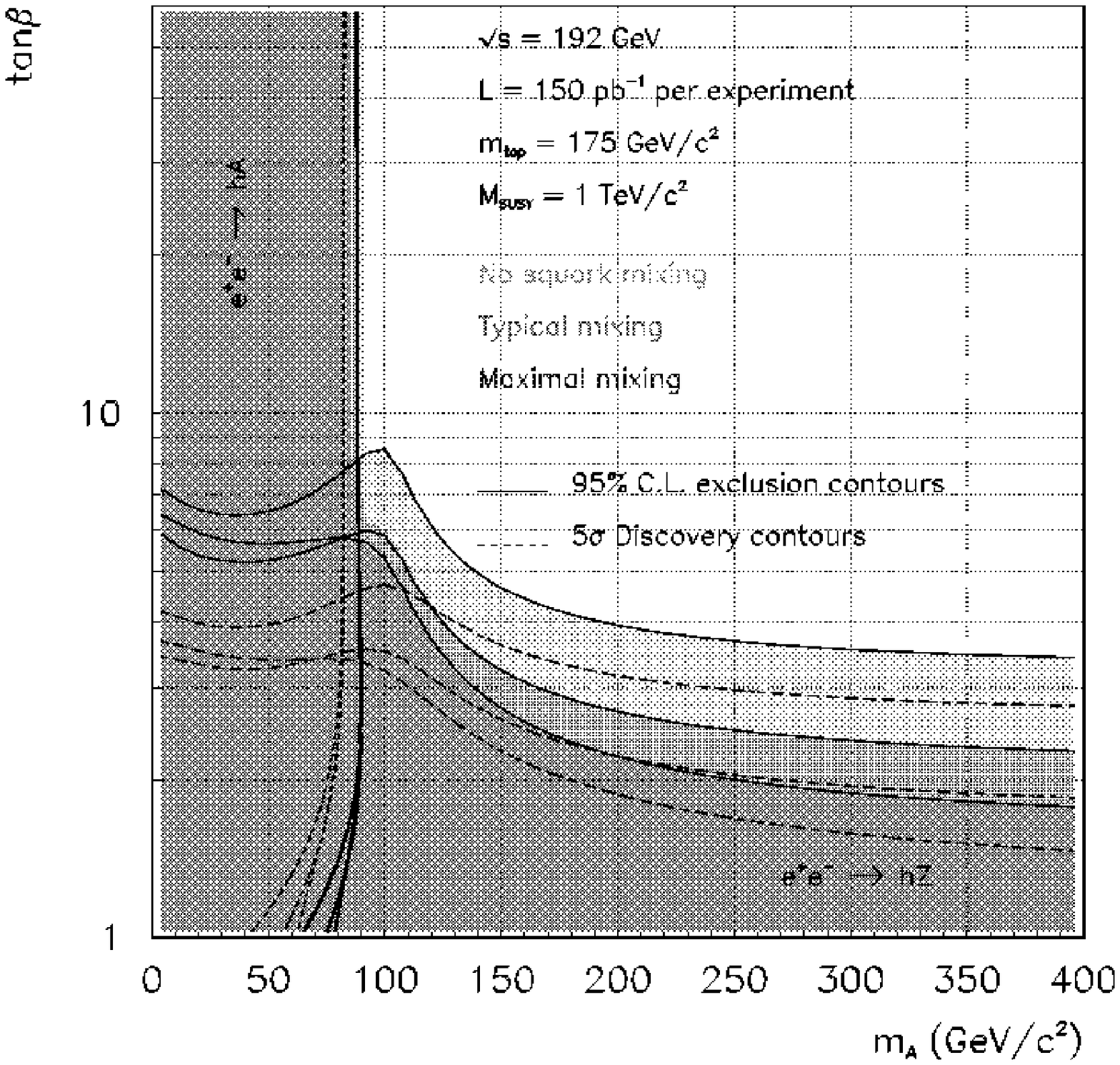}}
\put(15,140){(c)}
\put(25,-5){\epsfxsize100mm\epsfbox{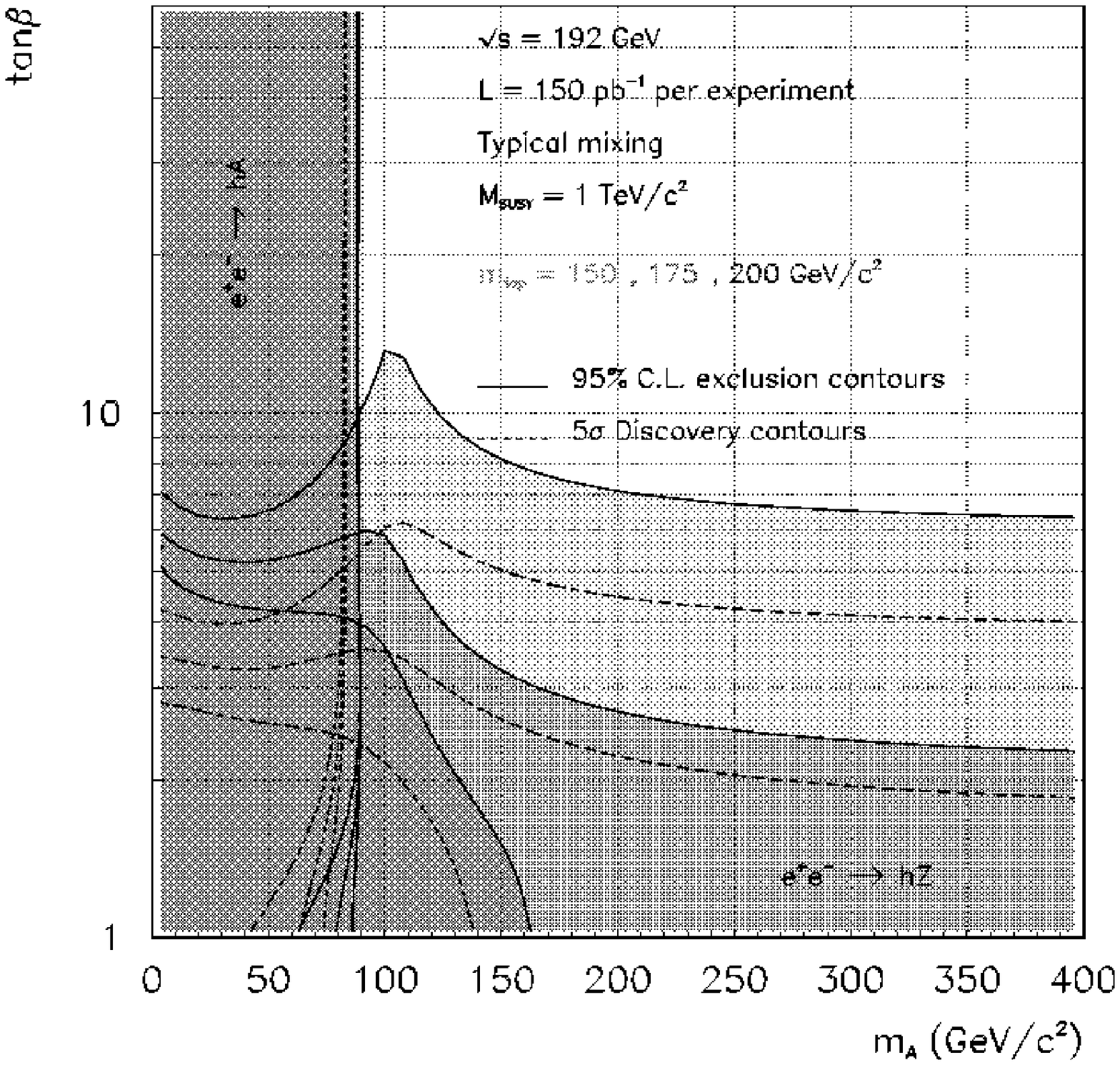}}
\put(15,45){(d)}
\end{picture}
\addtocounter{figure}{-1}
\caption{\it (cont'd)}
\end{figure}

\begin{figure}[p]
\begin{picture}(140,190)
\put(25,90){\epsfxsize100mm\epsfbox{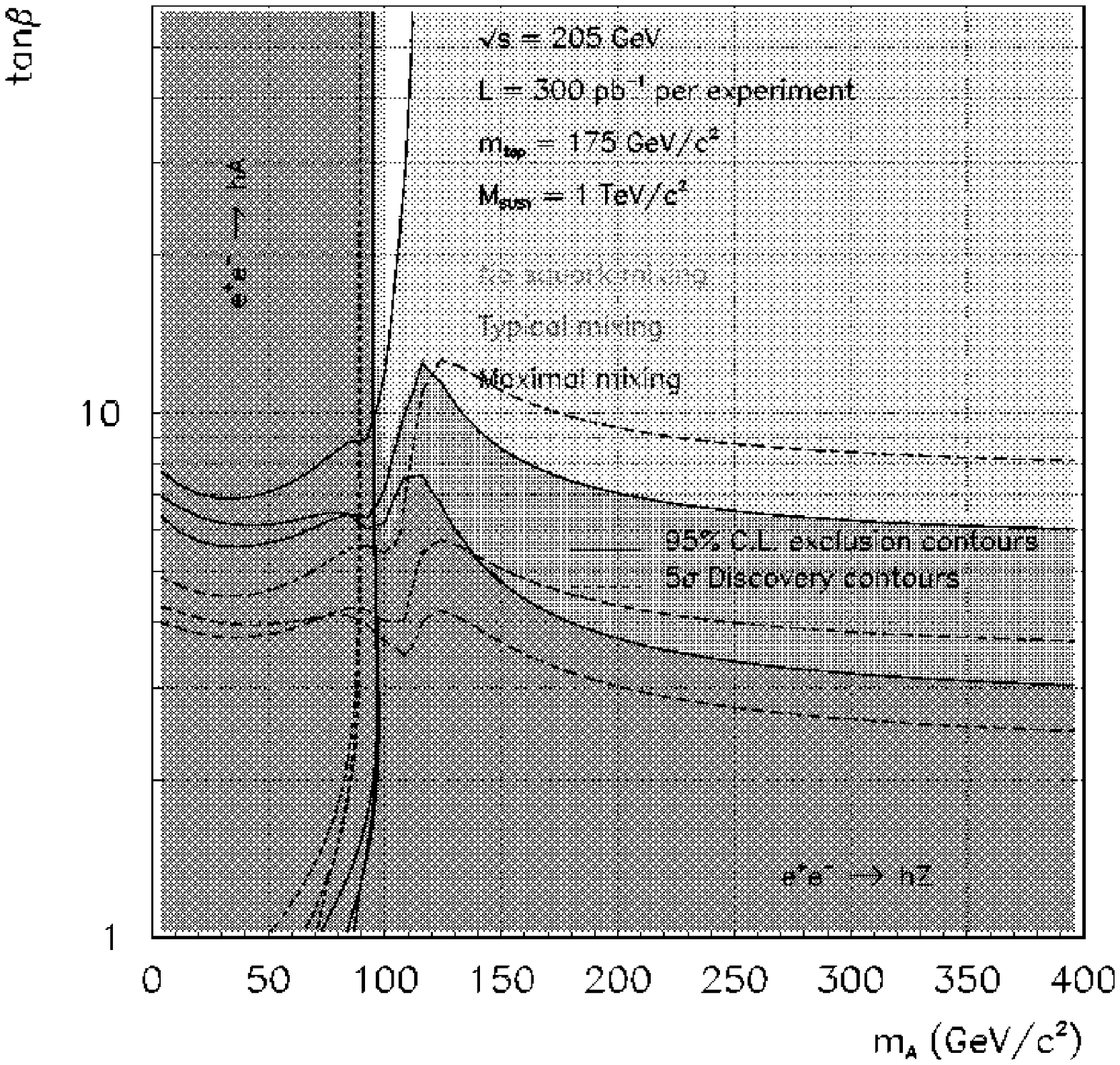}}
\put(15,140){(e)}
\put(25,-5){\epsfxsize100mm\epsfbox{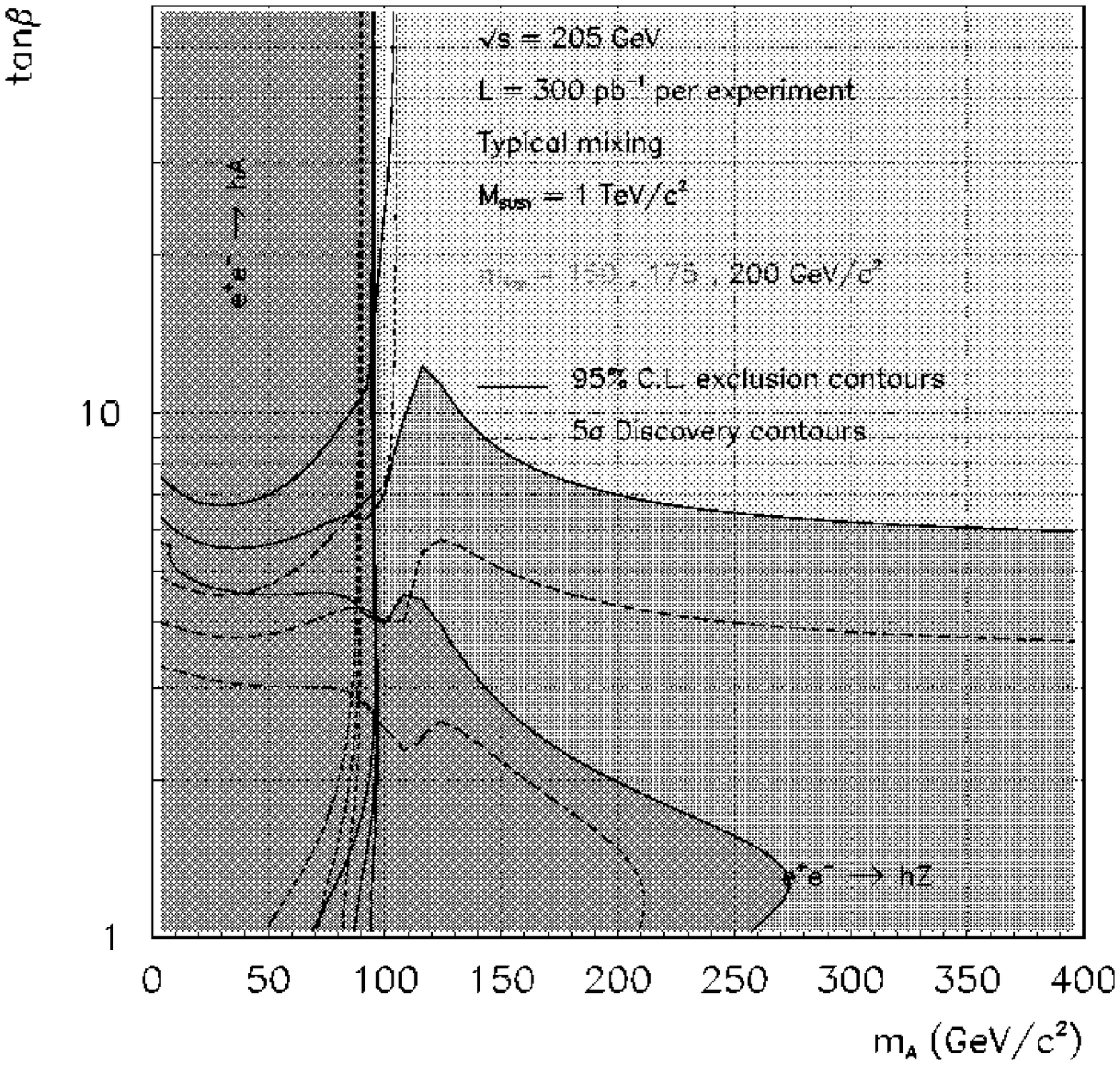}}
\put(15,45){(f)}
\end{picture}
\addtocounter{figure}{-1}
\caption{\it (cont'd)}
\end{figure}

\begin{figure}[p]
\begin{picture}(140,190)
\put(25,90){\epsfxsize100mm\epsfbox{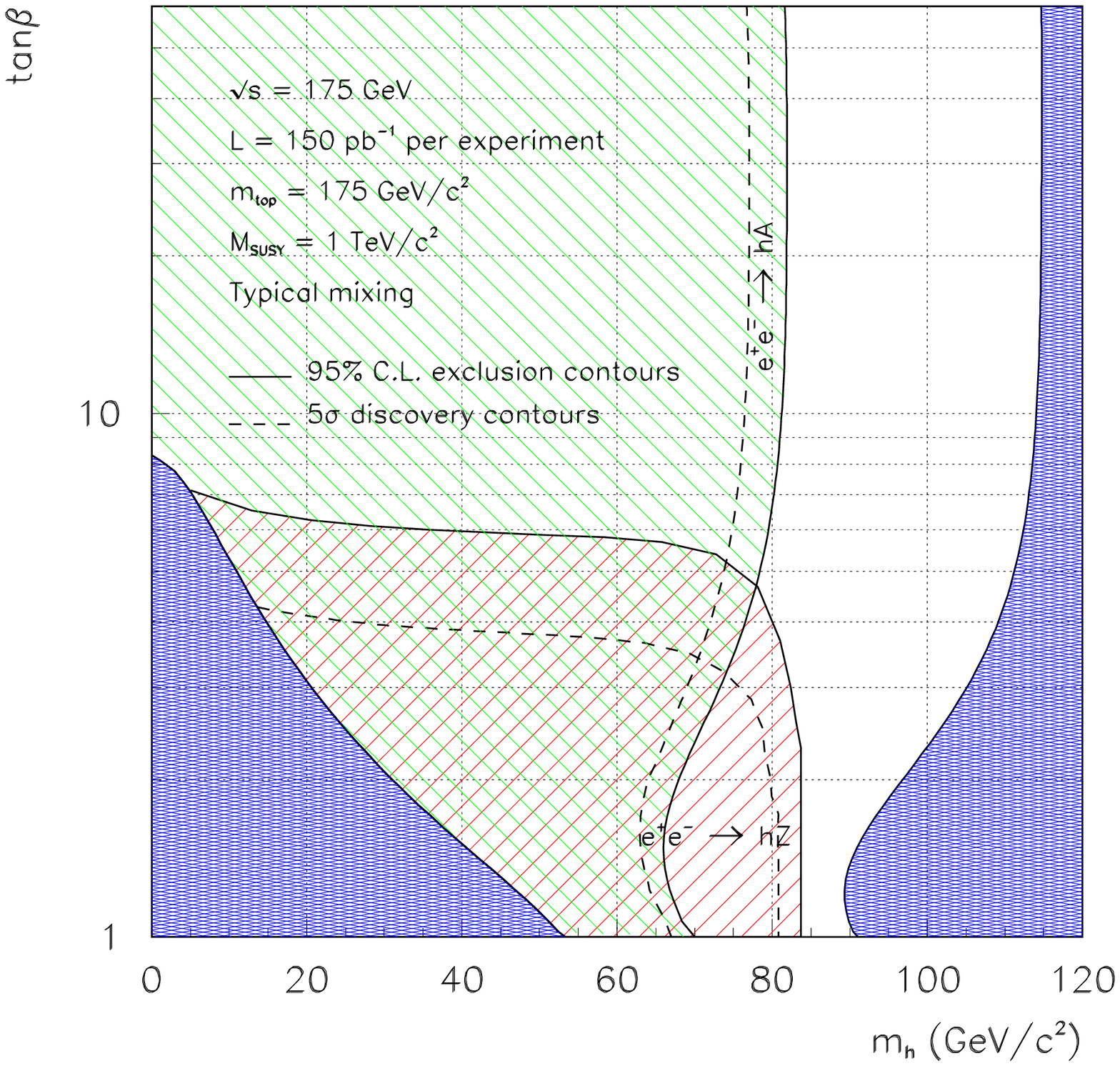}}
\put(15,140){(a)}
\put(25,-5){\epsfxsize100mm\epsfbox{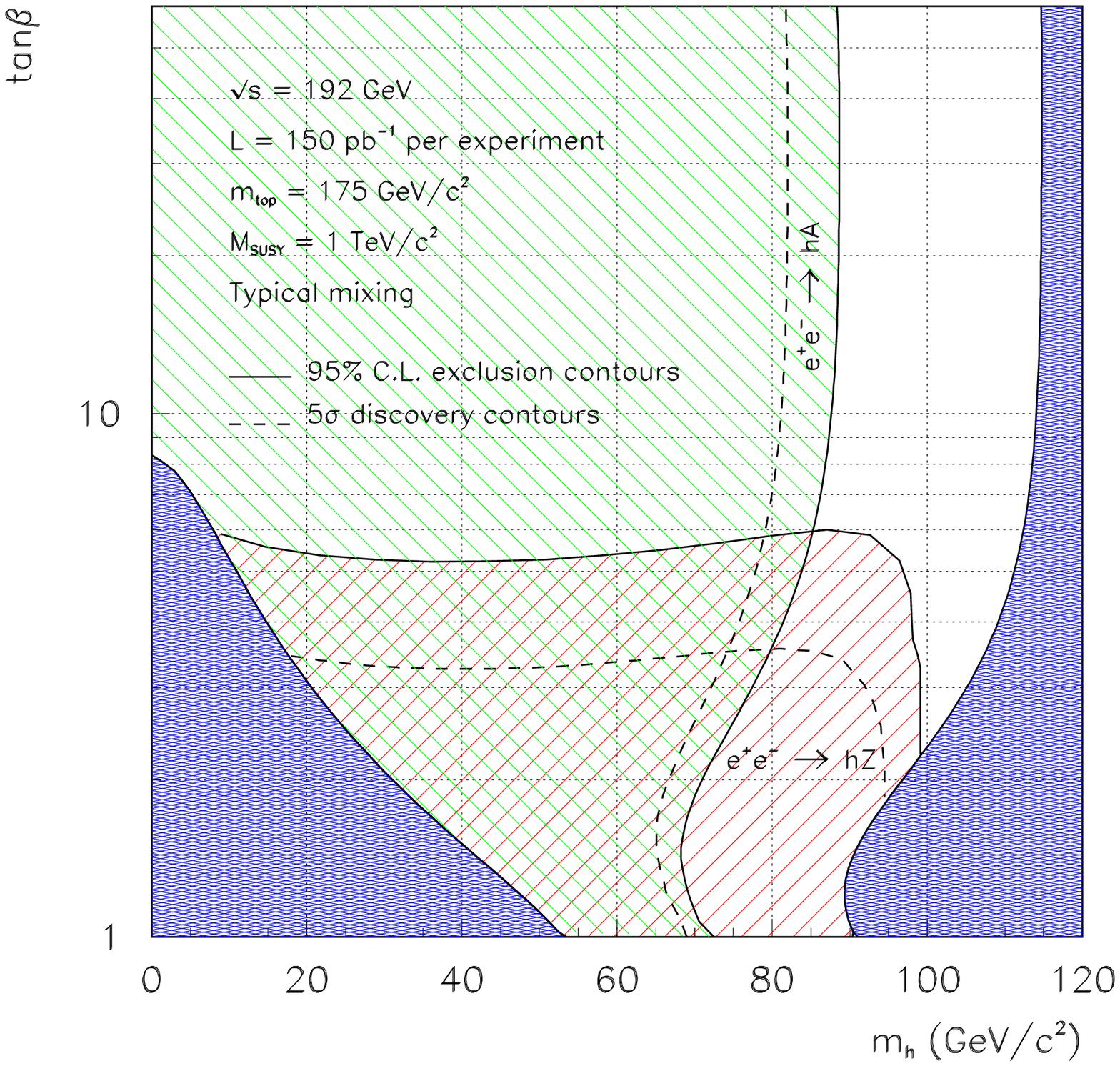}}
\put(15,45){(b)}
\end{picture}
\caption{\it 
  Exclusion and discovery limits in the $[\mh, \tanb]$ plane, for
  $M_t$ = 175 GeV and $A= - \mu = M_S = 1$ TeV for each of the
  center-of-mass energies. The dark shaded areas are excluded
  theoretically.}
\label{fig:dix}
\end{figure}

\begin{figure}[tbp]
\begin{picture}(140,90)
\put(25,-5){\epsfxsize100mm\epsfbox{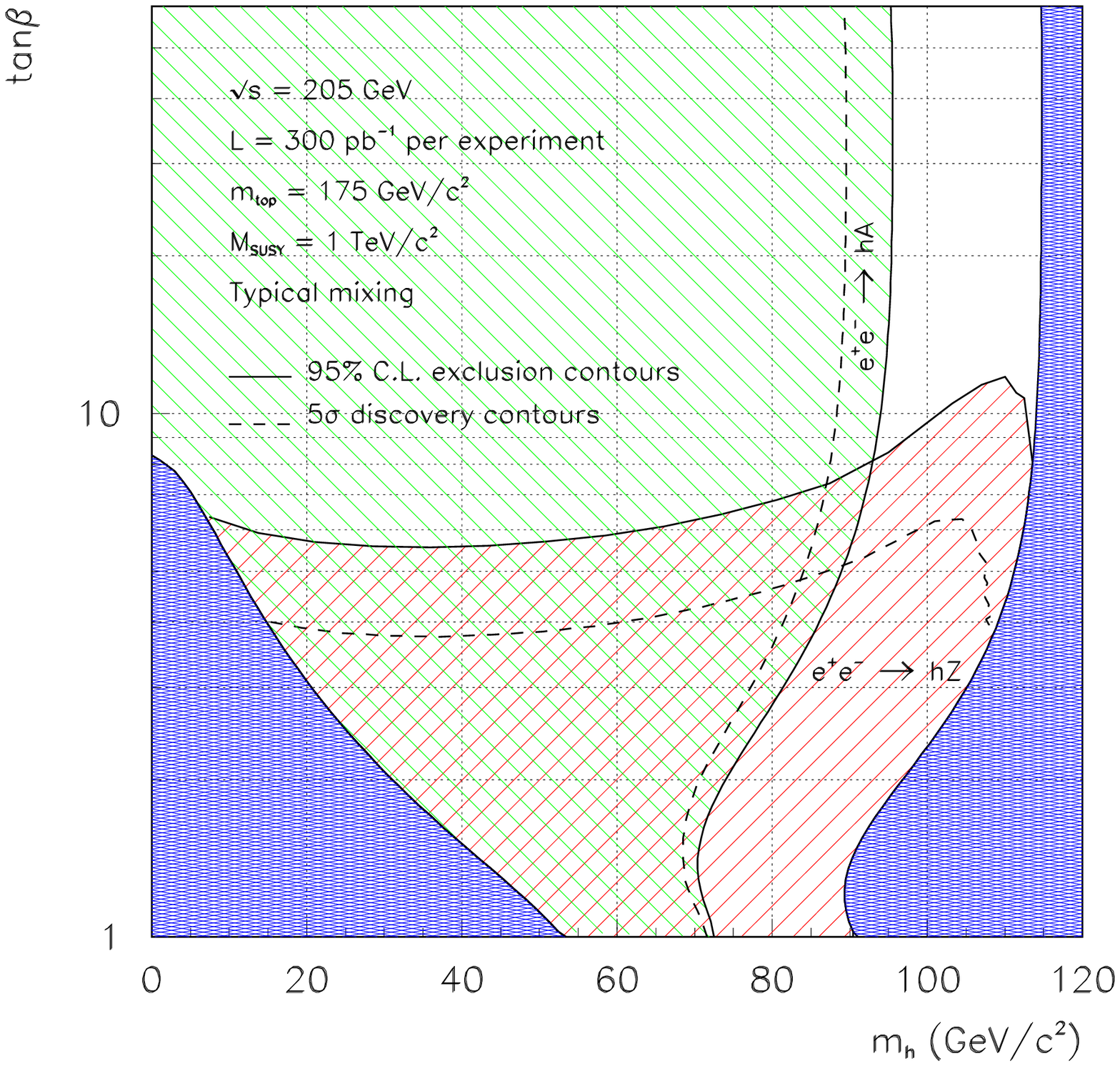}}
\put(15,45){(c)}
\end{picture}
\addtocounter{figure}{-1}
\caption{\it (cont'd)}
\end{figure}

Comparing the experimental limits for the center-of-mass energies
175 GeV and 192 GeV, Fig.\ref{fig:neuf} a/b and c/d,
it is clear that the higher energy value
allows a remarkably larger part of
the parameter space to be covered.
For $M_t = 175$~\Gcs\ and  in particular for
$\sqrt{s} = 175$ GeV, the potential of \LEPII\ is rather
limited at large \mA, while for 192~GeV it is possible to
cover the moderate and large $m_A$ for small $\tanb$,
$\tanb \leq 2-3$ (which is a difficult region for LHC).
This  is especially obvious if non-negligible mixing in the stop sector 
is considered.
For large values of \mA, the limit of the standard model Higgs boson 
is approached (light gaugino channels are not considered to be open)
and the lightest CP-even Higgs boson acquires its
maximal value. Since the maximal value of \mh\ increases with \tanb,
the range in \tanb\ covered by LEP2 directly reflects 
the range of standard model Higgs boson masses accessible to the
experiments.

\begin{table}[hbt]
\setlength{\tabcolsep}{0.45pc}
\caption{\it Maximal \mh\ (\h\Z\ and \h\A\ combined) and \mA\ (\h\A\ only), 
  in \Gcs, that can be directly discovered and excluded in the MSSM
  for $M_t = 175$~\Gcs\ and typical mixing at various \LEPII\ energies
  for two representative values of $\tan\beta$ (* These values of \mh\ 
  are already excluded theoretically in the MSSM for typical mixing,
  $M_S$ = 1 TeV and for the values of $M_t$ and $\tan \beta$
  considered here).}
\label{tab:MSSM}
\begin{center}
\begin{tabular}{|@{}c@{}|@{}c@{}|@{}c@{}|@{}c@{}|@{}c@{}|@{}c@{}|}
\hline\hline \rule[0mm]{0mm}{3ex}
$\sqrt{s}$ & $L_{min}$ & 
$\tanb=2$ & $\tanb=30$ & $\tanb=2$ & $\tanb=30$  \\[.5ex]
  \begin{tabular}{c}
   (GeV) \\ \hline \\[-1ex] 175 \\[.5ex] 192 \\[.5ex] 205 \\[1ex] 
  \end{tabular} &
  \begin{tabular}{c}
   (\inpb) \\ \hline \\[-1ex] 150 \\[.5ex] 150 \\[.5ex] 300 \\[1ex]     
  \end{tabular} &
  \begin{tabular}{cc}
   $\mh^{disc.}$ & $\mA^{disc.}$ \\ \hline \\[-1ex] 
   80 & 75 \\[.5ex] 95 & 78 \\[.5ex] 108$^*$ & 80 \\[1ex] 
  \end{tabular} &
  \begin{tabular}{cc}
   $\mh^{disc.}$ & $\mA^{disc.}$ \\ \hline \\[-1ex]
   77 & 77 \\[.5ex] 82 & 82 \\[.5ex] 88 & 88 \\[1ex] 
  \end{tabular} & 
  \begin{tabular}{cc}
   $\mh^{excl.}$ & $\mA^{excl.}$ \\ \hline \\[-1ex]
   84 & 82 \\[.5ex] 98 & 87 \\[.5ex] 113$^*$ & 93 \\[1ex] 
  \end{tabular} & 
  \begin{tabular}{cc}
   $\mh^{excl.}$ & $\mA^{excl.}$ \\ \hline \\[-1ex]
   83 & 83 \\[.5ex] 88 & 88 \\[.5ex] 95 & 95 \\[1ex]
  \end{tabular} \\ 
\hline\hline
\end{tabular}
\end{center}
\end{table}

It is  interesting to compare the upper limits on
\mh\ which can be reached experimentally,
Table~\ref{tab:MSSM} and Fig.\ref{fig:dix}, with the maximal
 \h\ masses expected in the MSSM. In
particular, if the experimental limits are compared to the \h\ mass
range preferred by gauge and b-$\tau$ Yukawa coupling unification,
for $M_t \simeq $ 175~\Gcs\,
it can be seen that a large part of the SUSY Higgs mass range is 
covered in the 192~GeV version of
\LEPII, 
in contrast to the lower energy of 175~GeV. In fact, the infrared fixed
point solution can be excluded at the 95 $\%$ C. L. at $\sqrt{s}$ = 192 GeV if
$M_t \leq $ 175~\Gcs. (For $M_t \simeq 185$~\Gcs\
the infrared fixed point solution can be
 excluded only at $\sqrt{s} = $205 GeV.)

\subsection{MSSM {\it vs.} SM}
\label{sec:disting}

No precise experimental analyses have yet been developed to cope with
the situation in which a Higgs boson would be discovered and thus
would need to be studied in detail. This can be partly explained by
the relatively low luminosity currently expected at \LEPII\ which will
not allow any accurate measurements to be performed in this field. It
could however be imagined that if a new particle were discovered, a
substantial extension of the \LEPII\ project could be decided upon
with the purpose of identifying this particle.

The angular distribution of the Higgs boson produced in the
Higgs-strahlung process is expected to be quite uniform at the \LEPII\ 
center-of-mass energies, and so is the angular distribution of the
decay products in the Higgs boson rest frame. Even in the most
difficult case $\mh\sim\mZ$, the numbers of events collected by the
four experiments with 1~\infb\ taken at 192~GeV (323 events from the
signal and 255 from the backgrounds, see Table~\ref{tab:192}) suffice
to exhibit the flatness of these distributions, and to characterize
unambiguously a $J^P = 0^+$ particle.

Such an excess of events after a b-tagging requirement is also
sufficient to claim that this object often decays into \bbbar, which
also characterizes a Higgs particle. However, a precise measurement of
the \bbbar\ branching fraction can only be done with a channel in
which the evidence for a signal can be claimed without b-tagging, {\it
  i.e.} the \H\lplm\ topology. This situation is particularly
favourable if the Higgs boson mass is not degenerate with the Z mass,
in which case the background is much reduced thanks to the excellent
recoil mass resolution.

For instance, the cross-section for a 80~\Gcs\ Higgs boson at
$\sqrt{s}=192$~GeV in the \H\lplm\ topology is 47 fb (after selection
cuts, but with no b-tagging requirement), to be compared to 14~fb for
the background.  This already allows a measurement of the \bbbar\ 
branching fraction (if close to 100\%) with a $\sim 20\%$ statistical
accuracy, if a luminosity of 1~\infb\ is given to each experiment.
Similarly, the quantity $\sigma(\epemto \h\Z) \times BR(\h\to\bbbar)$
can be determined with the same luminosity from the events collected
in all the topologies with a statistical accuracy of $\sim 5\%$ (resp.
10\%) for a 80~\Gcs (resp. 90~\Gcs) Higgs boson.

Unfortunately, even if the systematics uncertainties were negligible
({\it e.g.} the errors related to the b-tagging efficiency
determination) this is not sufficient to distinguish the standard
model and the MSSM in the region where only the Higgs-strahlung
process plays a role. In this region, a statistical accuracy better
than 1 or 2\% is indeed required \cite{500gev} to achieve this goal.
The extension of the standard model would be manifested in this
parameter range only if non-standard Higgs boson decays were observed.
It is otherwise only in the region where the $\epemto\ \h\A$ process
or, less likely, charged Higgs bosons (see Section~\ref{sec:charged})
can be discovered, that this distinction is  possible at \LEPII.

\subsection{Search for Charged Higgs Bosons}
\label{sec:charged}

In the MSSM the mass of the charged Higgs bosons \Hpm\ is expected to
be larger than \mW.  In general, for non-extreme stop mixing
configuration, it is of the order of $\mHpm^2 = \mA^2 + \mW^2$, {\it i.e.}
larger than $\sim 90~\Gcs$ within a few $\Gcs$, rendering rather
difficult the discovery of such heavy objects at \LEPII. However, this
does not hold necessarily in other, e.g.\ non-minimal supersymmetric
extensions of the Standard Model where light charged Higgs bosons --
although heavier than 44~\Gcs\ as shown by \LEPI\ data -- cannot be
ruled out. Their discovery would unambiguously signal the existence of
an extended Higgs sector.

\subsubsection{Production and Decays}
\label{sec:prodcha}

Charged Higgs bosons are produced in pairs in the process $\epemto\ 
\H^+\H^-$ with a rate depending only on \mHpm\ in the general
two-doublet model. About 100 such events are expected to be produced
with an integrated luminosity of 500~\inpb\ for $\mHpm \sim 70~\Gcs$,
irrespective of the center-of-mass energy from 175 to 205~GeV, and
this rate decreases rapidly with increasing mass due to the $\beta^3$
kinematic suppression factor.

Furthermore, if it is assumed that the cascade decay modes~\cite{DKZ}
like $\H^+ \to \W^{+\ast} \h$ are kinematically suppressed, the
charged Higgs bosons are expected to decay predominantly into the
heaviest kinematically accessible fermion pair provided it is not
suppressed by a small CKM matrix element, {\it i.e.} $\H^+ \to
\tau^+\nu_\tau$ or $\csbar$. Therefore the expected final states are
$\tau^+\nu_\tau\tau^-\bar\nu_\tau$, $\csbar\tau^-\bar\nu_\tau$ and
$\csbar\cbars$, thus leading to an important irreducible background
from $\epemto\ \W^+\W^-$ in addition to the low expected signal rate.
This renders almost hopeless the discovery of charged Higgs bosons
with mass around and above the W mass.

Searches for these final states have been developed by L3 \cite{sopchaque}
and DELPHI \cite{delphinote}, using 
full detector simulation of signal and background processes for an integrated luminosity of 
500~\inpb. A search for the four-jet final state $\csbar\cbars$ has so far
 been 
developed by L3.
\vvs1

\noindent {\bf a) The $\epemto\ \HpHm \to \csbar\cbars$ Channel}\pss{0.5}
The process $\epemto\ \HpHm \to \csbar\cbars$ leads to four-jet hadronic events.
In order to distinguish a Higgs signal from the main background of
\epemto\ \qqbar\ and $\W^+\W^-$, use is made of the different
topological properties. The simulated hadronic energy deposits are clustered
into four jets. For example at $\sqrt{s}=175$~GeV, for 60 and 70~\Gcs\ Higgs signals,
75\% and 69\% selection efficiencies are expected respectively.
The numbers of expected background events are: 3140 \qqbar, 4664 \W\W,
and 90 \Z\Z\ for 500~\inpb.

The four jets can be combined into two jet-pairs in three possible ways, and their energies
and directions are fitted to the energy-momentum conservation and to the 
$\mHp = \mHm$ constraint. The combination with the smallest $\chi^2$ 
is chosen, provided that the measured mass difference between the
two jet-pairs is smaller than 5~\Gcs, and a mass resolution of about 1~\Gcs\ is achieved.
Unlike the $\h\A \to \bbbar\bbbar$ channel, no b-tagging requirement can be applied 
to reject the $\W\W\ \to$ four-jet background, but the signal-to-noise ratio is
improved by removing events where one jet pair combination is consistent with a 
W pair, at the expense of a suppression of the signal efficiency when $\mHpm\simeq\mW$.

For $\mHpm \sim 60~\Gcs$, the expected signal efficiency is about 7\% and the number 
of background events is about 2 \qqbar\ and 3 \W\W\ events, for 
an integrated luminosity of 500~\inpb\ at $\sqrt{s} = 175$~GeV.
\vvs1

\noindent {\bf b) The $\epemto\ \HpHm \to \csbar\tau^-\bar\nu_\tau$ Channel}\pss{0.5}
The signature of an $\epemto\ \HpHm \to \csbar\tau^-\bar\nu_\tau$ signal is 
one isolated slim jet with missing energy coming from the $\tau$ decay  
recoiling against a hadronic system. In the DELPHI analysis, a preselection of hadronic final 
states is performed, and the events are clustered into three jets.
The lowest multiplicity jet, {\it i.e.} the $\tau$ candidate, is required 
to have at most three charged particle tracks. 
Further cuts on the mass and on the angle between the two most energetic jets 
are applied to reject events without a clear 3-jet topology. 
A kinematic fit of the neutrino direction and the $\tau$ momentum (four unknowns)
is performed by constraining the $\tau\nu$ and cs systems to have the same invariant 
mass and the total energy-momentum to be conserved (five equations). A $\chi^2$ cut 
is applied to improve the signal-to-noise ratio. 
This retains from 16\% to 29\% of the signal events,
depending on the H$^{\pm}$ mass and approximately 38 background events
remain for 500~\inpb\ at 175~GeV. 

\begin{figure}[htbp]
\begin{picture}(150,80)
\put(-7,14){\epsfxsize85mm\epsfbox{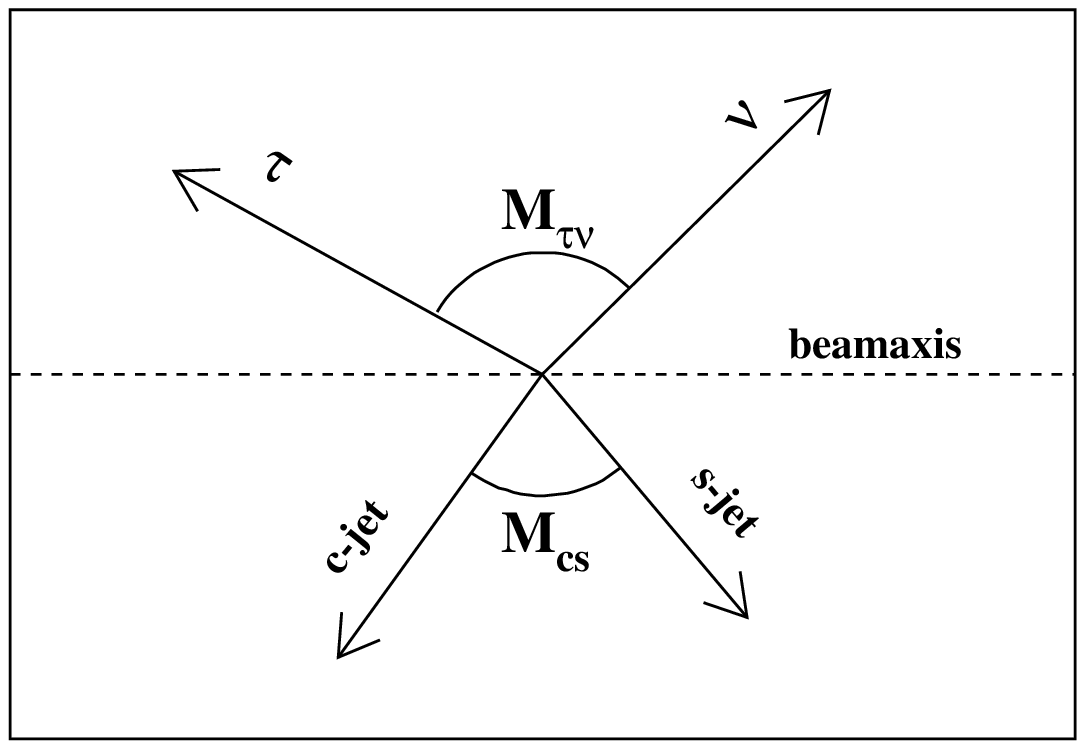}}
\put(82,-2){\epsfxsize85mm\epsfbox{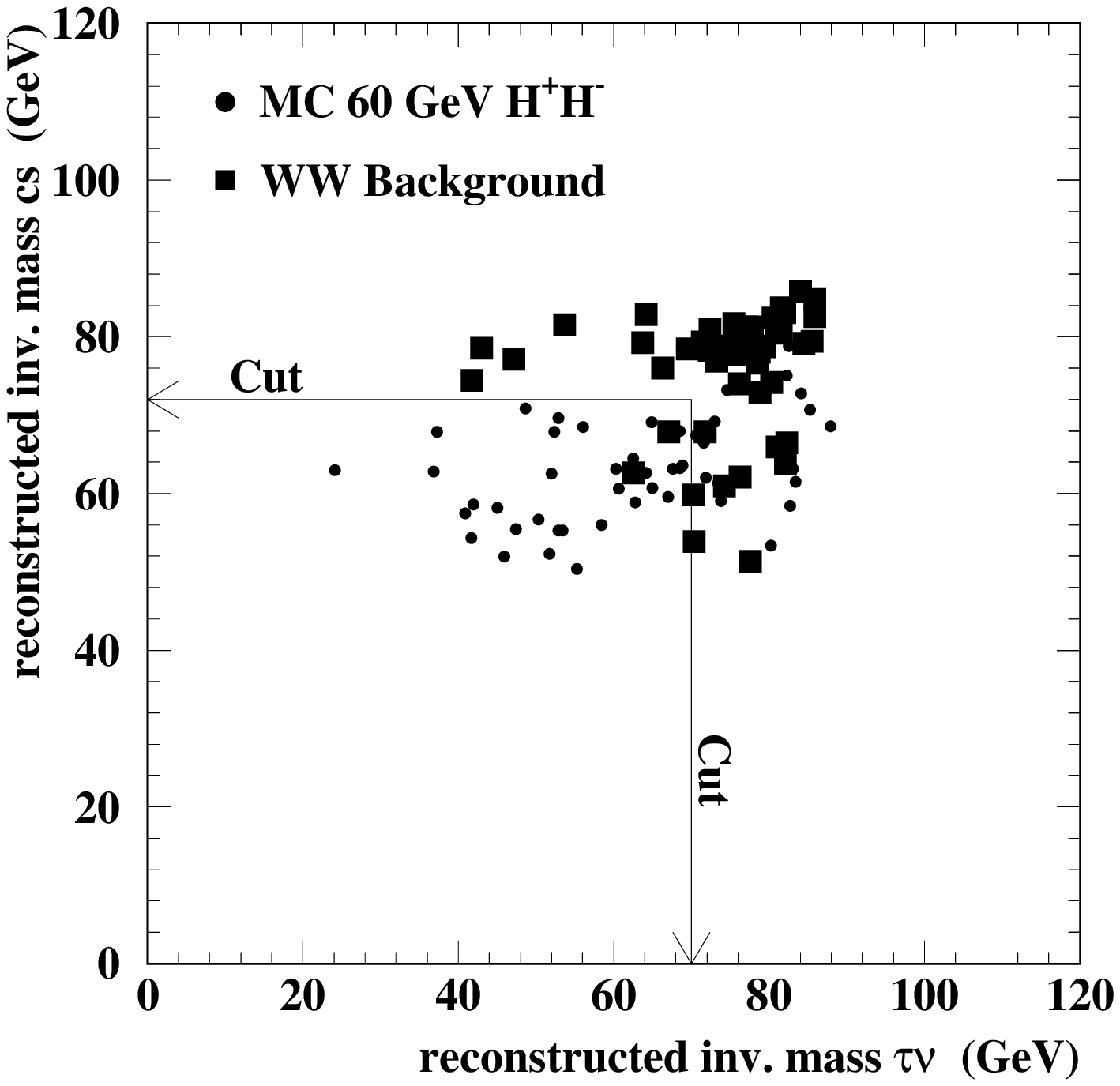}}
\end{picture}
\caption{\it Left: Schematic reconstruction of {\sl both} invariant 
  masses $M_{\mathrm{cs}}$ and $M_{\tau\nu}$ in the cs$\tau\nu$
  channel. For details see text.  Right: Reconstructed invariant
  masses of $M_{\mathrm{cs}}$ and $M_{\tau\nu}$ in the L3 analysis for
  a 60~\Gcs\ Higgs signal and background events in the cs$\tau\nu$
  channel for 500~\inpb\ at 175~GeV.}
\label{fig:cstnfinal}
\end{figure}

In the L3 analysis, the kinematic fit is replaced by the following approximate 
method: 

\begin{itemize}
\item[{\it (i)}] the missing momentum, $\vec{p}_{\mathrm{miss}}$,
      the missing energy, $E_{\mathrm{miss}}$,
      and the invariant mass of the two most energetic jets, $M_{\mathrm{cs}}$,
      are calculated. In order to improve the mass resolution,
      this mass is rescaled by the factor 
      $E_{\mathrm{beam}}/(E_{\mathrm{jet1}}+E_{\mathrm{jet2}})$;
\item[{\it (ii)}] for the other hemisphere, the invariant mass
      $M^2_{\tau\nu} = (E_{\mathrm{miss}} + E_\tau)^2 -
      (\vec{p}_{\mathrm{miss}} + \vec{p}_\tau)^2$ is calculated, where
      $\vec{p}_\tau$ is the visible $\tau$ momentum.
      \end{itemize}
The reconstructed masses of $M_{\mathrm{cs}}$ and $M_{\tau\nu}$
are shown in Fig.\ref{fig:cstnfinal}. Cuts as given in the figure are applied.
The expected signal efficiencies are about 5.6\% and about 2 \W\W\
background events are expected for 500~\inpb\ at 175~GeV. 
\vvs1

\noindent {\bf c) The $\epemto\ \HpHm \to
  \tau^+\nu_\tau\tau^-\bar\nu_\tau$ Channel}\pss{0.5}
\label{sec:tntn}
The $\epemto\ \HpHm \to \tau^+\nu_\tau\tau^-\bar\nu_\tau$ events are
characterized by a low particle multiplicity and large missing energy.
The background processes $\epemto\ \tptm(\gamma)$, $\qqbar$, $\W\W$
and $\gamma\gamma\to \ffbar$ are relevant in this channel.

After applying selection criteria based on the acoplanarity, the total
energy and the event thrust axis angle, the main remaining background
comes from leptonic \W\W\ decays. Unlike the two other channels, a
reconstruction of the Higgs boson masses is not possible because of
too many unknowns due to the numerous missing energy sources. However,
the energies of the decay products of the two taus can be measured,
and part of the W decays into e$\nu_e$ and $\mu\nu_\mu$ can be removed
by a cut on these two energies, which are expected to be larger in
that case than for $\tau\nu_\tau$ final states due to the additional
neutrinos from the $\tau$ decay.

The expected signal efficiencies are about 12\% and about 1 \W\W\
background event is expected in the L3 analysis for 500~\inpb\ at 175~GeV. 
For DELPHI, a 23\% signal efficiency is achieved, for a total of 85 background events 
expected.

\subsubsection{Results}
\label{sec:hphm}

\begin{figure}
\begin{picture}(150,100)
\put(15,5){\epsfxsize130mm\epsfbox{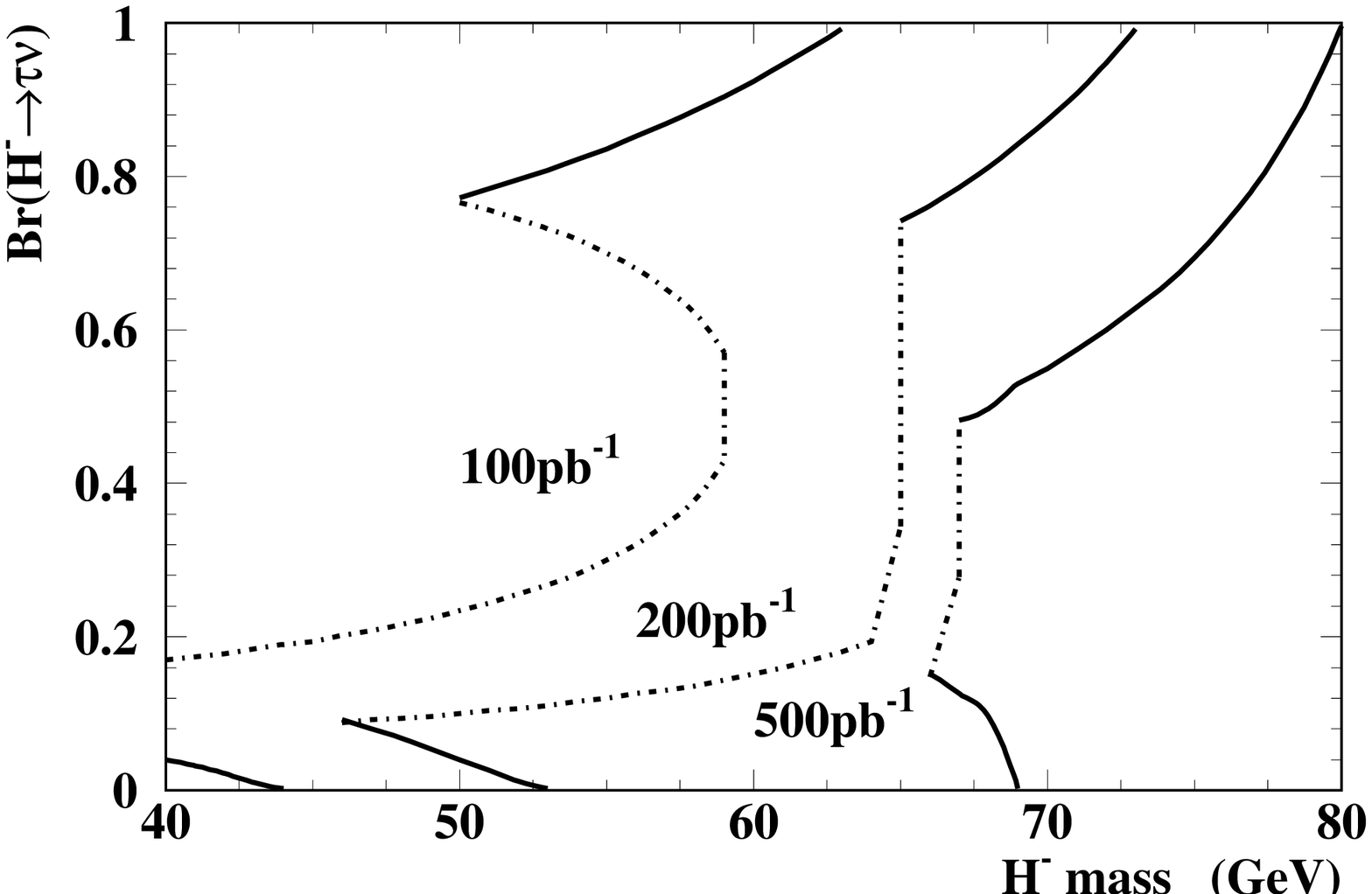}}
\end{picture}
\caption{\it 3$\sigma$ sensitivity
  regions as a function of the charged Higgs boson mass and its
  leptonic branching ratio at 175~GeV and for a total luminosity of
  100, 200, and 500~\inpb. The upper solid lines indicate the
  sensitivity limits from the $\tau\nu\tau\nu$ channel, the dashed
  lines come from the cs$\tau\nu$ channel, and the lower solid lines
  come from the cscs channel.  (For simplicity, combined contour lines
  in overlapping sensitivity regions are not shown.)}
\label{fig:hpmsens}
\end{figure}

In the framework of a general two Higgs doublet model, the results 
of this analysis can be expressed as a function of Br($\H^+\to\tau^+\nu_\tau$) 
(if it is assumed that off-shell decay modes are suppressed) and of \mHpm.
Combining the studies in the \csbar\cbars, $\csbar\tau^-\bar\nu_\tau$, and 
$\tau^+\nu_\tau\tau^-\bar\nu_\tau$ channels described in the previous sections,
the regions which can be explored with $\sqrt{s} = 175$~GeV are shown in 
Fig.\ref{fig:hpmsens} for luminosities of 100, 200, and 500~\inpb\ taken 
by one experiment. (These numbers would be roughly divided by four if the four
LEP experiments were combined.) 

Except for branching ratios into $\tau\nu_\tau$  near 0 or 1,
all three channels contribute simultaneously which extends the
sensitivity range by few~\Gcs. Charged Higgs bosons with masses up to 
about 70~\Gcs\ should be detectable independently of their decay branching ratios 
assuming a total luminosity of 500~\inpb\ at 175~GeV.
The region where a 99.73\%~CL ($3\sigma$) effect due to $\HpHm$ production 
can be detected is strongly dependent on the assumed
luminosity. The boundary lines also depend strongly on the
detection sensitivities due to the small change of the charged
Higgs boson production cross-section with different \mHpm.
The expected variations of the number of background events for
different Higgs masses is taken into account in the figure.
The effect of increasing the center-of-mass energy to about 200~GeV
is small since only a small change of the production cross section
is expected.

The background at \LEPII\ from \W\W\ production with identical decay modes
as for the charged Higgs bosons can be controlled. Irreducible
background only occurs if $\mHpm \sim \mW$. For leptonically decaying Higgs 
bosons, masses can however be explored up to about \mW\ due to the small 
branching fraction of the W boson into $\tau\nu_\tau$.

To summarize, \LEPII\ has a good  potential for a charged Higgs boson
discovery already in its first phase at 175~GeV [well beyond the current 
mass limit of \LEPI\ of 44~\Gcs] 
as soon as a sufficient luminosity of about 200~\inpb\ is collected.
Even with an increase of the machine energy to around 200~GeV, it
is extremely difficult  to explore the kinematic region around and above \mW\
because of the large irreducible background from \W\W\ production
and the low production cross section.

\subsection{Complementarity between LEP2 and LHC}

As with the SM Higgs boson, it is an important task to compare the
Higgs discovery potential of LEP2 with the potential of
LHC for the minimal supersymmetric
standard model MSSM \cite{0}. The comparison will be based again on the
LEP energy $\sqrt{s}=192 \rm{\;GeV}$ with an integrated luminosity of
$\int\!{\cal L}=150 \rm{\;pb}^{-1}$ per experiment [yet all four experiments
combined] and the presumably ultimate integrated luminosity
$\int\!{\cal L}=3\!\times\!10^5 \rm{\;pb}^{-1}$ of LHC, with the results from
ATLAS and CMS combined.

\begin{figure}[htb]
\vspace{-1cm}
\centerline{
\epsfig{figure=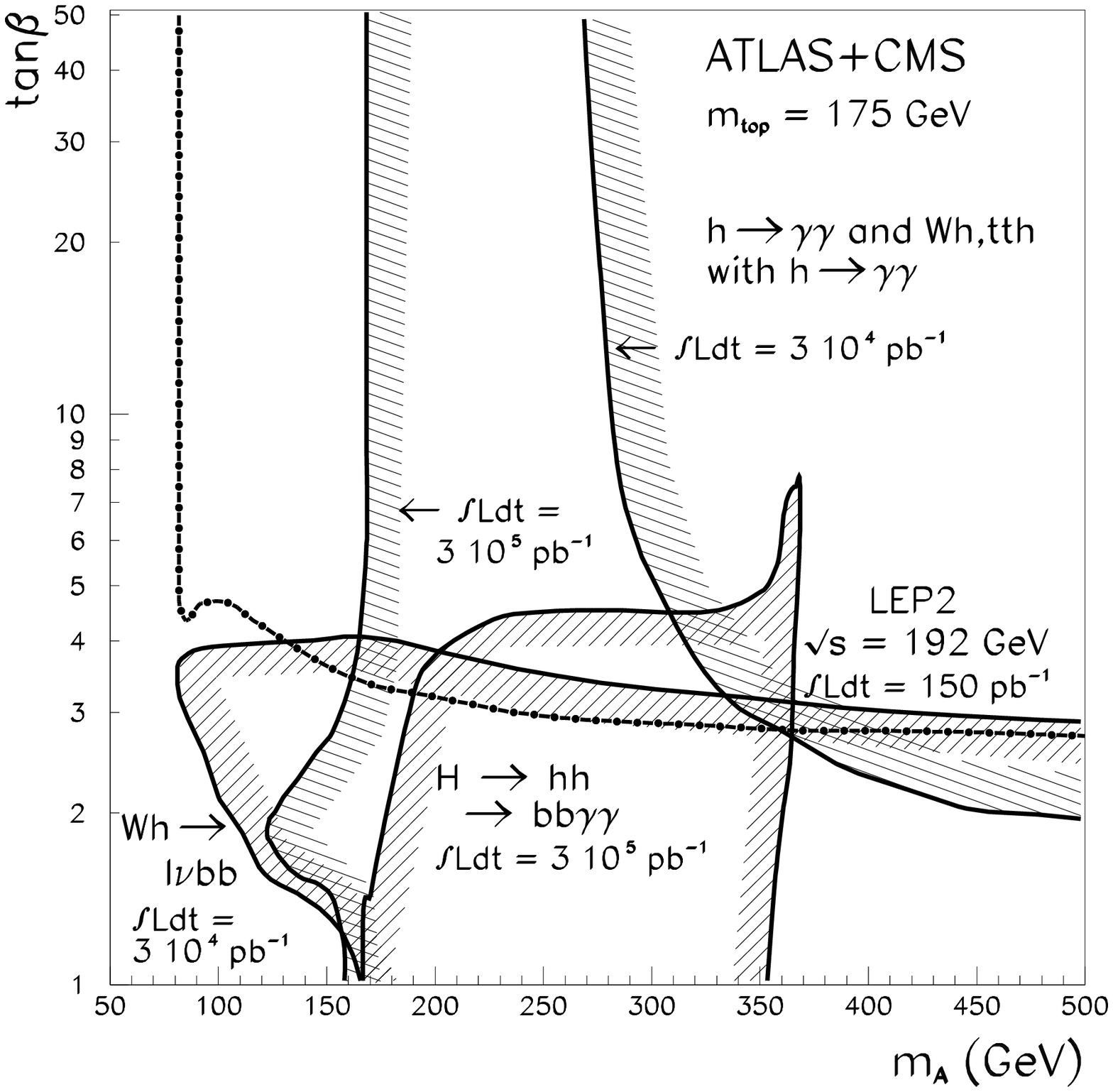,height=12cm,angle=0}}
\vspace{-.5cm}
\caption[0]{{\it 
    The regions in the [$m_A, \tan \beta$] parameter space in which
    the lightest CP-even Higgs boson $h$ can be covered at LEP2
    (Higgs--strahlung and associated production for 150 $pb^{-1}$ at
    $\sqrt{s}$ = 192 GeV) and at LHC ($\gamma\gamma$ decay channel in
    direct production and associated $W h, t \bar{t} h$ production for
    $3 \times 10^4 pb^{-1}$ and $3 \times 10^5 pb^{-1}$, $b\bar b$
    decay channel in associated $Wh,t\bar th$ production for $3 \times
    10^4 pb^{-1}$, and $H\to hh \to b\bar b\gamma\gamma$ decay channel
    for $3 \times 10^5 pb^{-1}$).  Parameters: $M_t$ = 175 GeV, $A=0$,
    $|\mu| \ll M_S= 1$ TeV, but the masses of all supersymmetric
    particles are set to $1$~TeV.} (Courtesy of D. Froidevaux and E.
  Richter-Was)}
\label{fdv1}
\end{figure}

\begin{figure}[htb]
\vspace{-1cm}
\centerline{
\epsfig{figure=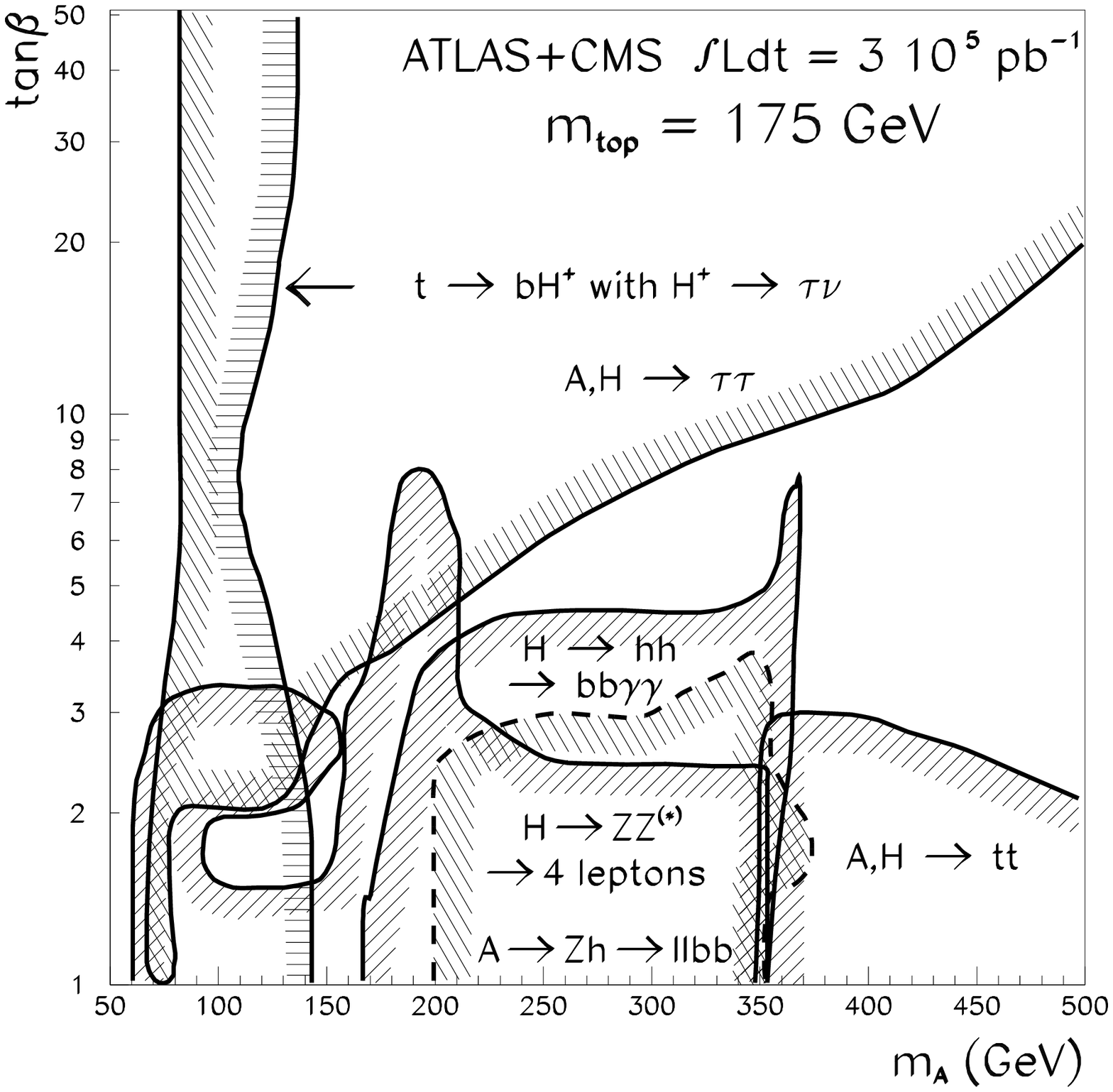,height=12cm,angle=0}}
\vspace{-.5cm}
\caption{{\it
    The regions in the [$m_A, \tan \beta$] parameter space which can
    be covered at LHC in all possible channels associated with the
    heavy scalar, the pseudoscalar and the charged Higgs bosons
    (parameters as in previous figure).} (Courtesy of D. Froidevaux
  and E. Richter-Was)}
\label{fdv2}
\end{figure}

The lightest of the neutral scalar Higgs bosons $h$ can be produced at
the LHC in gluon--gluon fusion \cite{smh18,s1a}, and through Higgs--strahlung
off $W$ bosons \cite{s2} and top quarks \cite{s3}.  This Higgs particle
will be searched for in the $\gamma\gamma$ and 
$b\bar{b}$ decay channels.  Tagging leptonic $W/Z$ and $t$ decays
provides an experimental trigger for the $b\bar b$ search, but also
reduces considerably the huge backgrounds from non--$b$ jets.
The area of the $[m_A,\rm{tan}\beta]$ parameter plane in which the light
scalar Higgs boson $h$ can be discovered in
this way at the LHC is shown in
Fig.\ref{fdv1} by the shaded regions
\cite{s4}. The boundary of the LEP2 discovery range is indicated
by the full line.
No method has yet been found which allows the discovery of $h$ in the
parameter range of large $\rm{tan}\beta > 5$ and $m_A$ between
$\sim 90$ and $\sim 170 \rm{\;GeV}$ at either of the two machines.

While the area in which the pseudoscalar Higgs boson $A$ can be
discovered at LEP is rather modest, a large domain is accessible at
LHC, Fig.\ref{fdv2}.  A clean signal of $A$ comes from $\tau^+\tau^-$
decays; in addition, cascade decays $A \to Zh$ with subsequent
leptonic $Z$ decays provide promising search channels. The search for
$A$ in $t\bar{t}$ decays requires the theoretical control of the top
background production at a level between 10 and 2$\%$ which is an
extremely difficult problem. A similar picture applies to the search
for the heavy scalar Higgs particle $H$ at the LHC, cf.
Fig.\ref{fdv2}. In particular, the classical four--lepton decay of $H$
via $ZZ$ intermediate states can be exploited. At LEP2 the heavy Higgs
$H$ might be produced only in a small region of the MSSM parameter
range. The search for charged MSSM Higgs particles is frustrated in
either machine. While the LEP2 energy is not sufficient to produce
these particles pairwise outside a tiny domain of the MSSM parameter
space, the search technique at the LHC is restricted so far to $t \to
bH^+$ decays with a rather limited range in the charged Higgs mass.

The predictions for the LHC have to be considered with some caution.
The computation of the Higgs spectrum and couplings have been treated
in analogy to the LEP2 simulations; however, the masses of all SUSY
particles have been assumed heavy.  Since the couplings in the $gg$
fusion cross sections as well as in the $\gamma\gamma$ branching
ratios for $h$ decays are generated by loops, this channel could be
affected strongly by light charginos and stop particles~\cite{s5}.
Depending on the point considered in the SUSY parameter space, the
variation of $\sigma\!\times\!$BR through SUSY-loop effects can go
either way, enhancing or spoiling the Higgs signal.  The problem of
SUSY loop corrections is much less severe in search channels which are
based on reactions realized already at the Born level, as $Wh$ at the
LHC, and all the search channels in $e^+e^-$ collisions.  A problem of
$pp$ collisions are the QCD corrections. They are known for the signal
in $gg$ fusion \cite{s1a}, but not for all background processes; the
assumption that significances are estimated in a conservative way by
setting $K$ factors to unity is expected to be fulfilled in large
parts of the SUSY parameter space, but this is not guaranteed yet for
large $\tan\beta$.  Moreover, if the Higgs bosons do not only decay to
SM particles but instead to invisible LSP and other
neutralino/chargino states with potentially large branching ratios
\cite{shgs3b,shgs9}, the analysis must be modified and the conclusions
would eventually be altered rather dramatically.

Combining the discovery potentials of LEP2 at $\sqrt{s}=192 \;{\rm
  GeV}$ and of LHC by summing up all Higgs production channels, the
entire $[m_A,\rm{tan}\beta]$ parameter plane of the MSSM is predicted
to be covered within the standard framework of non--SUSY Higgs decays
[based on the parameter set $M_t = 175$ GeV, $M_S=1$~TeV and $A,\mu\ll
M_S$].  The discovery potential of LEP (LHC) in the search for $h$
increases for smaller (larger) values of $M_t$, $A$, $\mu$ and $M_S$
which are associated with smaller (larger) values of $m_h$.  In the
region of $m_A$ values less than about 150$\rm{\;GeV}$, the search for
$h$ can be performed by LEP2 while the other heavy Higgs particles,
$H$, $A$ and $H^\pm$, can be searched for at the LHC.  As discussed
before, the observation of at least two different Higgs states, at
LEP2 or LHC, is a crucial step in disentangling the supersymmetric
theory from the Standard Model.  Moreover, the channels exploited in
the search for $h$ are different at LEP2 and LHC. This implies that
the couplings involved will be different and hence the physics tested
in both cases will be complementary.  \vvs2

\vfill

\newpage

\section{Non--Minimal Extensions}

\subsection{The Next--to--Minimal Supersymmetric Standard Model}

In this section we shall augment the MSSM by introducing a single
gauge singlet superfield $N$ leading to a model
which is referred to as the NMSSM\cite{NMSSM}.

The classic motivation for
singlets is that they can solve  the so-called $\mu$-problem of the
MSSM \cite{mu} by eliminating the $\mu$-term and replacing its effect
by the vacuum expectation value (vev) $<N>=x$, which may be naturally
related to the usual Higgs vev's $<H_i>=v_i$. However such models in
which the superpotential contains only trilinear terms, possess a
${\hbox{Z$\!\!$Z}}_3$ symmetry which is spontaneously broken at the
electroweak breaking scale. This results in cosmologically stable
domain walls \cite{walls} which make the energy density of the universe
too large. This cosmological catastrophe can be avoided by allowing
explicit and non-renormalizable ${\hbox{Z$\!\!$Z}}_3$ breaking  terms
suppressed by powers of the Planck mass which will ultimately dominate
the wall evolution \cite{NMSSMwalls} without affecting the
phenomenology of the model. However such terms induce a destabilisation
of the gauge hierarchy \cite{stabtwo} due to tadpole contributions to
the $N$ mass in supergravity models with supersymmetry breaking in the
hidden sector.

Where does all this leave the NMSSM? This depends on one's point of
view. If there might be some [yet unknown] solution 
to the domain
wall problem, then one can consider models with 
${\hbox{Z$\!\!$Z}}_3$ symmetry which is broken spontaneously. 
Another
approach is to avoid the domain wall problem by considering more general
NMSSM models {\em without} a ${\hbox{Z$\!\!$Z}}_3$ symmetry.
Note that ${\hbox{Z$\!\!$Z}}_3$ violating terms,
such as a $\mu$ term, large enough to avoid the domain wall problem, can
still be sufficiently small as to have no impact on collider
phenomenology. These more general models allow arbitrary renormalizable
mass terms in the superpotential including the $\mu$ parameter, and
linear and quadratic terms in $N$. The question of Planck scale tadpole
contributions arises in this case. However, such contributions depend on
supersymmetry breaking in a hidden sector of a supergravity theory, and
are hence model dependent. 

According to this discussion 
we shall consider two quite different versions
of the NMSSM:

\noindent 
(i) The ``General NMSSM'' is defined by 
the following superpotential:
\begin{equation}\label{nmssmdef}
W =-\mu H_1H_2 + \lambda NH_1H_2 - \frac{k}{3}N^3
  + \frac{1}{2}\mu^{\prime}N^2 + \mu^{\prime\prime}N + \ldots
\label{Wgeneral}
\end{equation}  
This version of the model is essentially a
generalization of the MSSM, and provides a more general realization of
low-energy SUSY which is equally consistent with  gauge coupling
unification and high precision measurements.
It  reduces to the MSSM in the limit in which the $N$ field
is removed, and since it does not have a ${\hbox{Z$\!\!$Z}}_3$ symmetry 
there is no domain wall problem. 

\noindent
(ii) The ``Constrained NMSSM''  is defined by the trilinear terms in 
eq.(\ref{Wgeneral}), i.e.\ 
$\mu=\mu^{\prime}=\mu^{\prime\prime}=0$, plus the constraints of
gauge coupling unification and universal soft SUSY-breaking parameters
imposed at the unification scale $M_{GUT} \approx 10^{16}$ GeV
\cite{esr,epic}. 

\subsubsection{The General NMSSM}

\noindent {\bf a) Masses and Couplings}\pss{0.5}
The superfield $N$ contains a singlet Majorana fermion,
plus a singlet complex scalar. The real part of the
complex scalar will be assumed to develop a vacuum expectation value.
The singlet yields one additional CP-even state and 
one additional CP-odd state which are gauge singlets but
can mix with the corresponding neutral Higgs states of the MSSM,
leading to three CP-even Higgs bosons $h_1,h_2,h_3$ and two
CP-odd Higgs bosons $A_1,A_2$.
Although there are more neutral Higgs bosons than in the MSSM,
they will have diluted couplings due to their singlet components,
making their production cross-sections smaller.

The tree-level CP-odd mass matrix, after 
rotating away the Goldstone mode as usual, reduces to the $2\times 2$
matrix in the basis $(A,N)$ 
where $A$ is the MSSM CP-odd field,
\be
M_{A}^2 =
\left(\begin{array}{ccc}  
 m_A^2 & . \\
 . & .
\end{array} \right)
\label{M_A}
\end{equation}
The entries represented by dots are complicated singlet
terms. Unlike the MSSM, the parameter
$m_A^2$ here is not a mass eigenvalue due
to singlet mixing. The CP-odd matrix is diagonalized
by rotating through an angle $\gamma$, leading to
two CP-odd eigenstates $A_1,A_2$ of mass $m_{A_1}\leq m_{A_2}$.

The tree-level 
CP-even mass squared matrix in the
basis $(H_1,H_2,N)$ is
\begin{equation}
M^2=
\left(\begin{array}{ccc}
m_Z^2\cos^2 \beta +m_A^2\sin^2 \beta &
-(m_Z^2+m_A^2-2\lambda^2v^2)\sin \beta \cos \beta   & .   \\
-(m_Z^2+m_A^2-2\lambda^2v^2)\sin \beta \cos \beta  &
m_Z^2\sin^2 \beta+m_A^2\cos^2 \beta & . \\
. & . & .
\end{array} \right)
\label{CPevenmass} 
\end{equation}
where, as usual, $v=174$ GeV, $\tan \beta= v_2 /v_1$;
again the dots correspond to singlet terms.
Apart from the terms involving $\lambda$, the upper 2$\times$2
block of this matrix is identical to the MSSM CP-even matrix.
However, whereas the MSSM matrix is diagonalized by rotation
through a single angle $\alpha$, the matrix in eq.\ref{CPevenmass}
is diagonalized by a 3$\times$3 unitary matrix $V$,
leading to three mass eigenstates $h_1,h_2,h_3$ with masses ordered
as $m_{h_1}\leq m_{h_2} \leq m_{h_3}$.

The singlets obviously cannot mix with charged scalars, and 
at tree-level the mass of the charged Higgs is
\be
m_{H^{\pm}}^2=m_A^2+m_W^2-\lambda^2v^2
\ee
Clearly a non-zero $\lambda$ tends to reduce the
charged scalar masses which can now be arbitrarily small,
and -- in contrast to the MSSM -- {\em below the W mass}.

We shall define the relative couplings
$R_i\equiv R_{ZZh_i}$ as the $ZZh_i$ coupling in units of the standard model
$ZZH$ coupling, and similarly we shall define a $Zh_iA_j$ coupling
factor $R_{Zh_iA_j}$.
For example $R_{ZZh_1}$ is a generalization of $\sin (\beta - \alpha)$
and the $R_{Zh_1A_i}$ are generalizations of $\cos (\beta - \alpha)$
in the MSSM. 
The $Zh_iA_j$ coupling factorises
into a CP-even factor $S_i$ and a CP-odd factor 
which depends only on the angle $\gamma$ 
which controls singlet mixing in the CP-odd sector.

It can be shown that the CP-even Higgs bosons in this
model respect the following relations \cite{NMSSMbound,Kam,kw}:
\bea
m_{h_1}^2 &\leq& \Lambda^2 =
m_Z^2\cos^2 2\beta + \lambda^2v^2\sin^22\beta\cr
m_{h_2}^2 &\leq& \frac{\Lambda^2-R_1^2m_{h_1}^2}{1-R_1^2}\cr
m_{h_3}^2 &=&
\frac{\Lambda^2-R_1^2m_{h_1}^2-R_2^2m_{h_2}^2}
{1-R_1^2-R_2^2}
\label{bounds}
\eea
In the case $\lambda=0$, 
$\Lambda$ is equal to the lightest 
CP-even upper mass bound in the MSSM.
The above results show that
if $R_1\approx 0$ we may simply ignore $h_1$  
and concentrate on $h_2$ which then becomes the lightest physically   
coupled CP-even state and must satisfy $m_{h_2}^2 \leq \Lambda^2$.
Similarly if both $R_1$ and $R_2$ are nearly zero, then $m_{h_3}^2=\Lambda^2$.

It can also be shown that 
\bea
m_{A_1}^2 & \leq & m_A^2 \cr
m_{A_2}^2 & = & \frac{m_A^2-m_{A_1}^2\cos^2 \gamma}
                       {1 - \cos^2 \gamma}
\eea
If the lighter CP-odd
state is weakly coupled ($\cos \gamma \approx 0$)
then it is mainly singlet,
and the heavier CP-odd state is then
identified with the MSSM state of mass
$m_A$.
\vvs1

\noindent {\bf b) Theoretical Upper Limit on $\Lambda$}\pss{0.5}
According to eq.(\ref{bounds}), $\Lambda$ is clearly a function of
$\tan \beta$ and $\lambda$, and to find the absolute upper bound on the
mass of the lightest CP-even Higgs boson we must maximize this
function ($\Lambda_{max}$). Radiative corrections, which 
drastically affect 
the bound \cite{boundRC}, 
are included using recently proposed methods \cite{RC4}.

\begin{figure}[htb]
\vspace{0.1cm}
\centerline{
\epsfig{figure=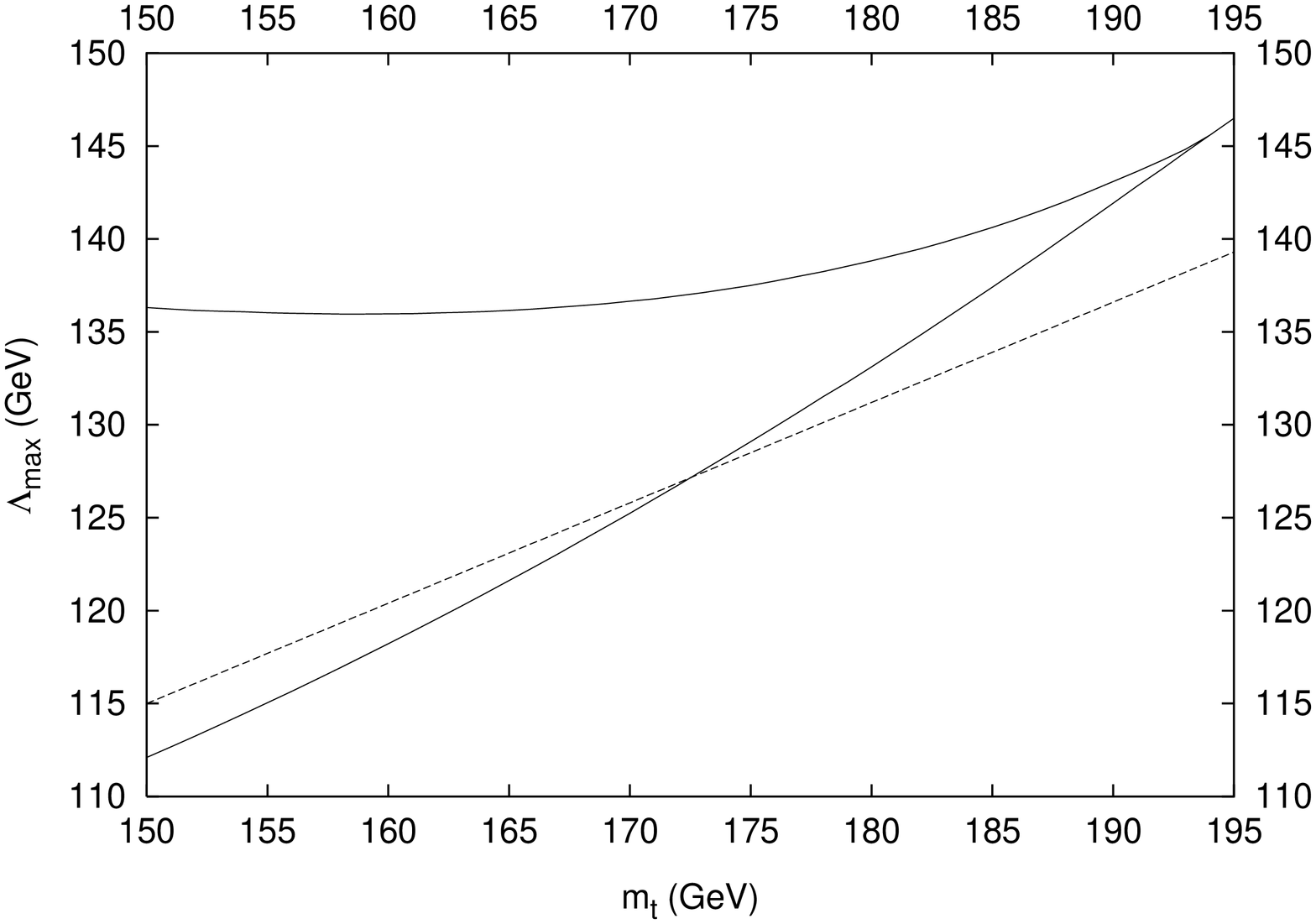,height=11cm,angle=270}
}
\caption{\it Theoretical upper bound
  on the mass of the lightest CP-even Higgs boson as a function of
  $M_t$. We have used $\alpha_3(m_Z)=0.12$, a mean squark mass of
  1~TeV, and varied $A_t$ and $\tan\beta$ in such a way as to maximize
  the upper bound.  The upper solid line is for the NMSSM and the
  lower solid line is for the MSSM. As $M_t$ becomes very large the
  two bounds become very close, since here $h_t$ is always close to
  triviality and hence $\lambda$ must be small.  The dashed line
  refers to the constrained NMSSM.  }
\label{boundmt}
\end{figure}

For a fixed $\tan \beta$, the maximum value of $\Lambda$ is given by
the maximum  value of $\lambda$ as derived from the triviality
requirement that none of the Yukawa couplings becomes non-perturbative
before the GUT scale of around $10^{16}$ GeV. Using the recently
calculated two-loop RGEs \cite{epic}, we find an upper limit on
$\lambda$ as a function of $\tan\beta$. The maximum value of $\lambda$
is typically in the range 0.6-0.7 for a wide range of $\tan \beta$,
depending on $M_t$ and $\alpha_3(m_Z)$, and falls off to zero for
$\tan\beta\rightarrow\approx$1.5 or 60 because  $h_t$ or $h_b$,
respectively, is very close to triviality. Having derived the maximum
value of $\lambda$ as a function of $\tan\beta$,  we can use this
information to obtain an $M_t$-dependent  maximum value of $\Lambda$
shown as the upper solid  curve in Fig.\ref{boundmt}. The MSSM bound
is also shown (lower solid curve) for comparison. The dashed line is
the corresponding upper mass bound in the constrained NMSSM (see
later).

As well as being the upper bound on the mass
of the lightest CP-even Higgs boson, the parameter $\Lambda$ plays an
important role in constraining all the CP-even Higgs boson masses and
couplings. 
Thus the value of $\Lambda_{max}$, corresponding to the upper solid
line in Fig.\ref{boundmt}, also constrains $h_2$ and $h_3$ according
to eq.(\ref{bounds}).
\vvs1

\noindent {\bf c)  Experimental Lower Limits on $\Lambda$}\pss{0.5}
$\Lambda$ has a
theoretical upper limit given by
$\Lambda_{max} \approx$ 146 GeV.
Now we shall discuss how  experiment may place
a {\em lower} limit on $\Lambda$ which we shall 
refer to as $\Lambda_{min}$. The meaning of $\Lambda_{min}$
is as follows. For each value of $\Lambda$ there are many
possible sets of parameters $(R_i,m_{h_i})$ 
subject to the bounds in eq.(\ref{bounds}).
Each of the three (CP-even) Higgs bosons in each set may 
or may not be discovered at LEP, depending on how light
it is and how strongly coupled to the $Z$ it is.
We can consider the present $R^2-m_h$ 95\% exclusion
plots at LEP \cite{sopchaque} and classify each of the three
Higgs bosons in each set (for a fixed $\Lambda$)
as excluded or not excluded. We may find, for some value of $\Lambda$,
that for {\em all} 
the allowed sets at least one out of the three Higgs bosons
is always excluded. In this case we classify this value of $\Lambda$ as
being excluded by experiment. We now define $\Lambda_{min}$ as
the largest value of $\Lambda$ which may be excluded by the LEP 
data. There will be a different $\Lambda_{min}$
for each of the expected LEP2 $R^2-m_h$ 95\% exclusion
plots (see Fig. \ref{fig:sbavis} and  Table~\ref{tab:sba}).
 If  $\Lambda_{min}$ exceeds
$\Lambda_{max}$ then the model is excluded. 

 $\Lambda_{min}$ is 
approximately determined by the ``worst case'' of 
all three CP-even Higgs bosons having equal masses
$m_{h_i}\approx \Lambda$, and equal
couplings ${R_i}^2\approx 1/3$. 
\footnote{This approximate result is exact in the limit that 95\%
exclusion is equated to 50 produced events.}
Using this simple approximation, together with current
$R^2-m_h$ exclusion limits \cite{sopchaque},
LEP1 already places a limit on $\Lambda$ of
$\Lambda> \Lambda_{min}=59$ GeV, which is just equal to the mass
limit for a CP-even Higgs boson with its
$ZZh$ coupling suppressed by $R^2=1/3$.

\begin{figure}[htb]
\vspace{0.1cm}
\centerline{
\epsfig{figure=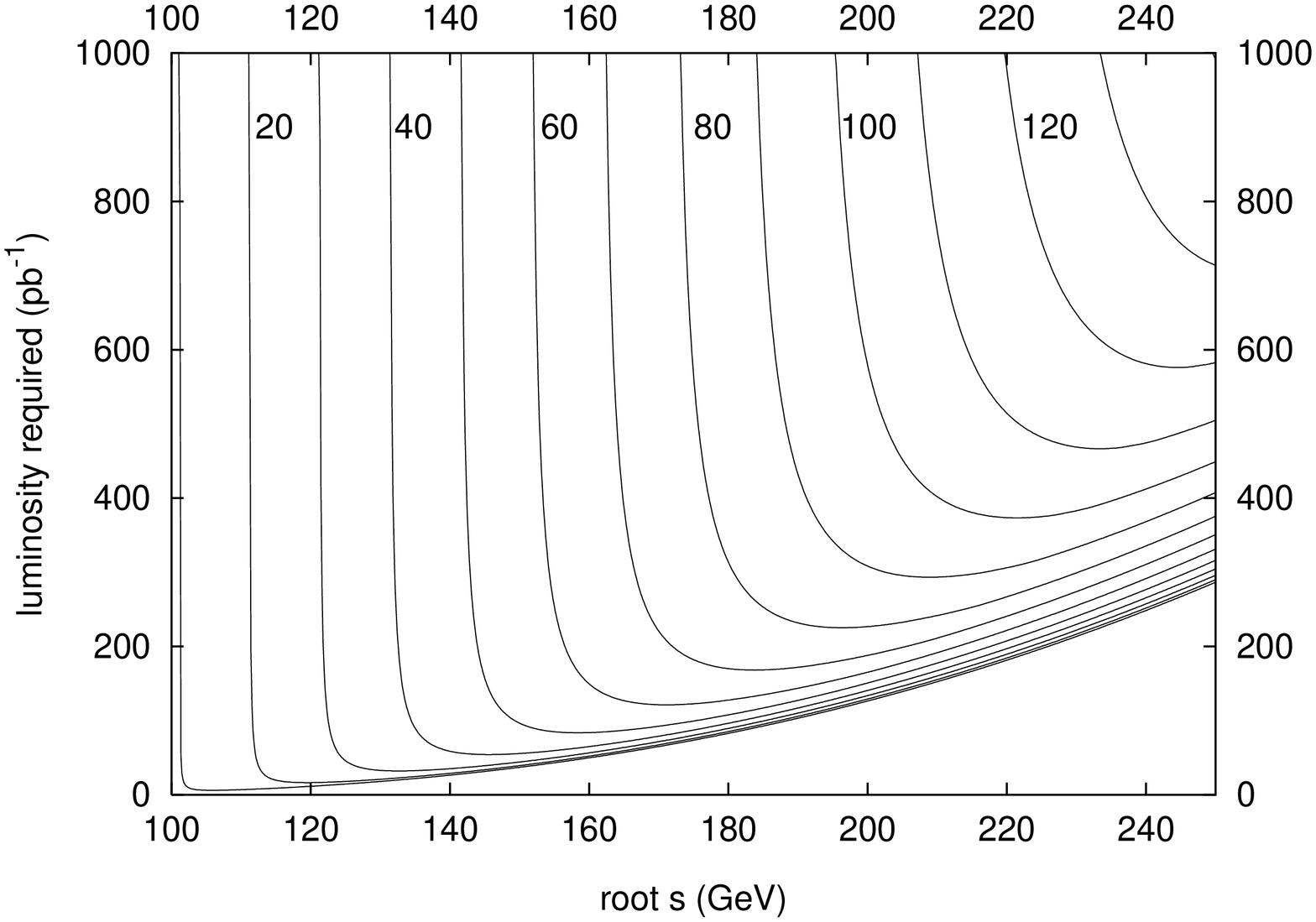,height=11cm,angle=270}
}
\caption[dum]{\it Excluded values of $\Lambda$ at $e^+e^-$ colliders
  with energy $\sqrt{s}$.}
\label{lum}
\end{figure}

The values of $\Lambda_{min}$ which may be excluded by a future 
$e^+e^-$ collider of a given energy and integrated luminosity [note
that this is total luminosity of all four experiments pooled]
are shown in Fig.\ref{lum} where exclusion is
approximately equated to 50 produced events.

Focusing specifically on LEP2, we
consider energies and
integrated luminosities per experiment of
$\sqrt{s}$  =  175, 192, 205   GeV and $ \int{\cal{L}}$=
 150, 150, 300  $pb^{-1}$, respectively.
Using the $R^2-m_h$ $95\%$ exclusion plots \cite{janot}, 
we find that LEP2 will yield the exclusion 
limits $\Lambda_{min}=81,93,105$~GeV respectively, for the three sets of
LEP2 machine parameters above.
These excluded values of $\Lambda$ (corresponding to $R^2=1/3$)
are not far from the
values of SM Higgs boson masses which may be excluded,
due to the steep rise of the exclusion curves in the $R^2-m_h$ plane.
\vvs1

\noindent {\bf d) Exclusion Limits in the $m_A-\tan\beta$ Plane}\pss{0.5}
It is possible to obtain 
exclusion limits in this model in the $m_A-\tan\beta$ plane,
rather similar to the familiar MSSM plots. 
The excluded regions are obtained from the following three
searches:

(i) For the processes $Z\rightarrow Zh_i$, we 
exploit the fact that the upper 2$\times$2
block of the CP-even mass squared matrix
is completely specified
(for fixed $\lambda$) in the $m_A$-$\tan\beta$ plane.
However, unlike the MSSM, the CP-even spectrum is not 
completely specified since it depends on
three remaining unknown real
parameters associated with singlet mixing
(i.e. the dots in eq.(\ref{CPevenmass})).
Nevertheless, since each choice of these parameters
completely specifies the
parameters $m_{h_i}$ and $R_i$, we can scan 
over the unknown parameters; 
if the resulting spectrum is always excluded, then
we conclude that this point in the $m_A$-$\tan\beta$ plane
(for fixed $\lambda$) is excluded. For these excluded regions
we use the available LEP1 and LEP2 $R^2-m_h$ $95\%$
exclusion lines.

(ii) An excluded lower limit on $m_A$,
as a function of $\tan \beta$ and $\lambda$ in this model comes
from the non-observation of $Z\rightarrow h_iA_j$.
The excluded lower limit on $m_A$ in this model is determined by
the ``worst-case'' 
values of $m_{h_i},m_{A_j},S_i,\gamma$
consistent with this value of $m_A$.
It turns out that the worst case experimentally
is when the three CP-even Higgs bosons all have equal masses
as heavy as possible 
\be
{m_{h_i}}^2={m_A}^2+({m_Z}^2-\lambda^2v^2)\sin^22\beta
\ee
and equal coupling factors, ${S_i}^2\approx 1/3$,
and the two CP-odd Higgs bosons both have masses equal
to $m_A$ and $\gamma = \pi/4$, leading to  
$R_{Zh_iA_j}^2=1/6$. For these excluded regions we use
the simple approximation that 50 events corresponds
to $95\%$ exclusion.

(iii) Finally we shall present excluded regions for
charged Higgs production, assuming detection up to
the kinematic limit. We note that the charged Higgs signal 
is the same as in the MSSM as considered in Section 3.

\begin{figure}
\begin{center}
\begin{picture}(150,85)
\put(-10,85){\epsfig{figure=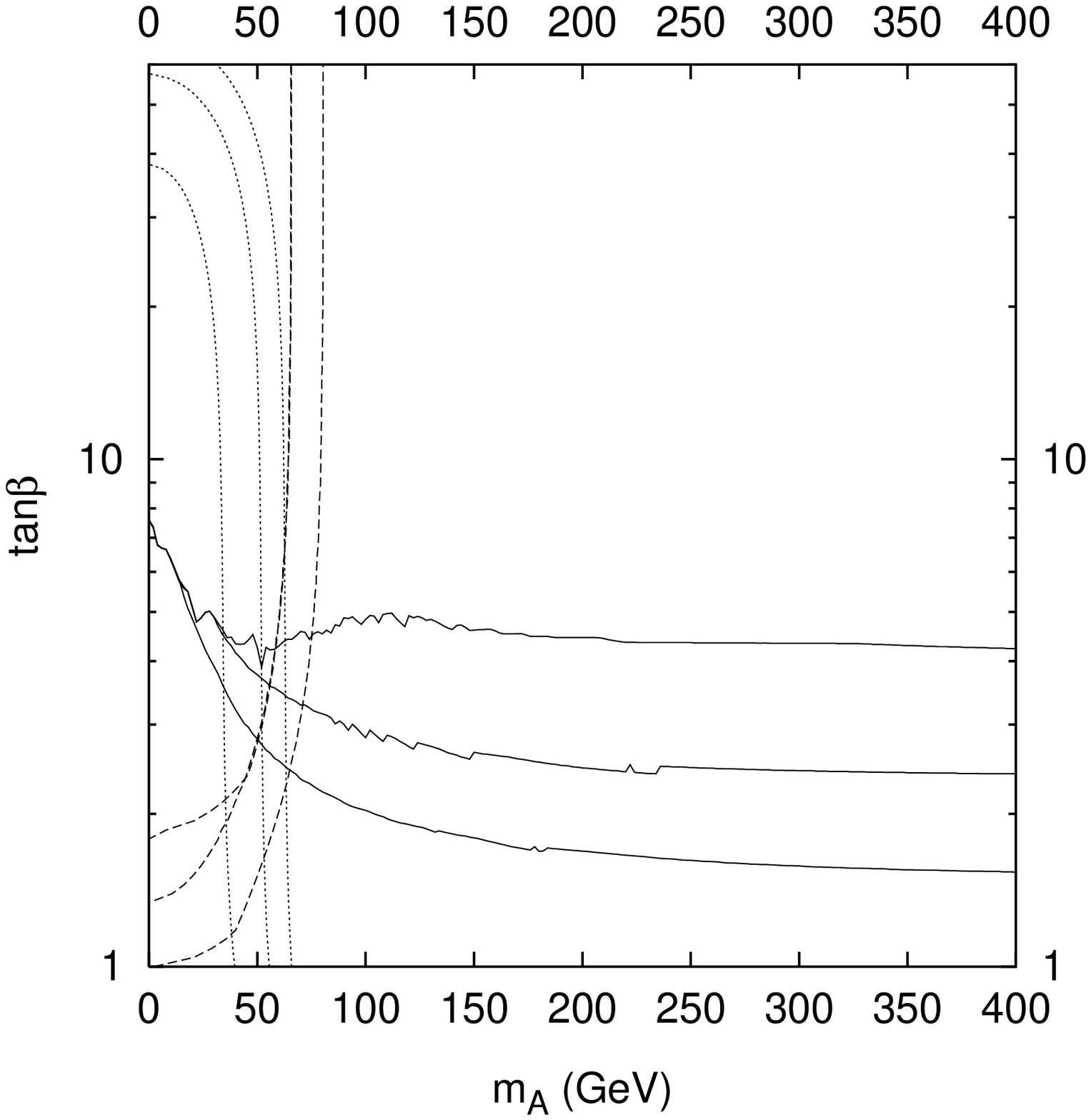,height=9cm,angle=270}}
\put(78,85){\epsfig{figure=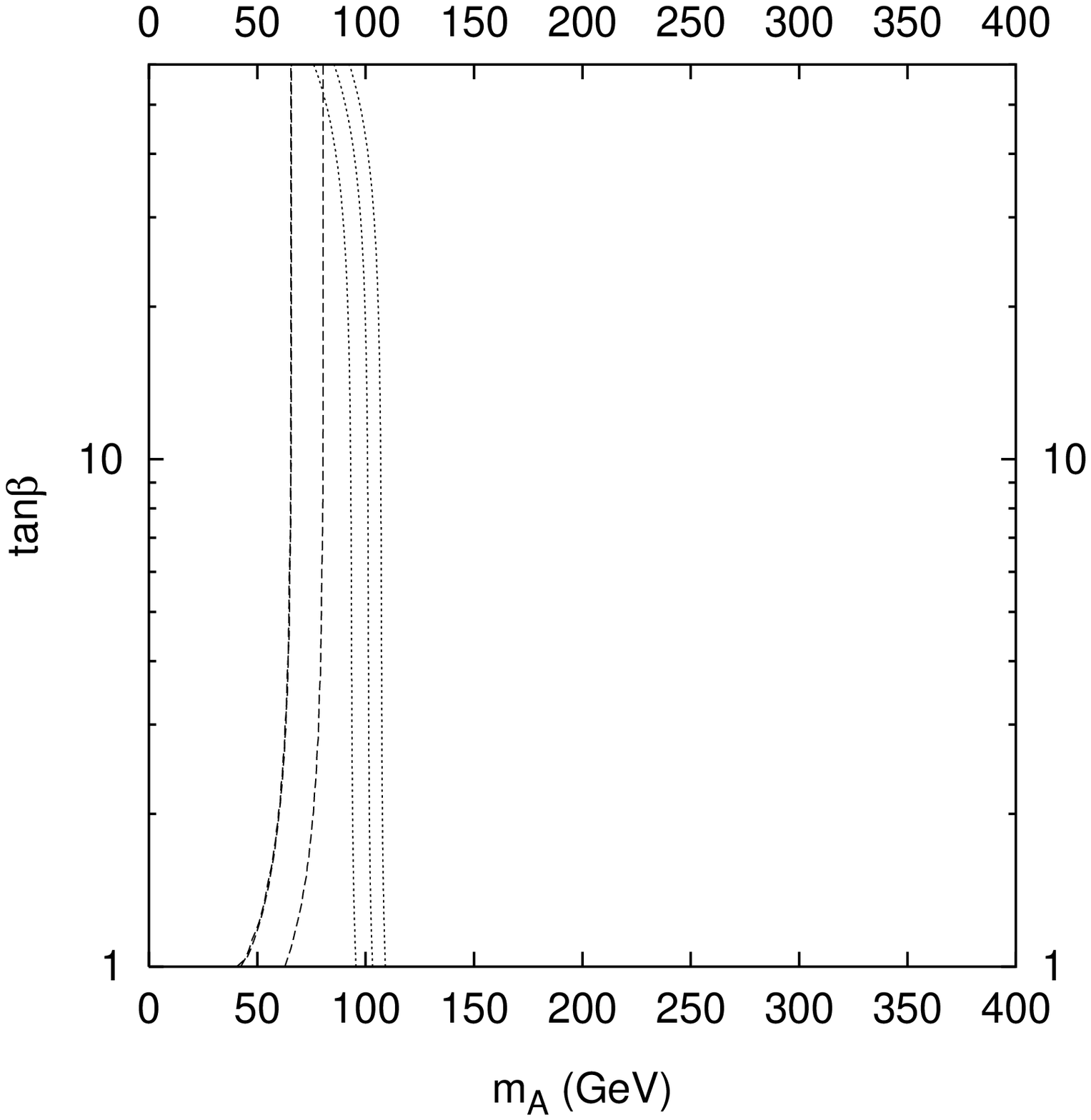,height=9cm,angle=270}}
\end{picture}
\end{center}
\caption{\it Excluded regions of the General NMSSM
  with $\lambda=0$ (left) and $\lambda=0.5$ (right). We have included
  radiative corrections by assuming degenerate squarks at 1 TeV, with
  no squark mixing and $M_t=175$ GeV. Note the influence of radiative
  corrections on the charged Higgs mass for large $\tan \beta$ since
  both $h_t$ and $h_b$ must be large for such corrections to be
  important.}
\label{plane}
\label{planel0.5}
\end{figure}
%

%
%

In Fig.\ref{plane} we show the excluded regions of this model
corresponding to the choice of parameter $\lambda=0$ \footnote{Strictly
speaking if $\lambda=0$ then there is no singlet mixing. However these
curves apply to the case where $\lambda$ is small (say less than 0.1)
and singlet mixing is possible. Such small values of $\lambda$ are
always found in the constrained NMSSM.} for the three sets of LEP2 machine
parameters $\sqrt{s}$  =  175, 192, 205   GeV and $ \int{\cal{L}}$=
 150, 150, 300  $pb^{-1}$, respectively,
 and using the LEP1 and LEP2  exclusion data, Fig. \ref{fig:sbavis}
 and  Table~\ref{tab:sba}. 
In each case the solid lines correspond to exclusion limits
from $Z\rightarrow Zh_i$ as obtained using procedure (i) above in the
NMSSM, the dashed lines correspond to exclusion limits from
$Z\rightarrow h_iA_j$ as obtained using procedure (ii) above in the
NMSSM, and the dotted lines correspond to the exclusion limits from
charged Higgs production using procedure (iii) above. These exclusion
plots should be compared to similar exclusion plots in the MSSM for
$M_t=175$ GeV and degenerate squarks at 1 TeV, which are the parameters
assumed in Fig.\ref{plane}. 

In Fig.\ref{planel0.5} we also show a similar plot but 
with $\lambda=0.5$. In this case the solid lines 
have disappeared beneath the $\tan \beta=1$ horizon,
because for larger $\lambda$ the bound $\Lambda$ may be
larger, which allows a heavier CP-even spectrum which
is consequently more difficult to exclude.
The charged Higgs and $h_iA_j$ channels now give
improved coverage, however, since larger $\lambda$ 
decreases the charged Higgs mass, and also decreases the
$h_i$ masses for a fixed $S_i$ coupling. 


\subsubsection{The Constrained NMSSM}

As noted in the introduction, the constrained NMSSM is defined by 
eq.(\ref{nmssmdef})
with $\mu$, $\mu'$, $\mu''$~$\to$~$0$ and the condition of universal
SUSY-breaking gaugino masses $M_{1/2}$, scalar masses $m_0$ and
trilinear couplings $A_0$ at $M_{GUT}$. In addition, the effective potential
has to have the correct properties, i.e. the $SU(2)\times U(1)$ symmetry
has to be broken by Higgs vev's, but the vev's of charged and/or colored
fields as sleptons, squarks and charged Higgs scalars have to vanish.
Finally, present lower limits on sparticle masses due to direct searches
have to be satisfied, and for the top-quark mass $M_t$ we require values
between 150~GeV and 200~GeV.

{\em A priori} the constrained NMSSM has six parameters at $M_{GUT}$, three
dimensionless couplings $\lambda$, $k$ and $h_t$ (the top-quark Yukawa
coupling) and three dimensionful ones $M_{1/2}$, $m_0$ and $A_0$. The
scale set by the known mass of the $Z$ boson reduces the number of free
parameters of the model to five. A scan of the parameter space of the
model, which is consistent with all the above constraints, has been
performed in \cite{esr} and \cite{epic}. Below we present results,
relevant for the Higgs search at LEP2, which are based on the data
obtained in \cite{esr}. We will discuss the allowed Higgs masses and 
couplings within the constrained NMSSM.

First  note that neither a CP-odd Higgs boson $A_j$ with sufficiently
large coupling $R_{Zh_iA_j}$ nor a charged Higgs boson
can be sufficiently light within the constrained NMSSM in order to be 
visible at LEP2. Thus we will concentrate on the neutral CP-even Higgs
bosons in the following. Concerning their decays, it follows that
neutralinos are too heavy to play a role, thus their branching ratios
are close to the standard model ones (essentially $b\bar b$) and
the same search criteria apply.

In general, the upper limit on the lightest Higgs mass
as a function of the top-quark mass
depends on the magnitude of the SUSY-breaking parameters due to
radiative corrections to the Higgs potential. 
For a gluino mass beyond 1~TeV more and more fine tuning
is required between these parameters; thus
Fig.\ref{boundmt} shows the upper 
limit on the lightest Higgs boson within the constrained NMSSM as a
dashed line, for gluinos lighter than 1~TeV.

\begin{figure}[htb]
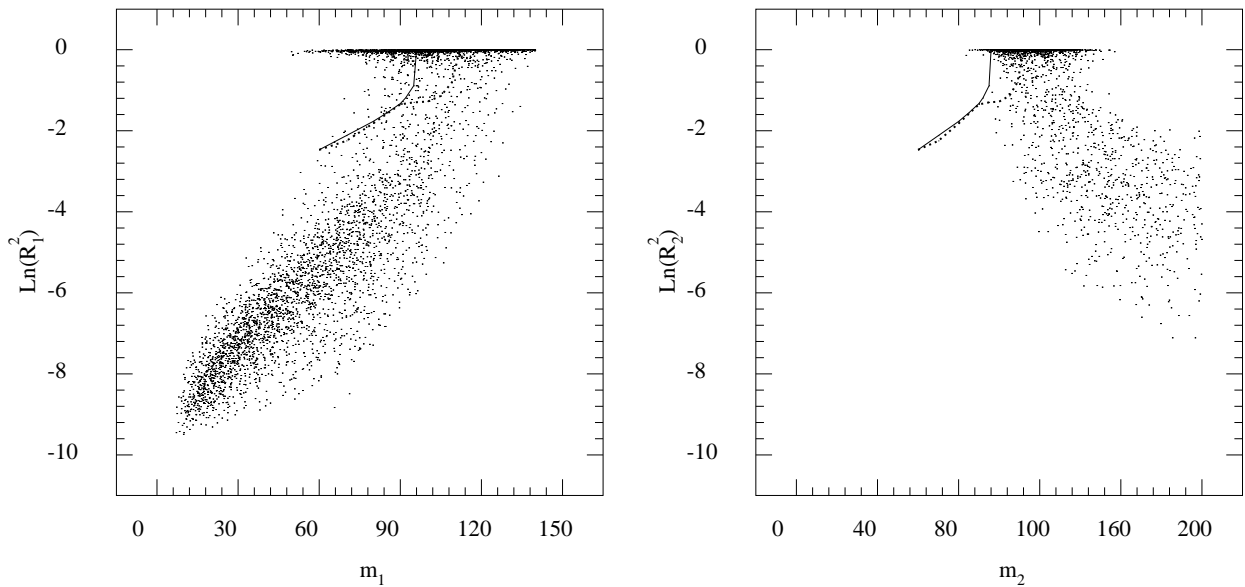
 
\begin{center}
\begin{picture}(150,80)
\put(-13,-5){\epsfig{figure=des1.ps,height=8cm} }
\put(72,-5){\epsfig{figure=des2.ps,height=8cm}}
\end{picture}
\end{center}
\caption{\it 
  The logarithm of the $ZZh_{1(2)}$ coupling $R_{1(2)}$ squared vs.
  the mass of the lightest CP-even state $h_{1(2)}$ in the constrained
  NMSSM.}
\label{des1} \label{des2}
\end{figure}

As noted before, the lightest Higgs boson within the NMSSM can 
essentially be a gauge singlet state and thus couple very weakly to the
$Z$ boson. Fig.\ref{des1} (left) shows the logarithm of the coupling
$R_1$
as a function of the mass $m_{1}$ of the lightest Higgs boson. Two
different regions exist within the constrained NMSSM: A densely
populated region with $R_1\sim 1$ and $m_{1} > 50$~GeV, and a tail
with $R_1 <$ (or $\ll$) $1$ and $m_{1}$ as small as $\sim$ 10~GeV.
Within the tail, the lightest Higgs boson is essentially a gauge
singlet state, which explains the small values of~$R_1$.

The solid line in Fig.\ref{des1} (left) indicates (for $m_{1} >
60$~GeV) the boundary of the region which can be tested at LEP2 with a
maximal c.m.  energy of 192~GeV and a luminosity of 150~pb$^{-1}$; the
dotted line corresponds to a maximal c.m. energy of 205~GeV and a
luminosity of 300~pb$^{-1}$ [both after combining all experiments
\cite{janot}]. A large part of the region with $R_1 \sim 1$, but only
a small part of the tail can be tested.

Fortunately, as noted above, the second lightest Higgs boson cannot be
too heavy if the lightest Higgs boson is essentially a gauge singlet
state \cite{NMSSMbound}, \cite{Kam}, \cite{kw}.  In the region of the
tail of Fig.\ref{des1} (left), within the constrained NMSSM, the mass of the
second Higgs boson $h_2$ varies between 80~GeV and the upper limit
indicated in Fig.\ref{boundmt} as a dashed line.  Its coupling to the
$Z$ boson $R_2$ is very close to 1 if $R_1 \ll 1$.  Fig.\ref{des2}
(right) shows the logarithm of the coupling $R_2$ as a function of the
mass $m_2$ of the second lightest Higgs. The tail of Fig.\ref{des1} (left)
corresponds to the region with $R_2 \sim 1$ in Fig.\ref{des2} (right) and
$vice$ $versa$.  Thus, a small part of the parameter space
corresponding to the tail of Fig.\ref{des1} (left) actually becomes visible
at LEP2 through the observation of $h_2$, which behaves like the
lightest Higgs boson of the MSSM in this case.


%
%

If a 
 Higgs boson is observed, it will generally be very difficult,
however, to distinguish the NMSSM from the MSSM \cite{epic}. This seems
only possible if a Higgs boson happens to contain a sizeable amount of
the singlet state (and hence a measurably reduced coupling to the $Z$
boson), but couples still strongly enough in order to be visible.
Finally, the constrained NMSSM could actually be ruled out at LEP2 if a
neutral CP-odd Higgs particle, a charged Higgs particle, or an invisibly 
decaying Higgs particle 
would be observed. 
\vvs1

\subsection{Non-linear Supersymmetry}

Most of the supersymmetric models investigated so far
are models based on linearly realized supersymmetry.  However, supersymmetry
may as well be realized nonlinearly.  Whereas the linear
supersymmetric models require supersymmetric partners for all
conventional particles in the standard model, the nonlinear models do not lead
to SUSY partners.  Most global nonlinear supersymmetric models require only
one new particle: the Akulov-Volkov field \cite{Vol72/73}, which is 
a Goldstone
fermion.  
This Goldstino can be removed by going over to curved space, i.e. to
supergravity, 
where it can be gauged away.  In the flat space limit, the
supergravity multiplet decouples from ordinary matter so
that supersymmetry can manifest itself only in the Higgs 
sector.

The formalism for extending the standard model to a 
supersymmetric theory in a nonlinear
way was developed in ref.\cite{Sam83/84}.  Recently, the general form of the 
nonlinear supersymmetric standard model has been constructed 
and the Higgs
potential in the flat space limit \cite{Kim95} has been derived. 

%
The Higgs sector of the nonlinear SUSY models is evidently larger than
that of the Standard Model.  It contains at least two dynamical Higgs
doublets and an auxiliary Higgs singlet.  In the case that both the dynamical 
and the auxiliary singlet are included in the theory, the spectra of Higgs
bosons in the nonlinear models resemble those of the linear model
with two Higgs doublets and one singlet 
(NMSSM).  Both models 
have three scalar,
two pseudoscalar Higgs bosons and one charged Higgs boson pair.
However, the structure of the Higgs potential is quite different between
nonlinear models and the NMSSM.\ \ For the general nonlinear model, the 
complete potential in the flat space limit is given by \cite{Kim95}
\begin{eqnarray}
\label{nlsusy.1}
        V &=& {\textstyle \frac18} (g_1^2 +g_2^2) (|H^1|^2 -|H^2|^2)^2
             + {\textstyle \frac12} g_2^2 |H^+_1H_2|^2
                       + |\mu_1 + \lambda_1 N|^2\,|H^1|^2     \cr
          & & \mbox{ } + |\mu_2 + \lambda_2 N|^2\,|H^2|^2
                       + |\lambda_0 H^{1T}\epsilon H^2 + \mu_0 N + k N^2|^2
\end{eqnarray}
involving novel types of interactions between the Higgs fields. 
The two Higgs doublets $H^1, H^2$ and the singlet $N$ develop the
vev's $v_1, v_2,\ \mbox{and}\ x$ respectively.
The three scalar Higgs bosons are the eigenstates of the scalar-Higgs
mass matrix. 
In a  way similar to  the NMSSM, an upper bound for the mass $m_1$ of the 
lightest scalar Higgs boson $S_1$ can be derived at the lowest 
order,\footnote{In the present 
exploratory analysis we have neglected the radiative corrections.}
\begin{equation}
\label{nlsusy.2}
        m^2_1 \le m_{1,max}^2 = 
                m_Z^2 (\cos^2 2\beta + 2\lambda_g^2 \sin^2 2\beta)
\end{equation}
where  $\lambda_g = \lambda_0^2/(g_1^2 +g_2^2)$.
Hence, $m_1 \le m_Z$ for $\lambda_0 \le \sqrt{(g_1^2+g_2^2)/2} \approx
0.52$ and $m_1 \le 1.92\lambda_0 m_Z$ for $\lambda_0 > 0.52$.
In the latter case, the upper bound of $\lambda_0$ determines the limit
of $m_1$. For $M_t = 175$ GeV and with the GUT scale
as cut-off scale, one obtains $m_1 \le 130$ GeV.
Even though at the c.m. energies of 175, 192, and 205 GeV for 
LEP2 the production of $S_1$ may be kinematically possible, 
the production rate is in general not large enough for 
$S_1$ to be detected.
The main contributions to the production cross sections come from (i) the
Higgs-strahlung
 process; (ii) the process 
where $S_1$ is radiated off leptons or quarks, 
and 
(iii) associated pair production $P_j S_1$, 
where $P_j\ (j=1,2)$ is a pseudoscalar Higgs
boson.

\begin{figure}[hbt]
\centering
\mbox{\epsfig{file=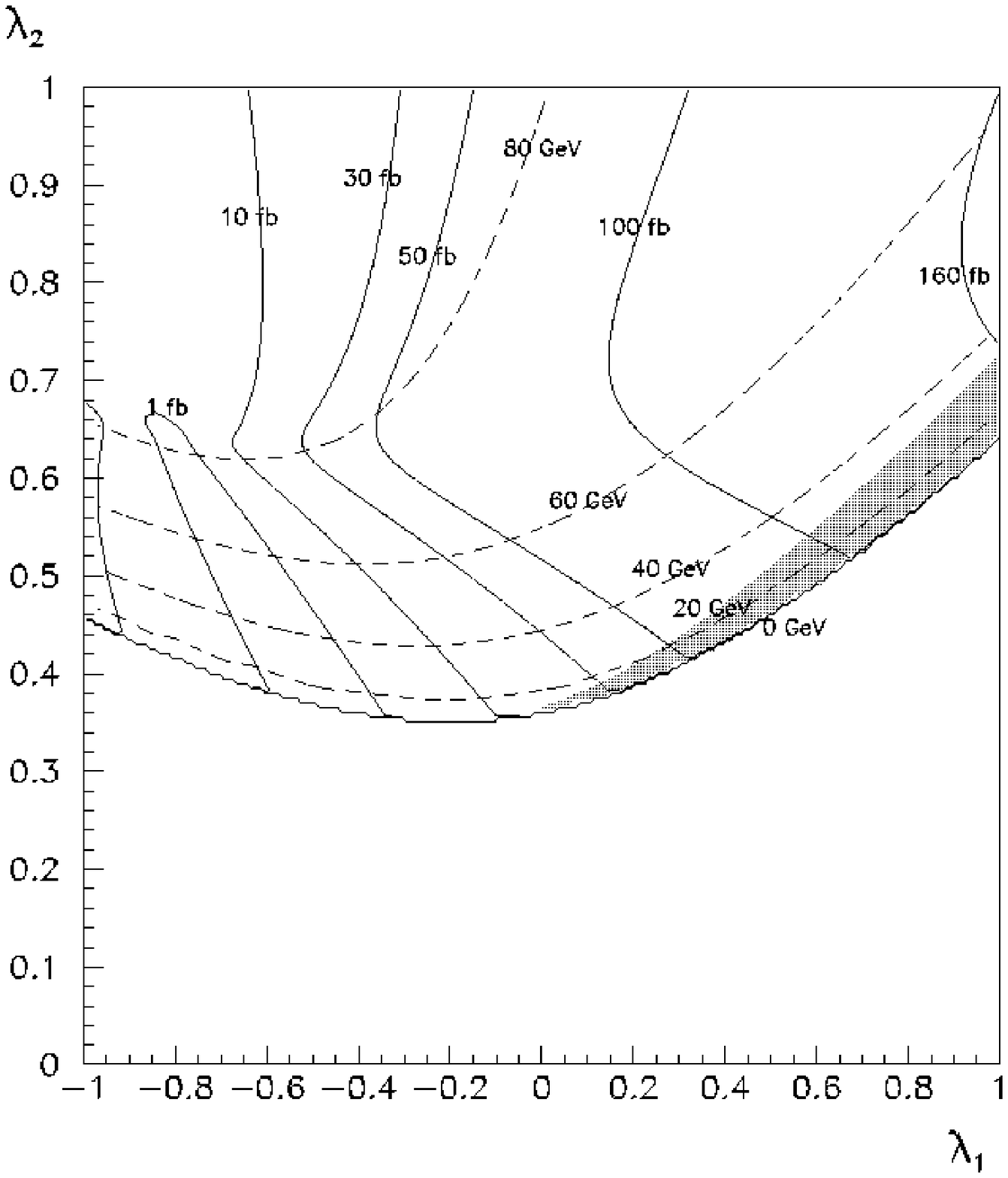,width=0.5\textwidth}}
\par
\caption[0]{\it 
  Contour lines of the lightest scalar Higgs mass $m_1$ (dashed) and
  of the production cross section $\sigma$ (full) at
  $\sqrt{s}=175$~GeV, as functions of $\lambda_1, \lambda_2$ for
  $\tan\beta=6,\ \lambda_0=0.3,\ k=-0.02,\ m_C=400\ \mbox{GeV}$.  The
  shaded area marks the parameter region excluded by LEP1, defined as
  the region where the production cross section at the $Z$ peak is
  greater than 1 pb.}
\label{nlsusy}
\end{figure}

We have first searched for parameter regions where the experimental
lower limit on $m_1$ given by the LEP1 data is minimal; it turned out
that there are regions where even $m_1 = 0$ is still allowed.
Fig.\ref{nlsusy} shows (dashed) the contour lines of $m_1$ and (full)
the contour lines of the cross section $\sigma$ for the production of
$S_1$ in $e^+ e^-$ collisions at $\sqrt s = \mbox{175 GeV}$, in the
$\lambda_1,\lambda_2$ plane for a representative set of the parameters
$\tan\beta,\ \lambda_0,\ k$ and $m_C$ ($m_C$ being the charged-Higgs
mass).  In the shaded region, the cross section $\sigma$ at $\sqrt s =
m_Z$ is greater than 1 pb, which corresponds to the discovery limit of
LEP1 for $m_1 \approx \mbox{65 GeV}$ \cite{janot}.  This region is
excluded by LEP1 since the discovery limit decreases with decreasing
$m_1$.  The discovery limit at $\sqrt s = \mbox{175 GeV}$ with a
luminosity of 500 $\mbox{pb}^{-1}$ is about 50 fb for $m_1 = \mbox{80
  GeV}$ (about 30 fb for $m_1 = \mbox{40 GeV}$) \cite{janot}.  Thus
the region accessible at LEP2 includes the area in Fig.\ref{nlsusy}
where $m_1 < \mbox{80 GeV}$ and $\sigma >$ 50 fb. As with LEP1, a
massless $S_1$ could be undetectable.

For some parameter sets, the nonlinear SUSY model may be tested even
if the lightest scalar is undetectable.  If the production cross
section of $S_1$ is smaller than the discovery limit, one can 
examine whether the production of the other Higgs bosons is
kinematically possible and whether their production rates are large
enough for discovery.  

For the higher energies 192 and 205 GeV, the effects are similar to the 
175~GeV case.
Though the accessible region increases, an undetectable massless
$S_1$ Higgs boson
is still possible.  Energies of 240 GeV and more are needed to test this
nonlinear supersymmetric 
model conclusively.


\subsection{Majoron Decays of Higgs Particles}

\noindent
There are a variety of well motivated extensions of the Standard 
Model (SM) with a spontaneously broken 
global symmetry. This symmetry could either be lepton number 
or a combination of family lepton numbers \cite{CMP,majoron}.
These models are characterized by a more complex symmetry
breaking sector which contains additional Higgs bosons. 
It is specially interesting for our purposes to consider models
where such symmetry is broken at the electroweak scale \cite{HJJ,MASI_pot3}.  
In general, these models contain a massless Goldstone boson, called majoron 
($J$), which interacts very weakly with normal matter.  In such models, 
the normal doublet Higgs boson is expected to have sizeable invisible decay 
modes to the majoron, due to the strong Higgs--majoron coupling.  
This can have a significant effect
on the Higgs phenomenology at LEP2.  In particular, the invisible
decay could contribute to the signal of two acoplanar jets and missing
momentum. This feature of majoron models allows one to strongly
constrain the Higgs mass in spite of the occurrence of extra
parameters compared to the SM.  In particular, the LEP1 limit on the
predominantly doublet Higgs mass is close to the SM limit irrespective
of the decay mode of the Higgs boson \cite{valle:1,dproy}.

We consider a model containing two Higgs doublets
($\phi_{1,2}$) and a singlet ($\sigma$) under the $SU(2)_L \times
U(1)_Y$ group.  The singlet Higgs field carries a non-vanishing 
$U(1)_L$ charge, which could be lepton number. Here we
only need to specify the scalar potential of the model:
\begin{eqnarray}
\label{N2}
V &=&\mu_{i}^2\phi^{\dagger}_i\phi_i+\mu_{\sigma}^2\sigma^{\dagger}
\sigma + \lambda_{i}(\phi^{\dagger}_i\phi_i)^2+
   \lambda_{3} (\sigma^{\dagger}\sigma)^2+
\nonumber\\
& &\lambda_{12}(\phi^{\dagger}_1\phi_1)(\phi^{\dagger}_2\phi_2)
+\lambda_{13}(\phi^{\dagger}_1\phi_1)(\sigma^{\dagger}\sigma)+
\lambda_{23}(\phi^{\dagger}_2\phi_2)(\sigma^{\dagger}\sigma)
\nonumber\\
& & +\delta(\phi^{\dagger}_1\phi_2)(\phi^{\dagger}_2\phi_1)+
    {\textstyle \frac{1}{2}}\kappa[(\phi^{\dagger}_1\phi_2)^2+\;\;h.c.]
\end{eqnarray}
where the sum over repeated indices $i$=1,2 is assumed.

Minimization of the above potential leads to the spontaneous $SU(2)_L
\times U(1)_Y \times U(1)_L$ symmetry breaking and allows us to identify
a total of three massive CP even scalars $H_{i} \:$ (i=1,2,3), plus a
massive pseudoscalar $A$ and the massless majoron $J$. We assume that
at the LEP2 energies only three Higgs particles can be produced: the
lightest CP-even scalar $h$, the CP-odd massive scalar $A$, and the
massless majoron $J$. Notwithstanding, our analysis is also valid for
the situation where the Higgs boson $A$ is absent \cite{valle:2}, 
which can be obtained
by setting the couplings of this field to zero.

At LEP2, the main production mechanisms of invisible Higgs bosons
are the Higgs-strahlung process ($e^+e^- \rightarrow h Z$) and the associated
production of Higgs bosons pairs ($e^+e^- \rightarrow A h$), which
rely upon the couplings $hZZ$ and $hAZ$ respectively. The important
feature of the above model is that, because of its singlet nature, 
the majoron is not sizeably coupled to the gauge bosons and cannot 
be produced directly, thereby evading strong LEP1 
constraints. The $hZZ$ and $hAZ$ couplings depend on the
model parameters via the appropriate mixing angles, but they can be
effectively expressed in terms of the two parameters $\epsilon_A$,
$\epsilon_B$:
\begin{eqnarray}
\label{HZZ3}
{\cal L}_{hZZ}
&=& \epsilon_B \left ( \sqrt{2} G_F \right )^{1/2}
m_Z^2 Z_\mu Z^\mu h\\
{\cal L}_{hAZ}&=& - \epsilon_A \frac{g}{\cos\theta_W} 
Z^\mu h \stackrel{\leftrightarrow}{\partial_\mu} A 
\end{eqnarray}
The couplings $\epsilon_{A(B)}$ are model dependent.  For instance,
the SM Higgs sector has $\epsilon_A=0$ and $\epsilon_B=1$, while a
majoron model with one doublet and one singlet leads to $\epsilon_A=0$
and $\epsilon_B^2 \leq 1$.

The signatures of the Higgs-strahlung process and the associated
production depend upon the allowed decay modes of the Higgs bosons $h$
and $A$. For Higgs boson masses $m_h$ accessible at LEP2 energies the
main decay modes for the CP-even state $h$ are $b \bar{b}$ and $JJ$.
We treat the branching fraction $B = BR(h \rightarrow JJ)$ as a free
parameter.  In most models $BR$ is basically unconstrained and can
vary from 0 to 1. Moreover we also assume that, as it happens in the
simplest models, the branching fraction for $A \rightarrow b \bar{b}$
is nearly one, and the invisible $A$ decay modes $A\rightarrow hJ$,
$A\rightarrow JJJ$ do not exist (although CP-allowed).  Therefore our
analysis depends upon five parameters: $m_h$, $m_A$, $\epsilon_A$,
$\epsilon_B$, and $B$. This parameterization is quite general and very
useful from the experimental point of view: limits on $m_h$, $m_A$,
$\epsilon_A$, $\epsilon_B$, and $B$ can be later translated into
bounds on the parameter space of many specific models.

The parameters defining our general parameterization can be constrained
by the LEP1 data. In fact Refs.\cite{valle:1,valle:asso} analyze
some signals for invisible decaying Higgs bosons, and conclude that
LEP1 excludes $m_h$ up to 60 GeV provided that $\epsilon_B > 0.4$.
Similar results are obtained in fig.(\ref{fig:sbainv}).


The $\bar{b}b + \ptmis$ topology is our main subject of investigation
and we carefully evaluate signals and backgrounds, choosing the cuts
that enhance the signal over the backgrounds.  Our goal is to evaluate
the limits on $m_h$, $m_A$, $\epsilon_A$, $\epsilon_B$, and $B$ that
can be obtained at LEP2 from this final state.  There are three
sources of signal events with the topology $\ptmis +$ 2 $b$-jets: one
due to the associated production and two due to the Higgs--strahlung.
\begin{eqnarray}
e^+  e^- &\rightarrow& ( Z  \rightarrow b   \bar{b} )~   
+ ~(h \rightarrow  J  J) 
\label{h:jj}
\\
e^+  e^- &\rightarrow& ( Z  \rightarrow \nu \bar{\nu} )~ 
+ ~(h \rightarrow  b \bar{b}) 
\label{h:sm}
\\
e^+  e^- &\rightarrow& ( A  \rightarrow b   \bar{b} )~   
+ ~(h \rightarrow  J  J) 
\label{h:a}
\end{eqnarray}
The signature of this final state is the presence of two jets
containing $b$ quarks and missing momentum ($\ptmis$).  It is
interesting to notice that for light $m_h$ and $m_A$, the associated
production dominates over the Higgs--strahlung mechanism \cite{valle:asso}.

There are several sources of background for this topology:
\begin{displaymath}
\begin{array}{ll}
e^+ e^-  \rightarrow  Z/\gamma~ Z/\gamma 
\rightarrow q \bar{q} ~ \nu \bar{\nu} &
\quad e^+ e^-  \rightarrow  (e^+e^-) \gamma\gamma 
\rightarrow [e^+e^-] q \bar{q}\\
e^+ e^-  \rightarrow  Z^*/\gamma^*~  
\rightarrow q \bar{q} [n \gamma ] &
\quad e^+ e^-  \rightarrow  W^+ W^- 
\rightarrow q \bar{q}^\prime ~ [\ell] \nu\\
e^+ e^-  \rightarrow W [e] \nu 
\rightarrow q \bar{q}^\prime ~ [e] \nu  & 
\quad e^+ e^-  \rightarrow Z  \nu  \bar{\nu} 
\rightarrow q \bar{q} ~ \nu \bar{\nu}
\end{array}
\end{displaymath}
where the particles in square brackets escape undetected and the jet
originating from the quark $q$ is identified (misidentified) as being
a $b$-jet. 

At this point the simplest and most efficient way to improve the
signal-over-background ratio is to use that the $A$ and
$h$ decays lead to jets containing $b$-quarks. So we require that the
events contain two $b$-tagged jets. Moreover, the background can be
further reduced requiring a large $\ptmis$. We therefore have imposed 
a set of cuts which is based on the cuts used by the
DELPHI collaboration for the SM Higgs boson search~\cite{delphione}.

\begin{figure}[p]
\begin{picture}(140,110)
\put(25,55){\epsfxsize110mm\epsfbox{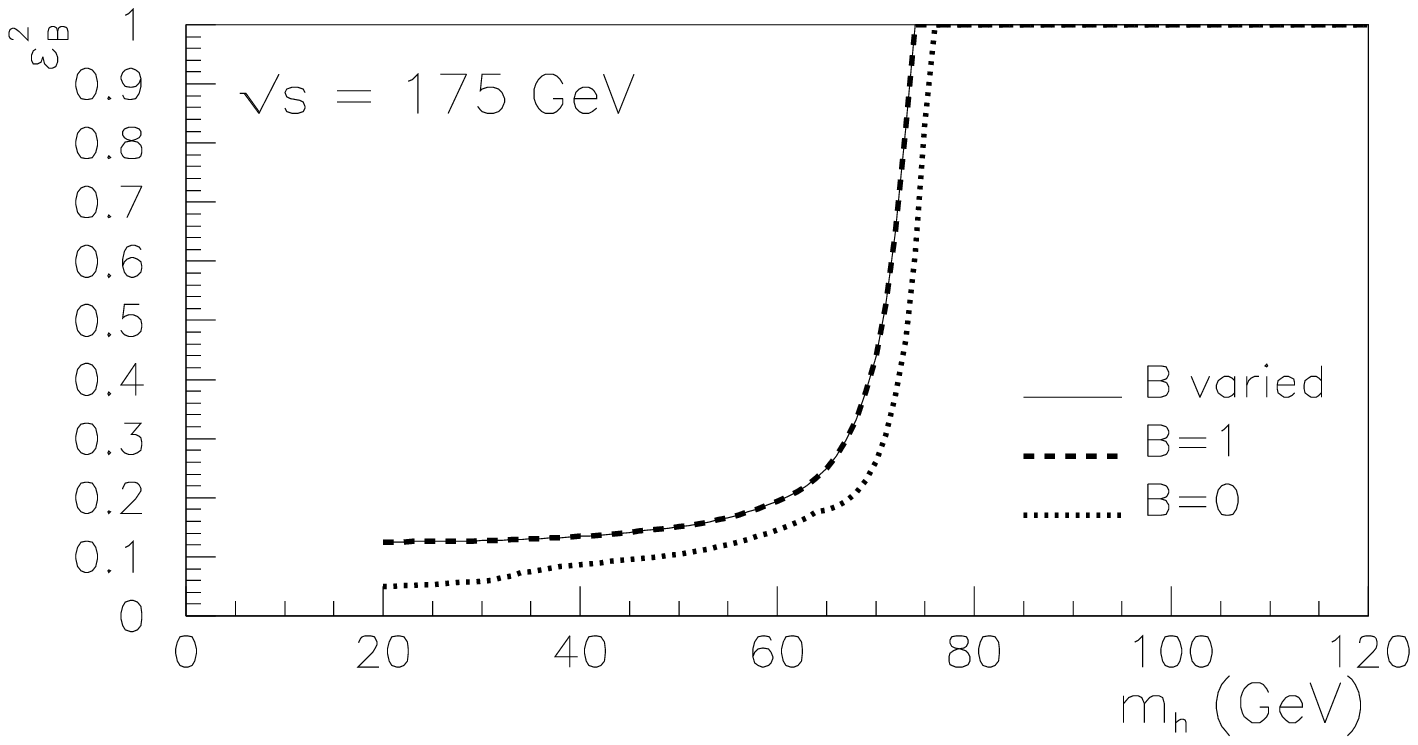}}
\put(25,0){\epsfxsize110mm\epsfbox{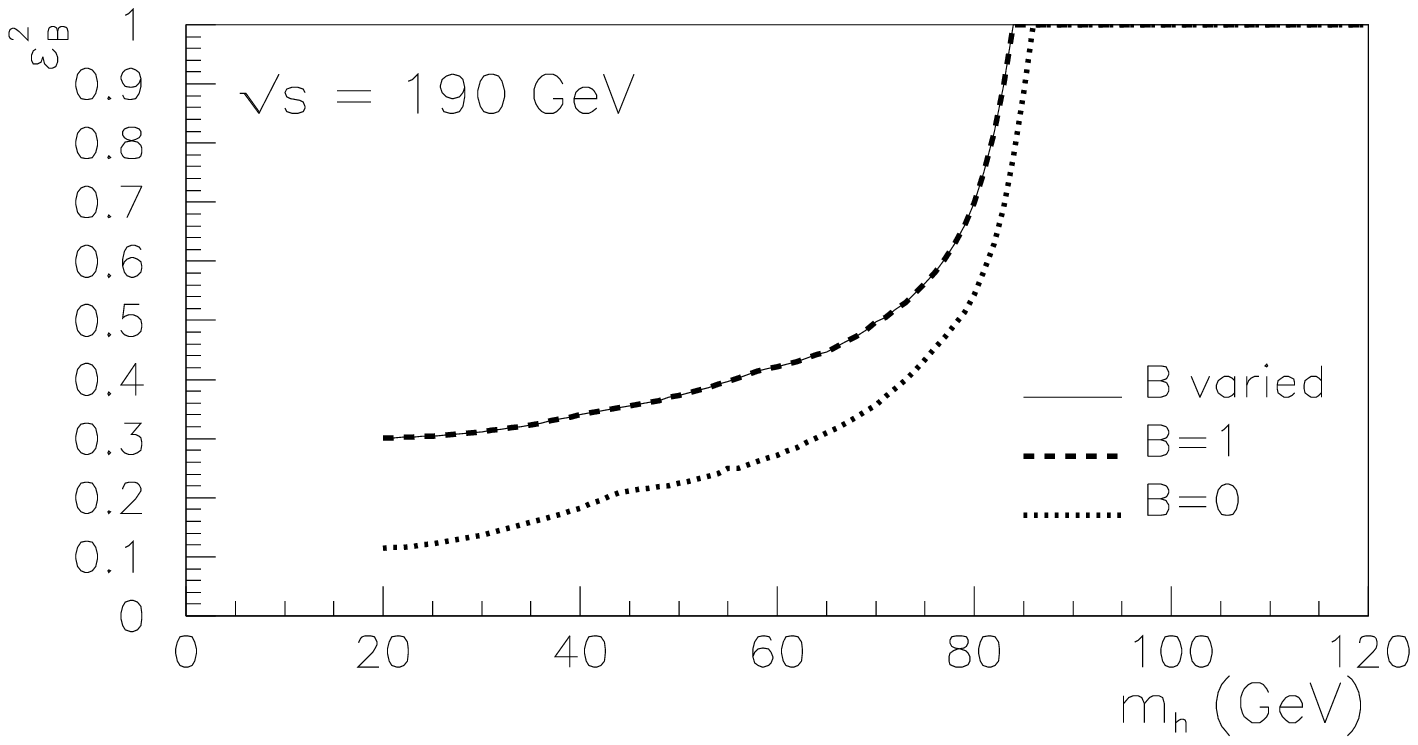}}
\end{picture}
\vspace{-.8cm}
\caption[0]{\it{
  Limits on $\epsilon_B^2$ as a function of $m_h$ for
  500 $pb^{-1}$ and $\sqrt{s}= $175 GeV and for 300 $pb^{-1}$ and 
$\sqrt{s}= 190 $ GeV;  for different values of
  $B=Br(h\rightarrow JJ)$}
\label{fig:zh}}
\vspace*{-8mm}
\begin{center}
\begin{tabular}{p{0.45\linewidth}p{0.45\linewidth}}
\begin{center}
\mbox{\epsfig{file=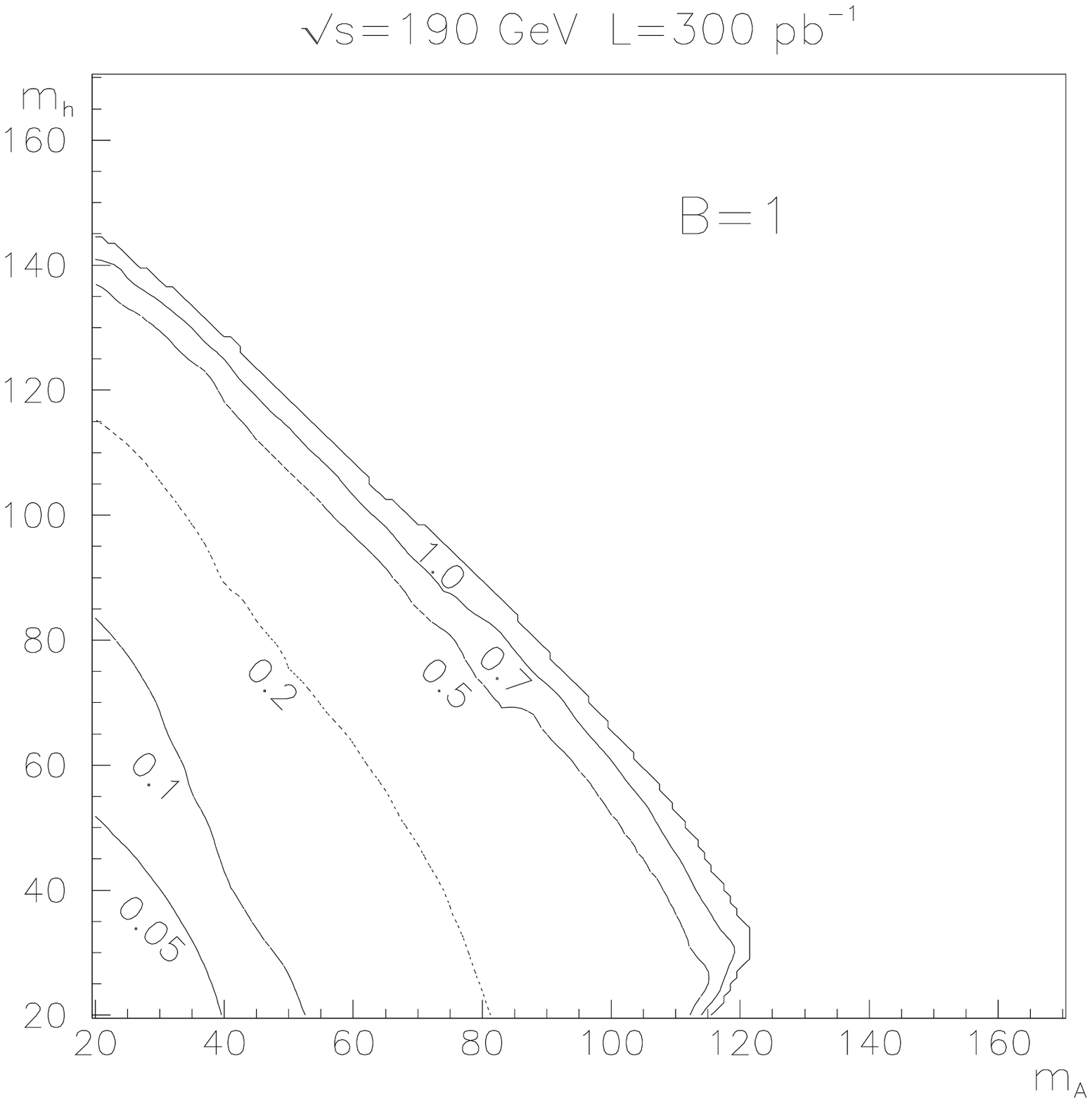,width=\linewidth}}
\end{center}&
\begin{center}
\mbox{\epsfig{file=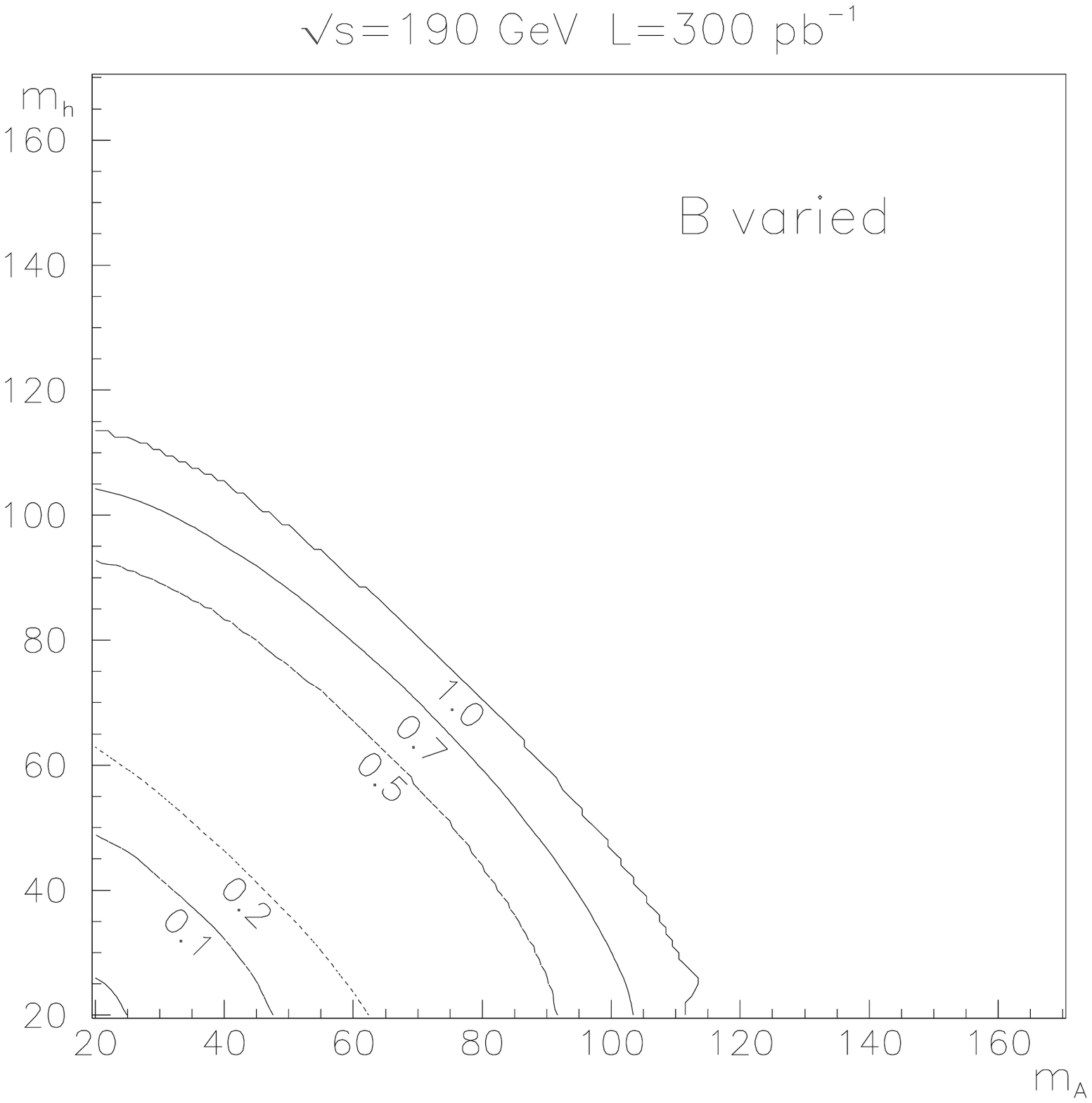,width=\linewidth}}
\end{center}\\
\end{tabular}
\vspace*{-10mm}
\caption[0]{\it 
  Limits on $\epsilon_A^2$ as a function of $m_h, m_A$ for $\sqrt{s} =
  190$ GeV. The left plot shows the limits obtained for
  $B=Br(h\rightarrow JJ)=1$, in the right plot $B$ is varied from 0 to
  1.}
\label{fig:ah}
\end{center}
\vspace*{-5mm}
\end{figure}

Depending on the $h$ and $A$ mass ranges, including or excluding an 
invariant mass cut $m\pm 10$~GeV [where $m$ is the mass of the particle 
decaying visibly] gives better or weaker limits on the $ZhA$ and
$ZZh$ couplings.  Therefore, for each mass combination four limits are
calculated (with or without invariant mass cut, with thrust cut or the
cut on the minimal two-jet energy) and the best limit is kept.

We denote the number of signal events for the three production processes
(\ref{h:jj} -- \ref{h:a}), after imposing all cuts, $N_{JJ}$, $N_{SM}$,
and $N_A$ respectively, assuming that $\epsilon_A = \epsilon_B = 1$. Then
the expected number of signal events when we take into account couplings
and branching ratios is
\begin{equation}
N_{exp} = \epsilon_B^2 \left [ B N_{JJ} + (1-B) N_{SM} \right ]
+ \epsilon_A^2 B N_A 
\end{equation} 
In general, this topology is dominated by the associated production,
provided it is not suppressed by small couplings $\epsilon_A$ or phase
space. The most important background after the cuts is $Z/\gamma Z/\gamma$ 
production. The total numbers of background events summed
over all relevant channels are 2.3, 2.8
and 5.9 for $\sqrt{s}=175$ , $190$ and $205$ GeV respectively.


In order to obtain the limits shown in
Figs.\ref{fig:zh}-\ref{fig:ah}, we assumed that only the background
events are observed, and we evaluated the 95 \% CL region of the
parameter space that can be excluded with this result.  By taking
the weakest bound, as we vary $B$, we obtained the absolute bound on
$\epsilon_A$, $\epsilon_B$, and $m_h$ independent of the $h$ decay
mode. The limits on $\epsilon_A$ obtained by searches for the 
$b\protect\bar{b} ~+~ \protect\ptmis$ final states are stronger 
than those obtained from the $b\protect\bar{b}b\protect\bar{b}$ 
topology.  Moreover, the bounds in the limiting case 
$\epsilon_A=0$ apply for the simplest model of invisibly 
decaying Higgs bosons, where just one singlet is added to the SM.
A more complete presentation of these results will be given
in Ref.\cite{inprep}.

\subsection{Strongly Interacting Higgs Particle}

The radiative corrections at LEP1 depend 
only logarithmically on the Higgs mass,
and the measurements, although very precise, are not sufficient  
to determine the structure of the Higgs sector. It is therefore necessary
to keep an open mind to the possibility that the Higgs sector is
more complicated than in the Standard Model. 
Beyond the Standard Model various extensions have been suggested.
One of the possibilities is supersymmetry which has been previously discussed. 
Another possibility is strong
interactions in the form of technicolor, which at
least in its simplest form is ruled out by the LEP1 data.
Strong interactions in the Standard Model itself imply a heavy Higgs boson
and can presumably be studied at the LHC.

However, 
the idea of strong interactions is more general. In particular
it is possible that strong interactions are present in the
singlet sector of the theory. In general the  choice of representations
in a gauge theory is arbitrary and presumably a clue to a deeper
underlying theory. Singlets do not have 
quantum numbers under the gauge group of the Standard Model.
They therefore do not feel the strong or electroweak forces,
but they can couple to the Higgs particle. As a consequence,
radiative corrections to weak processes are not sensitive to the
presence of singlets in the theory, because no Feynman graphs containing
singlets  appear
at the one--loop level. Because effects at the two--loop level
are below the experimental precision,
the presence of a singlet sector is not ruled out by any 
of the LEP1 precision data.

It is therefore not unreasonable to
assume that there exists a hidden sector that affects Higgs
physics only. Such an extension of the Standard Model involving
singlet fields preserves the essential simplicity of the model,
while at the same time acting as a realistic model for non--standard Higgs
properties. Here we will study the coupling
of a Higgs boson to an $O(N)$ symmetric set of scalars, which  
is one of  the simplest possibilities introducing only a few extra 
parameters in the theory. The extra scalars may give rise to 
 large invisible decay width of the Higgs particle.
When the coupling is large enough, the Higgs resonance can become
wide even for a light Higgs boson. This has led to the conclusion that
this Higgs particle becomes undetectable at the LHC \cite{vladimir}. 
As one can measure missing energy
more precisely at $e^+e^-$ colliders than at a hadron machine, 
LEP2 can give important constraints
on the parameters of the model. However, it is clear that there
will be a range of parameters where this Higgs boson can be seen neither
at LEP nor at the LHC.
\vfill\newpage

\begin{figure}[tb]
\begin{center}
\begin{picture}(170,80)
\put(0,0){\epsfig{figure=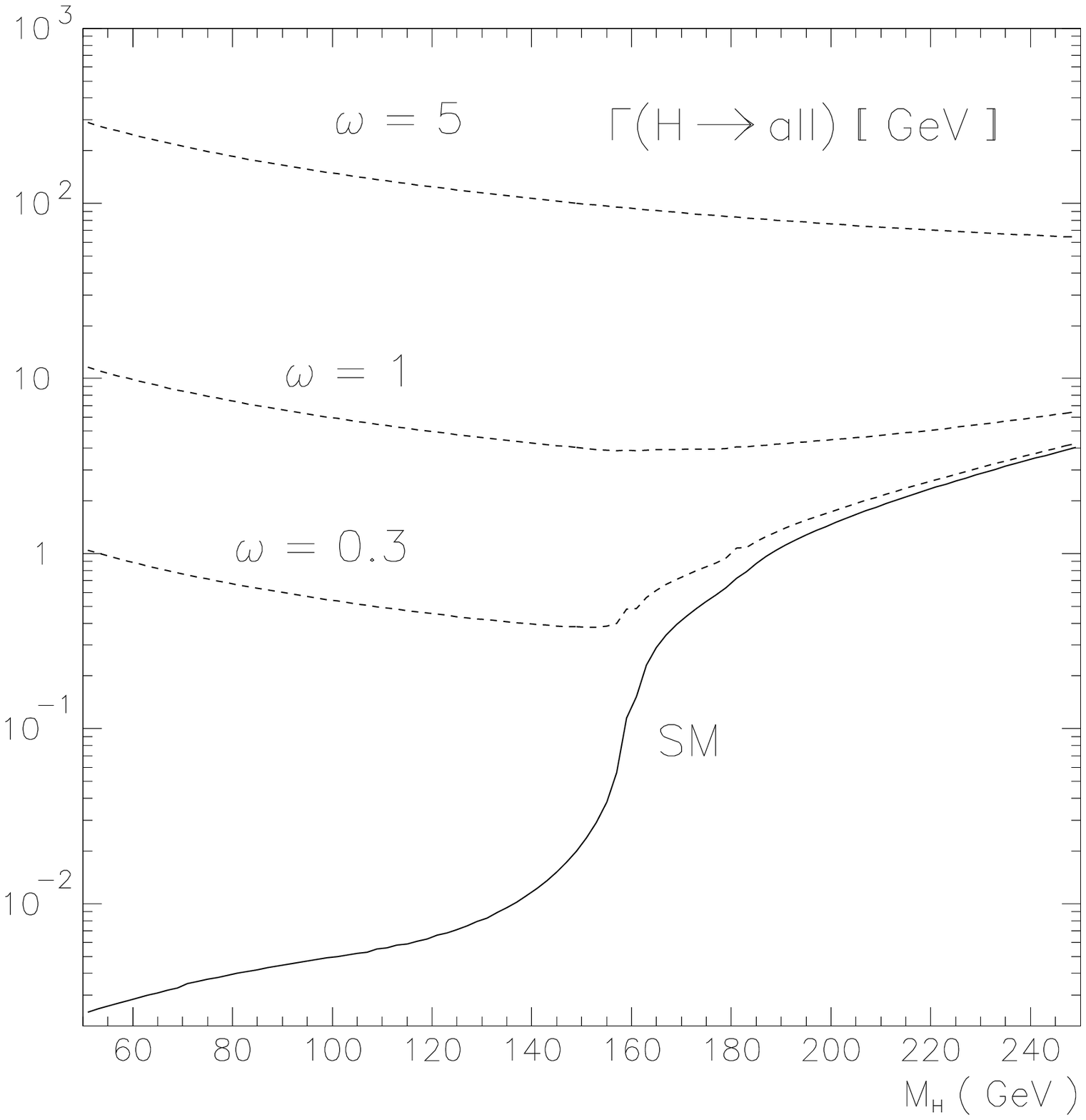,height=8cm,angle=0}}
\put(85,0){\epsfig{figure=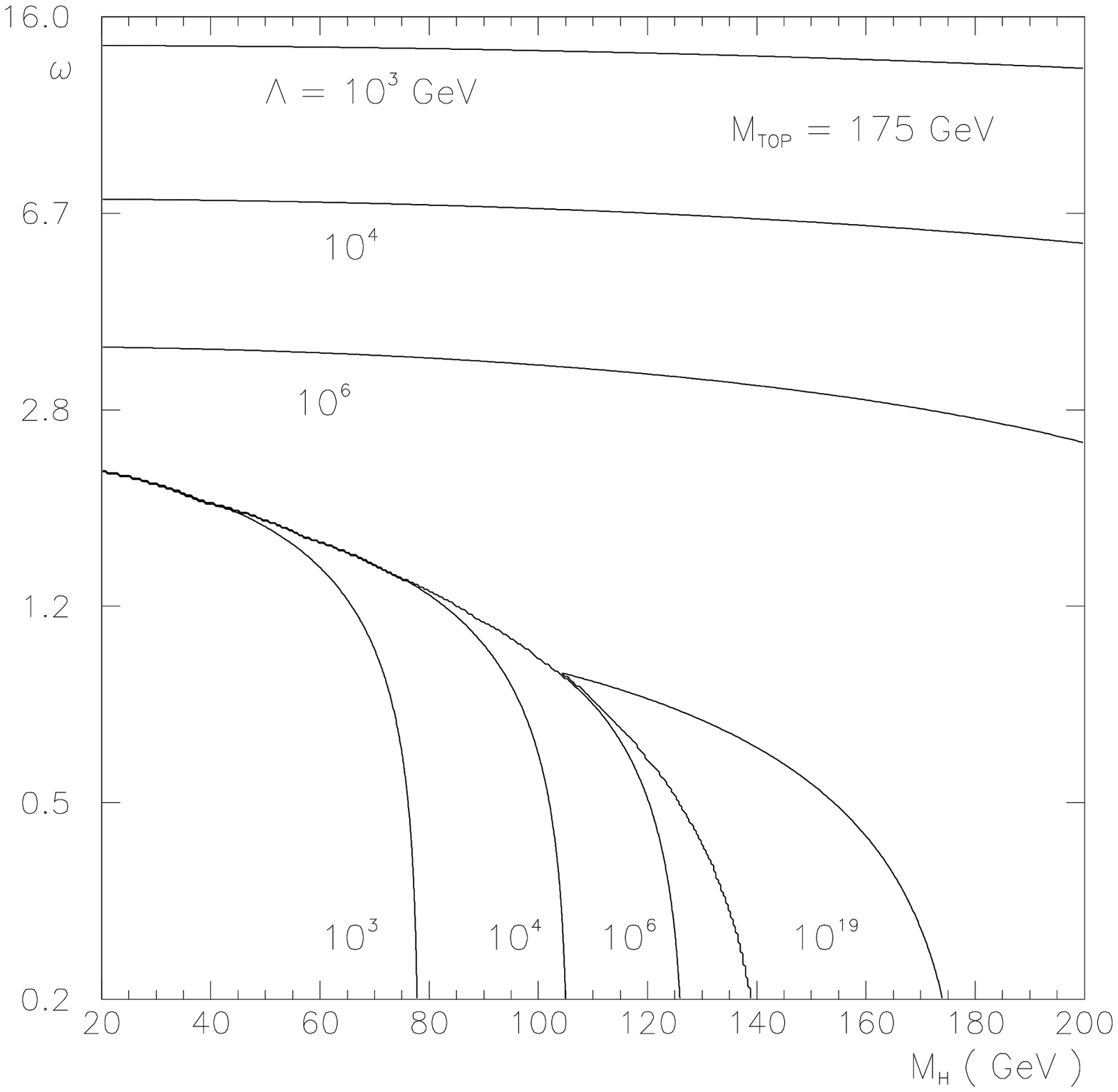,height=8cm,angle=0}}
\end{picture}
\end{center}
\caption{\it 
  {\rm (left)} Higgs width in the phion model, in comparison with the
  Standard Model.}
\label{width}
\caption{\it 
  {\rm (right)} Theoretical limits on the parameters of the model in
  the $\omega$ vs. $M_H$ plane. For a given scale $\Lambda$, the
  physical region is below the upper lines and to the right of the
  lower lines.}
\label{stability}
\end{figure}

\noindent {\bf a) The Model}\pss{0.5}
The Higgs sector of the model is described by the following Lagrangian,
\begin{eqnarray}
\label{definition}
 {\cal L}  &=&
 - \partial_{\mu}\phi^+ \partial^{\mu}\phi -\lambda (\phi^+\phi - v^2/2)^2
\nonumber \\
& - &    1/2\,\partial_{\mu} \vec\varphi \partial^{\mu}\vec\varphi
     -1/2 \, m^2 \,\vec\varphi^2 - \kappa/(8N) \, (\vec\varphi^2 )^2
    -\omega/(2\sqrt{N})\, \, \vec\varphi^2 \,\phi^+\phi \nonumber
\end{eqnarray}
where $\phi$ is the normal Higgs doublet and the vector $\vec\varphi$
is an N--component real vector of scalar fields, which we call phions.
Couplings to fermions and vector bosons are the same as in the
Standard Model.  The ordinary Higgs field acquires the vacuum
expectation value $v/\sqrt{2}$.  We assume that the
$\vec\varphi$--field acquires no vacuum expectation value, which can
be assured by taking $\omega$ positive. After the spontaneous symmetry
breaking one is left with the ordinary Higgs boson, coupled to the
phions into which it decays. Also the phions receive an induced mass
from the spontaneous symmetry breaking.  The factor $N$ is taken to be
large, so that the model can be analyzed in the $1/N$ expansion.  By
taking this limit, the phion mass stays small, but because there are
many phions, the decay width of the Higgs boson can become large.
Therefore the main effect of the presence of the phions is to give a
large invisible decay rate to the Higgs boson. The invisible decay
width is given by
\be \Gamma_H =\frac {\omega^2 v^2}{32 \pi m_H} \nonumber \ee
The Higgs width is compared with the width in the Standard Model for various choices
of the coupling $\omega$ in Fig.\ref{width}.
The model is different
from Majoron models, since the width is not necessarily small.
The model is similar to the technicolor--like model of Ref.\cite{chivukula}.

Consistency of the model requires two conditions.  One condition is
the absence of a Landau pole below a certain scale $\Lambda$. The
other follows from the stability of the vacuum up to a certain scale.
An example of such limits is given in Fig.\ref{stability}, where
$\kappa=0$ was taken at the scale $2m_Z$, which allows for the widest
range. For the model to be valid beyond a scale $\Lambda$ one should
be below the indicated upper lines in the figure.  One should be to
the right of the indicated lower lines to have stability of the
vacuum.

For the search for the Higgs boson there are basically
two channels; one is the standard decay, which is reduced in branching
ratio due to the decay into phions.
The other is the invisible decay, which rapidly becomes dominant,
eventually making the Higgs resonance very  wide, Fig.\ref{width}. 
In order to find the bounds we 
neglect the coupling $\kappa$ as this is a small effect. We
also neglect the phion mass. For non--zero values of the phion mass
the bounds can be found by rescaling the decay widths
with the appropriate phase space factor.  
The present bounds, coming from LEP1 invisible search,
are included as a dashed curve in Fig.\ref{exclusion} below.
\vvs1

\noindent {\bf b) LEP2 Bounds}\pss{0.5}
In the case of LEP2 the limits on the Higgs mass and couplings in the
present model come essentially from the invisible decay, as the
branching ratio into $\bar bb$ quarks drops rapidly with increasing
$\varphi$--Higgs coupling. To define the signal we look at events
around the maximum of the Higgs resonance, with an invariant mass $m_H
\pm \Delta$ for $\Delta=5$ GeV, which corresponds to a typical mass
resolution.  Exclusion limits are determined by Poisson statistics as
defined in Appendix \ref{sec:app}. The results are given by
the full lines in Fig.\ref{exclusion}. One notices the somewhat
reduced sensitivity for a Higgs mass near the Z boson mass and a
looser bound for small Higgs masses because there the effect of the
widening of the resonance prevails.  The small $\omega$ region is
covered by visible search.  There is a somewhat better limit on the
Higgs mass for moderate $\omega$ in comparison with the $\omega=0$
case; this is due to events from the extended tail of the Higgs boson
which is due to the increased width.

We conclude from the analysis that LEP2 can put significant limits on
the parameter space of such a model. However there is a range where
the Higgs boson will not be discovered, even if it does exist in this
mass range.  This also holds true when one considers the search at the
LHC.  Assuming moderate to large values of $\omega$, i.e. in the
already difficult intermediate mass range, it is unlikely that
sufficient signal events are left at the LHC. In that case the only
information can come directly from high--energy $e^+e^-$ colliders or
indirectly from higher precision experiments at LEP1.
\begin{figure}[p]
\vspace{0.1cm}
\centerline{\epsfig{figure=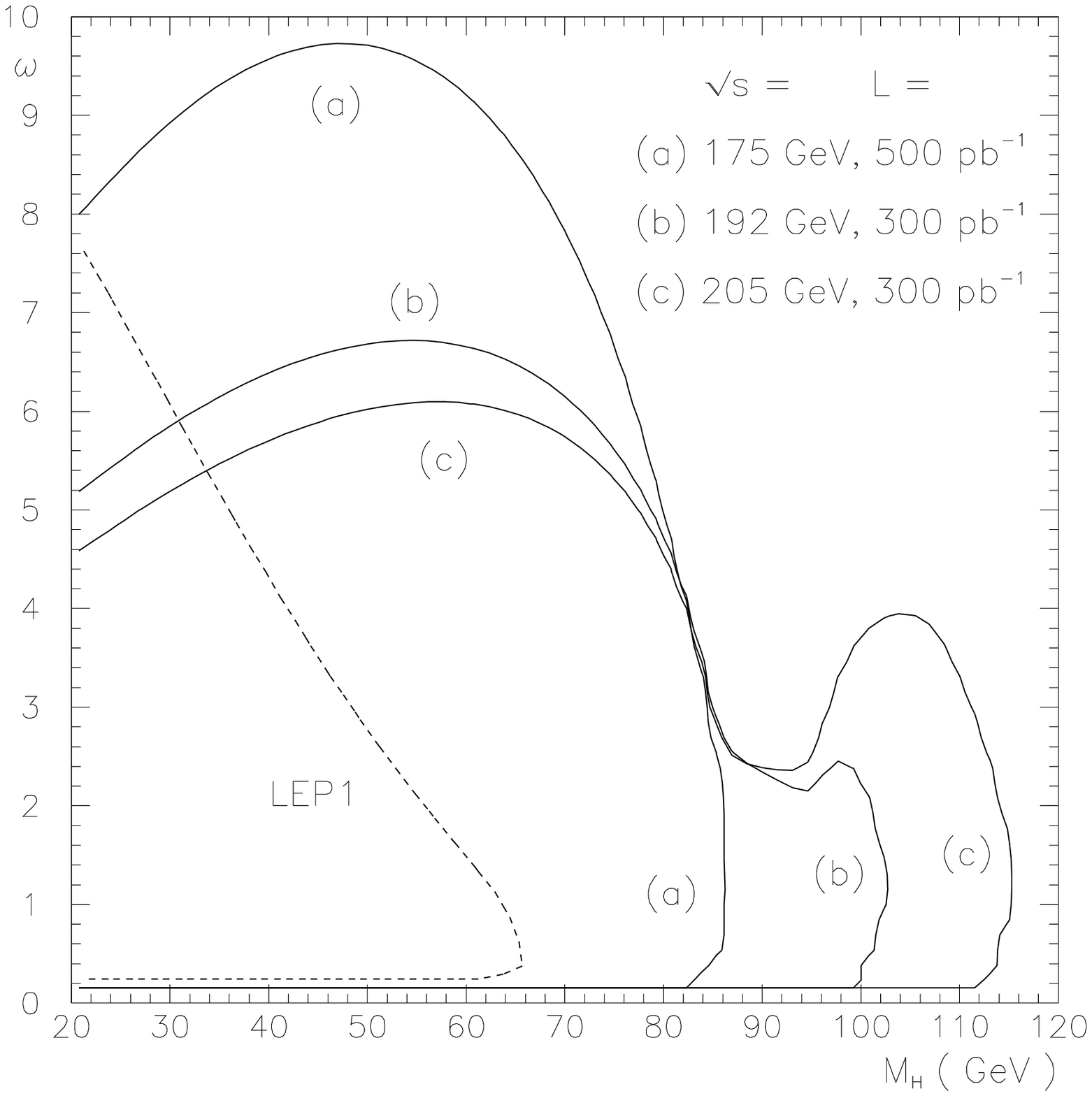,height=9.0cm,angle=0}}
\caption{\it Exclusion limits at LEP2 (full lines), and LEP1 (dashed). The region
  where $\omega$ is small can be covered by the search for visible
  Higgs decays.}
\label{exclusion}

\vspace*{11cm}

\end{figure}

\clearpage

 {\small
\section{Appendices} 

\subsection{Higgs-strahlung and $WW$ Fusion
\label{app:HSWW}}

Compact forms can be derived for the cross section of the 
process~\cite{R11C,R11B}
\begin{equation}
  e^+e^- \rightarrow H + \bar{\nu}\nu
\end{equation}
by choosing the energy $E_H$ and the polar angle $\theta$ of the Higgs
particle as the basic variables in the $e^+e^-$ c.m.\ frame.  The
overall cross section that will be observed experimentally, receives
contributions $3\times{\cal G}_S$ from Higgs-strahlung with $Z$ decays
into three types of neutrinos, ${\cal G}_W$ from $WW$ fusion, and
${\cal G}_I$ from the interference term between fusion and
Higgs-strahlung for $\bar{\nu}_e \nu_e$ final states.  We find:
\begin{equation}\label{totalx}
  \frac{d\sigma(H\bar\nu\nu)}{dE_H\,d\cos\theta}
  = \frac{G_F^3 m_Z^8p}{\sqrt2\,\pi^3s}
  \left(3{\cal G}_S + {\cal G}_I + {\cal G}_W \right)
\end{equation}
with
\begin{eqnarray}
  {\cal G}_S &=& \frac{v_e^2+a_e^2}{96}\;
    \frac{ss_\nu + s_1s_2}{\left(s-m_Z^2\right)^2
               \left[(s_\nu-m_Z^2)^2 + m_Z^2\Gamma_Z^2\right]}\\
  {\cal G}_I &=& \frac{(v_e+a_e)c_W^4}{8}\;
    \frac{s_\nu-m_Z^2}{\left(s-m_Z^2\right)
                \left[(s_\nu-m_Z^2)^2 + m_Z^2\Gamma_Z^2\right]}
    \nonumber\\
    &&\times
    \left[ 2 - (h_1+1)\log\frac{h_1+1}{h_1-1} 
             - (h_2+1)\log\frac{h_2+1}{h_2-1}
           +\, (h_1+1)(h_2+1)\frac{{\cal L}}{\sqrt{r}}\right]
    \\
  {\cal G}_W &=& \frac{c_W^8}{s_1 s_2 r}\,
    \Bigg\{(h_1+1)(h_2+1)
        \left[
        \frac{2}{h_1^2-1} + \frac{2}{h_2^2-1} - \frac{6s_\chi^2}{r}
        + \left(\frac{3t_1t_2}{r}-c_\chi\right)
          \frac{{\cal L}}{\sqrt{r}}\right]
    \nonumber\\
    &&\qquad\quad
    - \left[\frac{2t_1}{h_2-1} + \frac{2t_2}{h_1-1}
      + \left(t_1+t_2+s_\chi^2\right)
        \frac{{\cal L}}{\sqrt{r}}\right]
    \Bigg\}
\end{eqnarray}
where $a_e=-1$ and $v_e = -1+4s_W^2$.  The cross section is written
explicitly in terms of the Higgs momentum $p=(E_H^2-m_H^2)^{1/2}$, and
the energy $\epsilon_\nu=\sqrt{s}-E_H$ and invariant mass squared $s_\nu =
\epsilon_\nu^2-p^2$ of the neutrino pair.  The expression for ${\cal
G}_W$ had first been obtained in Ref.\cite{R11C}.  The following
abbreviations have been introduced:
\begin{displaymath}
  \begin{array}{r@{\;=\;}l}
  s_{1,2} & \sqrt{s}(\epsilon_\nu\pm p\cos\theta)\\[1mm]
  h_{1,2} & 1 + 2m_W^2/s_{1,2}\\[1mm]
  c_\chi & 1 - 2 s s_\nu /(s_1s_2)\\[1mm]
  s_\chi^2 & 1 - c_\chi^2
  \end{array}
  \qquad
  \begin{array}{r@{\;=\;}l}
  t_{1,2} & h_{1,2} + c_\chi h_{2,1}\\[1mm]
  r & h_1^2 + h_2^2 + 2c_\chi h_1h_2 - s_\chi^2\\[1mm]
  {\cal L} & {\displaystyle \log\frac{h_1h_2 + c_\chi + \sqrt{r}}
                        {h_1h_2 + c_\chi - \sqrt{r}}}
  \end{array}
\end{displaymath}
To derive the total cross section $\sigma(e^+e^-\to H\bar\nu\nu)$, the
differential cross section must be integrated over the region
$-1<\cos\theta<1$ and $m_H < E <
\frac12\sqrt{s}\left(1+m_H^2/s\right)$.

\subsection{Higgs Mass Computation: analytical approximation in the
limit of a common scale $M_S$ and restricted mixing parameters 
\label{app:H1}}

In this appendix we present the results of the analytical approximation 
which reproduces the two--loop RG improved effective potential results 
in the case of two light Higgs doublets below $M_S$ ($m_A \leq M_S$)
 \cite{RC4}.
The two CP-even and the charged Higgs masses
read

\begin{eqnarray}
m^2_{h(H)} &=& \frac{Tr M^2 \mp \sqrt{(TrM^2)^2 - 4 \det M^2}}{2}
\label{mhH}
\end{eqnarray}   
\begin{eqnarray}
 m^2_{H^{\pm}} &=& m_A^2 + (\lambda_5 - \lambda_4) v^2,
\label{mhch}
\end{eqnarray}   
where
\begin{equation}  
         TrM^2 =  M_{11}^2 + M_{22}^2 \;\; ; \;\;\;\;\;
     \det M^2 = M_{11}^2 M_{22}^2 - \left( M_{12}^2 \right)^2,
\label{detm2}
\end{equation}
with
 \begin{eqnarray}
\label{m2ij}
  M^2_{12} &=&  2 v^2 [\sin \beta \cos \beta (\lambda_3 + \lambda_4) + 
     \lambda_6 \cos^2 \beta + \lambda_7 \sin^2 \beta ] -
     m_A^2 \sin \beta \cos \beta
 \nonumber\\    
     M^2_{11} &=& 2 v^2 [\lambda_1  \cos^2 \beta + 2 
     \lambda_6  \cos \beta \sin \beta 
     + \lambda_5 \sin^2 \beta] + m_A^2 \sin^2 \beta \\
     M^2_{22} &=&  2 v^2 [\lambda_2 \sin^2 \beta +2 \lambda_7  \cos \beta
     \sin \beta + \lambda_5 \cos^2 \beta] + m_A^2 \cos^2 \beta . \nonumber
 \end{eqnarray}  
The mixing angle $\alpha$ is equally determined by
\begin{eqnarray}
      \sin 2\alpha = \frac{2M_{12}^2}
{\sqrt{\left(Tr M^2\right)^2-4\det M^2}}
\;\;\;\;\;\;\;\;
      \cos 2\alpha = \frac{M_{11}^2-M_{22}^2}
{\sqrt{\left(Tr M^2\right)^2-4\det M^2}}
\label{sincos}
\end{eqnarray}

The above quartic couplings are given by    
\begin{eqnarray}
  \lambda_1 &=& \frac{g_1^2 + g_2^2}{4} \left(1-\frac{3}{8 \pi^2} \;h_b^2 \; t
                 \right)    
      +   \frac{3}{8 \pi^2}\; h_b^4\; \left[
         t + \frac{X_{b}}{2} + \frac{1}{16 \pi^2}
        \left( \frac{3}{2} \;h_b^2 + \frac{1}{2}\;h_t^2       
     - 8\; g_3^2 \right) \left( X_{b}\;t + t^2\right) \right]
    \nonumber\\                                                 
  &&{} -   \frac{3}{96\pi^2} \; h_t^4\;\frac{\mu^4}{M_{_S}^4}
        \left[ 1+ \frac{1}{16 \pi^2} \left( 9\;h_t^2 -5  h_b^2
     -  16 g_3^2 \right) t  \right]
\label{lambda1}  
\end{eqnarray}

\begin{eqnarray}
      \lambda_2 &=& \frac{g_1^2 + g_2^2}{4} \left(1-
\frac{3}{8 \pi^2} \;h_t^2
       \; t\right)                      
        + \frac{3}{8 \pi^2}\; h_t^4\; \left[
         t + \frac{X_{t}}{2} + \frac{1}{16 \pi^2}
        \left( \frac{3 \;h_t^2}{2} + \frac{h_b^2}{2}       
     - 8\; g_3^2 \right) \left( X_{t}\;t + t^2\right) \right]
    \nonumber\\
  &&{}-  \frac{3}{96\pi^2} \; h_b^4\;\frac{\mu^4}{M_{_S}^4}
        \left[ 1+ \frac{1}{16 \pi^2} \left(9\;h_b^2 -5  h_t^2
     -  16 g_3^2 \right) t  \right]
\label{lambda2}   
\end{eqnarray}
    
\begin{eqnarray}
    \lambda_3 &=& \frac{g_2^2 - g_1^2}{4} \left(1-
\frac{3}{16 \pi^2}(h_t^2 + h_b^2) \; t \right)                      
 + \frac{6}{16 \pi^2}\; h_t^2\, h_b^2\; \left[
         t + \frac{A_{tb}}{2} + \frac{1}{16 \pi^2}
        \left( h_t^2 + h_b^2       
    - 8\; g_3^2 \right) \left( A_{tb}\;t + t^2\right) \right]
  \nonumber\\
  &&{}+ \frac{3}{96\pi^2} \; h_t^4\; \left[\frac{3 \mu^2}{M_{_S}^2}
                - \frac{\mu^2 A_t^2}{M_{_S}^4} \right]
        \left[ 1+ \frac{1}{16 \pi^2} \left (6\;h_t^2 -2  h_b^2
     -  16 g_3^2 \right) t  \right]
 \nonumber\\
  &&{}+ \frac{3}{96\pi^2} \; h_b^4\; \left[\frac{3 \mu^2}{M_{_S}^2}
              - \frac{\mu^2 A_b^2}{M_{_S}^4} \right]
        \left[ 1+ \frac{1}{16 \pi^2} \left (6\;h_b^2 -2  h_t^2
     -  16 g_3^2 \right) t  \right]
\label{lambda3}   
\end{eqnarray}

\begin{eqnarray}
       \lambda_4 &=&  -\; \frac{g_2^2}{2} \left(1
-\frac{3}{16 \pi^2} 
       (h_t^2 + h_b^2) \; t\right)
        - \frac{6}{16 \pi^2}\; h_t^2\; h_b^2\; \left[
         t + \frac{A_{tb}}{2} + \frac{1}{16 \pi^2}
        \left( h_t^2 + h_b^2       
     - 8\; g_3^2 \right) \left( A_{tb}\;t + t^2\right) \right]
   \nonumber\\
  &&{}+\frac{3}{96\pi^2} \; h_t^4\; \left[\frac{3 \mu^2}{M_{_S}^2}
                         - \frac{\mu^2 A_t^2}{M_{_S}^4} \right]
        \left[ 1+ \frac{1}{16 \pi^2} \left (6\;h_t^2 -2  h_b^2
     -  16 g_3^2 \right) t  \right]
 \nonumber\\
  &&{}+\frac{3}{96\pi^2} \; h_b^4\; \left[\frac{3 \mu^2}{M_{_S}^2}
                         - \frac{\mu^2 A_b^2}{M_{_S}^4} \right]
        \left[ 1+ \frac{1}{16 \pi^2} \left (6\;h_b^2 -2  h_t^2
     -  16 g_3^2 \right) t  \right]
\label{lambda4}   
\end{eqnarray}

\begin{eqnarray}
 \lambda_5 &=& -\; \frac{3}{96\pi^2} \; h_t^4\; 
                   \frac{\mu^2 A_t^2}{M_{_S}^4} 
        \left[ 1- \frac{1}{16 \pi^2} \left (2  h_b^2 - 6\;h_t^2 
     +  16 g_3^2 \right) t  \right]
\nonumber \\  
&&{}- \frac{3}{96\pi^2} \; h_b^4 \;
                         \frac{\mu^2 A_b^2}{M_{_S}^4} 
        \left[ 1- \frac{1}{16 \pi^2} \left ( 2  h_t^2 -6\;h_b^2
     +  16 g_3^2 \right) t  \right]
\label{lambda5}   
\end{eqnarray}
       
\begin{eqnarray}
         \lambda_6 &=& \frac{3}{96 \pi^2}\; h_t^4\;
       \frac{\mu^3 A_t}{M_{_S}^4}
   \left[1- \frac{1}{16\pi^2}
   \left(\frac{7}{2} h_b^2  -\frac{15}{2} h_t^2 + 16 g_3^2 \right)
        t \right]  
\label{lambda6}  
\\  &&{}+ 
\frac{3}{96 \pi^2}\; h_b^4\;  
\frac{\mu}{M_{_S}} \left(\frac{A_b^3}{M_{_S}^3}
                                            - \frac{6 A_b}{M_{_S}}\right)
   \left[1- \frac{1}{16\pi^2}
\left(\frac{1}{2} h_t^2  -\frac{9}{2} h_b^2 + 16 g_3^2 \right)
       t  \right]  
\nonumber 
\end{eqnarray}

\begin{eqnarray}
\lambda_7 &=& \frac{3}{96 \pi^2}\; h_b^4\;\frac{\mu^3 A_b}{M_{_S}^4}
   \left[1- \frac{1}{16\pi^2}
\left(\frac{7}{2} h_t^2  -\frac{15}{2} h_b^2 + 16 g_3^2 \right)
       t  \right]  
\label{lambda7}  
\\ && {}+  
\frac{3}{96 \pi^2}\; h_t^4\;\frac{\mu}{M_{_S}} 
\left(\frac{A_t^3}{M_{_S}^3}
         - \frac{6 A_t}{M_{_S}}\right)
   \left[1- \frac{1}{16\pi^2}
\left(\frac{1}{2} h_b^2  -\frac{9}{2} h_t^2 + 16 g_3^2 \right)
       t  \right]  
\nonumber
\end{eqnarray}
They contain the same kind of corrections as eq.(\ref{mhsm}),
including the leading $D$-term contributions, and 
we have defined,

\begin{equation}
X_{t(b)}  =  \frac{2 A_{t(b)}^2}{M_{_S}^2}
                  \left(1 - \frac{A_{t(b)}^2}{12 M_{_S}^2} \right) ; \;\;\;\;\;
    \;\;\;\;
      A_{tb} = \frac{1}{6} \left[-\frac{ 6 \mu^2}{M_{_S}^2}
        - \frac{(\mu^2 - A_b A_t)^2}{ M_{_S}^4}
     + \frac{3 (A_t + A_b)^2}{M_{_S}^2}\right]. \nonumber
\end{equation}

All quantities in the approximate formulae are defined at the scale
$M_t$.
and
$h_t =  m_t(M_t)/ (v \sin \beta)$  
$h_b  =  m_b(M_t)/ (v \cos \beta)   $
are the top and bottom Yukawa couplings in the two-Higgs doublet model.

For $m_A\leq M_t$,  $\tan\beta$ is fixed at the scale $m_A$,
while for $m_A\geq M_t$,  $\tan\beta$ is given by~\cite{B}
\be
\label{tanbeta}
\tan\beta(M_t)=\tan\beta(m_A)\left[1+\frac{3}{32\pi^2}(h_t^2 - h_b^2)
\log\frac{m_A^2}{M_t^2}\right].
\ee

For the case in which the CP-odd Higgs mass $m_A$ is lower than
$M_{_S}$, but still  larger than the top-quark mass scale,
we decouple, in the numerical computations, the heavy Higgs doublet
and define an effective quartic coupling for the light Higgs,
which is related to the running  Higgs mass at the scale $m_A$
through
$\lambda(m_A) = (m_h(m_A)/2 v^2)$.
The low energy value of the quartic coupling is then obtained by
running the SM renormalization-group equations from the scale
$m_A$ down to the scale $M_t$. In the analytical approximation, 
for simplicity  the effect of decoupling of the 
heavy Higgs doublet at an intermediate scale is ignored but is 
partially  compensated by relating
the value of $\tan\beta$ at the scale $M_t$ with its
corresponding value at the scale $m_A$ through its renormalization-group 
running, eq.(\ref{tanbeta}). A subroutine implementing the
above computations is available \cite{subh}.

\subsection{Deriving $5\sigma$ Discovery and 95\% C.L. Exclusion
Contours}
\label{sec:app}

The minimum luminosity needed to assess the discovery or to
exclude the existence of a Higgs boson with mass \mH\ can directly be
determined from the numbers of events expected from the signal and from the
background processes at the three different center-of-mass energies.
Given the rather small numbers of events involved in this process, it is
preferable to use Poisson statistics to derive the result.

Several definitions for the ``minimum luminosity needed'' were proposed.
For instance, the minimum luminosity needed for a $5\sigma$ discovery can be
defined either {\it (i)} as the luminosity needed by the {\it typical}
experiment, {\it i.e.} by an experiment that would actually observe the
number of events expected; or {\it (ii)} as the
luminosity for which 50\% of the experiments would make the discovery
at the requested $5\sigma$ level, where the {\it a priori}
unknown numbers of events observed are properly generated according to
a Poisson distribution around their expected values.
Although both definitions lead to the same numerical result, a
preference was given to the second one, which allows in addition
the proportion of the experiments required to make the discovery to
be varied.

In detail, let $b$ and $s$ be the numbers of background
and signal events expected with a luminosity of 1~\infb, and $\alpha$ be
the fraction needed for the discovery.
The first definition corresponds to finding the smallest
value of $\alpha$ that fulfills the condition
$$1 - \exp(-\alpha b) \sum_{i=0}^{N-1}
{(\alpha b)^i \over i!} \le 5.7 \times 10^{-7},\eqno (1)$$
where $N = \alpha(s+b)$, {\it i.e.} that renders the probability of a
background fluctuation smaller than the probability of a $5\sigma$ effect
in the case of Gaussian distributions. The second requirement
consists in finding the smallest value of $\alpha$ for which the number
of events $N_1$ that would correspond to a $5\sigma$ (high) fluctuation of
the background alone is smaller than the numbers of events $N_2$ that
would correspond to a 50\% probability (low) fluctuation of the total
number of events (signal included). This amounts to finding a value of $N$
 which fulfills, in addition to (1), the following condition
$$ \exp\left[-\alpha(s+b)\right] \sum_{i=0}^{N-1}
{\left[\alpha (s+b)\right]^i \over i!} \le 0.5. \eqno (2)$$

As to the exclusion of the existence of a signal at the 95\% confidence level,
the minimum luminosity needed has been similarly defined as the luminosity
for which 50\% of the experiments would actually exclude it in the case of the
absence of signal. Again, this is equivalent to the luminosity needed by
the {\it typical} experiment, which is given by the value of
$\alpha$ such that
$${\displaystyle \exp\left[-\alpha (s+b)\right] \sum_{i=0}^N
  {\left[\alpha (s+b)\right]^i \over i!} \over \displaystyle
  \exp(-\alpha b) \sum_{i=0}^N {(\alpha b)^i \over i!}} \le 0.05,$$
where $N=\alpha b$.

In both instances, when deriving the result, $b$ and $s$ were conservatively
increased (resp. reduced) by their systematic uncertainties, mainly coming
from the yet limited Monte Carlo statistics. The numbers of events expected
by each of the four experiments were then added together, and the individual
uncertainties were added in quadrature.

However, one caveat should be  mentioned. Even if it is legitimate to
compute the minimum luminosity needed by each
of the four individual experiments by requiring only 50\% of the
Gedanken-experiments to make the discovery/exclusion, this becomes unclear
for the combined experiment: this minimum luminosity would not suffice in 50\%
of the cases, and this would not be ``compensated'' by having two
(or four) such combined experiments. Since a choice for this
fraction cannot be uniquely defined,  the
combined results have been presented with a fraction of 50\% too.


}

\end{document}